\documentclass[preprint,tightenlines,superscriptaddress,nofootinbib]{revtex4}
\usepackage{amsmath,amssymb,url}
\usepackage{graphicx,tabularx,axodraw}
\usepackage{color,colordvi}
\usepackage{fancybox,multirow}
\usepackage{pst-plot,epsfig,wrapfig,pstricks}
\usepackage{pifont}
\usepackage[latin1]{inputenc}

\def\bq{\begin{equation}}
\def\eq{\end{equation}}
\def\ba{\begin{eqnarray}}
\def\ea{\end{eqnarray}}

\newcommand{\sla}[1]{/\!\!\!#1}
\newcommand{\wt}{\widetilde}

\newcommand{\lpm}{\ell^\pm}
\newcommand{\pp}{p\bar{p}}
\newcommand{\uu}{u\bar{u}}
\newcommand{\dd}{d\bar{d}}
\newcommand{\bb}{b\bar{b}}
\newcommand{\cc}{c\bar{c}}
\newcommand{\qq}{q\bar{q}}
\newcommand{\tops}{t\bar{t}}
\newcommand{\nn}{\nu\bar\nu}
\newcommand{\mm}{\mu^+\mu^-}
\newcommand{\ells}{\ell^+\ell^-}
\newcommand{\taus}{\tau^+\tau^-}
\newcommand{\ww}{W^+W^-}

\newcommand{\etal}{\textit{et al.}}

\newcommand{\eg}{{\sl e.g. }}

\begin{document}

%\preprint{}

\date{\today}

\title{Searching for the Higgs boson}

\author{D.~Rainwater}
\email{rain@pas.rochester.edu}
\affiliation{Dept.~of Physics and Astronomy,
             University of Rochester, Rochester, NY, USA}

\begin{abstract}
These lectures on Higgs boson collider searches were presented at TASI
2006.  I first review the Standard Model searches: what LEP did,
prospects for Tevatron searches, the program planned for LHC, and some
of the possibilities at a future ILC.  I then cover in-depth what
comes after a candidate discovery at LHC: the various measurements one
has to make to determine exactly what the Higgs sector is.  Finally, I
discuss the MSSM extension to the Higgs sector.
\end{abstract}

\maketitle

\tableofcontents

%%%%%%%%%%%%%%%%%%%%%%%%%%%%%%%%%%%%%%%%%%%%%%%%%%%%%%%%%%%%%%%%%%%%%%%%
%%%%%%%%%%%%%%%%%%%%%%%%%%%%%%%%%%%%%%%%%%%%%%%%%%%%%%%%%%%%%%%%%%%%%%%%
%%%%%%%%%%%%%%%%%%%%%%%%%%%%%%%%%%%%%%%%%%%%%%%%%%%%%%%%%%%%%%%%%%%%%%%%

\section{Introduction}

Despite all the remarkable progress made early in the 21$^{\rm st}$
century formulating possible explanations for the weakness of gravity
relative to the other forces, the nature of dark matter (and dark
energy), what drove cosmological inflation, why neutrino masses are so
small, and what might unify the gauge forces, we still have not yet
answered the supposedly more readily accessible problem of electroweak
symmetry breaking.  Just what, exactly, gives mass to the weak gauge
bosons and the known fermions?  Is it weakly-coupled and spontaneous,
involving fundamental scalars, or strongly-coupled, involving
composite scalars?  Is the flavor problem linked?  Do we discover the
physics behind dark matter (and {\it its} mass), gauge unification and
flavor at the same time?  Or are those disconnected problems?

Our starting point is unitarity, the conservation of probability: the
weak interaction of the Standard Model (SM) of particle physics
violates it at about 1~TeV~\cite{LQT}.  The theory demands at least
one new propagating scalar state with gauge coupling to weak bosons to
keep this under control.  The same problem holds for fermion--boson
interactions~\cite{Appelquist:1987cf,Maltoni:2001dc,Dicus:2004rg,Dicus:2005ku},
only at much higher energy, so is generally less often
discussed\footnote{The original study~\cite{Appelquist:1987cf} was
clearly incorrect, but the correct line of reasoning is a work in
progress~\cite{Maltoni:2001dc,Dicus:2004rg,Dicus:2005ku}.}.  While the
variety of explanations for electroweak symmetry breaking (EWSB) is
vast, what we call the Standard Model (SM) assumes the existence of a
single fundamental scalar field which spontaneously acquires a vacuum
expectation value to generate all fermion and boson masses.  It is a
remarkably compact and elegant explanation, simple in the extreme.
Yet while it tidies up the immediate necessities of the SM, it suffers
from glaring theoretical pathologies that drive much of the
model-building behind more ambitious explanations.

Numerous lectures and review articles already exist, covering the SM
Higgs sector and the minimal supersymmetric (MSSM)
extension~\cite{HHG,Djouadi,Spira:1997dg,gunion_haber}, which are
useful both for learning nitty-gritty theoretical details and serving
as formulae references.  These lectures are instead a crash-course
tour of theory in practical application: previous, present and planned
Higgs searches, what happens after a candidate Higgs discovery, and an
overview of MSSM Higgs phenomenology as a perturbation of that for SM
Higgs.  They are not comprehensive, but do provide a solid grounding
in the basics of Higgs hunting.  They should be read only after one
has become intimate with the SM Higgs sector and its underlying
theoretical issues.  Within TASI 2006, this means you should already
have studied Sally Dawson's lectures.  After both of these you should
also be able to explain to your friends how we look for a Higgs boson
at colliders (if they care), how to confirm it's a Higgs and figure
out what variety it is (since we care), and describe how some basic
extensions to the SM Higgs sector behave as a function of their
parameter space (nature might not care for the SM).

Herein I'll assume that nature prefers fundamental scalars and
spontaneous symmetry breaking.  This is a strong bias, but one that
provides a solid framework for phenomenology.  The ambitious student
who wants to really learn all the varieties of EWSB should also study
strong dynamics~\cite{SD}, dimensional
deconstruction~\cite{Arkani-Hamed:2001nc}, extra-dimensional Higgsless
constructions~\cite{Simmons:2006iw} and the Little
Higgs~\cite{LH-reviews} and Twin Higgs
mechanisms~\cite{Chacko:2005pe}.  In many of these classes of theories
the Higgs sector appears to be very SM-like, but in some no Higgs
appears and one instead would pay great attention to weak boson
scattering around a TeV.

%%%%%%%%%%%%%%%%%%%%%%%%%%%%%%%%%%%%%%%%%%%%%%%%%%%%%%%%%%%%%%%%%%%%%%%%
%%%%%%%%%%%%%%%%%%%%%%%%%%%%%%%%%%%%%%%%%%%%%%%%%%%%%%%%%%%%%%%%%%%%%%%%
%%%%%%%%%%%%%%%%%%%%%%%%%%%%%%%%%%%%%%%%%%%%%%%%%%%%%%%%%%%%%%%%%%%%%%%%

\section{Collider searches for the Standard Model Higgs}
\label{sec:SM}

Even though the SM Higgs sector doesn't explain flavor (why all the
fermion masses are scattered about over 12 orders of magnitude in
energy) and has a disconcerting radiative stability problem that
surely must involve new physics beyond the SM, it's a suitable
jumping-off point for formulating Higgs phenomenology.  That is, the
study of physical phenomena associated with a theory, exploring the
connection between theory and experiment.  Without this connection,
experiments would not make sense and theory would flail about,
untested.  To survey SM Higgs collider physics we need to recall a few
fundamentals about the SM Higgs boson.
\begin{itemize}
\item[1.] The Higgs boson unitarizes weak boson scattering, $VV\to VV$,
          so its interaction with weak bosons is very strictly defined
          to be the electroweak gauge coupling times the vacuum
          expectation value (vev); i.e., proportional to the weak
          boson masses.
\item[2.] The Higgs also unitarizes $VV\to f\bar{f}$ scattering, so
          its fermion couplings (except $\nu_i$) are proportional to
          the fermion mass, with a strictly defined universal
          coefficient.
\item[3.] Because of the coupling strengths, the Higgs is dominantly
          produced by or in association with massive particles
          (including loop-induced processes, as we'll see in
          Sec.~\ref{sub:Hdecay}), and prefers to decay to the most
          massive particles kinematically allowed.
\item[4.] The Higgs boson mass itself is a free parameter\footnote{We
          know it is not massless, due to the absence of additional
          long-range forces.}, but influences EW observables, so we
          can fit EW precision data to make a prediction for its mass.
\end{itemize}
We may thus define the SM Higgs sector by its vacuum expectation
value, $v$, measured via $M_W$, $G_F$, etc., and the known electroweak
gauge couplings; 9 Yukawa couplings (fermion mass parameters, ignoring
neutrinos and CKM mixing angles); and one free parameter, $M_H$.

Prior to the Large Electron Positron (LEP) collider era starting
around 1990, Higgs searches involved looking for resonances amongst
the low energy hadronic spectra in $e^+e^-$ collisions.  These were in
fact non-trivial searches, mostly involving decays of hadrons to Higgs
plus a photon, but are generally regarded as comprehensive and set a
lower mass bound of $M_H\gtrsim 3$~GeV.

Higgs hunting in the 1990s was owned by LEP, an $e^+e^-$ collider at
CERN which steadily marched up in energy over the decade.  It found no
Higgs bosons\footnote{This may be a somewhat controversial statement,
depending on what lunch table you're sitting at.  See
Sec.~\ref{sub:LEP-H}.}.  Attention then turned to the long-delayed
Tevatron Run~II program, proton--antiproton collisions at 2~TeV, which
got off to a shaky start but is now performing splendidly.  It so far
sees nothing Higgs-like, either, but has not yet gathered enough data
to be able to say much.  The proton--proton Large Hadron Collider
(LHC) at CERN is also many years behind schedule, but its construction
is now nearing completion and we may expect physics data within a few
years.

Our survey begins with LEP from a historical perspective and some
general statements about Higgs boson behavior as a function of its
mass.  Next we turn our attention to the ongoing Tev2 search, for
which the prospects hinge critically on machine performance.  Then we
delve into the intricacies of LHC Higgs pheno, which is far more
complicated than either LEP or Tevatron, yet essentially guarantees an
answer to our burning questions.

%%%%%%%%%%%%%%%%%%%%%%%%%%%%%%%%%%%%%%%%%%%%%%%%%%%%%%%%%%%%%%%%%%%%%%%%
%%%%%%%%%%%%%%%%%%%%%%%%%%%%%%%%%%%%%%%%%%%%%%%%%%%%%%%%%%%%%%%%%%%%%%%%

\subsection{The LEP Higgs search}
\label{sub:LEP}

An obvious question to ask is, can we produce the Higgs directly in
$e^+e^-$ collisions?  We could then probe Higgs masses up to our
machine energy, which for LEP-II eventually reached 209~GeV.
Recalling that the Higgs--electron coupling is proportional to the
electron mass, which is quite a bit smaller than the electroweak vev
of 246~GeV, the coupling strength is about $1.5\times 10^{-6}$, or
teeny-tiny in technical parlance.  A quick calculation reveals that it
would take about 4 years running full-tilt to produce just one Higgs
boson.  This one event would have to be distinguished from the general
scattering cross section to fermion pairs in the SM, which is beyond
hopeless.

\begin{figure}[hb!]
\includegraphics[width=5cm]{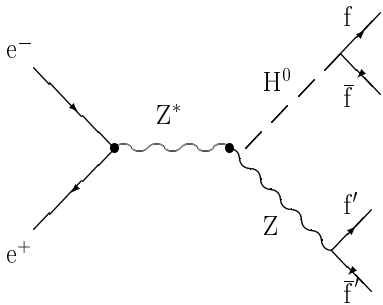} 
\caption{Feynman diagram for the process $e^+e^-\to ZH$ with subsequent
Higgs and $Z$ boson decays to fermion pairs.  All LEP Higgs searches
were based primarily on this process, with various fermion
combinations in the final state composing the different search
channels.}
\label{fig:ee_ZH}
\end{figure}

Instead, we think of what process involves something massive, with
vastly larger Higgs coupling, so that the interaction rate is large
enough to produce a statistically useful number of Higgs bosons.  The
two obvious possibilities are $e^+e^-\to\ww H$ (two $W$'s required for
charge conservation) and $e^+e^-\to ZH$.  The first process will
obviously have less reach in $M_H$ as the two $W$ bosons require far
more energy than a single $Z$ boson to produce.  LEP Higgs searches
therefore focused on the latter process, shown as a Feynman diagram in
Fig.~\ref{fig:ee_ZH}: the electron and positron annihilate to form a
virtual $Z$, far above its mass shell, which returns on-shell by
spitting off a Higgs boson.  This process is generically known as
Higgsstrahlung, analogous to bremsstrahlung radiation.  Both the Higgs
and $Z$ immediately decay to an asymptotic final state of SM
particles.  For the Higgs this is preferentially to the most massive
kinematically-allowed pair, while $Z$ decays are governed by the
fermion gauge couplings\footnote{See the PDG~\cite{Yao:2006px} for $Z$
boson branching ratios, which you should memorize.}.  In brief, the
$Z$ decays $70\%$ of the time to jets, $20\%$ of the time invisibly
(to neutrinos, which the detectors can't see), and about $10\%$ to
charged leptons, which are the most distinctive, ``clean'' objects in
a detector.

%

%%%%%%%%%%%%%%%%%%%%%%%%%%%%%%%%%%%%%%%%%%%%%%%%%%%%%%%%%%%%%%%%%%%%%%%%

\subsubsection{Momentary diversion: Higgs decays}
\label{sub:Hdecay}

What, precisely, are the Higgs branching ratios (BRs)?  To find these,
we first need the Higgs partial widths; that is, the inverse decay
rates to each final state kinematically allowed.  Everyone should
calculate these once as an exercise.

Let's start with the easiest case: Higgs decay to fermion pairs, which
is a very simple matrix element. The general result at tree-level is:
\bq\label{eq:H_ff}
\Gamma_{f\bar{f}} \;=\;
\frac{N_c \, G_F \, m^2_f \, M_H}{4\sqrt{2}\,\pi} \, \beta^3
\qquad {\rm where} \;\; \beta\,=\,\sqrt{1-\frac{4m^2_f}{M^2_H}}
\eq
One factor of the fermion velocity $\beta$ comes from the matrix
element and two factors come from the phase space.  I emphasize that
this is at tree-level because there are significant QCD corrections to
colored fermions.  The bulk of these corrections are absorbed into a
running mass (see Ref.~\cite{Spira:1997dg}).  For calculations we
should always use $m_q(M_H)$, the quark mass renormalized to the Higgs
mass scale, rather than the quark pole mass.  Programs such as {\sc
hdecay}~\cite{Djouadi:1997yw} will calculate these automatically given
SM parameter inputs, greatly simplifying practical phenomenology.

\begin{figure}[ht!]
\includegraphics[scale=0.85]{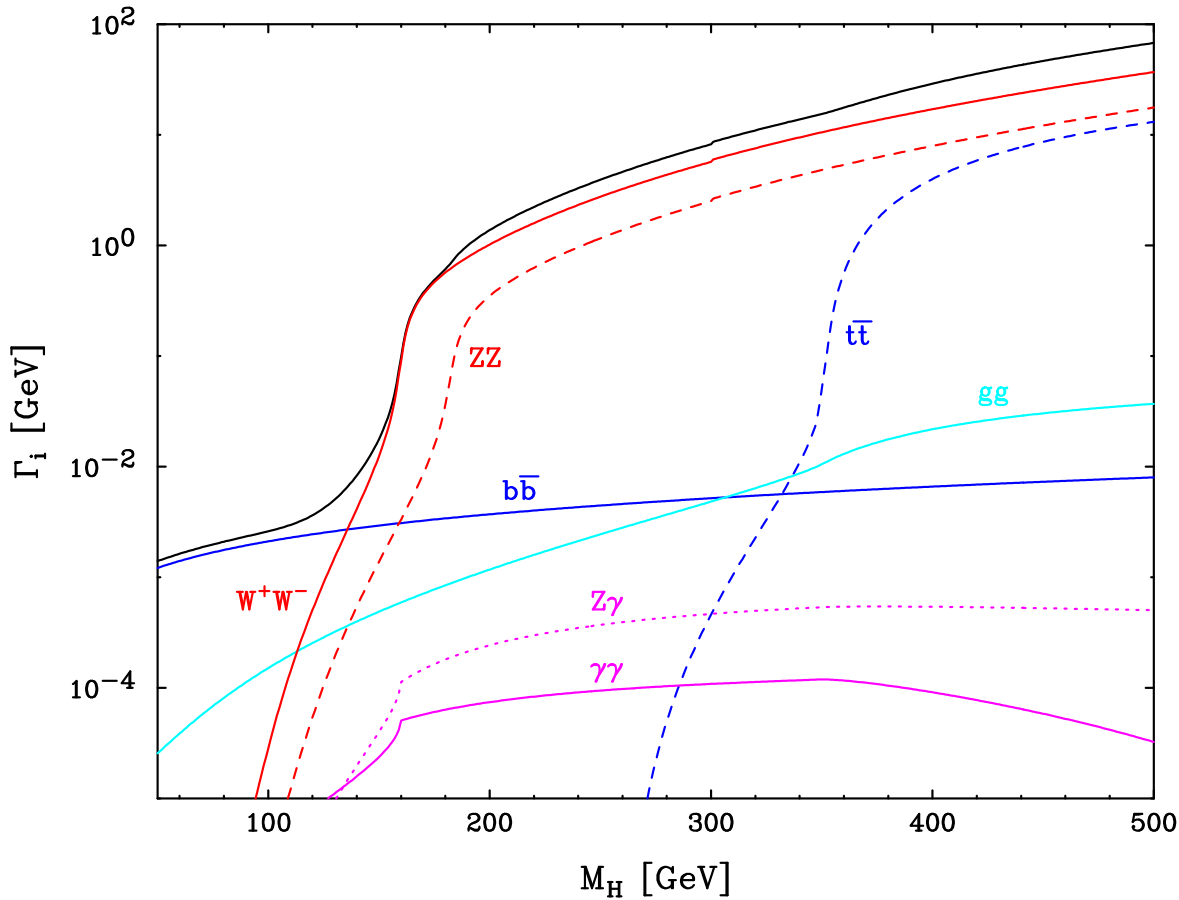}
\caption{Select Standard Model Higgs boson partial widths, as a 
function of mass, $M_H$.  Individual partial widths are labeled, while
the total width (sum of all partial widths, some minor ones not shown)
is the black curve.  Widths calculated with {\sc
hdecay}~\protect\cite{Djouadi:1997yw}.}
\label{fig:SM_Hwid}
\end{figure}

Note that the partial width to fermions is linear in $M_H$, modulo the
cubic fermion velocity dependence, which steepens the ascent with
$M_H$ near threshold.  Partial widths for various Higgs decays are
shown in Fig.~\ref{fig:SM_Hwid}.  While the total Higgs width above
fermion thresholds grows with Higgs mass, Higgs total widths below $W$
pair threshold are on the order of tens of MeV -- quite narrow.  The
only complicated partial width to fermions is that for top quarks, for
which we must treat the fermions as virtual (at least near threshold)
and use the matrix elements to the full six-fermion final state,
integrated over phase space.  This is slightly more complicated, but
easily performed numerically.

Before the decay to top quarks is kinematically allowed, however, the
decays to weak bosons turn on.  A few $W$/$Z$ widths above threshold
the $W$ and $Z$ may be treated as on-shell asymptotic final states,
making the partial width calculation easier.  We find:
\bq\label{eq:H_VV}
\Gamma_{VV} \;=\; \frac{G_F M^3_H}{16\sqrt{2}\pi}\;\delta_V\beta \,
                  \biggl( 1 - x_V + \frac{3}{4}x^2_V \biggr)
\;\; {\rm where} \;
\begin{cases} \delta_{W,Z}\,=\,2,1         \\
              \beta\,=\,\sqrt{1-x_V}       \\
              x_V\,=\,\frac{4M^2_V}{M^2_H}
\end{cases}
\eq
The factor of $\beta$ comes from phase space, while the matrix
elements give the more complicated function of $x_V$.  The partial
width is dominantly cubic in $M_H$, although the factors of beta and
$x_V$ enhance this somewhat near threshold, as in the fermion case.
We can see this in Fig.~\ref{fig:SM_Hwid}: the partial widths to $VV$
gradually flatten out to cubic behavior above threshold.  The reason
for this stronger $M_H$ dependence compared to fermions is that a
longitudinal massive boson wavefunction is proportional to its energy
in the high-energy limit, which enhances the coupling by a factor
$E/M_V$.  (Recall that it is this property of massive gauge bosons
that requires the Higgs, lest their scattering amplitude rise as
$E^2/M_V^2$, violating unitarity.  The Higgs in fact generates the
longitudinal modes.)  This much stronger dependence on $M_H$ leads to
a very rapid total width growth with $M_H$, which reaches 1~GeV around
$M_H=190$~GeV.  We'll return to this when discussing Higgs couplings
measurements in Sec.~\ref{sub:LHC-coup}.  The bottom line is that
bosons ``win'' compared to fermions.  Thus, even though the top quark
has a larger mass than $W$ or $Z$, it cannot compete for partial width
and thus BR.  Note that the partial widths to $VV$ are non-trivial
below threshold: the $W$ and $Z$ are unstable and therefore have
finite widths; they may be produced off-shell.  The Higgs can decay to
these virtual states because its coupling is proportional to the
daughter pole masses (or, in the case of quarks, the running masses),
not the virtual $q^2$, which can be much smaller.  Below threshold the
analytical expressions are known~\cite{H_VV_OFS} (see
Ref.~\cite{Djouadi} for a summary), but are not particularly
insightful to derive as an exercise.

The astute reader will have noticed by now that Fig.~\ref{fig:SM_Hwid}
contains curves for Higgs partial widths to {\it massless} final
states!  (Have another look if you didn't notice.)  We know the Higgs
couples to particles proportional to their masses, so this requires
some explanation.  Recall that loop-induced transitions can occur at
higher orders in perturbation theory.  Such interactions typically are
important to calculate only when a tree-level interaction doesn't
exist.  They are responsible for rare decays of various mesons, for
instance, and are in some cases sensitive to new physics which may
appear in the loop.  Here, we consider only SM particles in the loop.
Which ones are important?  Recall also once again that the Higgs boson
couples proportional to particle mass.  Thus, the top quark and EW
gauge bosons are most important.  For $H\to gg$, then, that means only
the top quark, while for $H\to\gamma\gamma$ it is both the top quark
and $W$ loops (there is no $ZZ\gamma$ vertex).  The $H\to gg$
expression (for the Feynman diagram of Fig.~\ref{fig:H_gg})
is~\cite{Rizzo:1979mf}:
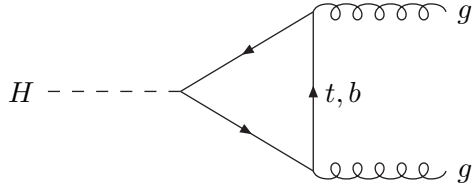
\begin{figure}[ht!]
\begin{center}
\setlength{\unitlength}{1pt}
\begin{picture}(180,100)(0,0)
\Gluon(100,20)(150,20){-3}{5}
\Gluon(100,80)(150,80){3}{5}
\ArrowLine(100,20)(100,80)
\ArrowLine(100,80)(50,50)
\ArrowLine(50,50)(100,20)
\DashLine(0,50)(50,50){5}
\put(-15,46){$H$}
\put(105,46){$t,b$}
\put(155,18){$g$}
\put(155,78){$g$}
\end{picture}
\setlength{\unitlength}{1pt}
\end{center}
\vspace*{-10mm}
\caption{Feynman diagram for the loop-induced process $H\to gg$ in the
SM.  All quarks enter the loop, but contribute according to their
Yukawa coupling squared (mass squared).  In the SM, only the top quark
is important.}
\label{fig:H_gg}
\end{figure}
\ba\label{eq:H_gg}
\Gamma_{gg} \; = \;
\frac{\alpha^2_s G_F M_H^3}{16\sqrt{2}\,\pi^3}
\bigg| \sum\limits_i \tau_i \bigl[ 1+(1-\tau_i)f(\tau_i) \bigr] \bigg|^2
&
\\ {\rm with} \quad
\tau_i \, = \, \frac{4m^2_f}{M^2_H}
\quad {\rm and} &
f(\tau)\,=\,\begin{cases}\bigl[\sin^{-1}\sqrt{1/\tau}\bigr]^2&\tau\geq 1\\
 -\frac{1}{4}\bigl[\ln\frac{1+\sqrt{1-\tau}}{1-\sqrt{1-\tau}}-i\pi\bigr]^2
 &\tau<1\end{cases}
\ea
which is for a general quark in the loop with SM Yukawa coupling.
It's easy to see that in the SM the $b$ quark contribution, which is
second in size to that of the top quark, is inconsequential.  Remember
to use the running mass $m_f(M_H)$ to take into account the largest
QCD effects.  When you derive this expression yourself as an exercise,
take care to solve the loop integral in $d>4$ dimensions, otherwise
you miss a finite piece.  The $H\to\gamma\gamma,Z\gamma$ expressions
have a similar form~\cite{H_gamgam}, but with two loop functions,
since it can also be mediated a $W$ boson loop (which interferes
destructively with the top quark loop!):
\bq\label{eq:H_gamgam}
\Gamma_{\gamma\gamma} \; = \;
\frac{\alpha^2 G_F M_H^3}{128\sqrt{2}\,\pi^3}
\bigg| \sum\limits_i N_{c,i} Q_i^2 F_i \bigg|^2
\eq
\bq
F_1     \, = \, 2 + 3\tau[1+(2-\tau)f(\tau)] \, , \;
F_{1/2} \, = \,   - 2\tau[1+(1-\tau)f(\tau)] \, , \;
F_0     \, = \, \tau[1-\tau f(\tau)]
\eq
where $N_{c,i}$ is the number of colors, $Q_i$ the charge, and $F_j$
the particle's spin.

Now look again more closely at Fig.~\ref{fig:SM_Hwid}.  The important
feature to notice is that these loop-induced partial widths are
ostensibly proportional to $M_H^3$, like the decays to gauge bosons.
However, the contents of the brackets, specifically the $f(\tau_i)$
function, can alter this in non-obvious ways.  For $H\to gg$,
Fig.~\ref{fig:SM_Hwid} shows a slightly more than cubic dependence at
low masses, leveling of to approximately $M_H^3$, and flattening out
to approximately quadratic a bit above the top quark pair threshold.
We see from Eq.~\ref{eq:H_gg} that the functional form changes at that
threshold, albeit fairly smoothly, by picking up a constant imaginary
piece when the top quarks in the loop can be on-shell.

The partial widths to $\gamma\gamma$ and $Z\gamma$ behave very
differently than $gg$.  For $M_H$ below $W$ pair threshold, the
interference between top quark and $W$ loops produces an extremely
sharp rise with $M_H$, which transitions to something slightly more
than linear in $M_H$ at $W$ pair threshold where the $W$ bosons in the
loop go on-shell.  There is is a smoother transition at the top quark
pair threshold, where they can similarly go on-shell.  The
$\gamma\gamma$ and $Z\gamma$ partial widths behave differently because
of the different $t\bar{t}\gamma$ and $t\bar{t}Z$ couplings: the
partial width to $Z\gamma$ at large $M_H$ is almost a constant, but
falls off for $\gamma\gamma$ almost inverse cubic in $M_H$.

\begin{figure}[ht!]
\includegraphics[scale=0.85]{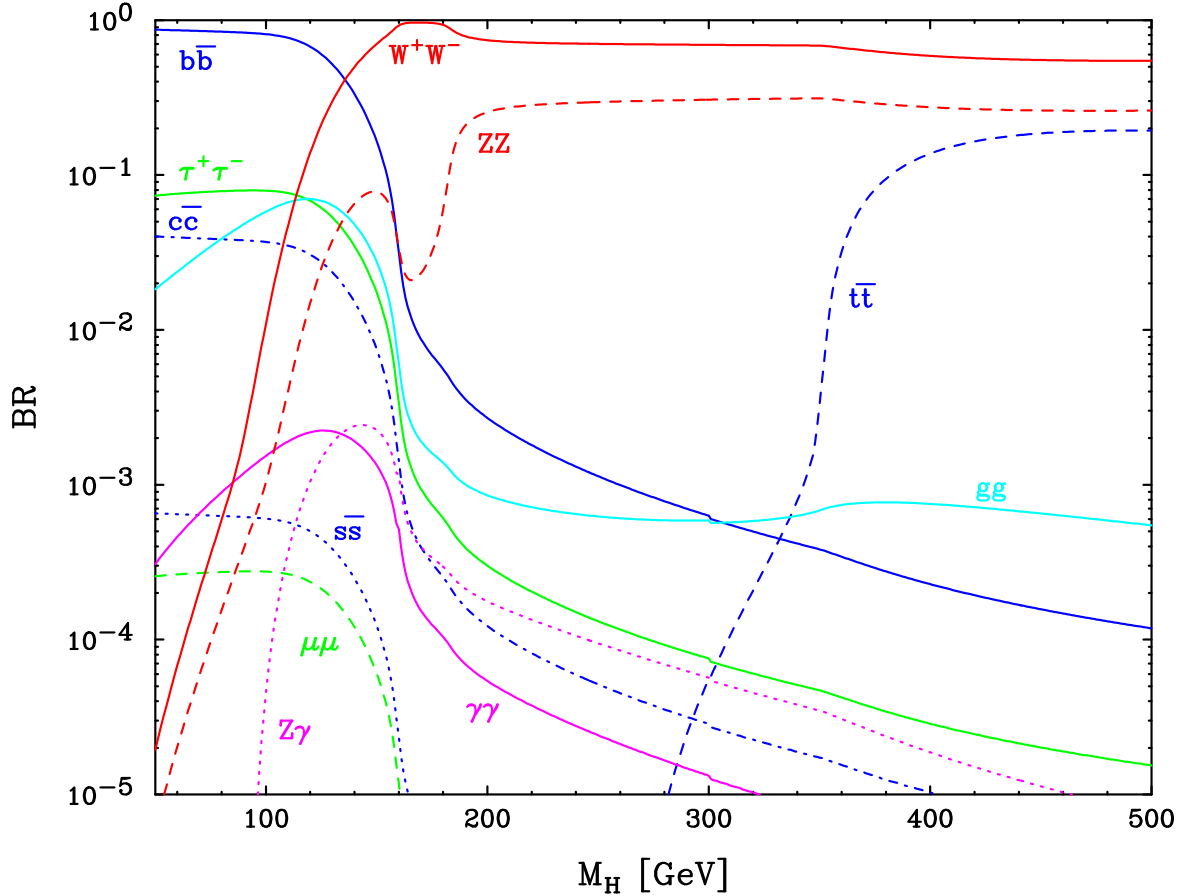}
\caption{Select Standard Model Higgs boson branching ratios as a 
function of mass, $M_H$~\protect\cite{Djouadi:1997yw}.  The Higgs
prefers to decay to the most massive possible final state.  The ratio
of fermionic branching ratios are proportional to fermion masses
squared, modulo color factors and radiative corrections.}
\label{fig:SM_BR}
\end{figure}

Once we've calculated all the various possible partial widths, we sum
them up to find the Higgs total width.  Each BR is then simply the
ratio $\Gamma_i/\Gamma_{tot}$.  These are shown in
Fig.~\ref{fig:SM_BR}; note the log scale.  If it wasn't obvious from
the partial width discussion, it should be now: near thresholds,
properly including finite width effects can be very important to get
the BRs correct.  Observe how the BR to $WW^*$ (at least one $W$ is
necessarily off-shell) is $50\%$ at $M_H=140$~GeV, 20~GeV below $W$
pair threshold.  BR($H\to\bb$)$\sim$BR($H\to\ww$) at $M_H=136$~GeV.

%

%%%%%%%%%%%%%%%%%%%%%%%%%%%%%%%%%%%%%%%%%%%%%%%%%%%%%%%%%%%%%%%%%%%%%%%%

\subsubsection{A brief word on statistics -- the simple view}
\label{sub:stats}

Now that we understand the basics of Higgs decay, and production in
electron-positron collisions, we should take a moment to consider
statistics.  The reason we must resort to statistics is that particle
detectors are imperfect instruments.  It is impossible to precisely
measure the energy of all outgoing particles in every collision.  The
calorimeters are sampling devices, which means they don't capture all
the energy; rather they're calibrated to give an accurate central
value at large statistics, with some Gaussian uncertainty about the
mean for any single event.  Excess energy can also appear, due to
cosmic rays, beam--gas or beam secondary interactions.  Quark final
states hadronize, resulting in the true final state in the detector (a
jet) being far more complicated and difficult even to identify
uniquely.  The electronics can suffer hiccups, and software {\it
always} has bugs, leading to imperfect analysis.  Thus, we would never
see two or three events at precisely the Higgs mass of, say,
122.6288...~GeV, and pop the champagne.  Rather, we'll get a
distribution of masses and have to identify the central value and its
associated uncertainty.

\begin{figure}[ht!]
\begin{center}
\includegraphics[scale=3]{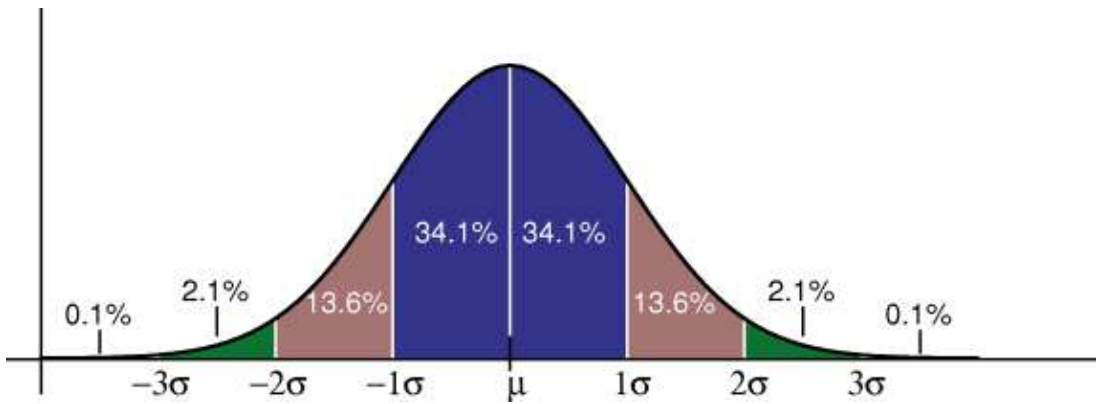}
\end{center}
\caption{Gaussian distribution about a mean $\mu$, showing the
fractional probability of events within one, two and three standard
deviations of the mean.}
\label{fig:gauss}
\end{figure}

In any experiment, event counts are quantum rolls of the dice.  For a
sufficient number of events, they also follow a Gaussian distribution
about the true mean:
\bq\label{eq:gauss}
f(x;\mu,\sigma) \; = \;
\frac{1}{\sigma\sqrt{2\pi}}
\:exp\biggl(-\frac{(x-\mu)^2}{2\sigma^2}\biggr)
\eq
The statistical uncertainty in the rate then goes as $1/\sqrt{N}$,
where $N$ is the number of events.  This is ``one sigma'' of
uncertainty: $68.2\%$ of identically-conducted experiments would
obtain $N$ within $\sigma\approx\pm\sqrt{N}$ about $\mu=N_{\rm true}$,
representing the true cross section.  Fig.~\ref{fig:gauss} shows the
fractional probabilities for various ``sigma'', or number of standard
deviations from the true mean.  To claim observation of a signal
deviating from our expected background, we generally use a $5\sigma$
criteria for discovery.  This means, if systematic errors have been
properly accounted for, that there is only a $0.00006\%$ chance that
the signal is due to a statistical fluctuation.  However, this
threshold is subjective, and you will often hear colleagues take
$4\sigma$ or even $3\sigma$ deviations seriously.  Since particle
physics has seen dozens of three sigma deviations come and go over the
decades, I would encourage you to regard $3\sigma$ as ``getting
interesting'', and $4\sigma$ as ``pay close attention and ask lots of
questions about systematics''.

Because SM processes can produce the same final state as any $ZH$
combined BR, we must know accurately what the background rate is for
each signal channel (final state) and how it is distributed in
invariant mass, then look for a statistically significant fluctuation
from the expected background over a fixed window region.  The size of
the window is determined by detector resolution: the better the
detector, the narrower the window, so the smaller the background,
yielding a better signal-to-background rate.  Generally, the window is
adjusted to accept one or two standard deviations of the hypothesized
signal ($68$--$95\%$).

Analyses are then defined by two different Gaussians: that governing
how many signal (and background) events were produced, and that
parameterizing the detector's measurement abilities.  The event count
$N$ in our above expression is the actual number of events observed,
in an experiment.  But in performing calculations ahead of time for
expected signal and background, it is variously taken as just $B$, the
number of background events expected, or $S+B$, expected signal
included, depending on the relative sizes of $S$ and $B$.  For doing
phenomenology, trying to decide which signals to study and calculate
more precisely, the distinction is often ignored.

The statistical picture I've outlined here is quite simplified.  Not
all experiments have sufficient numbers of events to describe their
data by Gaussians -- Poisson statistics may be more appropriate.  (An
excellent text on statistics for HEP is Ref.~\cite{Lyons:1986em}.)
Not all detector effects are Gaussian-distributed.  Nevertheless, it
gets across the main point: multiple sources of randomness introduce a
level of uncertainty that must be parameterized by statistics.  Only
when the probability of a random background fluctuation up or down to
the observed number of events is small enough, perhaps in some
distribution, can signal observation be claimed.  Exactly where this
line lies is admittedly a little hazy, but there's certainly a point
of several sigmas at which everybody would agree.

%%%%%%%%%%%%%%%%%%%%%%%%%%%%%%%%%%%%%%%%%%%%%%%%%%%%%%%%%%%%%%%%%%%%%%%%

\subsubsection{LEP Higgs data and results}
\label{sub:LEP-H}

Now to the actual LEP search.  Electrons and positrons have only
electroweak interactions, so backgrounds and a potential Higgs signal
are qualitatively of the same size.  (We'll see shortly in
Sec.~\ref{sub:Tev2} how this is not so at a hadron collider, which has
colored initial states.)  LEP thus had the ability to examine almost
all $Z$ and $H$ decay combinations: $\bb jj$, $\bb\ell^+\ell^-$,
$\bb\nu\bar\nu$, $\tau^+\tau^-jj$, $jjjj$, etc.  The largest of these
is $\bb jj$, as it combines the largest BRs of both the $Z$ and $H$.
It's closely followed by $\bb\nn$, since a $Z$ will go to neutrinos
$20\%$ of the time.  Neutrinos are missing energy, however, so not
precisely measured, making it possible that any observed missing
energy didn't in fact come from a $Z$.  Jets are much less
well-measured than leptons, so a narrower mass window can be used for
the $Z$ in $\bb\ell^+\ell^-$ events than $\bb jj$; the smaller
backgrounds in the narrower window might beat the smaller statistics
of the leptonic final state.

The exact details of each LEP search channel are not so important, as
lack of observation means we're more interested in channels' signal
and background attributes at hadron colliders.  For these lectures I
just present the final LEP result combining all four experiments.  The
interested student should read Eilam Gross' ``Higgs Statistics for
Pedestrians'', which goes into much more depth, and with wonderful
clarity~\cite{Gross:2002wg}.

\begin{figure}[ht!]
\begin{center}
\includegraphics[scale=0.9]{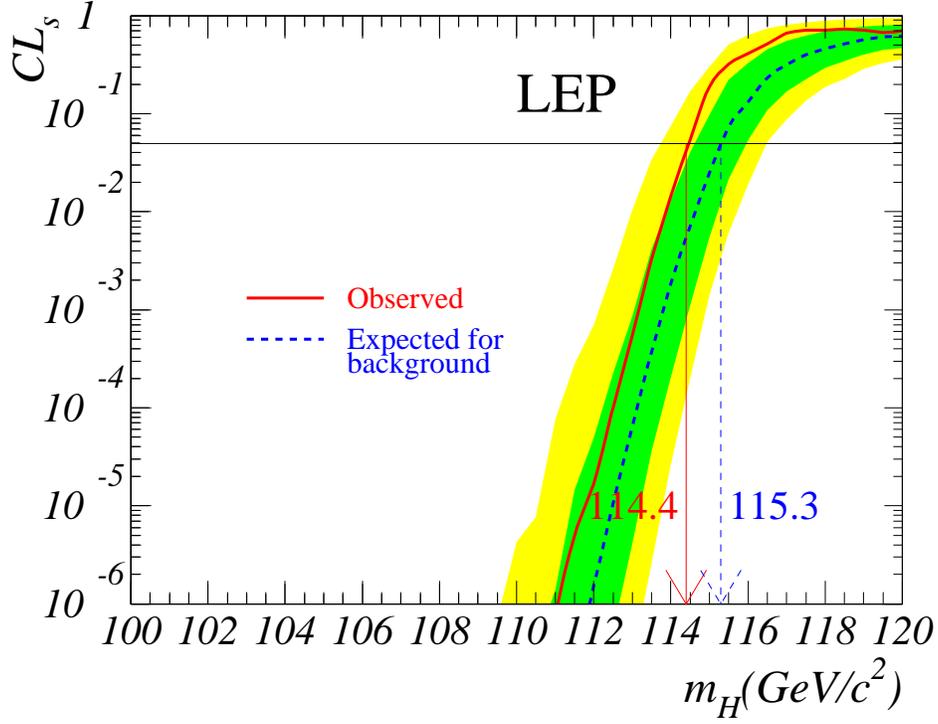}
\end{center}
\vspace*{-5mm}
\caption{Four-experiment combined result of the LEP Standard Model
Higgs search.  No signal was observed, establishing a lower limit of
114.4~GeV.  See text of Ref.~\protect\cite{Gross:2002wg} for
explanation.}
\label{fig:LEP}
\end{figure}

The money plot is shown in Fig.~\ref{fig:LEP}.  It shows the expected
confidence level (CL) for the signal+background hypothesis as a
function of Higgs mass.  The thin solid horizontal line at CL=0.05
signifies a $5\%$ probability that a true signal together with the
background would have fluctuated down in number of events to not be
discriminated from the expected background.  The green and yellow
regions are the $1\sigma$ and $2\sigma$ {\it expected} uncertainty
bands as a function of $M_H$, taking into account all sources of
uncertainty, calculational as well as detector effects.  Where the
central value (dashed curve) crosses 0.05 defines the $95\%$~CL
expected exclusion (lower mass limit).  This is essentially the
available collision energy minus the $Z$ mass minus a few extra GeV to
account for the $Z$ finite width -- it may be produced slightly
off-shell with some usable rate.  The solid red curve is the actual
experimental result, which is slightly above the experimental result
everywhere, meaning that the experiments gathered a couple more events
than expected in the 115-116~GeV mass bin.

\begin{figure}[hb!]
\begin{center}
\includegraphics[width=9cm,angle=-90]{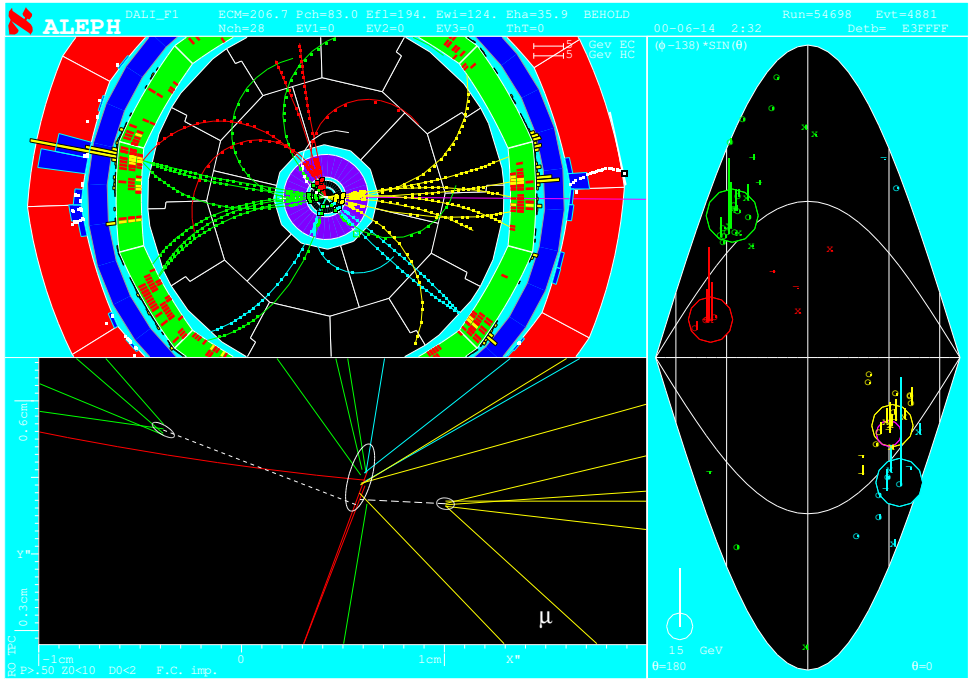} 
\end{center}
\vspace*{-5mm}
\caption{Event display of an interesting candidate $Z\to jj$,
$H\to\bb$ event in the Aleph detector at the end of LEP-II running and
at the machine's kinematic limit~\protect\cite{ALEPH_H-cand_event}.}
\label{fig:aleph}
\end{figure}

The end of LEP running involved a certain amount of histrionics.  At
first, the number of excess event at the kinematic machine limit was a
few, but more careful analyses removed most of these.  For example,
one particularly notorious event originally included in one
experiment's analysis had more energy than the beam delivered.
Another experiment removed a candidate event because some of the
outgoing particles traveled down a poorly-instrumented region of the
detector which was not normally used in analysis.  The final, most
credible enumeration was one candidate event in one experiment, show
in Fig.~\ref{fig:aleph}.

%

%%%%%%%%%%%%%%%%%%%%%%%%%%%%%%%%%%%%%%%%%%%%%%%%%%%%%%%%%%%%%%%%%%%%%%%%
%%%%%%%%%%%%%%%%%%%%%%%%%%%%%%%%%%%%%%%%%%%%%%%%%%%%%%%%%%%%%%%%%%%%%%%%

\subsection{Prospects at Tevatron}
\label{sub:Tev2}

With the end of the LEP era, all eyes turned to Run~II of the upgraded
Fermilab Tevatron.  Its energy increased from 1.8 to 1.96~GeV, and is
expected to gather many tens of times the amount of data in Run~I.
Higgs-hunting hopes were high~\cite{Carena:2000yx}, although it was
clear that the machine and both detectors have to perform
exceptionally well to have a chance, as Tevatron's Higgs mass reach
will not be all that great, and will have significant observability
gaps in the mass region expected from precision EW data.

\begin{figure}[hb!]
\begin{center}
\includegraphics[width=14cm,height=9cm]{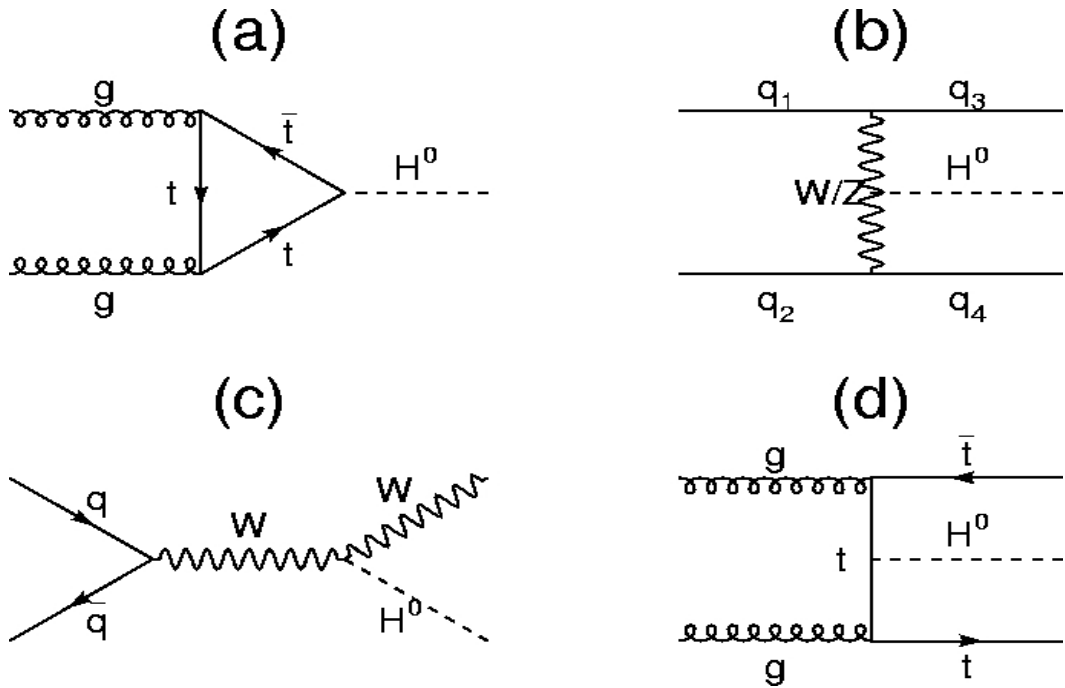}
\end{center}
\vspace*{-5mm}
\caption{Feynman diagrams for the four dominant Higgs production
processes at a hadron collider.}
\label{fig:4proc}
\end{figure}

\begin{figure}[ht!]
\begin{center}
\includegraphics[scale=0.62]{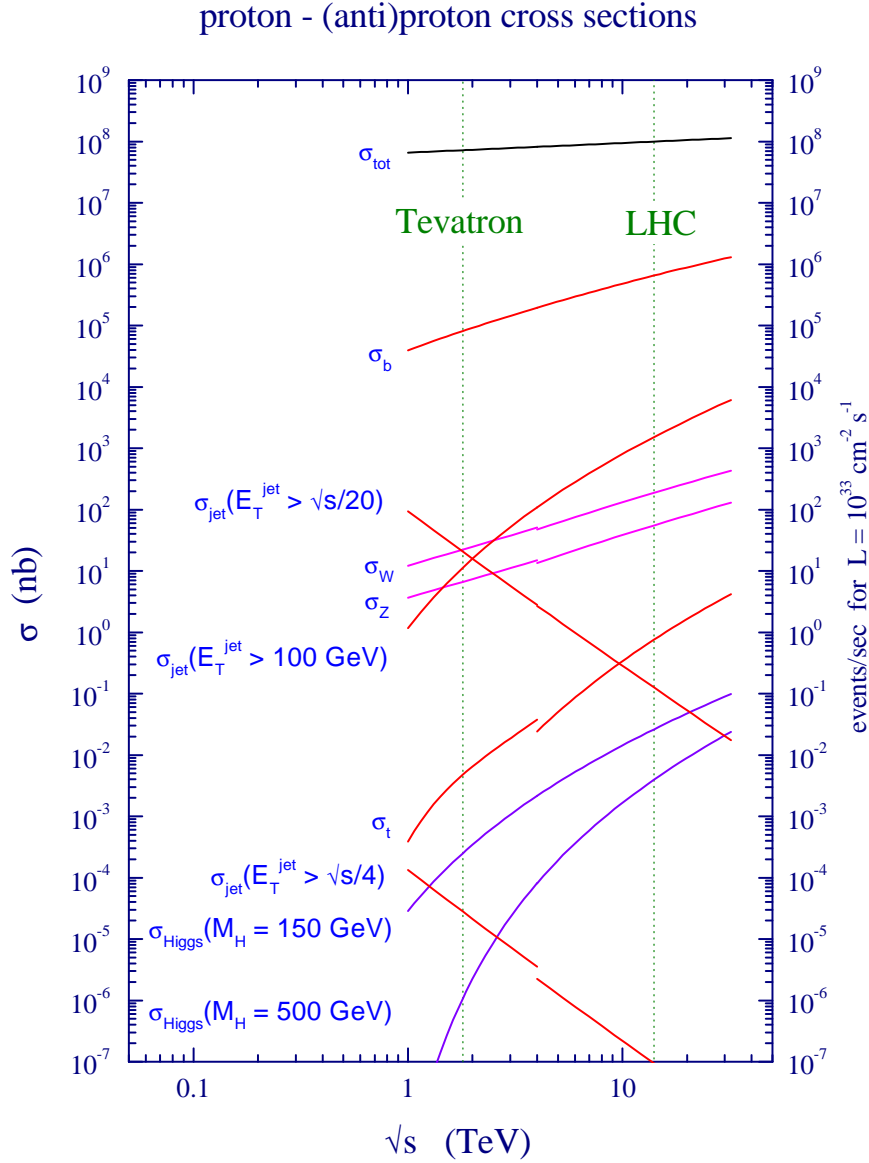}
\end{center}
\vspace*{-5mm}
\caption{Various SM ``standard candle'' cross sections at hadron
colliders of varying energy, with Tevatron and LHC marked in
particular.  Note the log scale.  Discontinuities are due to the
difference between $p\bar{p}$ for Tevatron and $pp$ for LHC.  Figure
from Ref.~\protect\cite{Gianotti:2004qs}.}
\label{fig:pp-xsecs}
\end{figure}

To understand the details and issues, we first need to identify how a
Higgs boson may be produced in proton-antiproton collisions.  Like the
electron, the light quarks have too small a mass (Yukawa coupling) to
produce a Higgs directly with any useful rate, discernible against the
large QCD backgrounds produced in hadron collisions\footnote{For
example, $H\to\bb$ is the dominant BR of a light Higgs, but QCD $b$
jet pair production in hadron collisions is many orders of magnitude
larger.  Cf. Fig.~\ref{fig:pp-xsecs}.}.  Quarks may annihilate,
however, to EW gauge bosons, which have large coupling to the Higgs;
and likewise to a top quark pair.  Incoming quarks may also emit a
pair of gauge bosons which fuse to form a Higgs, a process known as
weak boson fusion (WBF).  But high energy protons also possess a large
gluon content; recall that gluons have a loop-induced coupling to the
Higgs.  Fig.~\ref{fig:4proc} displays Feynman diagrams for all four of
these processes at hadron colliders.  The questions are, what are
their relative sizes, and what are their backgrounds?  Because of the
partonic nature of hadron collisions, the Higgs couplings are not
enough to tell us the relative sizes; we also need to take into
account incoming parton fluxes and final state phase space -- single
Higgs production is much less greedy than $t\bar{t}H$ associated
production, for instance.  In addition, the internal propagator
structure of the processes is important: $WH$,$ZH$ bremsstrahlung are
$s$-channel suppressed, but no other process is.

\begin{figure}[ht!]
\begin{center}
\includegraphics[scale=0.60,angle=270]{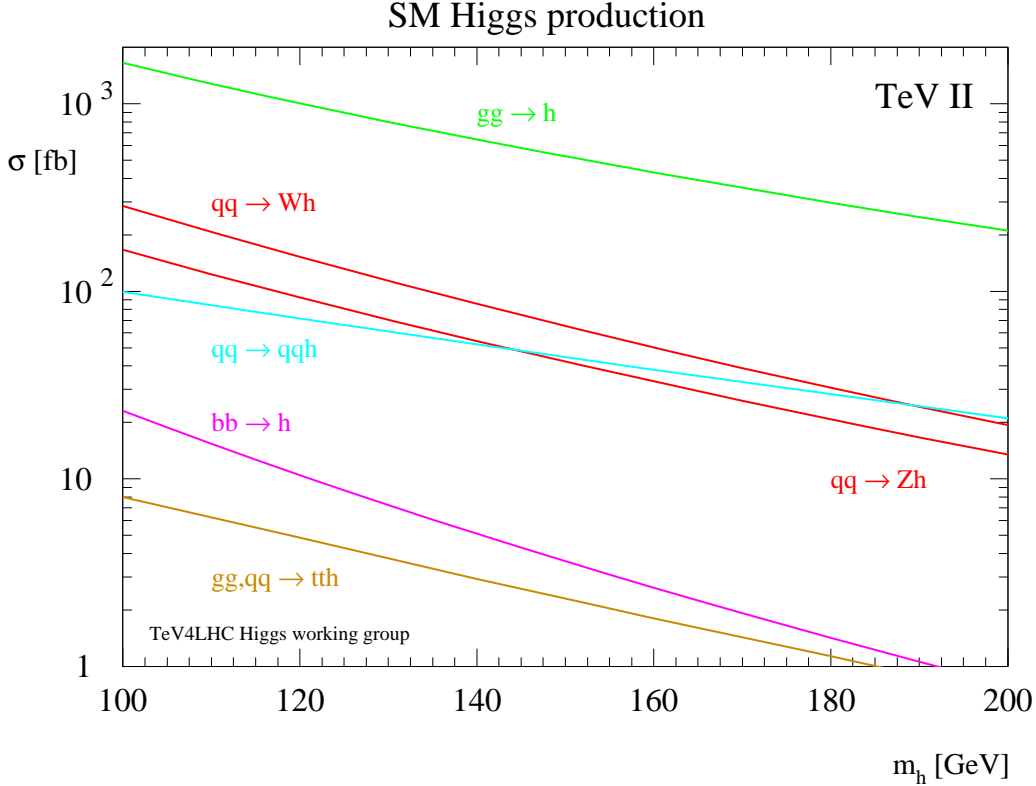}
\end{center}
\vspace*{-5mm}
\caption{Cross sections for Higgs production in various channels at
Tevatron Run~II ($\sqrt{s}=2$~TeV).  Note the log scale.  Figure from
the Tev4LHC Higgs working group~\protect\cite{Hahn:2006my}.}
\label{fig:Tev2-xsecs}
\end{figure}

The various rates, updated in 2006 with the latest theoretical
calculations~\cite{Hahn:2006my,Aglietti:2006ne}, are shown in
Fig.~\ref{fig:Tev2-xsecs} for a light SM Higgs boson.  Students not
already familiar with hadron collider Higgs physics will probably be
surprised to learn that $gg\to H$, gluon fusion Higgs production,
dominates at Tevatron energy.  This is partly because the coupling is
actually not all that small, partly because high-energy protons
contain a plethora of gluons, and partly because there is no
propagator suppression, and much less phase space suppression,
compared to other processes.  Higgsstrahlung (Fig.~\ref{fig:4proc}(c))
is still important at Tevatron, analogous to LEP.  Note that the
smaller cross sections have more complicated final states, therefore
potentially less background, and possibly distinctive kinematic
distributions that could assist in separating a signal from the
background.  It's not obvious that the largest rate is the most useful
channel!  Considering that the Higgs decays predominantly to different
final states as a function of its mass, it's also not obvious that the
optimal channel at one mass is optimal for all masses.  In fact,
that's definitely not the case.

Not knowing the answer, we naturally start by considering the largest
cross section times branching ratio, $gg\to H\to\bb$.  Just how large
is the background, QCD $pp\to\bb$ production?  Fig.~\ref{fig:pp-xsecs}
shows a variety of SM cross section for hadron collisions of various
energy, and marks off in particular Tevatron and LHC.  (The
discontinuity in some curves is because Tevatron is $\pp$ and LHC is
$pp$.)  We immediately notice that the $\bb$ inclusive rate is almost
nine orders of magnitude larger than inclusive $H\to\bb$.  Of course
the background will be smaller in a finite window about the Higgs
mass.  But jets are not so well-measured, necessitating a fairly large
window, $\sim$15--20~GeV either side of the central value.  We lose
only a few orders of magnitude of the background, taking us from
``laughable'' to just terminally hopeless.

The general rule of thumb at hadron collider experiments is to require
a final state with at least one high-energy lepton.  This means lower
backgrounds because the event had at least some EW component, such as
a $W$ or $Z$, or came from a massive object, such as the top quark,
which is not produced in such great abundance due to phase space
suppression.

Tevatron's Higgs search is rate-limited.  We can see this by
multiplying the 150~GeV Higgs cross section from
Fig.~\ref{fig:Tev2-xsecs} by the expected integrated luminosity of
4--8~fb$^{-1}$ during Run~II.  Because of this, and the very low
efficiency of identifying final-state taus in a hadron collider
environment (unlike at LEP), Tevatron's experiments CDF and D\O\ focus
on the $H\to\bb$ final state where that decay dominates the BR, and
Higgsstrahlung to obtain the lepton tag.  For larger Higgs masses,
where $H\to\ww$ dominates, gluon fusion Higgs production is the
largest rate, but Higgsstrahlung has some analyzing power.  To
summarize~\cite{Carena:2000yx}:
\begin{itemize}
\item[$M_H\lesssim$] 140~GeV: $H\to\bb$ dominates, so we use:
\vspace{-1mm}
\begin{itemize}
\item[$\cdot$] $WH\to\ell^\pm\nu\bb$
\vspace{-1mm}
\item[$\cdot$] $ZH\to\ell^+\ell^-\bb$
\vspace{-1mm}
\item[$\cdot$] $WH$, $ZH\to jj\bb$
\vspace{-1mm}
\item[$\cdot$] $ZH\to\nu\bar\nu\bb$
\end{itemize}
\item[$M_H\gtrsim$] 140~GeV: $H\to\ww$ dominates, so we use:
\vspace{-1mm}
\begin{itemize}
\item[$\cdot$] $gg\to H\to\ww$ (dileptons)
\vspace{-1mm}
\item[$\cdot$] $WH\to W^\pm\ww$ ($2\ell$ and $3\ell$ channels)
\end{itemize}
\end{itemize}

%%%%%%%%%%%%%%%%%%%%%%%%%%%%%%%%%%%%%%%%%%%%%%%%%%%%%%%%%%%%%%%%%%%%%%%%

\subsubsection{$VH,H\to\bb$ at Tevatron}
\label{sub:Tev2-VH}

While a lepton tag gets rid of most QCD backgrounds, it doesn't
automatically eliminate top quarks: they decay to $Wb$, thus the event
often contains one lepton and two jets, or two leptons and missing
energy, in addition to the $b$ jet pair.  This is the same final state
as our Higgs signal, with either extra jets or transverse energy
imbalance.  Kinematic cuts help, but because the detectors are
imperfect some top quark events will leak through.  Jet mismeasurement
gives fake missing energy, for example (and is one of the most
difficult uncertainties to quantify in a hadron collider experiment).
In addition, QCD initial-state radiation from the incoming partons can
give extra jets.  Thus top quark and Higgs signal events qualitatively
become very similar.  To control this further the experiments have to
look at other observables, such as angular distributions of the $b$
jets and leptons.  Other backgrounds to consider are QCD $W\bb$
production, weak bosons pairs where one decays to $\bb$ (and thus has
invariant mass close to the Higgs signal window).

\begin{figure}[hb!]
\begin{center}
\includegraphics[scale=0.75]{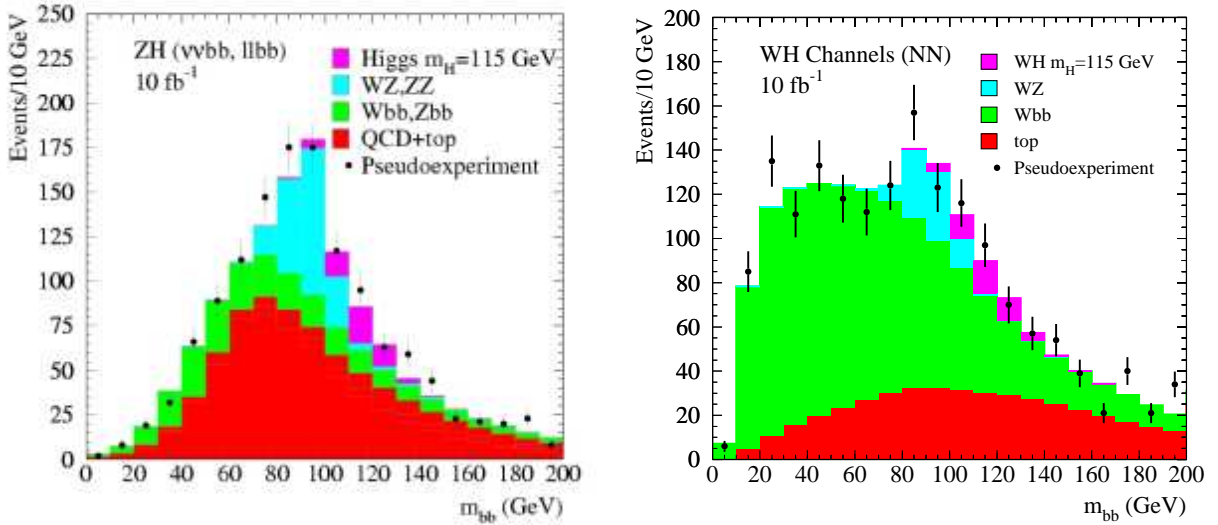}
\includegraphics[scale=0.42]{Higgs.figs/CDF/Tev2-CDF_WH_mbb.eps}
\end{center}
\vspace*{-8mm}
\caption{CDF simulations of a 115~GeV signal at Tevatron Run~II in
$ZH$ (left) and $WH$ (right) Higgsstrahlung production with Higgs
decays $H\to\bb$ and assuming 10~fb$^{-1}$ is
collected~\protect\cite{Babukhadia:2003zu}.}
\label{fig:Tev2-VH}
\end{figure}

Fig.~\ref{fig:Tev2-VH} shows the results of a CDF simulation study of
$WH$ and $ZH$ Higgsstrahlung events at Run~II for $M_H=115$~GeV (right
at the LEP Higgs limit)~\cite{Babukhadia:2003zu}.  First note how the
top quark pair and diboson backgrounds peak very close to the Higgs
mass.  Eyeballing the plots and simplistically applying our knowledge
of Gaussian statistics, we could easily believe that this could yield
a four or five sigma signal, perhaps combined with D\O\ results.
However, carefully observe that the shape of the invariant mass
distribution for background alone and with signal are extremely
similar: they are both steeply falling; the Higgs signal is not a
stand-out peak above a fairly flat background.  Therein lies a hidden
systematic!  This means that we must understand the
kinematic-differential shape of the QCD backgrounds to a very high
degree of confidence.  This is not just knowing the SM background at
higher orders in QCD, differentially, but also the detector response.
This criticality is not appreciated in most discussions of a potential
discovery at Tevatron.  It should be obvious that an excess in one of
these channels would cause a scramble of cross-checking and probably
further theoretical work to ensure confidence, in spite of the
statistics alone.  We'll run into this feature again with one of the
LHC channels in Sec.~\ref{sub:LHC-ttH}, but quantified.

\begin{figure}[hb!]
\begin{center}
\includegraphics[scale=0.7,angle=270]{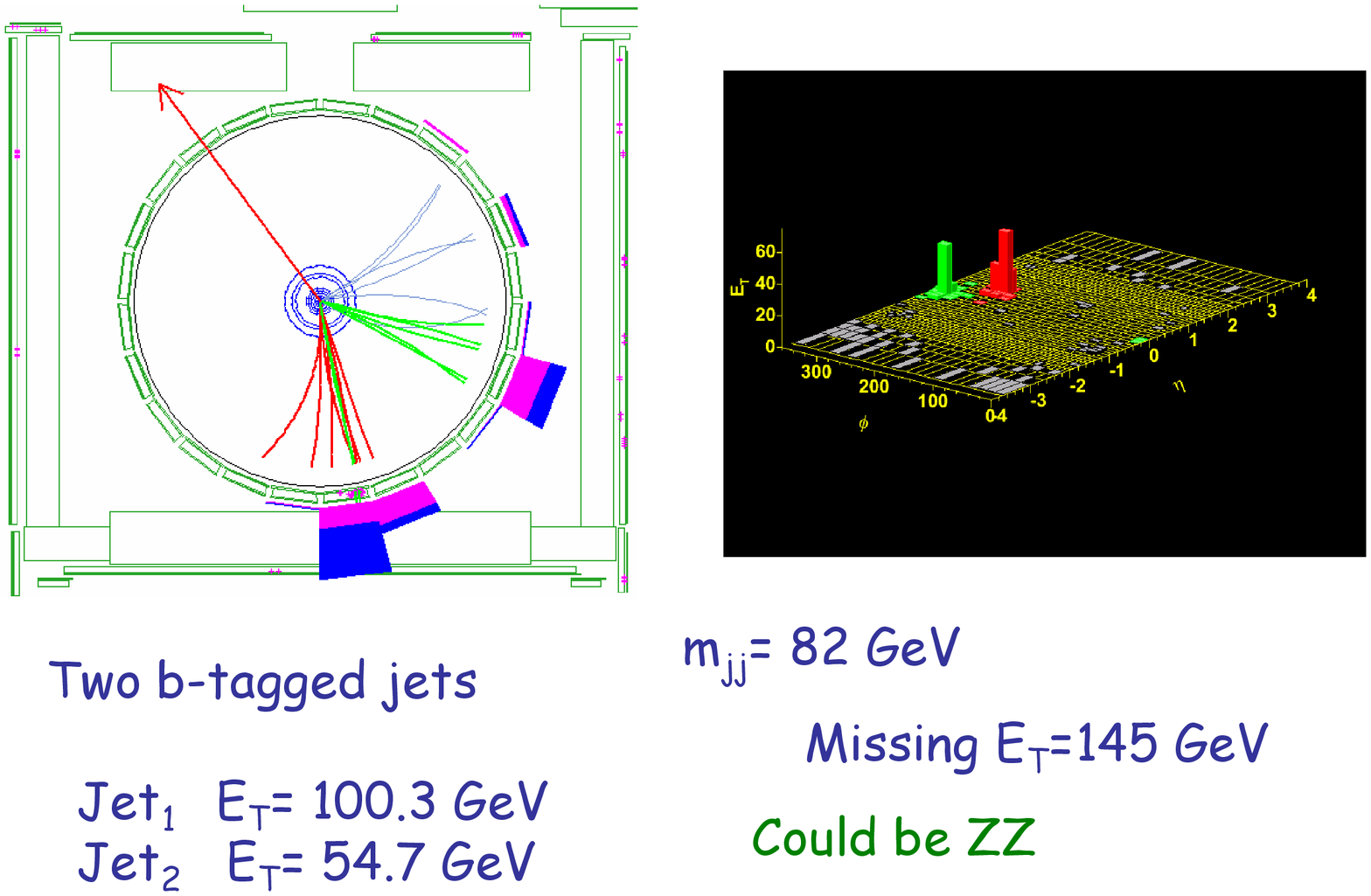}
\end{center}
\vspace*{-3mm}
\caption{Interesting $bb\sla{p}_T$ event at CDF in Tevatron 
Run~II~\protect\cite{CDF8842}.}
\label{fig:CDF-bbvv}
\end{figure}

CDF has in fact already observed an interesting candidate Higgs event
in Run~II, in the first few hundred pb$^{-1}$.  It is in the
$ZH\to\nn\bb$ channel (a $b$ jet pair plus missing transverse energy).
The event display and key kinematic information are shown in
Fig.~\ref{fig:CDF-bbvv}.  Given the very low $b$ jet pair invariant
mass, it's much more likely that the event came from EW $ZZ$ or QCD
$Z\bb$ production (cf. Fig.~\ref{fig:Tev2-VH}).  It therefore doesn't
generate the kind of excitement that the handful of events at LEP did.
Nevertheless, finding this event was a milestone, showing that CDF
could perform such an analysis and find Higgs-like events with good
efficiency.

\begin{table}[ht!]
\begin{center}\begin{tabular}{llccccc}
& & \multicolumn{5}{c}{Higgs Mass (GeV/$c^2$)} \\ \cline{3-7}
\noalign{\vskip3pt}
\multicolumn{1}{c}{Channel} &\multicolumn{1}{c}{Rate}& 
                               90   &  100  &  110  &  120  &  130 \\ \hline\hline
%---------------------------------------------------------------------------------
             & $S$          &  8.7  &  9.0  &  4.8  &  4.4  &  3.7 \\[-1mm]
$\lpm\nu\bb$ & $B$          &  28   &  39   &  19   &  26   &  46  \\[-1mm]
             & $S/\sqrt{B}$ &  1.6  &  1.4  &  1.1  &  0.9  &  0.5 \\ \hline\hline
%---------------------------------------------------------------------------------
             & $S$          &  12   &   8   &  6.3  &  4.7  &  3.9 \\[-1mm]
$\nn\bb$     & $B$          & 123   &  70   &  55   &  45   &  47  \\[-1mm]
             & $S/\sqrt{B}$ &  1.1  &  1.0  &  0.8  &  0.7  &  0.6 \\ \hline\hline
%---------------------------------------------------------------------------------
             & $S$          &  1.2  &  0.9  &  0.8  &  0.8  &  0.6 \\[-1mm]
$\ells\bb$   & $B$          &  2.9  &  1.9  &  2.3  &  2.8  &  1.9 \\[-1mm]
             & $S/\sqrt{B}$ &  0.7  &  0.7  &  0.5  &  0.5  &  0.4 \\ \hline\hline
%---------------------------------------------------------------------------------
             & $S$          &  8.1  &  5.6  &  3.5  &  2.5  &  1.3 \\[-1mm]
$\qq\bb$     & $B$          & 6800  & 3600  & 2800  & 2300  & 2000 \\[-1mm]
             & $S/\sqrt{B}$ & 0.10  & 0.09  & 0.07  & 0.05  & 0.03 \\ \hline 
%---------------------------------------------------------------------------------
\end{tabular}\end{center}
\vspace*{-5mm}
\caption{Predicted signal significances at Tevatron Run~II, for one
detector and 1~fb$^{-1}$, for various $VH,H\to\bb$ searches, taken
from Ref.~\protect\cite{Carena:2000yx}.}
\label{tab:Tev2-VH}
\end{table}

Table~\ref{tab:Tev2-VH} summarizes the 2000 Tevatron Higgs Working
Group Report predictions for Higgsstrahlung reach in
Run~II~\cite{Carena:2000yx}.  The results are quoted for one detector
and per fb$^{-1}$, hence the rather small significances.  CDF and D\O\
will eventually combine results, giving a factor of two in statistics.
However, it's not known how much data they'll eventually collect by
2009 or 2010, when LHC is expected to have first physics results and
CDF \& D\O\ detector degradation becomes an issue.  Fairly low Higgs
masses are shown, because when the report was written nobody expected
LEP to perform as well as it did, greatly exceeding its anticipated
search reach.  It should be obvious that a clear discovery would
require a large amount of data, combining multiple channels, and the
Higgs boson happening to be fairly light; not to mention the QCD shape
systematic concern I described earlier (but is not quantified).  In
spite of this apparent pessimism, however, CDF and D\O\ seem to be
performing modestly better than expected -- higher efficiencies for
$b$ tagging and phase space coverage, better jet resolution, etc.
There is as yet no detailed updated report with tables such as this,
but there are some newer graphically-presented expectations I'll show
as a summary.

%%%%%%%%%%%%%%%%%%%%%%%%%%%%%%%%%%%%%%%%%%%%%%%%%%%%%%%%%%%%%%%%%%%%%%%%

\subsubsection{$gg\to H\to\ww$ at Tevatron}
\label{sub:Tev2-H_VV}

For $M_H\gtrsim 140$~GeV, a SM Higgs will decay mostly to $W$ pairs
(cf. Fig.~\ref{fig:SM_BR}), which has a decent rate to dileptons and
has very little SM background -- essentially just EW $W$ pair
production, with some background from top quark pairs where both $b$
jets are lost.  This channel has some special characteristics due to
how the Higgs decay proceeds.  There is a marked angular correlation
between the outgoing leptons which differs from the SM backgrounds:
they prefer to be emitted together, that is close to the same flight
direction in the center-of-mass frame~\cite{Dittmar:1996ss}.

\begin{figure}[ht!]
\begin{center}
\includegraphics[scale=0.8]{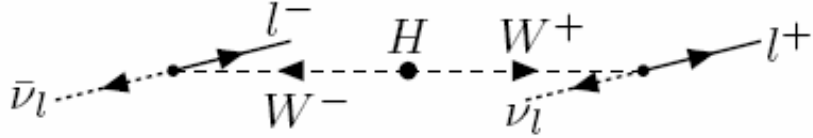}
\end{center}
\vspace*{-3mm}
\caption{Diagram showing the preferred flight direction of charged
leptons in $H\to\ell^+\nu\ell^-\bar\nu$.}
\label{fig:H-WW-llcorr-1}
\end{figure}

\begin{figure}[hb!]
\begin{center}
\includegraphics[scale=0.6]{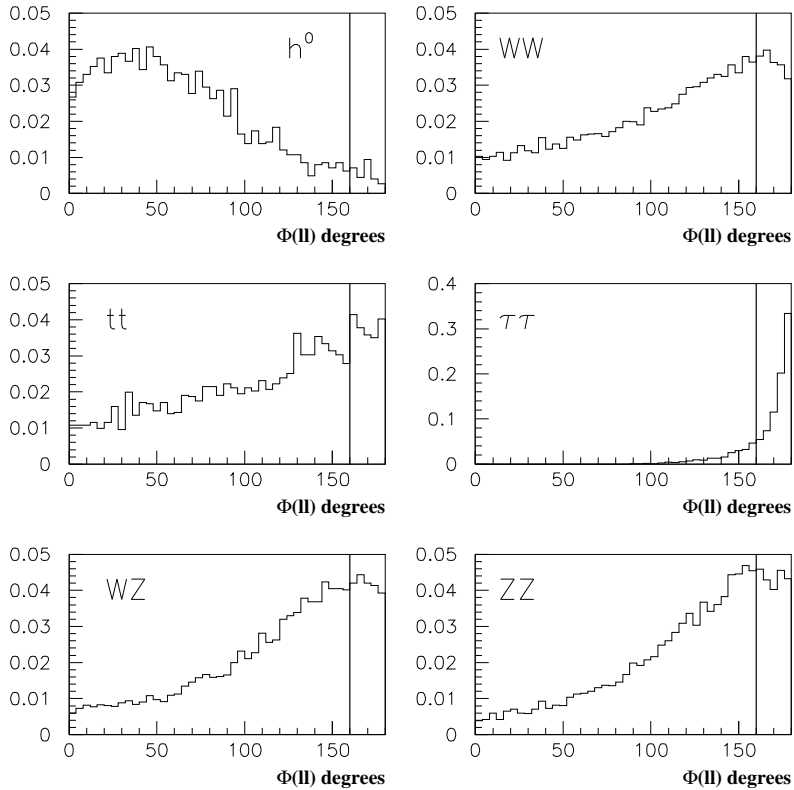}
\end{center}
\vspace*{-3mm}
\caption{Dilepton azimuthal angular correlation for a 
$H\to\ww\to\ell^+\nu\ell^-\bar\nu$ signal and its backgrounds.  The
efficacy of the cut (vertical line) can easily be estimated visually.
From the Tevatron Run~II Higgs Working Group
Report~\protect\cite{Carena:2000yx}.}
\label{fig:H-WW-llcorr-2}
\end{figure}

To understand this correlation, consider what happens if the Higgs
decays to a pair of transversely-polarized $W$ bosons.  For $W$
decays, the lepton angle with respect to the $W^\pm$ spin follows a
$(1\pm\cos\theta_{\ell^\pm})^2$ distribution.  That is, the
positively-charged lepton prefers to be emitted with the $W$ spin,
while the negatively-charged lepton prefers to be emitted opposite the
$W$ spin.  Since the Higgs is a scalar (spin-0), the $W$ spins are
anti-correlated, thus the leptons are preferentially emitted in the
same direction.  For longitudinal $W$ bosons, the lepton follows a
$\sin^2\theta_\ell$ distribution.  The $W$ spins are still correlated,
however, and the matrix element squared (an excellent exercise for the
student) is proportional to $(p_{\ell^-}\cdot p_\nu)(p_{\ell^+}\cdot
p_{\bar\nu})$.  Since a charged lepton and neutrino are emitted
back-to-back in the $W$ rest frame, this is again maximized for the
charged leptons emitted together.  This correlation is shown visually
by the schematic of Fig.~\ref{fig:H-WW-llcorr-1}.  Projected onto the
azimuthal plane (transverse to the beam), its efficacy is shown in
Fig.~\ref{fig:H-WW-llcorr-2} by comparison to various
backgrounds~\cite{Carena:2000yx,Han_H_WW_Tev}.


In addition to this angular correlation, we may also construct a
transverse mass ($M_T$) for the system, despite the fact that two
neutrinos go missing~\cite{Rainwater:1999sd}.  We first write down the
transverse energy ($p_T$) of the dilepton and missing transverse
energy ($\sla{E}_T$) systems,
\bq\label{eq:ET-def}
E_{T_{\ells}} \; = \; \sqrt{\vec{p}_{T_{\ells}}^2 + m_{\ells}^2}
\; , \qquad
\sla{E}_T  \; = \; \sqrt{\sla{\vec{p}}_T^2 + m_{\ells}^2}
\eq
where I've substituted the dilepton invariant mass $m_{\ells}^2$ for
$m_{\nn}^2$.  This is exact at $H\to WW$ threshold, and is a very
good approximation for Higgs masses below about 200~GeV and where this
decay mode is open.  The $W$ pair transverse mass is now
straightforward:
\bq\label{eq:M_T_WW}
M_{T_{WW}} \; = \; \sqrt{  (\sla{E}_T + E_{T_{\ells}})^2
                         - (\vec{p}_{T_{\ells}} + \sla{\vec{p}}_T)^2}
\eq
This gives a nice Jacobian peak for the Higgs signal, modulo detector
missing-transverse-energy resolution, whereas the SM backgrounds tend
to be comparatively flat.

Utilizing these techniques gives Tevatron some reach for a heavier
Higgs boson, mostly in the mass range $150\lesssim M_H\lesssim
180$~GeV, where the BR to $WW$ is significant and the Higgs production
rate is not too small.

%%%%%%%%%%%%%%%%%%%%%%%%%%%%%%%%%%%%%%%%%%%%%%%%%%%%%%%%%%%%%%%%%%%%%%%%

\subsubsection{Tevatron Higgs summary expectations}
\label{sub:Tev2-summ}

\begin{figure}[hb!]
\begin{center}
\includegraphics[scale=0.65]{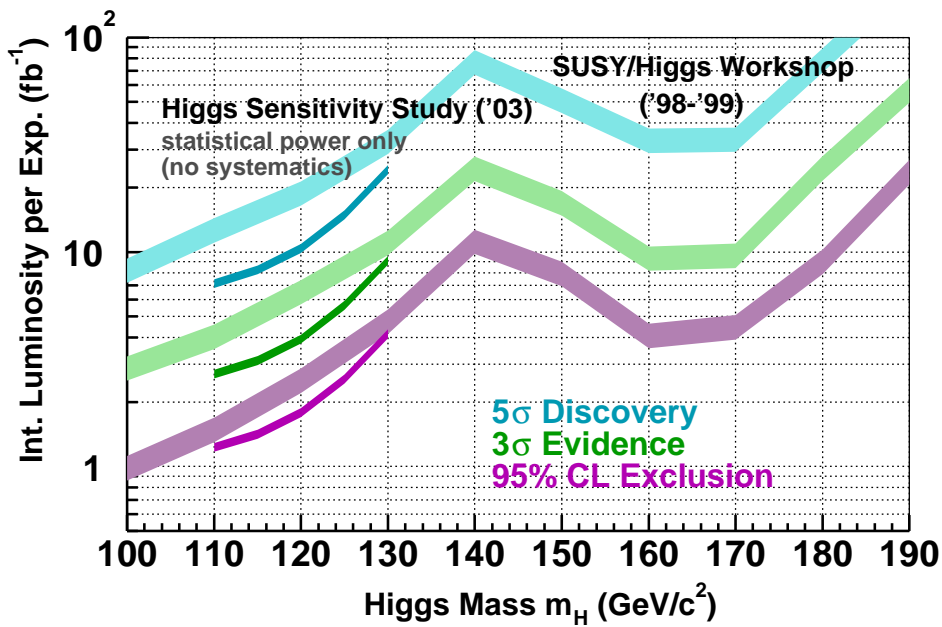}
\end{center}
\vspace*{-8mm}
\caption{Expected required integrated luminosity per experiment 
required in Run~II to observe a SM Higgs as a function of
$M_H$~\protect\cite{Babukhadia:2003zu}.}
\label{fig:Tev2-summ}
\end{figure}

Tevatron Higgs physics expectations have changed since the 2000
Report, as D\O\ and CDF have better understood their detectors and
made analysis improvements.  As yet, the only progress summary is from
2003, shown in Fig.~\ref{fig:Tev2-summ}.  It compares the original
Report's findings, shown by the thick curves, with improved findings
for the low-mass region, shown by the thinner lines.  However, the new
results do not yet include systematic uncertainties, which may be
considerable.  We should expect some form of a new summary expectation
sometime in 2007.  A final note on the undiscussed WBF production
mode: some study has been done (see Sec.~II.C.4 of
Ref.~\cite{Carena:2000yx}), but D\O\ and CDF both lack sufficient
coverage of the forward region to use this mode.  This is not the case
at LHC.

Run~II now has about 1~fb$^{-1}$ of analyzed data, and a Higgs search
summary progress report is available in Ref.~\cite{Bernardi:2006fd},
which updates each channel's expectations.

\medskip

%

%%%%%%%%%%%%%%%%%%%%%%%%%%%%%%%%%%%%%%%%%%%%%%%%%%%%%%%%%%%%%%%%%%%%%%%%
%%%%%%%%%%%%%%%%%%%%%%%%%%%%%%%%%%%%%%%%%%%%%%%%%%%%%%%%%%%%%%%%%%%%%%%%

\subsection{Higgs at LHC}
\label{sub:LHC}

\begin{figure}[hb!]
\includegraphics[scale=0.6,angle=270]{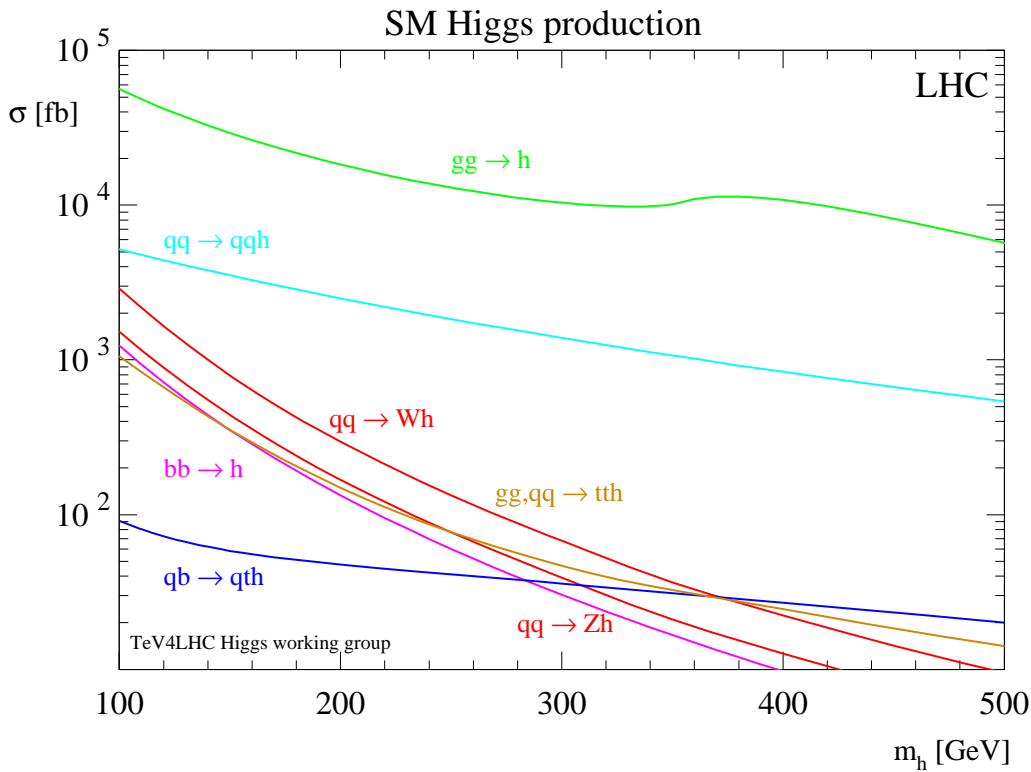}
\caption{Cross sections for Higgs production in various channels at
LHC ($\sqrt{s}=14$~TeV)~\protect\cite{Hahn:2006my}.}
\label{fig:LHC-xsecs}
\end{figure}

Higgs physics at LHC will be similar to that at Tevatron.  There is
the slight difference that LHC will be $pp$ collisions rather than
$p\bar{p}$.  The biggest difference, however, is the increased energy,
from 2 to 14~TeV.  Particle production in the 100~GeV mass range will
be at far lower Feynman $x$, where the gluon density is much larger
than the quark density.  In fact, it's useful (for Higgs physics) to
think of the LHC as a gluon collider to first order.  The ratio
between gluon fusion Higgs production and Higgsstrahlung is thus
larger than at Tevatron.  Fig.~\ref{fig:LHC-xsecs} displays the
various SM Higgs cross sections, only over a much larger range of
$M_H$ -- at LHC, large-$M_H$ cross sections are not trivially small,
compared to at the Tevatron.  There are huge QCD corrections to the
$gg\to H$ rate (also at Tevatron), but these are now known at NNLO and
under control~\cite{H-NNLO} (and included in
Fig.~\ref{fig:LHC-xsecs}).  They don't affect the basic phenomenology,
however.  Knowing that LHC is plans to collect several hundred
fb$^{-1}$ of data, a quick calculation reveals that the LHC will truly
be a Higgs factory, producing hundreds of thousands of light Higgs
bosons, or tens of thousands if it's heavy.

Looking back at Fig.~\ref{fig:pp-xsecs}, we see that while the Higgs
cross section rises quite steeply with collision energy ($gg\to H$ is
basically a QCD process), so do important backgrounds like top quark
production.  The inclusive $b$ cross section is still too large to
access to $gg\to H\to\bb$, but note that the EW gauge boson cross
sections do not rise as swiftly with energy.  Immediately we realize
that channels like $gg\to H\to\ww$ should have a much better
signal-to-background (S/B) ratio.  (In fact it suffers from
non-trivial single-top quark~\cite{Kauer:2004fg} and
$gg\to\ww$~\cite{Binoth_gg-WW} backgrounds, but is still an excellent
channel for $M_H\gtrsim 150$~GeV.)  The figure does not show cross
sections like $W\bb$ or $Z\bb$, which grow QCD-like and thus become a
terminal problem for $WH$ and $ZH$ channels.

Obviously there are a few significant differences between Tevatron and
LHC with implications for Higgs physics.  We'll lose access to $WH$
and $ZH$ at low mass, at least for Higgs decay to $b$ jets.  What
about rare decays, since the production rate is large?  The
$t\bar{t}H$ cross section is large and would yield a healthy event
rate.  It's complexity is distinctive, so one might speculate that
perhaps it could be useful.  WBF production is also accessible due to
better detectors, and likewise its more complex signature is worthy of
a look.  It will in fact turn out to be perhaps the best production
mode at LHC.

As with Tevatron, we need to understand both the signal and background
for each Higgs channel we wish to examine.  As a prelude to
Chapter~\ref{sec:meas}, Higgs measurements, at LHC we won't want to
just find the Higgs in one mode.  Rather, we'll want to observe it in
as many production and decay modes as possible, to study all its
properties, such as couplings.

%%%%%%%%%%%%%%%%%%%%%%%%%%%%%%%%%%%%%%%%%%%%%%%%%%%%%%%%%%%%%%%%%%%%%%%%

\subsubsection{$t\bar{t}H,H\to\bb$}
\label{sub:LHC-ttH}

\begin{figure}[hb!]
\includegraphics[scale=0.35]{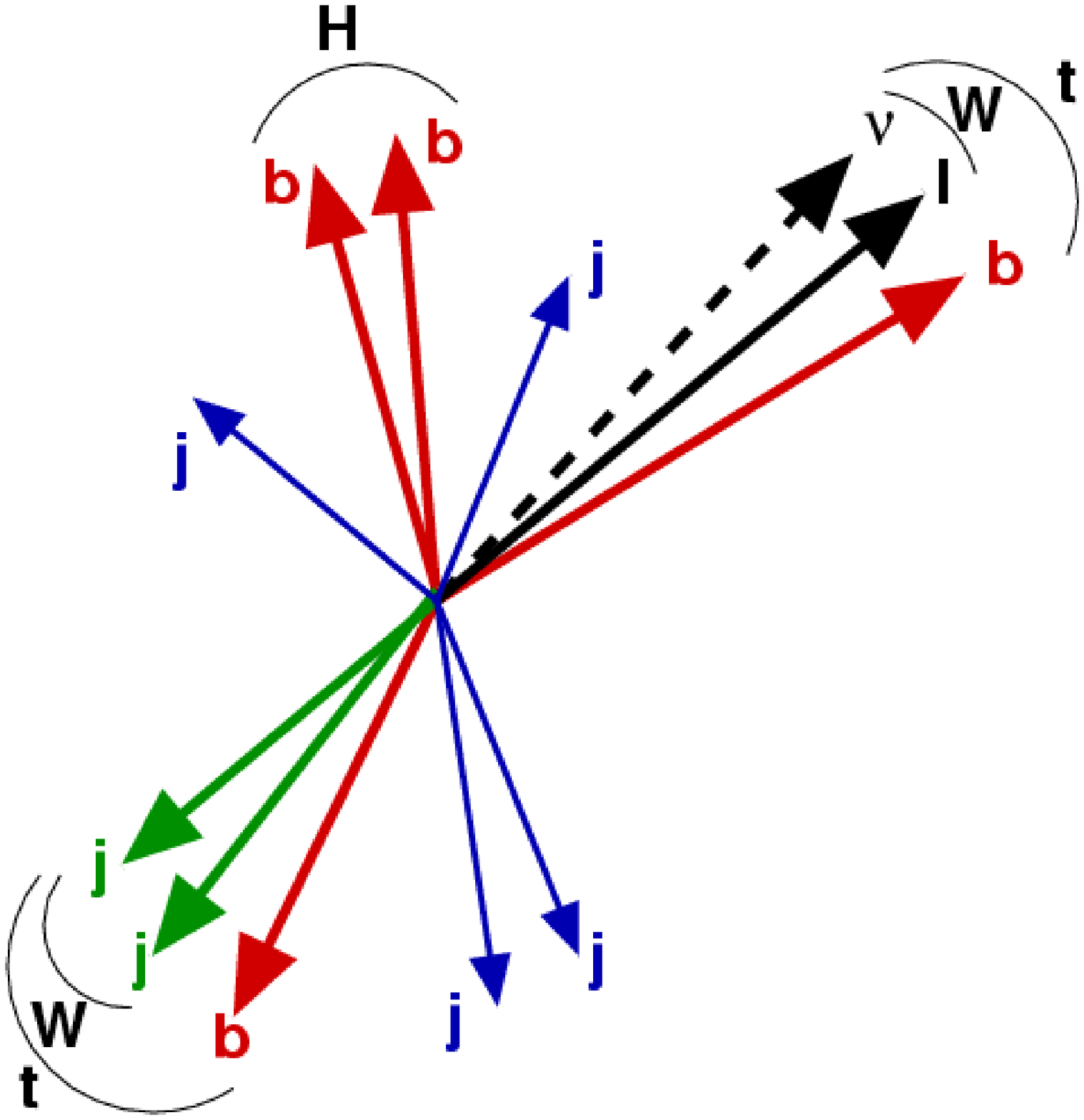}
\hspace*{5mm}
\includegraphics[scale=0.38]{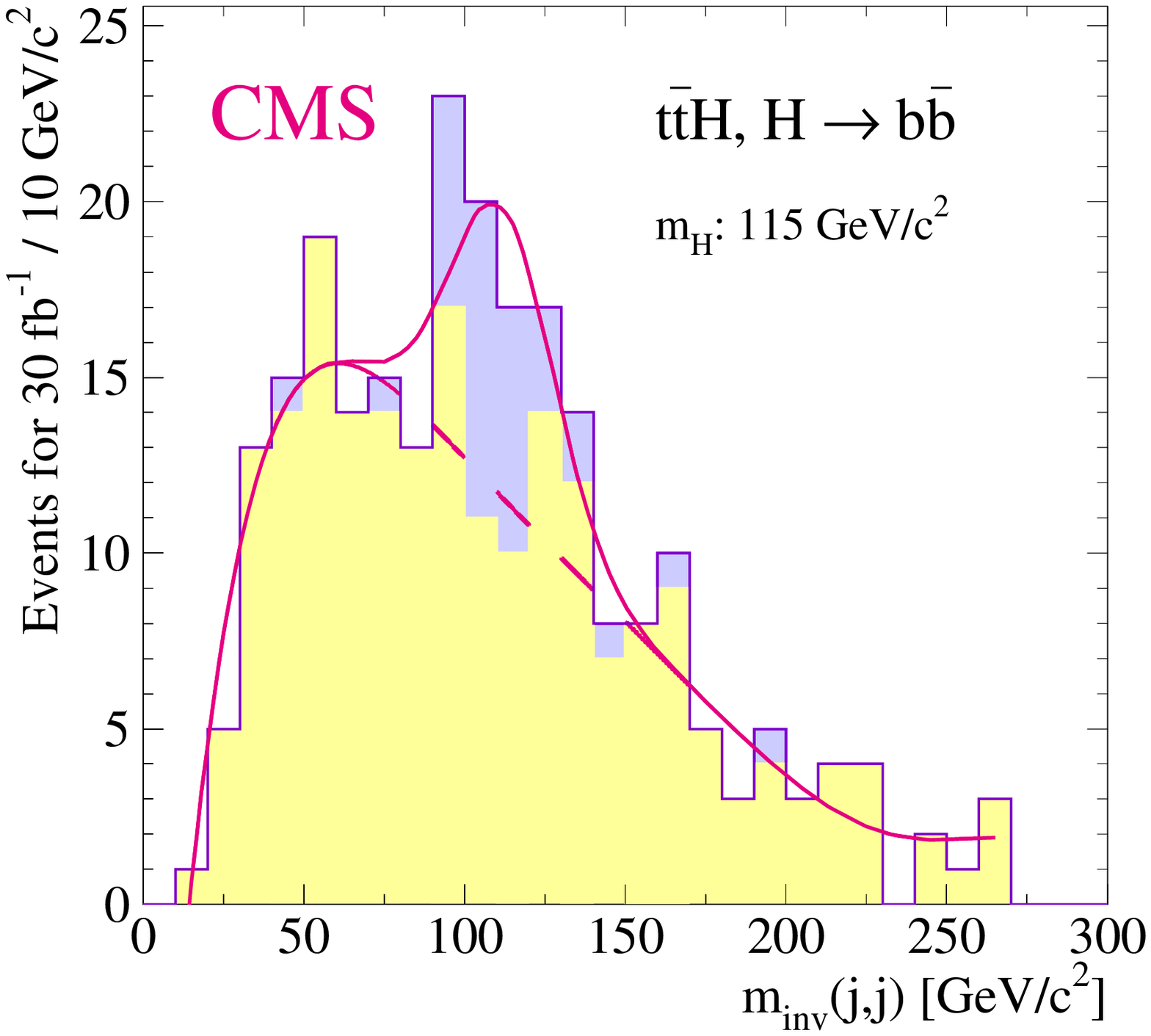}
\caption{Left: schematic of the outgoing particles in a typical
$t\bar{t}H,H\to\bb$ event at LHC~\protect\cite{Cammin:2004sz}.  Right:
early CMS study expectations for a $\bb$ mass peak in such events, for
$M_H=115$~GeV~\protect\cite{Drollinger:2001ym,Abdullin:2005yn}.}
\label{fig:ttH-event}
\end{figure}

Let's begin by discussing a very complex channel, top quark associated
production at low mass, $t\bar{t}H,H\to\bb$.  This was studied early
on in the ATLAS TDR~\cite{ATLAS_TDR} and in various obscure CMS notes,
and found to be a sure-fire way to find a light Higgs.
Fig.~\ref{fig:ttH-event} shows a schematic of such an event, with
multiple $b$ jets from both top quarks and the Higgs, at least one
lepton from a $W$ for triggering, and possibly extra soft jets from
QCD radiation.  The schematic is a bit fanciful in the neatness of
separation of the decay products, but is useful to get an idea of
what's going on.

These early
studies~\cite{ATLAS_TDR,Richter-Was:1999sa,Drollinger:2001ym,Abdullin:2005yn}
were too ambitious, however.  The backgrounds to this signal are
$\tops\bb$ and $t\bar{t}jj$\footnote{Non-$b$ jets can fake $b$ jets
with a probability of about $1\%$ or a little less.} production, pure
QCD processes.  The extra ($b$) jets must be fairly energetic, or
hard, because the signal is a 100+~GeV-mass object which decays to
essentially massless objects.  Despite this being a known
problem~\cite{Giele:1990vh}, these backgrounds were calculated using
the soft/collinear approximation for extra jet emission implemented in
standard Monte Carlo tools such as {\sc pythia} or {\sc herwig}.  This
greatly underestimated the backgrounds.

\begin{figure}[ht!]
\includegraphics[scale=0.48]{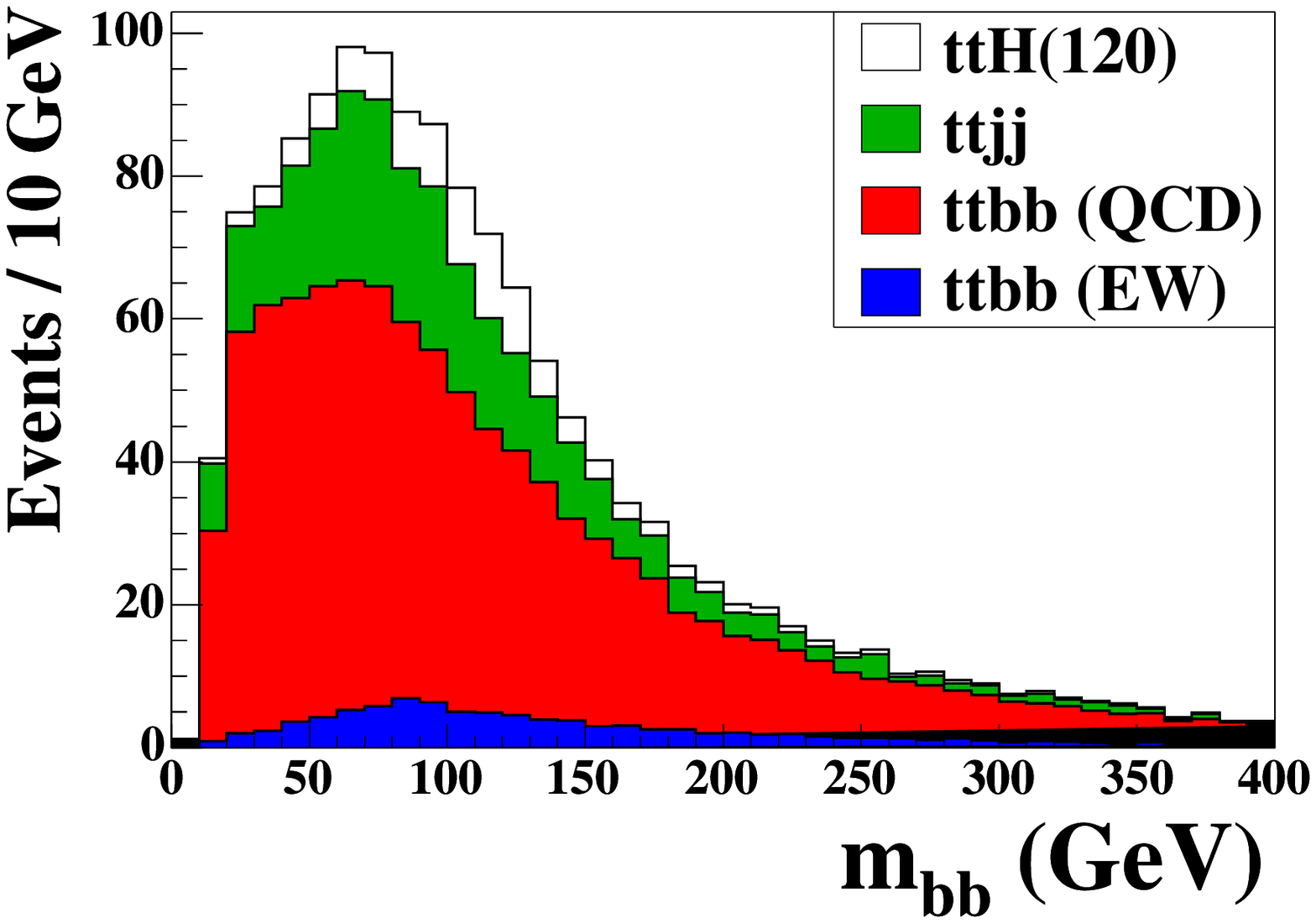}
\includegraphics[scale=0.38]{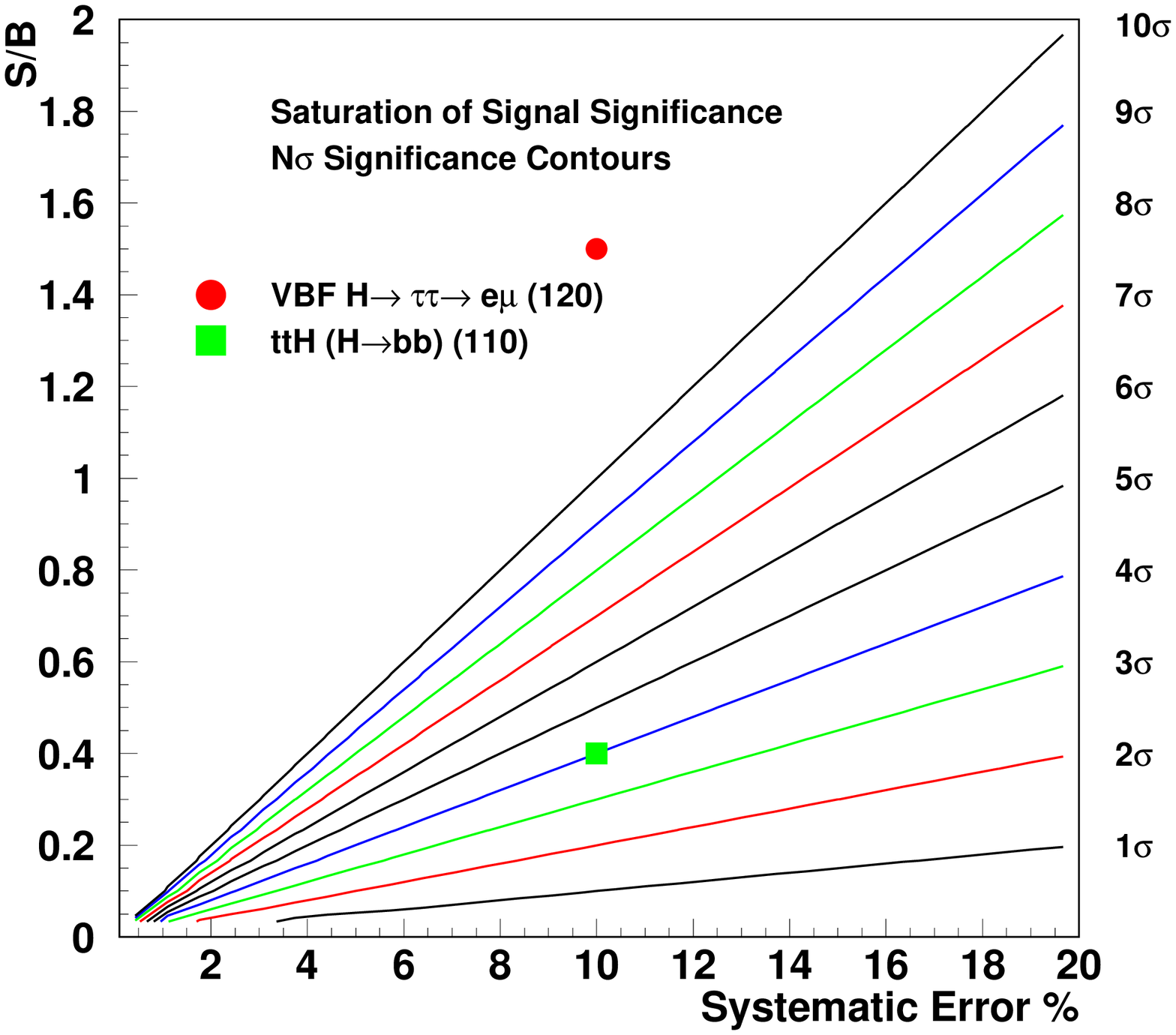}
\vspace*{-6mm}
\caption{Left: results of a more up-to-date ATLAS studying of 
$t\bar{t}H,H\to\bb$ production at LHC, for 30~fb$^{-1}$ of data and a
Higgs mass of $M_H=120$~GeV~\protect\cite{Cammin:2004sz}.  The QCD
backgrounds were calculated with exact matrix elements rather than in
the soft/collinear approximation.  Right: maximum achievable signal
significance for two LHC Higgs channels as a function of $S/B$ and
shape systematic uncertainty $\triangle$~\protect\cite{Cranmer}, as
discussed in the text.}
\label{fig:ATLAS-ttH}
\end{figure}

The left panel of Fig.~\ref{fig:ATLAS-ttH} shows the results of a
repeated study by ATLAS using a proper background
calculation~\cite{Cammin:2004sz}.  (Recent CMS studies found similar
results, and the new CMS TDR~\cite{CMS_TDR} does not even bother to
discuss this channel.)  There is no longer any clearly-visible mass
peak, and S/B is now about 1/6, much poorer.  While the figure
reflects only 1/10 of the expected total integrated luminosity at LHC,
statistics is not the problem.  Rather, it is systematic: uncertainty
on the exact shape of the QCD backgrounds.

Therein lies the sleeping dragon.  Now is a good time to explain how
systematic errors may enter our estimate of signal significance.
Our simple formula is modified:
\bq\label{eq:error}
\frac{S}{\sqrt{B}}
\;\; \to \;\;
\frac{S}{\sqrt{B(1+B\triangle^2)}}
\;\; \stackrel{{\cal L}\to\infty} \longrightarrow \;\;
\frac{S/B}{\triangle}
\eq
where $\triangle$ is the shape uncertainty in the background, a kind
of normalization uncertainty.  In the limit of infinite data, if $S/B$
is fixed (which it is), signal significance saturates.  The only way
around this is to perform higher-order calculations of the background
to reduce $\triangle$ (and hope you understand the residual
theoretical uncertainties).  The right panel of
Fig.~\ref{fig:ATLAS-ttH} shows the spectrum of
possibilities~\cite{Cranmer}.  For the known $10\%$ QCD shape
systematic for $t\bar{t}H$, even an infinite amount of data would
never be able to grant us more than about a $3\sigma$ significance.
This could still potentially be useful for a coupling measurement,
albeit poorly, but will not be a discovery channel unless higher-order
QCD calculations can improve the situation.  Calculating even just
$\tops\bb$ at NLO is currently beyond the state of the art, but is
likely to become feasible within a few years.

While I don't discuss it here, top quark associated Higgs production
does show some promise for the rare Higgs decays to photons.  Photons
are very clean, well-measured, and the detectors have good rejection
against QCD jet fakes.  The final word probably hasn't been written on
this, but the CMS TDR~\cite{CMS_TDR} does have updated simulation
results which the interested student may read up on.

%%%%%%%%%%%%%%%%%%%%%%%%%%%%%%%%%%%%%%%%%%%%%%%%%%%%%%%%%%%%%%%%%%%%%%%%

\subsubsection{$gg\to H\to\gamma\gamma$}
\label{sub:LHC-H2gg}

We've just seen that QCD can be a really annoying problem for Higgs
hunting at LHC.  A logical alternative for a low-mass Higgs is to look
for its rare decays to EW objects, \eg photons.  The BR is at about
the two per-mille level for a light Higgs, $110\lesssim M_H\lesssim
140$~GeV.  The LHC will certainly produce enough Higgses, but what are
the backgrounds like?

It turns out that the loop-induced QCD process $gg\to\gamma\gamma$ is
a non-trivial contribution, but we also have to worry about single and
double jet fakes from QCD $j\gamma$ and $jj$ production.  This occurs
when a leading $\pi^0$ from jet fragmentation goes to photons,
depositing most of the energy in the EM calorimeter, thereby looking
like a real photon.  Fortunately, because photons and jets are
massless, the invariant mass distribution obeys a very linear
$1/m_{\gamma\gamma}$ falloff in our region of interest.  The
experiments can in that case normalize the background very precisely
from the sidebands, where we know there is no Higgs signal.  Shape
systematics are not much of a concern, thus avoiding the pitfalls of
the $t\bar{t}H,H\to\bb$ case.

\begin{figure}[hb!]
\includegraphics[scale=0.85,angle=270]{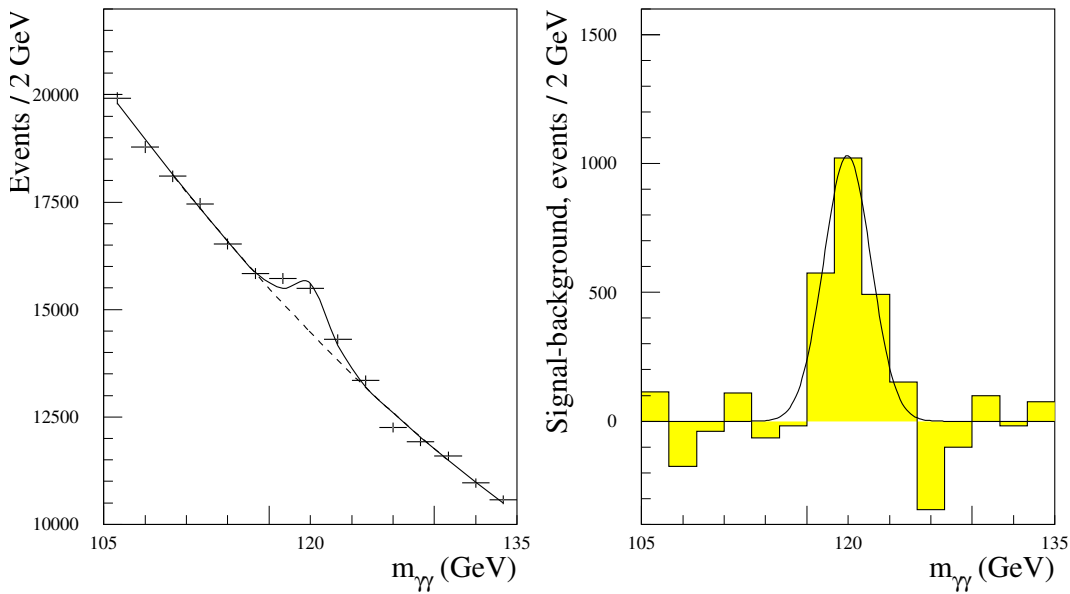}
\caption{ATLAS simulation of $gg\to H\to\gamma\gamma$ at LHC for
$M_H=120$~GeV and 30~fb$^{-1}$ of data~\protect\cite{ATLAS_TDR}.  The
right panel is the mass distribution after background subtraction,
normalized from sidebands.}
\label{fig:ATLAS-H2gg}
\end{figure}

Fig.~\ref{fig:ATLAS-H2gg} shows the results of an ATLAS study for this
channel using 30~fb$^{-1}$ of data~\cite{ATLAS_TDR}, 1/10 of the LHC
run program or 3 years at low-luminosity running.  The exact
expectations are still uncertain, mostly due to an ongoing factor of
two uncertainty in the fake jet rejection efficiency.  A conservative
estimate shows that this channel isn't likely to be the first
discovery mode, but would be crucial for measuring the Higgs mass
precisely at low $M_H$, to about $1\%$~\cite{ATLAS_TDR,CMS_TDR}.
Photon energy calibration nonlinearity in the detector may be an issue
for the ultimate precision, but is generally regarded as minor.  We'll
come back to this point in Chapter~\ref{sec:meas} on Higgs property
measurements.

While I focus here on the SM, keep in mind that because
$H\to\gamma\gamma$ is a rare decay, it can be very sensitive to new
physics.  Recall that the coupling is induced via both top quark and
$W$ loops which mostly cancel.  Depending on how the new physics
alters couplings, or what new particles appear in the loop, the
partial width could be greatly suppressed or enhanced.  (Anticipating
Chapter~\ref{sec:BSM}, the interested student could peruse
Ref.~\cite{Kane:1995ek} and references therein to see how this can
happen in supersymmetry.)

%

%%%%%%%%%%%%%%%%%%%%%%%%%%%%%%%%%%%%%%%%%%%%%%%%%%%%%%%%%%%%%%%%%%%%%%%%

\subsubsection{Weak boson fusion Higgs production}
\label{sub:WBF}

\begin{figure}[hb!]
\includegraphics[scale=1.1]{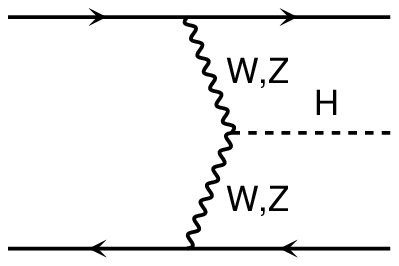}
\hspace*{10mm}
\includegraphics[scale=0.45]{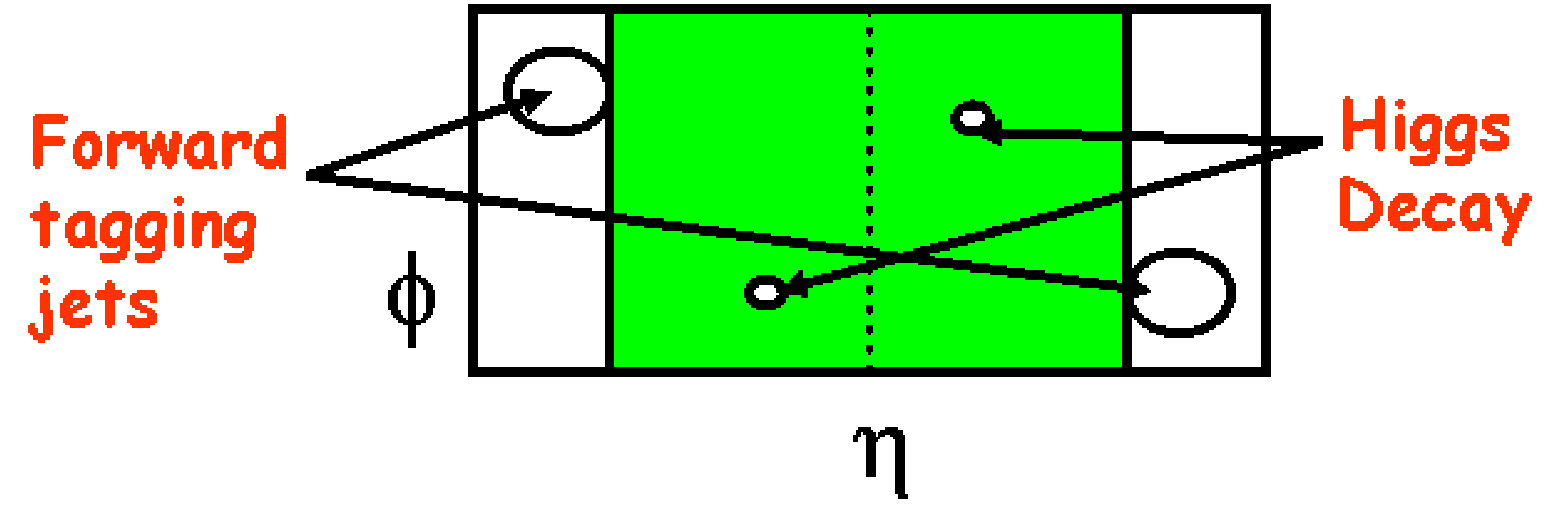}
\caption{WBF Higgs production Feynman diagram and lego plot schematic
of a typical event.}
\label{fig:WBF}
\end{figure}

Let's explore this other production mechanism I said isn't accessible
at Tevatron, weak boson fusion (WBF).  It was long ignored for LHC
light Higgs phenomenology because its rate is about an order of
magnitude smaller than $gg\to H$ there.  However, it has quite
distinctive kinemattics and QCD properties that make it easy to
suppress backgrounds, for all Higgs decay channels.  The process
itself is described by an incoming pair of quark partons which brem a
pair of weak gauge bosons, which fuse to produce a Higgs; see
Fig.~\ref{fig:WBF}.

\begin{figure}[hb!]
\includegraphics[scale=0.27]{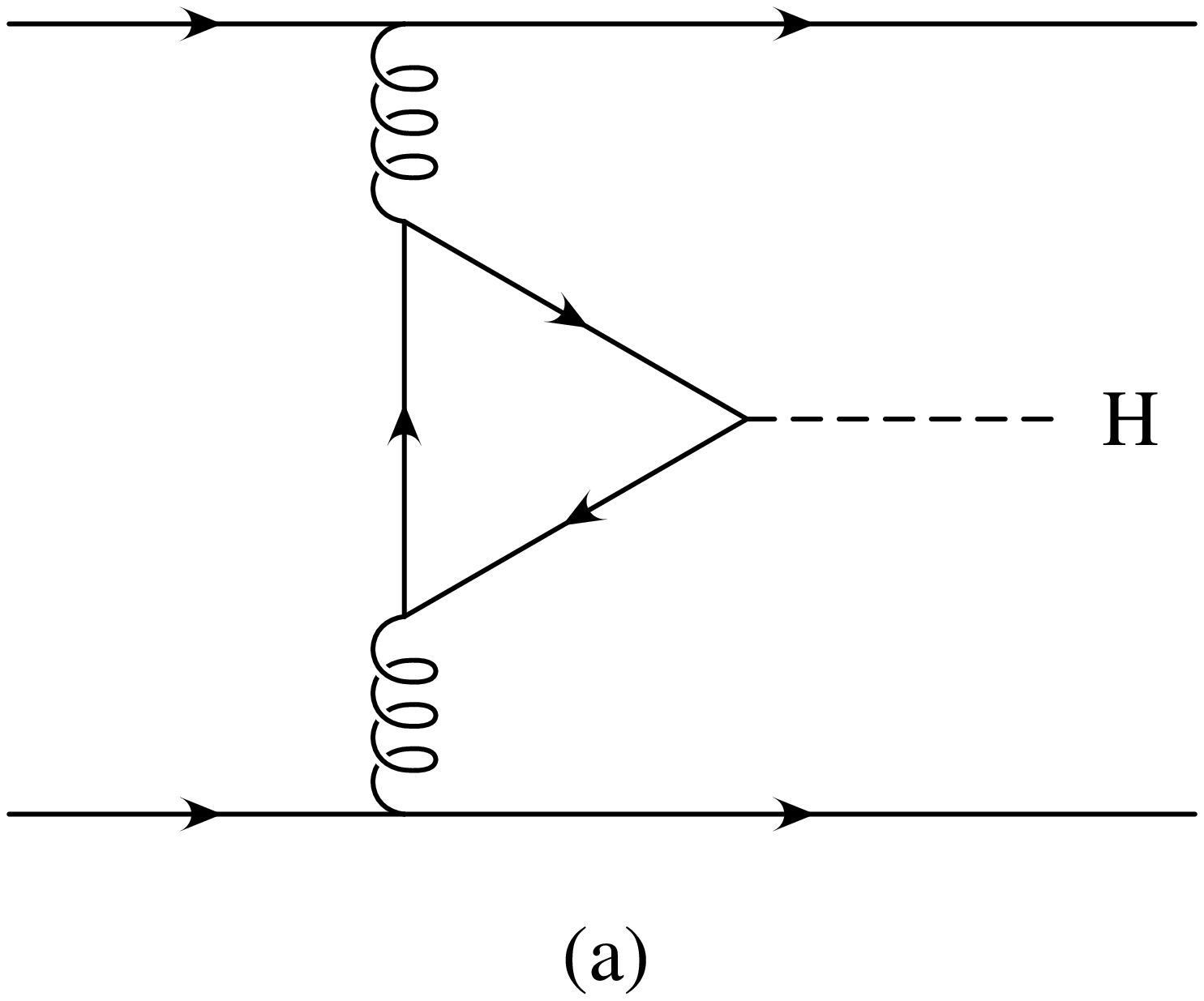}
\hspace*{3mm}
\includegraphics[scale=0.27]{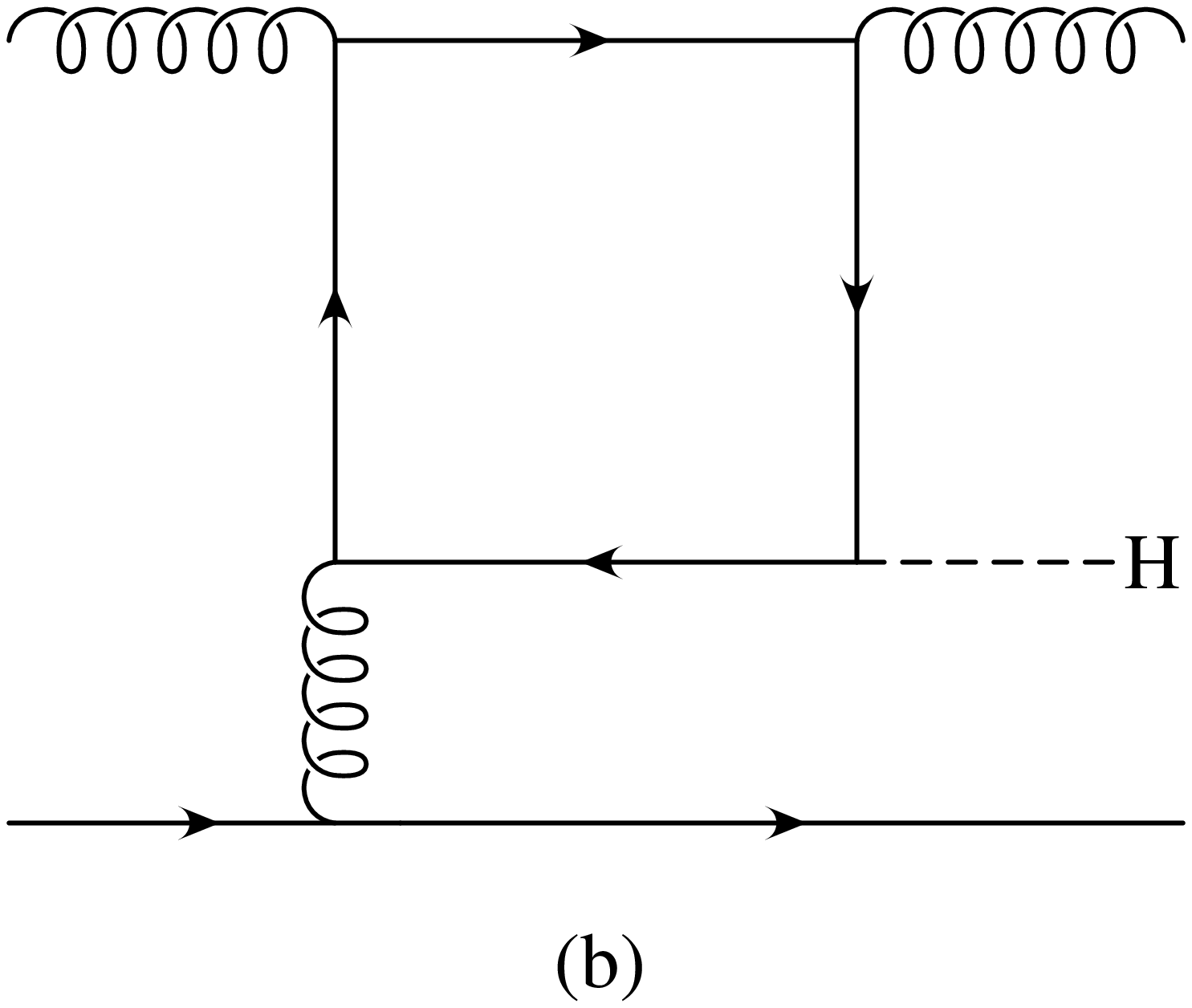}
\hspace*{3mm}
\includegraphics[scale=0.27]{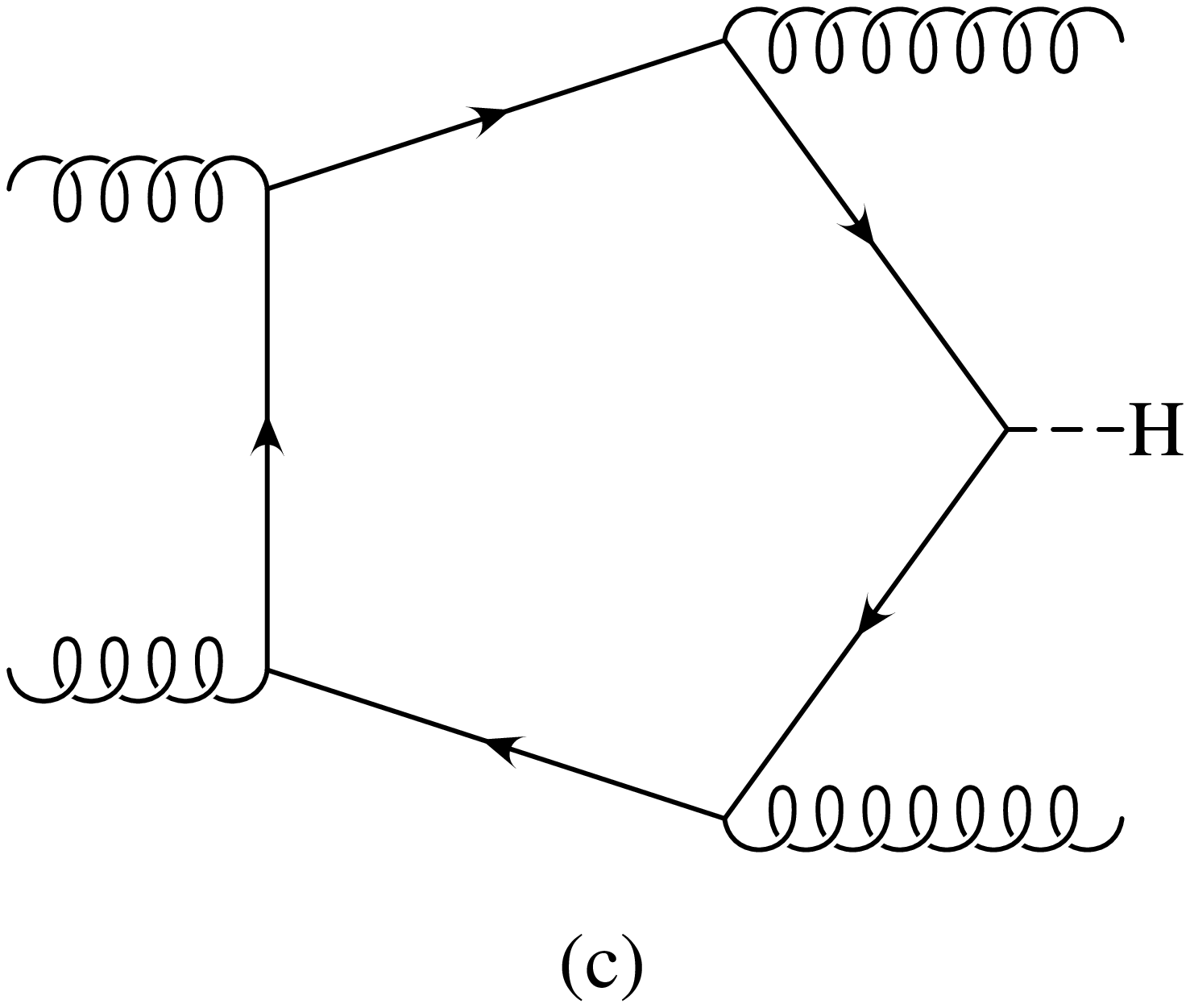}
\vspace*{-3mm}
\caption{Representative Feynman diagrams for gluon fusion Higgs plus
two jets production~\protect\cite{gg_Hjj}.}
\label{fig:GF-Hjj-Feyn}
\end{figure}

The first distinctive characteristic of WBF\footnote{Some
experimentalists refer to this as vector boson fusion (VBF), even
though the vector QCD boson (gluon) process of
Fig.~\ref{fig:GF-Hjj-Feyn} is not included.  This will cause
increasing confusion as time goes by.} is that the quarks scatter with
significant transverse momentum, and will show up as far forward and
backward jets in the hadronic calorimeters of CMS and ATLAS.  The
Higgs boson is produced centrally, however, so its decay products,
regardless of decay mode, typically show up in the central detector
region.  This is shown in the lego plot schematic in the right panel
of Fig.~\ref{fig:WBF}\footnote{The angle $\phi$ is the azimuthal angle
perpendicular to the beam axis.  Pseudorapidity $\eta$ is a
boost-invariant description the polar scattering angle,
$\eta=-\log(\tan\frac{\theta}{2})$.  The lego plot is a Cartesian map
of the finite-resolution detector in these coordinates, as if the
detector had been sliced lengthwise and unrolled.}.

\begin{figure}[ht!]
\includegraphics[scale=1]{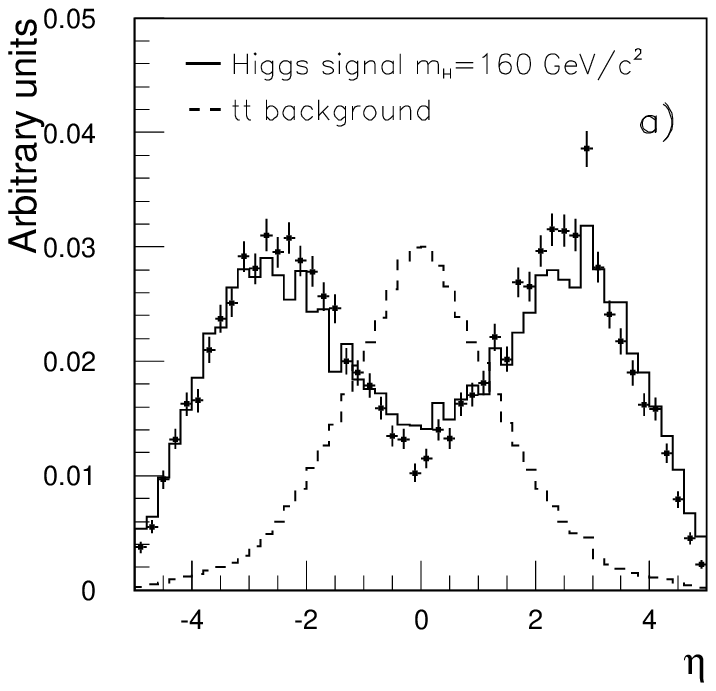}
\hspace*{10mm}
\includegraphics[scale=1]{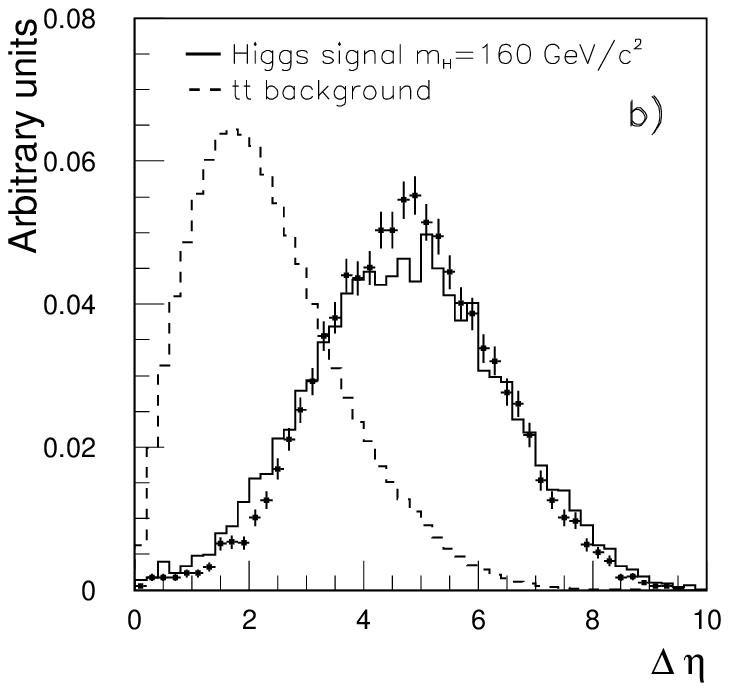}
\vspace*{-3mm}
\caption{Tagging jet rapidity (left) and separation (right) for WBF
Higgs production v. QCD $\tops$
production~\protect\cite{Asai:2004ws}.}
\label{fig:WBF-eta}
\end{figure}

The reason for this scattering behavior comes from the $W$ (or $Z$)
propagator, $1/(Q^2-M^2)$.  For $t$-channel processes, $Q^2$ is
necessarily always negative.  Thus the propagator suppresses the
amplitude least when $Q^2$ is small.  For small $Q^2$, we have
$Q^2=(p_f-p_i)^2\approx E_q^2(1-x)\theta^2$, where $x$ is the fraction
of incoming quark energy the weak boson takes with it, and is small.
Thus $\theta$ prefers to be small, translating into large
pseudorapidity.  One quark will be scattered in the far forward
detector, the other far backward, and the pseudorapidity separation
between them will tend to be large.  We call these ``tagging'' jets.
QCD processes with an extra EW object(s) which mimics a Higgs decay,
on the other hand, have a fundamentally different propagator structure
and prefer larger scattering
angles~\cite{early_WBF-H,Rainwater:1996ud}, including at
NLO~\cite{Campbell:2003hd}.  The differences between the two are shown
in Fig.~\ref{fig:WBF-eta}~\protect\cite{Asai:2004ws}.

The second distinctive characteristic is QCD radiation~\cite{QCDrad}.
Additional jet activity in WBF prefers to be forward of the scattered
quarks.  This is because it occurs via bremsstrahlung off color
charge, which is scattered at small angles, with no connection between
them.  In contrast, QCD production always involves color charge being
exchanged between the incoming partons: acceleration through 180
degrees.  QCD bremsstrahlung thus takes place over large angles,
covering the central region.  Central jet activity can be vetoed,
giving large background suppression~\cite{Barger:1994zq}.  We won't
discuss it further, due to theoretical uncertainties; the interested
student may learn more from Ref.~\cite{Rainwater:1999gg}.

We'll see in the next few subsections that WBF Higgs channels are
extremely powerful even without a central jet (minijet)
veto\footnote{A technical topic outside our present scope: see
Refs.~\cite{Barger:1994zq,Rainwater:1996ud,Rainwater:1999gg,Barger:1990py}
and the literature they reference.}.  Eventually a veto will be used,
after calibration from observing EW v. QCD $Zjj$ production in the
early running of LHC~\cite{Rainwater:1996ud}.  There is however
another lingering theoretical uncertainty, coming from Higgs
production itself!

\begin{figure}[ht!]
\includegraphics[scale=0.5]{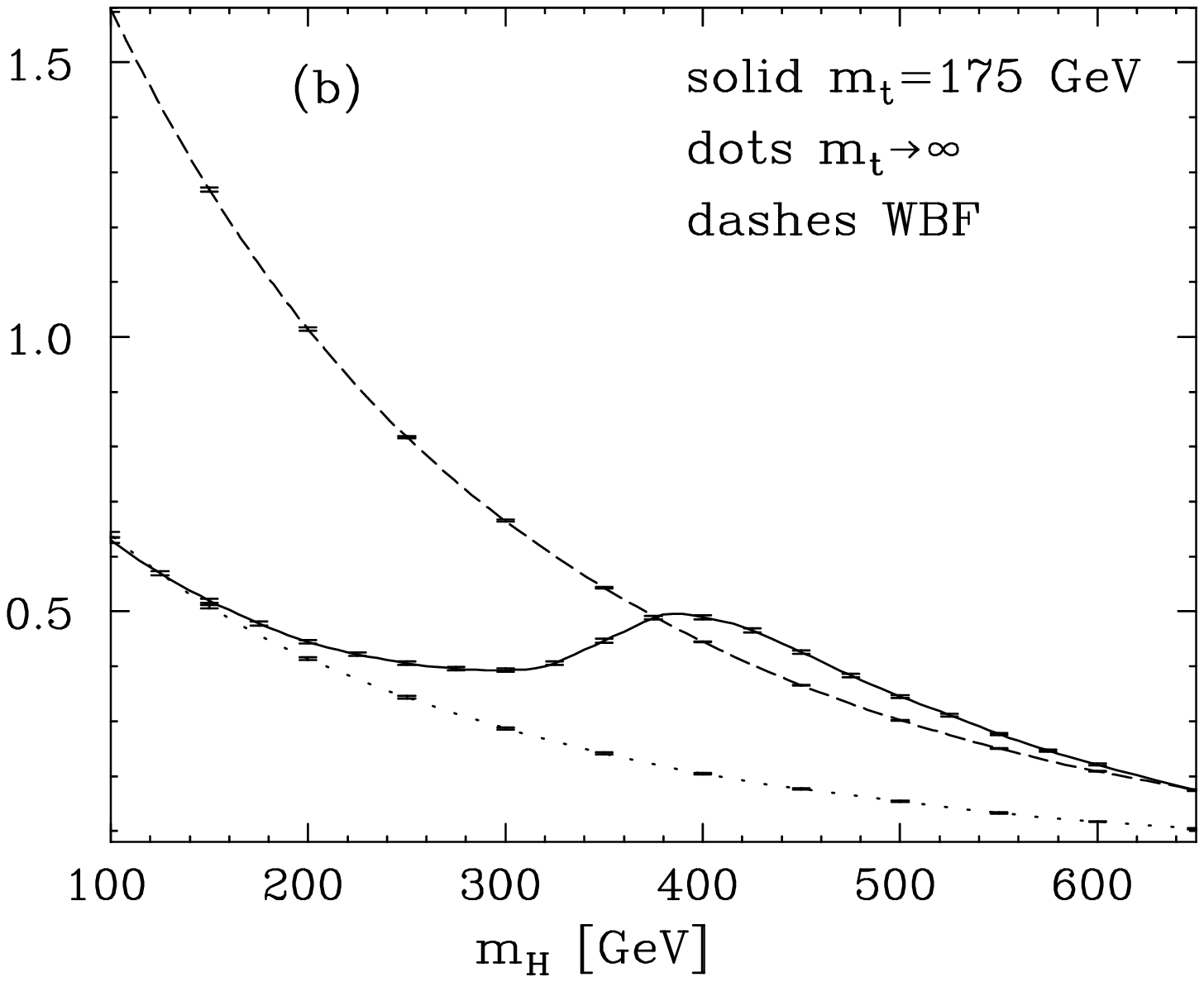}
\hspace*{1cm}
\includegraphics[scale=0.5]{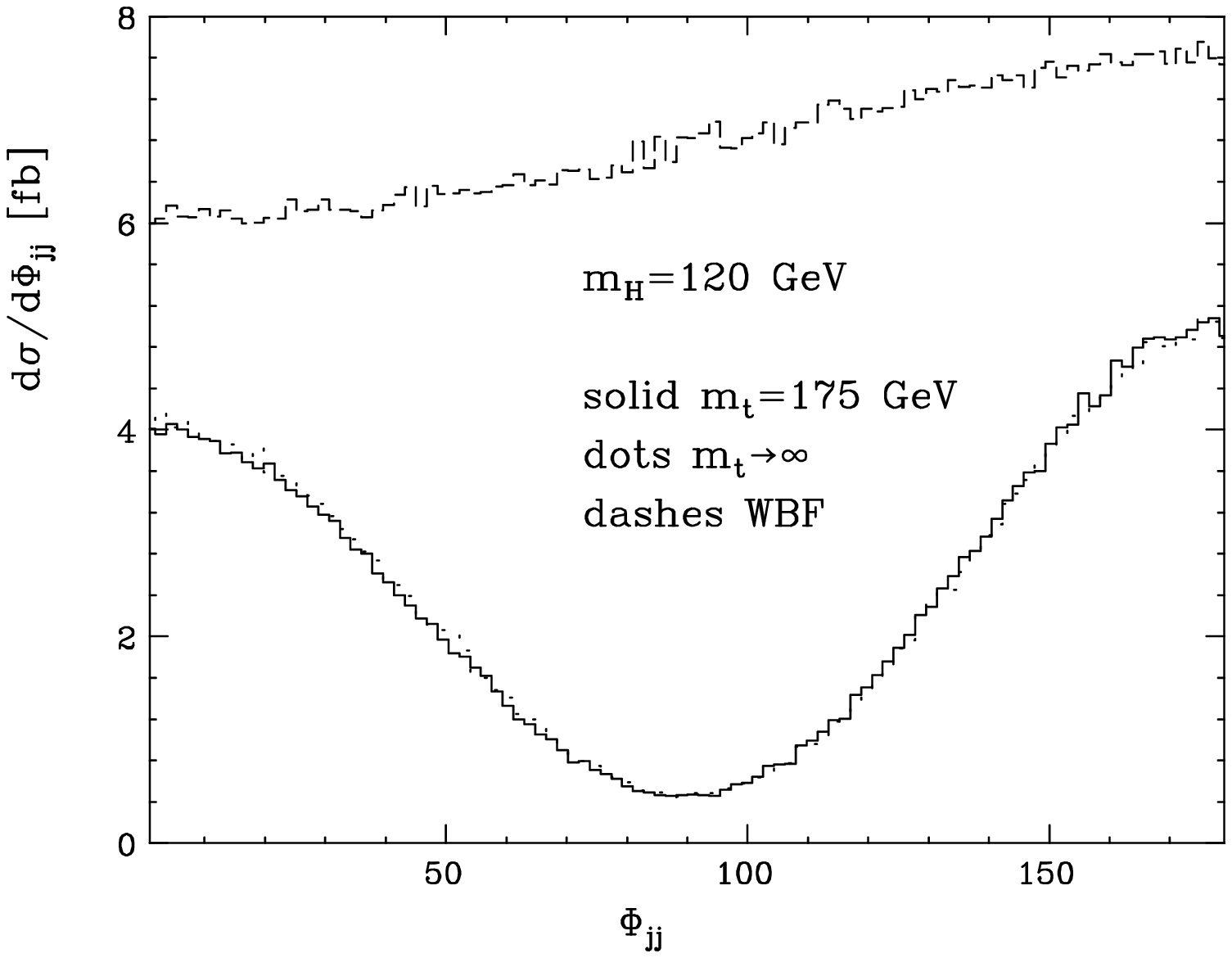}
\caption{Left: WBF and gluon fusion contributions to the forward-tagged
$Hjj$ sample at LHC.  Right: azimuthal angular distributions for the
same two processes, showing distinctive differences.  Figures taken
from Ref.~\protect\cite{gg_Hjj}.}
\label{fig:GF-Hjj-dists}
\end{figure}

QCD Higgs production via loop-induced couplings may itself give rise
to two forward tagging jets, which would then fall into the WBF Higgs
sample~\cite{gg_Hjj}.  Some representative Feynman diagrams for this
process are shown in Fig.~\ref{fig:GF-Hjj-Feyn}.  After imposing
WBF-type kinematic cuts (far forward/backward, well-separated jets,
central Higgs decay products), this contribution to the WBF sample
adds about another third for a light Higgs, or doubles it for a very
heavy Higgs, $M_H\gtrsim 350$~GeV, as shown in the left panel of
Fig.~\ref{fig:GF-Hjj-dists}.  The residual QCD theoretical cross
section uncertainty is about a factor of two, however, and being QCD
it will produce far more central jets, which will be vetoed to reject
QCD backgrounds.  Na\"ively, then, gluon fusion $Hjj$ is an $\sim
10\%$ contribution to WBF, but with a huge uncertainty.

This contribution is a mixed blessing.  It's part of the signal, so
would hasten discovery.  Yet it creates confusion, since at some point
we want to measure couplings, and the WBF and gluon fusion components
arise from different couplings.  Fortunately, there is a difference!
WBF produces an almost-flat distribution in $\phi_{jj}$, the azimuthal
tagging jet separation, but gluon fusion has a suppression at 90
degrees~\cite{gg_Hjj}; cf. right panel of Fig.~\ref{fig:GF-Hjj-dists}.


\vspace*{-5mm}

%%%%%%%%%%%%%%%%%%%%%%%%%%%%%%%%%%%%%%%%%%%%%%%%%%%%%%%%%%%%%%%%%%%%%%%%

\subsubsection{Weak boson fusion $H\to\taus$}
\label{sub:WBF-taus}

Now we know that the WBF signature can strongly suppress QCD
backgrounds because of its unique kinematic characteristics.  We
expect that $H\to\gamma\gamma$ is visible in
WBF~\cite{Rainwater:1997dg,Buscher:2005re,CMS_TDR}, but being a rare
decay in a smaller-rate channel, it's not expected to lead to
discovery.  Rather, it would be a useful additional channel for
couplings measurements.  Let's now instead discuss a decay mode we
haven't yet considered, $H\to\taus$.  This is sub-dominant to
$H\to\bb$ in the light Higgs region, $M_H\lesssim 150$~GeV, but the
backgrounds are more EW than QCD.  We thus have some hope to see it,
whereas $H\to\bb$ remains frustratingly hopeless.

We first have to realize that taus decay to a variety of final states:
\begin{itemize}
\vspace{-2mm}
\item[$\cdot$] $35\%$ $\tau\to\ell\nu_\ell\nu_\tau$,
               ID efficiency $\epsilon_\ell\sim90\%$
\vspace{-2mm}
\item[$\cdot$] $50\%$ $\tau\to h_1\nu_\tau$ ``1-prong'' hadronic
               (one charged track), ID efficiency $\epsilon_h\sim 25\%$
\vspace{-2mm}
\item[$\cdot$] $15\%$ $\tau\to h_3\nu_\tau$ ``3-prong'' hadronic
               (three charged tracks), which are thrown away
\end{itemize}
The obvious problem is that with at least two neutrinos escaping, the
Higgs cannot be reconstructed from its decay products.  Or can it?

Let's assume the taus decay collinearly.  This is an excellent
approximation: since 50+~GeV energy taus have far more energy than
their mass, so their decay products are highly collimated.  We then
have two unknowns, $x_+$ and $x_-$, the fractions of tau energy that
the charged particles take with them.  What experiment measures is
missing transverse energy in the $x$ and $y$ directions.  Two unknowns
with two measurements is exactly solvable.  For our system this
gives~\cite{Ellis:1987xu}:
\bq\label{m_taus}
m_{\taus}^2 \; = \; \frac{m_{\ells}^2}{x_+x_-} + 2m_\tau^2
\eq
(an excellent exercise for all students to get a grip on kinematics
and useful tricks at hadron colliders).  An important note is that
this doesn't work for back-to-back taus (the derivation will reveal
why), but WBF Higgses are typically kicked out with about 100~GeV of
$p_T$, so this almost never happens in WBF.  This trick can't be used
in the bulk of $gg\to H$ events because there it is produced mostly at
rest with nearly all taus back-to-back.

\begin{figure}[hb!]
\includegraphics[scale=0.6]{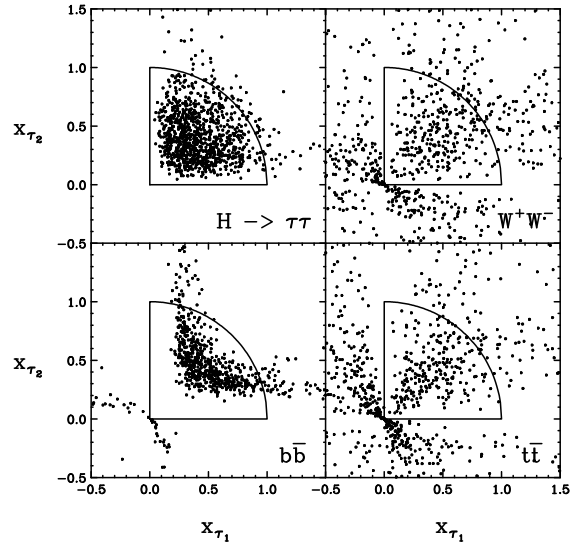}
\vspace*{-5mm}
\caption{Reconstructed $x_+$ v. $x_-$ ($x_1$, $x_2$) for a WBF
$H\to\taus$ signal v. non-$Z$
backgrounds~\protect\cite{Plehn:1999xi}.}
\label{fig:x1x2}
\end{figure}

We need a lepton trigger, so consider two channels: $\taus\to\lpm h$
and $\taus\to\ell^+\ell^{\prime-}$ ($\ell=e,\mu$).  The main
backgrounds are EW and QCD $Zjj$ production (really $Z/\gamma^*$), top
quark pairs, EW \& QCD $WWjj$ and QCD $\bb jj$ production.  But after
reconstruction, the non-$Z$ backgrounds look very different than the
signal in $x_+$--$x_-$ space, as shown in Fig.~\ref{fig:x1x2}.

\begin{figure}[hb!]
\includegraphics[scale=0.9]{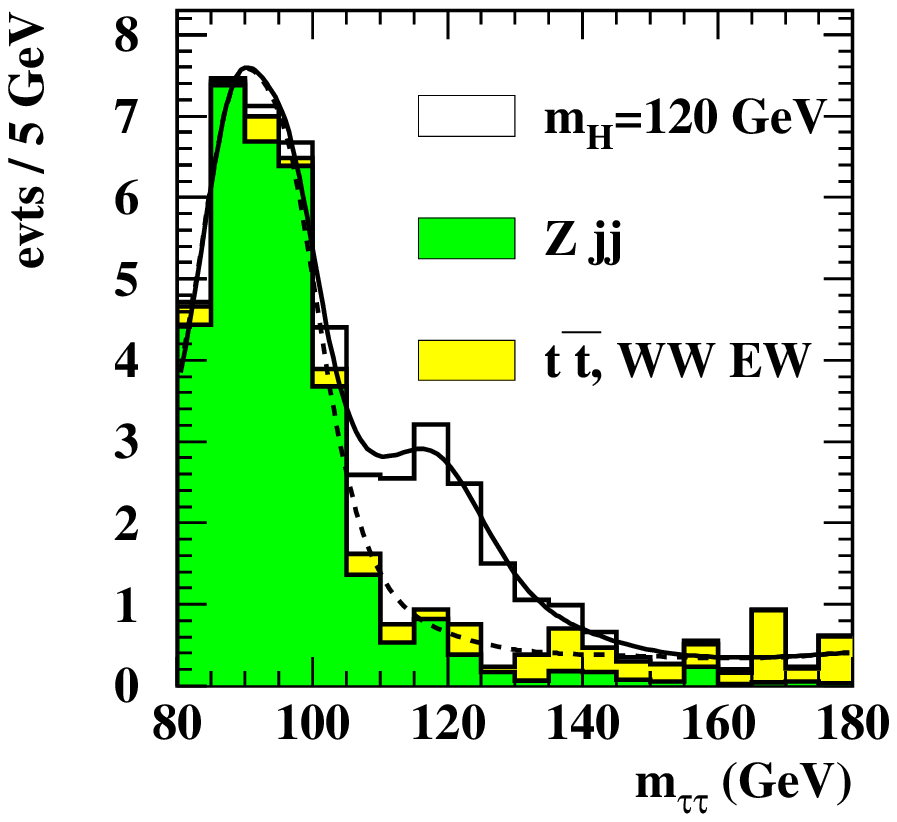}
\includegraphics[scale=0.45]{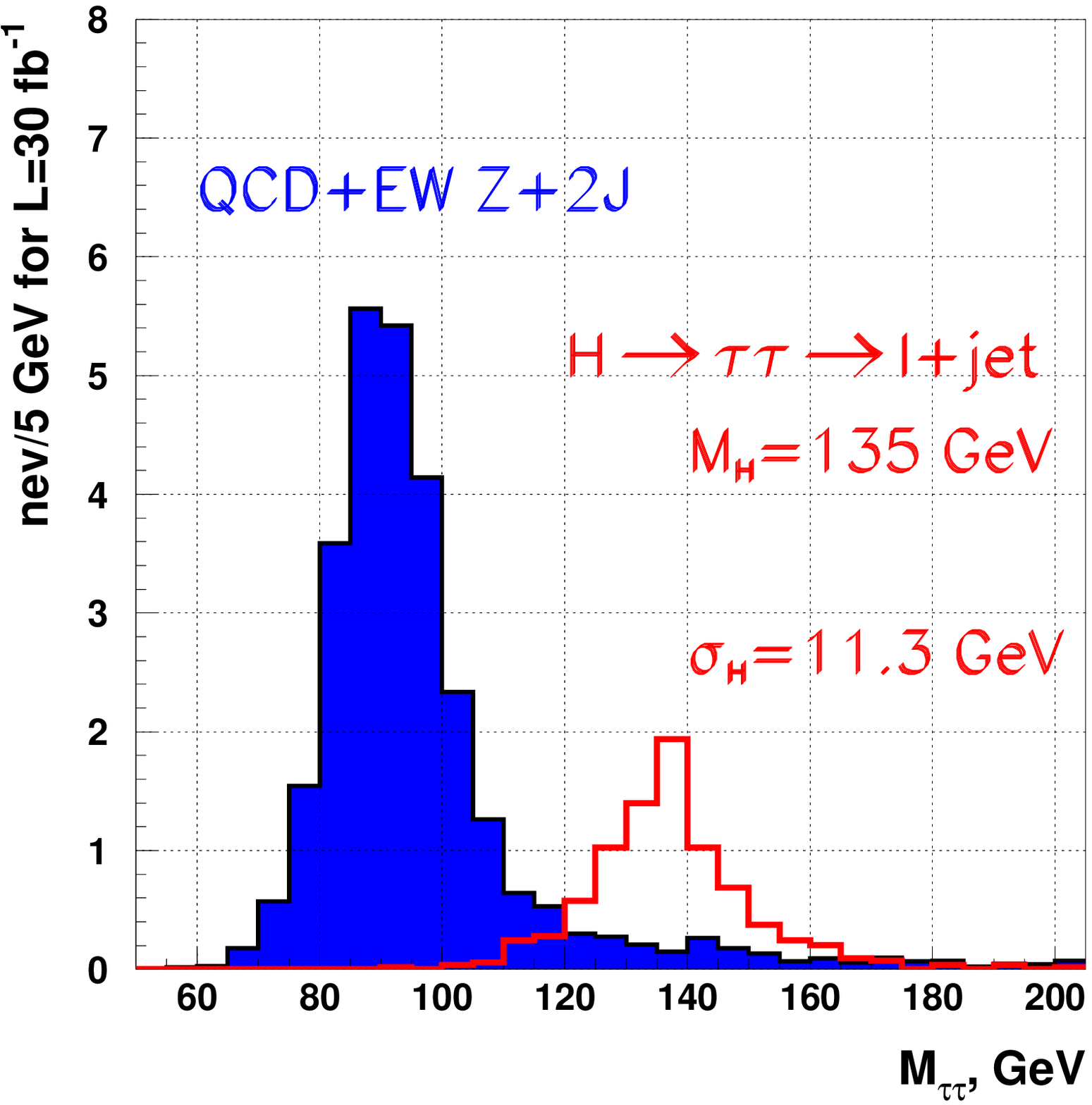}
\vspace*{-3mm}
\caption{ATLAS (left) and CMS (right) simulations of WBF $H\to\taus$
events after 30~fb$^{-1}$ of data at LHC.  The Higgs resonance clearly
stands out from the background.  Figures from
Ref.~\protect\cite{Asai:2004ws}.}
\label{fig:WBF-taus}
\end{figure}

ATLAS and CMS have both studied these channels with full detector
simulation and WBF kinematic cuts, but no minijet veto, and found
extremely promising results~\cite{Asai:2004ws}.
Fig.~\ref{fig:WBF-taus} shows invariant mass distributions for a
reconstructed Higgs in the two different decay channels, assuming only
30~fb$^{-1}$ of data.  The Higgs peak is easily seen above the
backgrounds and away from the $Z$ pole.  Mass resolution is expected
to be a few GeV.

\vspace*{-8mm}

But this joint study by CMS and ATLAS~\cite{Asai:2004ws} is not the
best we can do.  The joint study ignored the minijet veto, for
instance.  While that will assuredly improve the situation further,
we're just not sure precisely how much.  Putting this aside for the
moment, there are yet further tricks to play to improve the situation.

\begin{figure}[hb!]
\includegraphics[scale=0.6]{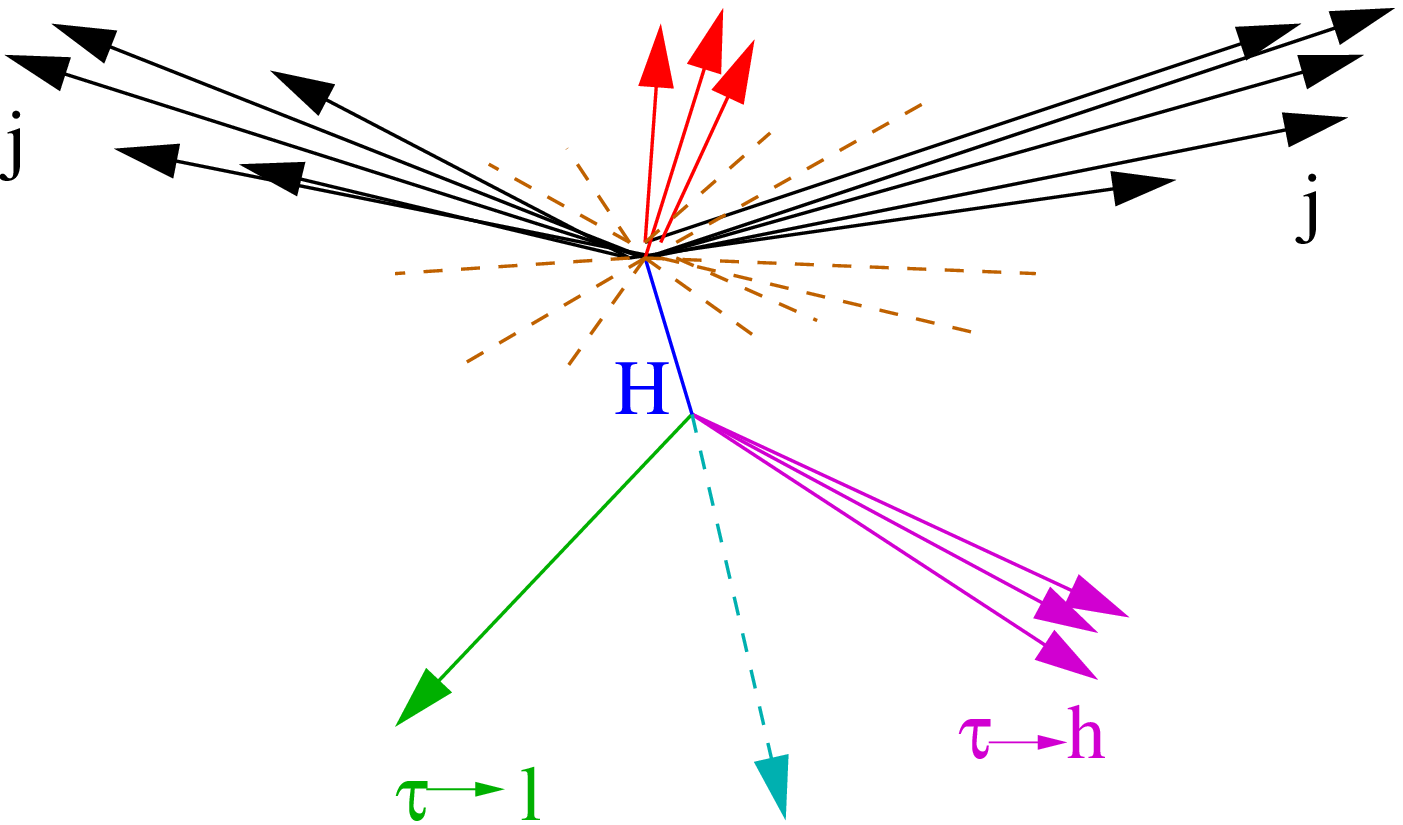}
\hspace*{5mm}
\includegraphics[scale=0.3]{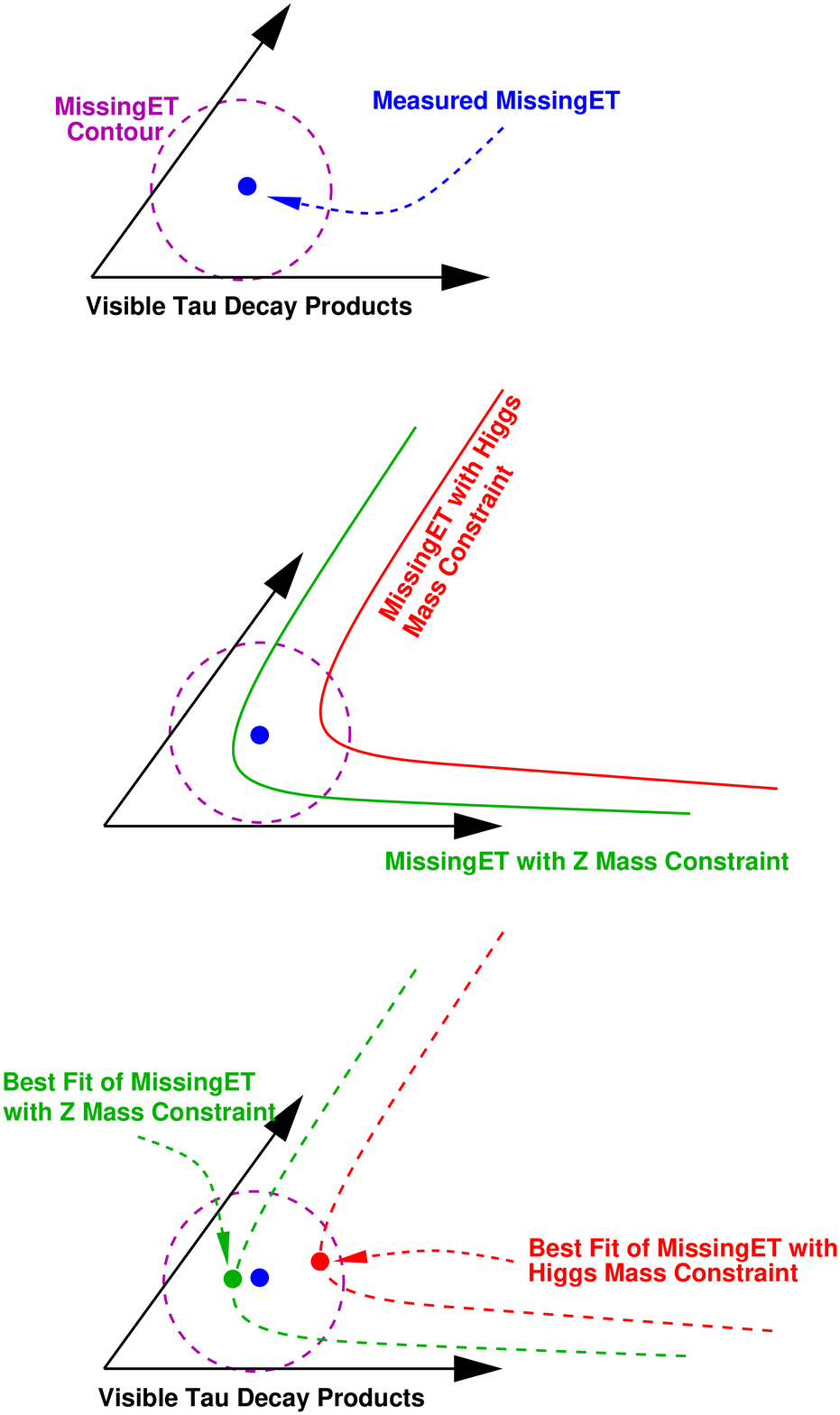}
\vspace*{-2mm}
\caption{Left: schematic azimuthal projection of WBF $H\to\taus$ events
at LHC.  Right: diagram illustrating the $1\sigma$ uncertainty region
(due to jet mismeasurement) of missing $p_T$, and how a $Z$ mass or
Higgs mass hypothesis can be best fit using a $\chi^2$ test.  Figures
from Ref.~\cite{Cranmer-taumod}.}
\label{fig:WBF-tautrick}
\end{figure}

The leading idea zeroes in on the fact that missing transverse
momentum ($\sla{p}_T$) has some uncertainty due to jet energy
mismeasurement (those imperfect detectors).  Using a $\chi^2$ test,
one determines which is more likely: $Z\to\taus$ or $H\to\taus$, {\it
using a fixed Higgs mass constraint}~\cite{Cranmer-taus}.  Examining
the schematics in Fig.~\ref{fig:WBF-tautrick}, we see this is
tantamount to deciding which fit is closer to the center of the
$\sla{p}_T$ uncertainty region.  Early indications are that this
technique would improve $S/B$ by about a factor {\it four}, in
addition to recovering some signal lost using more traditional strict
kinematic cuts on $x_+$ and $x_-$ (recall Fig.~\ref{fig:x1x2}).  This
would approximately halve the data required to discover a light SM
Higgs boson using this channel.  Keep it in mind when we see the
current official discovery expectations in Sec.~\ref{sub:LHC-sum}.
Further improvements might also be expected from neural-net type
analyses, which are coming to the fore now that Tevatron has
demonstrated their viability.

A final word on systematic uncertainties.  Unlike the tortuous case of
$t\bar{t}H,H\to\bb$, we don't have to worry about shape systematics
here.  The dominant background is $Zjj$ production.  We can separately
examine $Z\to ee,\mu\mu$, which produces an extremely sharp, clean
peak, precisely calibrating $Zjj$ production in Monte Carlo.  The only
uncertainty then is tau decay modeling, which is very well understood
from the LEP era.

%

%%%%%%%%%%%%%%%%%%%%%%%%%%%%%%%%%%%%%%%%%%%%%%%%%%%%%%%%%%%%%%%%%%%%%%%%

\subsubsection{Weak boson fusion $H\to\ww$}
\label{sub:WBF-WW}

A natural question to ask is, how well does WBF Higgs hunting work for
$M_H\gtrsim 140$~GeV, where $H\to\ww$ dominates?  We should expect
fairly well, since it's the production process characteristics that
supply most of the background suppression, leaving us only to look for
separated reconstructed mass peaks.

For $H\to\ww$ we'll consider only the dilepton channel, as it has
relatively low backgrounds, while QCD gives a large rate for the other
possible channel, one central lepton plus two central jets (and the
minijet veto will likely not work).  We'll therefore rely on exactly
the same angular correlations and transverse mass variable we
encountered in the Tevatron case~\cite{Rainwater:1999sd}
(cf. Eqs.~\ref{eq:ET-def},\ref{eq:M_T_WW}).  The only critical
distinction is then $e\mu$ v. $ee$, $\mu\mu$ samples, as the latter
have a continuum background ($Z^*/\gamma^*$).  These are not too much
of a concern, however.

Without going too much into detail, I'll simply say that top quarks
are a major background, and they have the largest uncertainty.  The
largest component comes from $t\bar{t}j$ production, where the extra
hard parton is far forward and ID'd as one tagging jet; a $b$ jet from
top decay gives the other tagging jet, and the other $b$ jet is
unobserved.  This background requires care to simulate, because the
soft/collinear approximation in standard codes is no good.  There is
also a significant contribution from single-top production, and
off-shell effects are crucial to simulate, which is not normally an
issue for backgrounds at LHC~\cite{OFS-tops}.  Work is still needed in
this area to be fully prepared for this particular search channel.
Fortunately, we may expect an NLO calculation of $t\bar{t}j$ before
LHC start~\cite{DUW}.

\begin{figure}[hb!]
\includegraphics[scale=0.95]{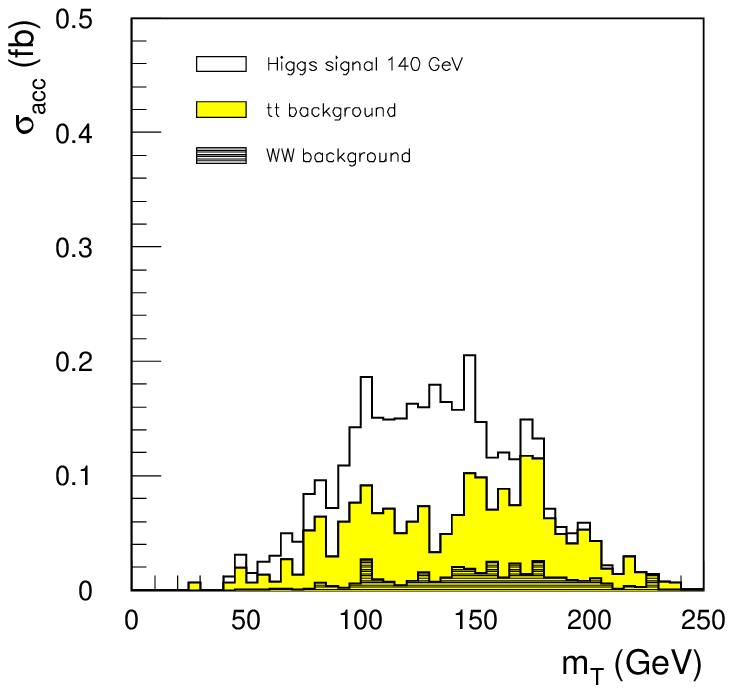}
\includegraphics[scale=0.95]{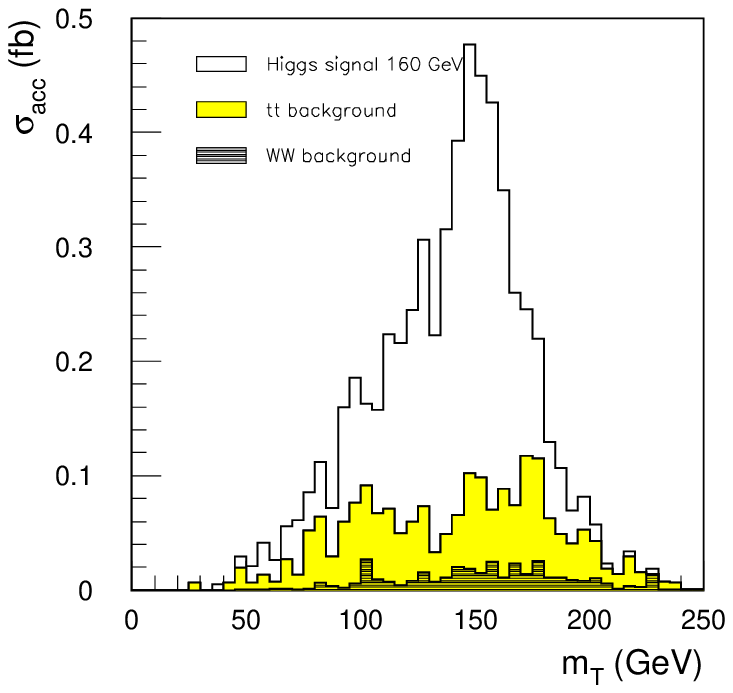}
\vspace*{-3mm}
\caption{ATLAS simulations of WBF $H\to\ww$ events after 30~fb$^{-1}$ 
of data at LHC for $M_H=140$~GeV (left) and 160~GeV (right).  The
Higgs signal clearly stands out from the background in both cases,
although the Jacobian peak is easier to identify closer to threshold.
Figures taken from Ref.~\protect\cite{Asai:2004ws}.}
\label{fig:WBF-WW}
\end{figure}

Fig.~\ref{fig:WBF-WW} shows the results of the same ATLAS/CMS joint
WBF Higgs study for this channel~\cite{Asai:2004ws}.  The results are
extremely positive, with $S/B>1/1$ without a minijet veto over a large
mass range; even for $M_H=120$~GeV, $S/B\sim 1/2$, allowing for Higgs
observation even down to the LEP limit in this channel.  The
transverse mass variable works extremely well for Higgs masses near
$WW$ threshold, and reasonably well for lower masses, where the $W$
bosons are off-shell.

%

%%%%%%%%%%%%%%%%%%%%%%%%%%%%%%%%%%%%%%%%%%%%%%%%%%%%%%%%%%%%%%%%%%%%%%%%

\subsubsection{$t\bar{t}H,H\to\ww$ at higher mass}
\label{sub:ttH-WW}

\begin{figure}[hb!]
\includegraphics[scale=0.25]{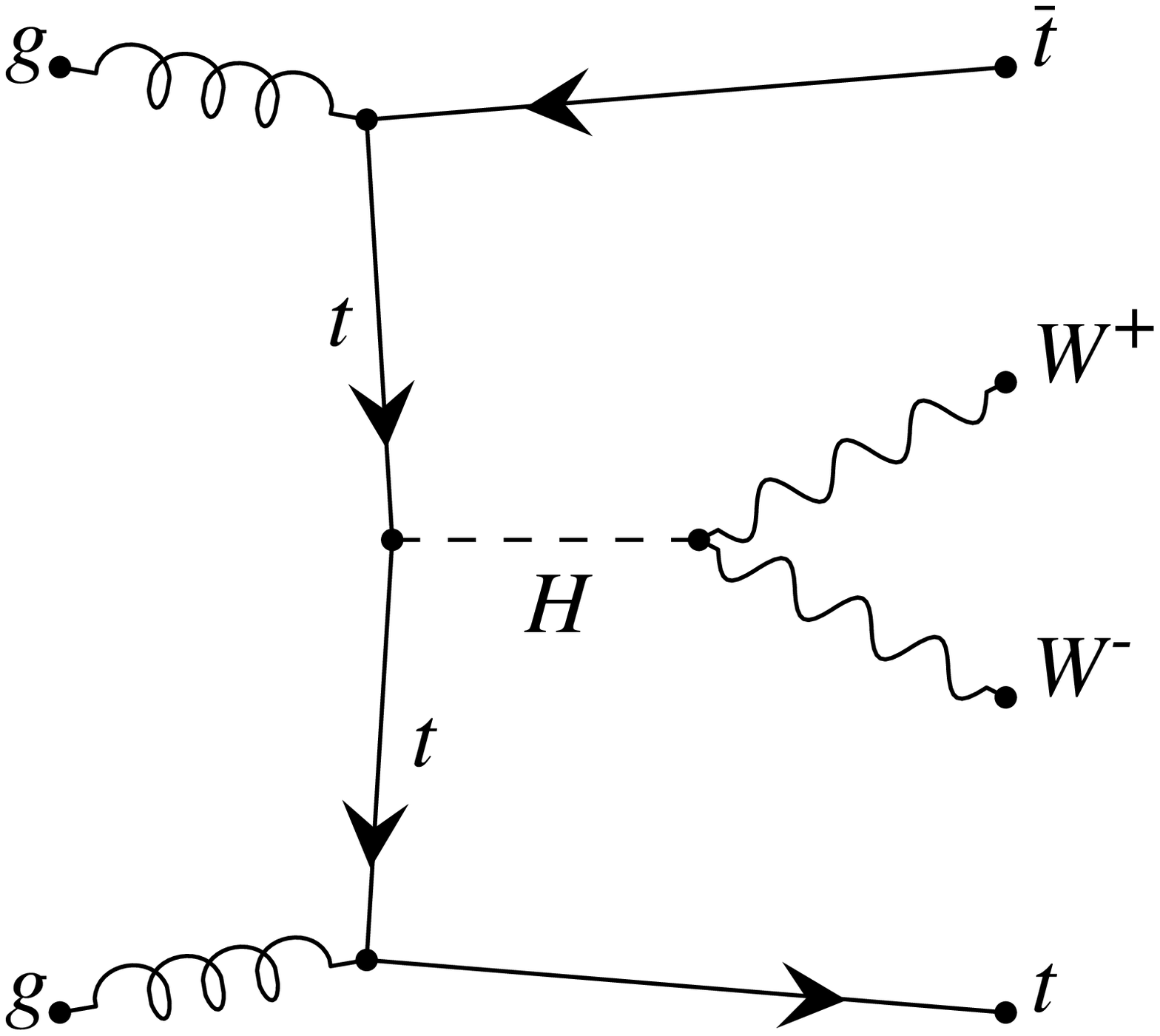}
\hspace*{5mm}
\includegraphics[scale=0.25]{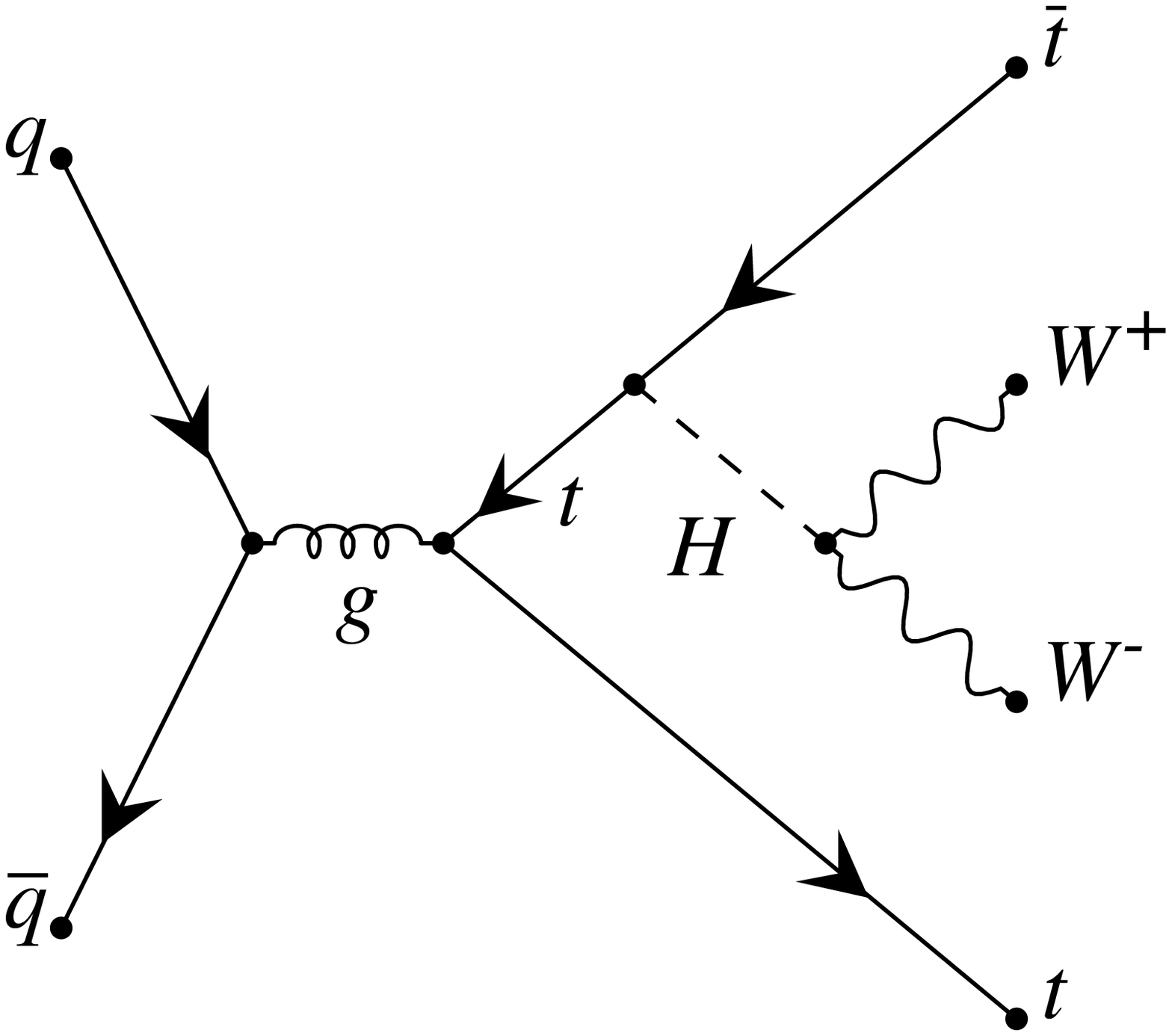}
\vspace*{-8mm}
\caption{Representative Feynman diagrams for $t\bar{t}H,H\to\ww$
production at LHC.}
\label{fig:ttH-WW-Feyn}
\end{figure}

A late entry to the Higgs game at LHC is top quark associated
production, but with Higgs decaying to $W$ bosons.  Representative
Feynman diagrams are shown in Fig.~\ref{fig:ttH-WW-Feyn}.  Obviously
this is intended to apply to larger Higgs masses, but turns out to
work fairly well even below $W$ pair
threshold~\cite{Maltoni:2002jr,ATL-PHYS-2002-019}.  The key is to use
same-sign dilepton and trilepton subsamples.  The backgrounds then
don't come from pure QCD production, rather from mixed QCD-EW top
quark pairs plus $W$, $Z/\gamma^*$, $W^+W^-$, etc.  We would be
especially eager to observe this channel because, if the $HWW$
coupling is measured elsewhere, it provides the only viable direct
measurement of the top quark Yukawa coupling.  More on this in
Chapter~\ref{sec:meas}.

\begin{figure}[hb!]
\includegraphics[scale=0.38]{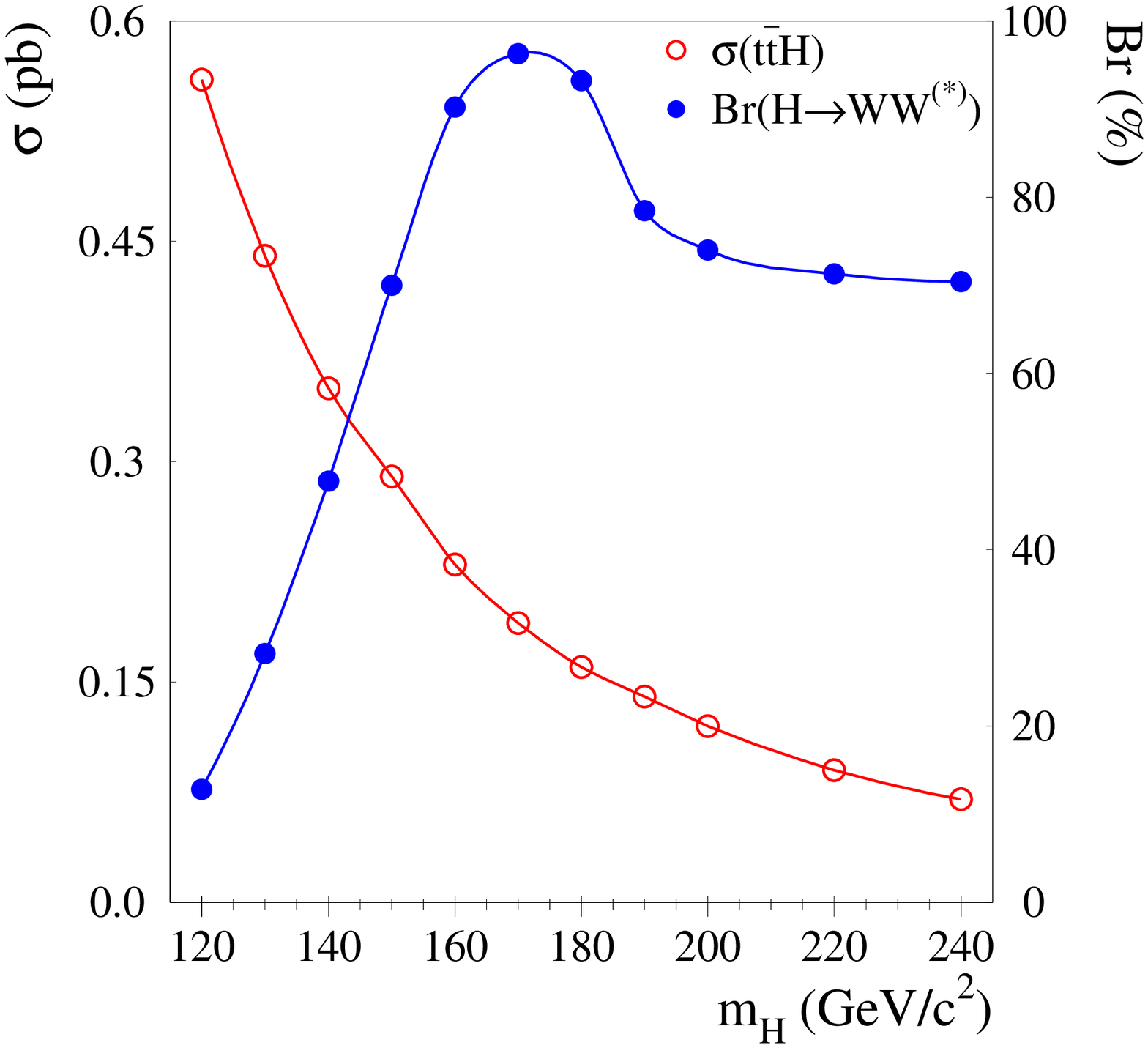}
\hspace*{5mm}
\includegraphics[scale=0.38]{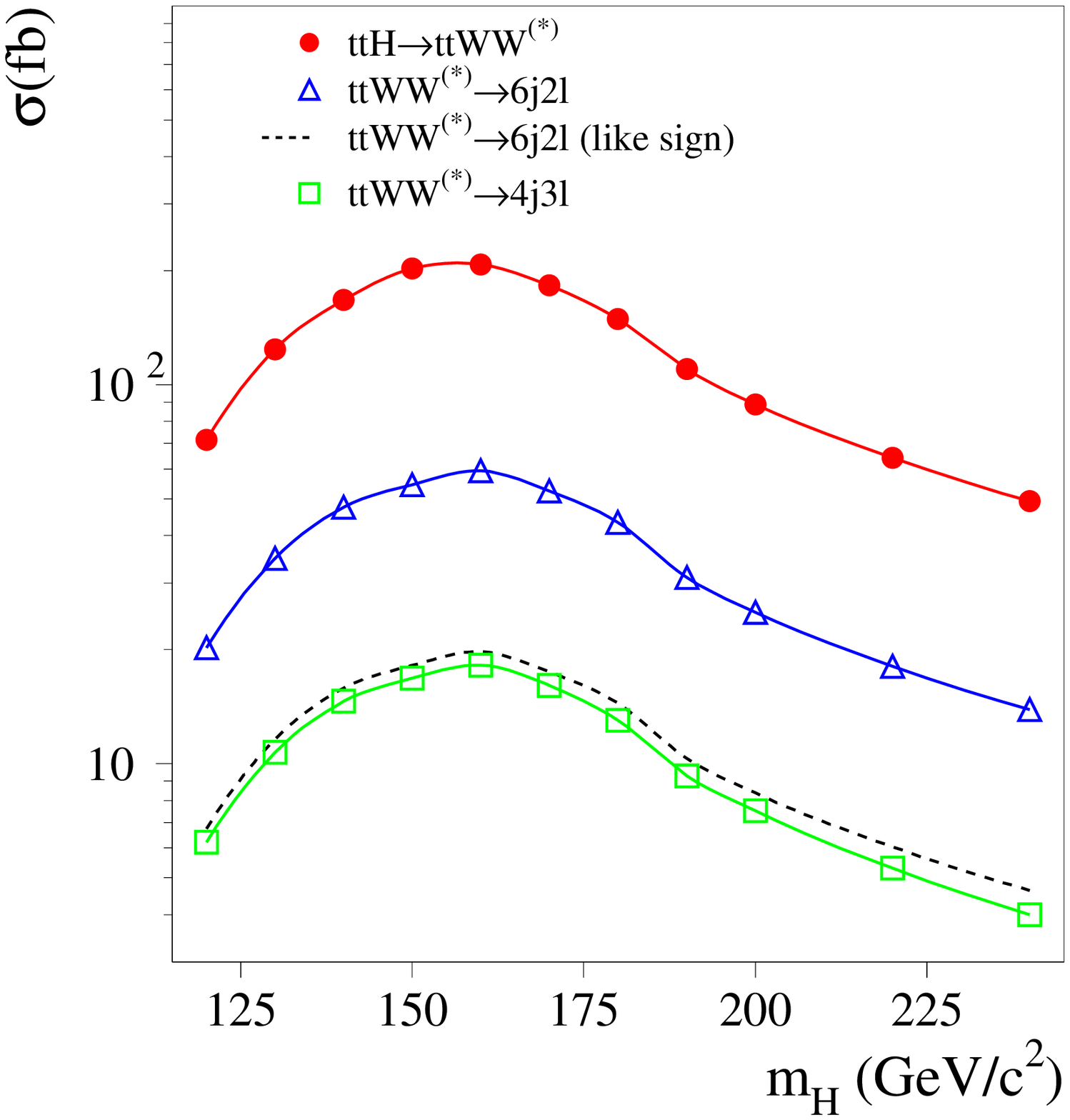}
\vspace*{-3mm}
\caption{Left: $t\bar{t}H$ cross section and BR($H\to\ww$) as a 
function of $M_H$.  Right: the cross section to three different final
states after top quark and Higgs decays.  Figures from
Ref.~\protect\cite{ATL-PHYS-2002-019}.}
\label{fig:ttH-WW-rate}
\end{figure}

\begin{figure}[ht!]
\includegraphics[scale=0.38]{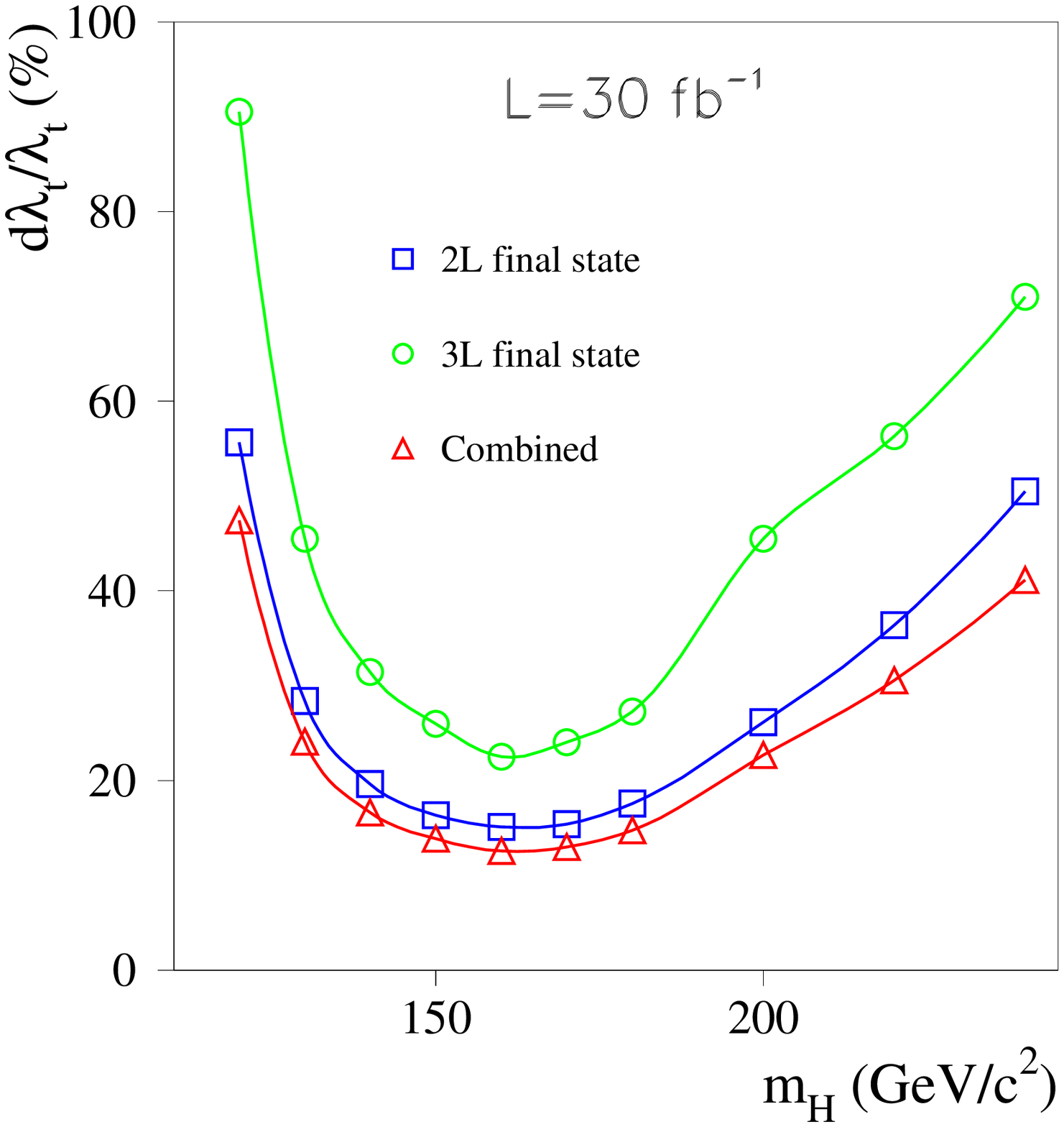}
\includegraphics[scale=0.38]{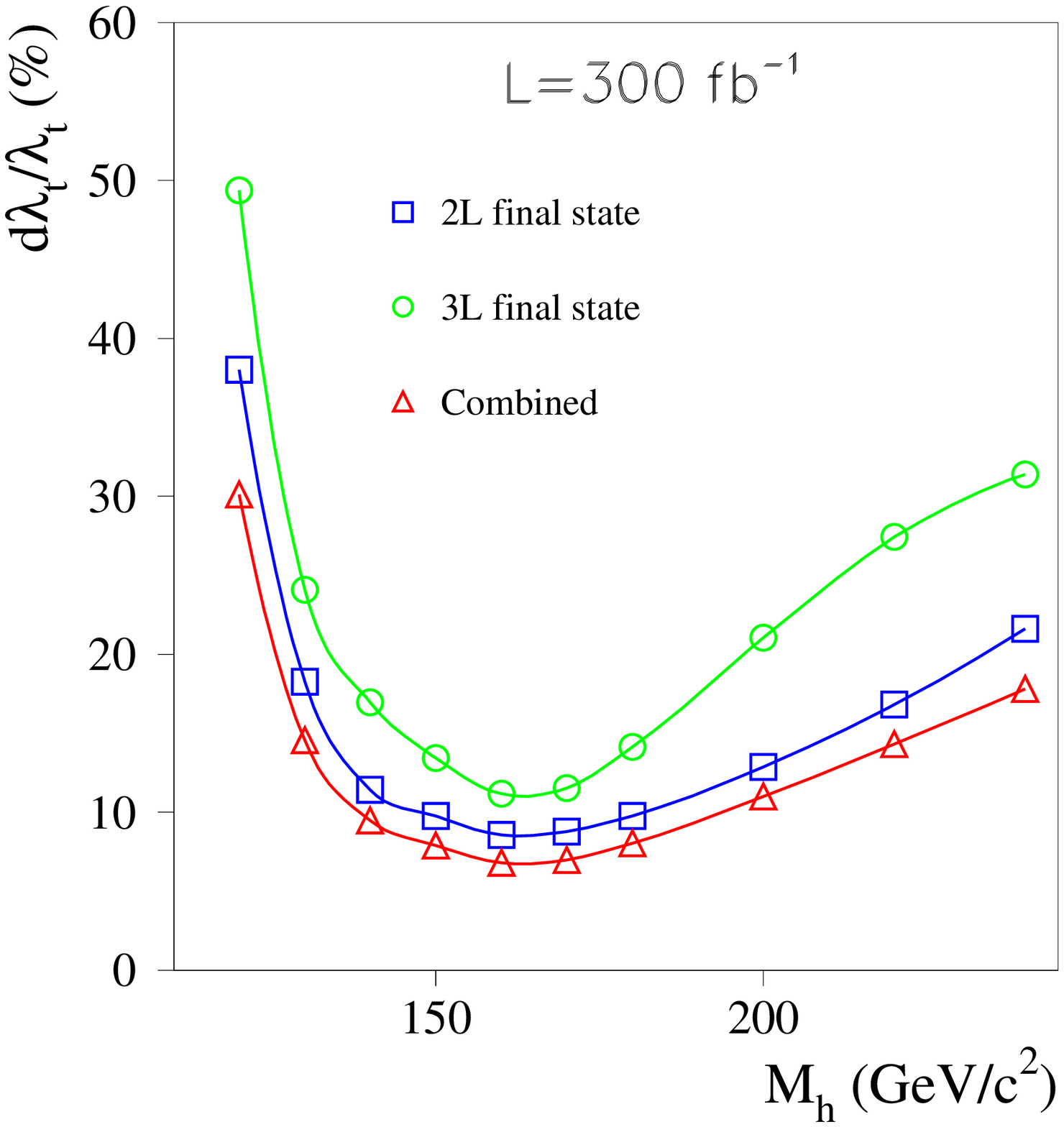}
\vspace*{-5mm}
\caption{ATLAS prediction~\protect\cite{ATL-PHYS-2002-019} for the top
quark Yukawa coupling measurement uncertainty (statistical only) from
$t\bar{t}H,H\to\ww$, for separate leptonic final-state channels and
combined.}
\label{fig:ttH-WW-coup}
\end{figure}

A noteworthy features of this channel is that while the $t\bar{t}H$
cross section falls with increasing $M_H$, BR($H\to\ww$) rises with
increasing $M_H$ in our mass region of interest, and the two trends
coincidentally approximately balance each other.  From a final-state
rate perspective, this channel is approximately constant over a wide
mass range, up to about 200~GeV.  Fig.~\ref{fig:ttH-WW-rate} shows
this numerically.  Fig.~\ref{fig:ttH-WW-coup} shows ATLAS's expected
statistical uncertainty on the top quark Yukawa coupling.  It ranges
from about $20\%$ over a broad mass range for 30~fb$^{-1}$ of data, to
about $10\%$ from the full LHC run.  Systematic uncertainties are
currently unexplored.

\vspace*{-5mm}
%

%%%%%%%%%%%%%%%%%%%%%%%%%%%%%%%%%%%%%%%%%%%%%%%%%%%%%%%%%%%%%%%%%%%%%%%%

\subsubsection{LHC Higgs in a nutshell}
\label{sub:LHC-sum}

\begin{figure}[hb!]
\includegraphics[scale=0.5]{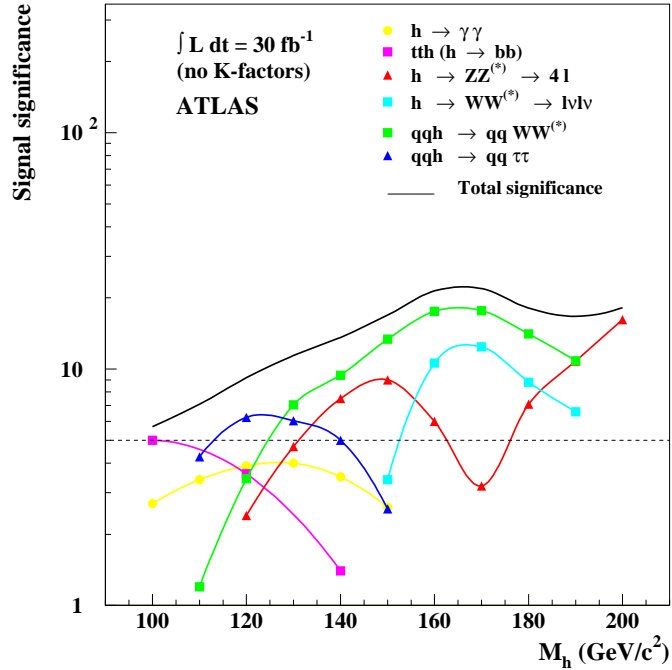}
\vspace*{-5mm}
\caption{ATLAS significance projections in multiple Higgs channels w/
30~fb$^{-1}$ of LHC data~\protect\cite{Asai:2004ws}.}
\label{fig:H-sum}
\end{figure}

LHC Higgs phenomenology has come a long way in the decade since the
first comprehensive studies were reported (e.g. the ATLAS
TDR~\cite{ATLAS_TDR}).  The old studies give a seriously misleading
picture of LHC capabilities.  Students should refer to newer ATLAS
Notes and the new CMS TDR~\cite{CMS_TDR}.  Solid grounds exist for
expecting even more improvements.  Fig.~\ref{fig:H-sum} summarizes
ATLAS's projections for multiple Higgs channels as a function of Higgs
mass.  Note especially the new dominance of WBF channels and
degradation of $t\bar{t}H$.

\vspace{-10mm}

%

%%%%%%%%%%%%%%%%%%%%%%%%%%%%%%%%%%%%%%%%%%%%%%%%%%%%%%%%%%%%%%%%%%%%%%%%
%%%%%%%%%%%%%%%%%%%%%%%%%%%%%%%%%%%%%%%%%%%%%%%%%%%%%%%%%%%%%%%%%%%%%%%%
%%%%%%%%%%%%%%%%%%%%%%%%%%%%%%%%%%%%%%%%%%%%%%%%%%%%%%%%%%%%%%%%%%%%%%%%

\section{Is it the Standard Model Higgs?}
\label{sec:meas}

Imagine yourself in 2010 (hey, we're optimists!), squished
shoulder-to-shoulder in the CERN auditorium, waiting for the speaker
to get to the punchline.  Rumors have been circulating for months
about excess events showing up in some light Higgs channels, but not
all that would be expected.  LHC has 40~fb$^{-1}$, after all.  Your
experimental friends tell you that both collaborations have been
scrambling madly, independent groups cross-checking the original first
analyses.  Then the null result slides start passing by.  No diphoton
peaks anywhere.  Nothing in the $WW$ or $ZZ$ channels.  Even CMS's
invisible Higgs search (WBF -- tagging jets with no central objects at
all) doesn't show anything.  Numerous standard MSSM Higgs results fly
by, invariant mass spectra fitting the SM predictions perfectly.  The
audience becomes restless, irritated.  People around you mutter that
there must not be a Higgs after all.  But you realize that the speaker
skipped mention of the WBF $H\to\taus$ channel.  Then suddenly it
appears, and there's a peak above the $Z$ pole, centered around
125~GeV, broader than you'd expect but the speaker says something
about resolution will improve with further refinement of the tau
reconstruction algorithms. It's also a too-small rate, less than half
what's expected.

So what is this beast?  The bump showed up in a Higgs search channel,
but at that mass it should have shown up in several others as well.
If it's Standard Model, that is.  At 125~GeV there should be $H\to\ww$
in WBF, and $H\to\gamma\gamma$ both inclusively and in WBF, although
maybe they're still marginal.  Photons turned out to be hard at first,
and QCD predictions weren't quite on the mark.  Quite a few people are
on their cell phones already.  You hear a dozen different
exclamations, ranging from ``We found the Higgs!'' to ``The Standard
Model is dead!''.  Quite obviously this is a new physics discovery,
but what exactly is going on?

By now you should get the point of this imaginary scenario: finding a
new bump is merely the start of real physics.  For numerous reasons
you've heard at this summer school, some better than others, finding a
SM Higgs really isn't very likely.  But as we'll see in
Chapter~\ref{sec:BSM}, SM Higgs phenomenology is a superb base for
beyond-the-SM (BSM) Higgs sectors.  They're variations on a theme in
some sense, with the occasional special channel thrown in, like the
invisible Higgs search alluded to above.  Our job will be to figure
out what any new resonance is.  But how do we go about doing that in a
systematic way that's useful to theorists for constructing the New
Standard Model?

For starters, we want to know the complete set of quantum numbers for
any Higgs candidate we find.  Standard Model expectations will
probably prejudice us as to what they are (roughly, at least) based on
which search channel a bump shows up in.  But for the scenario above,
I can envision at least three very reasonable yet completely different
models that would give that kind of a result in early LHC running.  We
should keep in mind that further data may reveal more resonances --
not everything is easy to see against backgrounds, or is produced with
enough rate to emerge with only 1/10 of the planned LHC data.  In some
cases we would have to wait much longer, using data from the planned
LHC luminosity upgrade (SLHC)~\cite{Gianotti:2002xx}.  New physics
could also mean new quantum numbers that we don't yet know about, so
we should be prepared to expand our list of measurements needed to
sort out the theory, and spend time {\it now} thinking about what
kinds of observables are even possible at the LHC.  Some measurements
will almost certainly require the clean environment of a future
high-energy electron-positron machine like an
ILC~\cite{Aguilar-Saavedra:2001rg,Gunion:2003fd}.  The most complete
picture would emerge only after combining
results~\cite{Weiglein:2004hn}, which could take than a decade.  In
the meantime we might get a good picture of the new physics, but not
its details.

Let's prepare a preliminary list of quantum numbers we need to measure
for a candidate Higgs resonance, which I'll generically call $\phi$.
In brackets is the SM expectation.  I'll order them in increasing
level of difficulty.  (See also the review article of
Ref.~\cite{Burgess:1999ha}.)
\begin{itemize}
\vspace{-2mm}
\item[$\cdot$] electric charge [neutral]
\vspace{-2mm}
\item[$\cdot$] color charge [neutral]
\vspace{-2mm}
\item[$\cdot$] mass [free parameter]
\vspace{-2mm}
\item[$\cdot$] spin [0]
\vspace{-2mm}
\item[$\cdot$] CP [even]
\vspace{-2mm}
\item[$\cdot$] gauge coupling ($g_{WWH}$) 
              [$SU(2)_L$ with tensor structure $g^{\mu\nu}$]
\vspace{-2mm}
\item[$\cdot$] Yukawa couplings [$m_f/v$]
\vspace{-2mm}
\item[$\cdot$] spontaneous symmetry breaking potential (self-couplings)
               {[fixed by the mass]}
\end{itemize}
Of course, the first two of those, electric and color charge, are
known immediately from the decay products.  (A non-color-singlet
scalar is a radically different beast than the SM Higgs and would have
dramatically different couplings and signatures.)  Mass is also almost
immediate, with some level of uncertainty that depends almost purely
on detector effects.  Spin and CP are related to some degree, and not
entirely straightforward if the Higgs sector is non-minimal and
contains CP violation.  Gauge and Yukawa couplings are generally
regarded as the most crucial observables, and in some sense I would
agree.  However, I would argue that the linchpin of spontaneous
symmetry breaking (SSB) is the existence of a Higgs potential, which
requires Higgs self-couplings.  Measuring these and finding they match
to some gauge theory with a SSB Higgs sector would to me be the most
definitive proof of SSB, and strongly suggest that the Higgs is a
fundamental scalar, not composite.  It is also the most difficult task
-- perhaps not even possible.

A cautionary note: the results I show in this section are in general
applicable only to the Standard Model Higgs!  This point is often lost
in many presentations highlighting the capabilities of various
experiments, but it is very easy to understand.  For example, if for
some reason the Higgs sector has suppressed couplings to colored
fermions, then any measurement of, say, the $b$ Yukawa coupling, will
be less precise, simply because the signal rate is lower, yet the
background remains fixed.  It's statistics!

%%%%%%%%%%%%%%%%%%%%%%%%%%%%%%%%%%%%%%%%%%%%%%%%%%%%%%%%%%%%%%%%%%%%%%%%
%%%%%%%%%%%%%%%%%%%%%%%%%%%%%%%%%%%%%%%%%%%%%%%%%%%%%%%%%%%%%%%%%%%%%%%%

\subsection{Mass measurement}
\label{sub:mass}

\begin{figure}[ht!]
\includegraphics[scale=0.5]{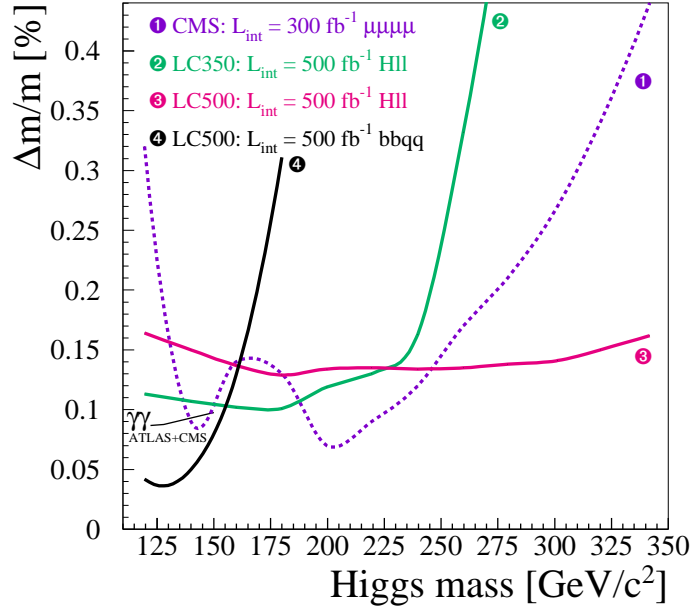}
\vspace*{-2mm}
\caption{Expected SM Higgs mass precision at LHC (for CMS; ATLAS will
be slightly different but comparable) and a future ILC, as a function
of Higgs mass~\protect\cite{Drollinger:2001bc}.}
\label{fig:mass}
\end{figure}

As already noted, our Higgs hunt pretty much gets us this quantum
number immediately, but with some slop driven by detector performance.
We want to measure it as accurately as possible, but in practice a GeV
or so is good enough, because theoretical uncertainties in parameter
fits tend to dominate for most BSM physics.  (This is a long-standing
problem in SUSY scenarios, for example.  It may be that we need to
know the Higgs mass theoretical prediction to four loops~\cite{Spira};
at present only a partial three-loop calculation is
known~\cite{Martin:2007pg}, and only two-loop results exist in usable
code~\cite{Hahn:2006np}.)  Fig.~\ref{fig:mass} shows the CMS and ILC
expected Higgs mass precision as a function of
$M_H$~\cite{Drollinger:2001bc}.  It varies, of course, because
different decay modes are accessible at different $M_H$, and detector
resolution depends on the final state.  In general, photon pairs
($H\to\gamma\gamma$) and four leptons coming from $Z$ pairs ($H\to
ZZ\to\ells\ell^{\prime+}\ell^{\prime-}$) will give the most precise
measurement.  As a rule of thumb, we may expect per-mille precision
over a broad mass range, translating typically to a few hundred MeV.

%

%%%%%%%%%%%%%%%%%%%%%%%%%%%%%%%%%%%%%%%%%%%%%%%%%%%%%%%%%%%%%%%%%%%%%%%%
%%%%%%%%%%%%%%%%%%%%%%%%%%%%%%%%%%%%%%%%%%%%%%%%%%%%%%%%%%%%%%%%%%%%%%%%

\subsection{Spin \& CP measurement}
\label{sub:spin-CP}

Spin and CP ($J^{PC}$) experimental measurements are linked, because
both require angular distributions to obtain.  Numerous techniques
have been proposed to address this, with significant overlap but also
some unique features with each method.  I'll highlight the leading
proposals which garner the most attention from LHC experimentalists
today.

From the observed final state we can tell that the Higgs candidate is
a boson.  We'll start by assuming that it may be spin 0, 1 or 2, but
no higher\footnote{$S\ge3$ fundamental particles are believed to have
deep problems in renormalizable field theory~\cite{S2+}.}.  Then we
recall that the Yang-Landau Theorem~\cite{Yang:1950rg} forbids a
coupling between three $S=1$ bosons if two of them are identical.
Thus, if we observe $\phi\to\gamma\gamma$, then our new object cannot
be spin-1, and $C=1$.  For the very curious student who wants to delve
deeper, there is a recent report on CP Higgs studies at
colliders~\cite{Accomando:2006ga}.

%%%%%%%%%%%%%%%%%%%%%%%%%%%%%%%%%%%%%%%%%%%%%%%%%%%%%%%%%%%%%%%%%%%%%%%%

\subsubsection{Nelson technique}
\label{sub:Nelson}

\begin{figure}[ht!]
\includegraphics[scale=0.8]{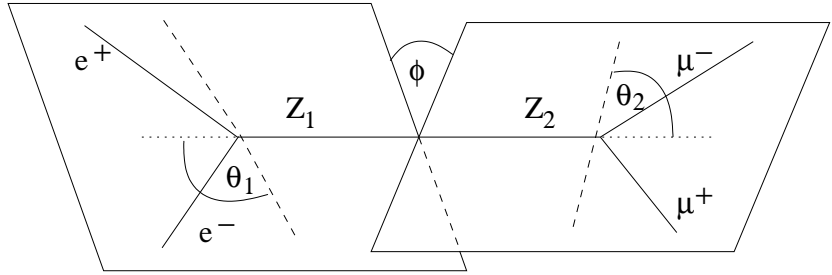}
\vspace*{-2mm}
\caption{Schematic of the azimuthal angle between the decay planes of
$Z$ bosons arising from massive scalar decay.  All angles are in the
scalar rest frame.  Figure from Ref.~\protect\cite{Buszello:2002uu}.}
\label{fig:planes}
\end{figure}

The first method is the oldest, developed by
Nelson~\cite{Dell'Aquila:1985vc}.  It assumes the object is a scalar
or pseudoscalar\footnote{A pseudoscalar doesn't couple at tree-level
to $W$ or $Z$, but can have a (large) loop-induced coupling.} and
relies on the decay angular distributions to a pair of EW gauge
bosons, which decay further.  The most practical aspect relevant for
LHC Higgs physics is in essence a measurement of the relative
azimuthal angle between the decay planes of two $Z$ bosons in turn
coming from the scalar decay, in the scalar particle's rest frame.
See Fig.~\ref{fig:planes} for clarity.  One bins the data in this
distribution and fits to the equation:
\bq\label{eq:Nelson}
F(\phi) \, = \, 1 + \alpha\cos(\phi) + \beta\cos(2\phi)
\eq
For a scalar, such as the SM Higgs, the coefficients $\alpha$ and
$\beta$ are functions of the scalar mass, and further we have the
constraint that $\alpha(M_\phi)>\frac{1}{4}$.  In contrast, for a
pseudoscalar, $\alpha=0$ and $\beta=-0.25$, independent of the mass.

\begin{figure}[ht!]
\includegraphics[width=12cm]{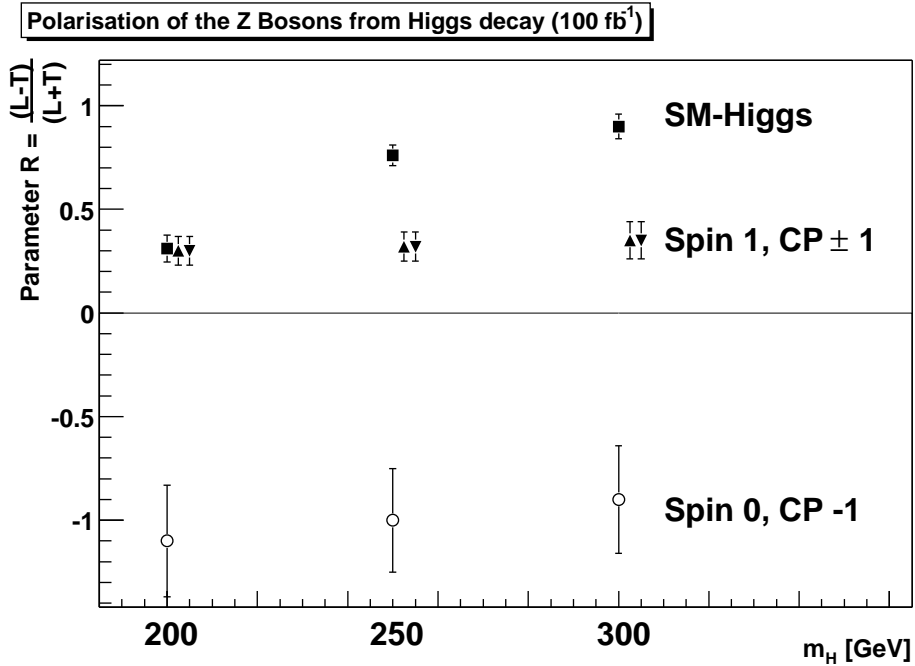}
\vspace*{-2mm}
\caption{Results of the LHC expectations spin/CP study of 
Ref.~\protect\cite{Buszello:2002uu}, showing how a SM Higgs could be
distinguished from a pseudoscalar or spin-1 boson as a function of
$M_H$.}
\label{fig:Bij-spin}
\end{figure}

Ref.~\cite{Buszello:2002uu} was the first to apply this to the LHC
Higgs physics program using detector simulation.  Assuming
100~fb$^{-1}$ of data, the study found that LHC could readily
distinguish a SM Higgs from a pseudoscalar for $M_H>200$~GeV, and from
a spin-1 boson of either CP state from a little above that, but not
right at 200~GeV; see Fig.~\ref{fig:Bij-spin}.  Applying this
technique to $M_H<200$ but above $ZZ$ threshold was not examined.

As a practical matter, $H\to ZZ^{(*)}$ observation is assured only for
both $Z$ bosons decaying to leptons ($e$ or $\mu$), where there is
essentially zero background.  Unfortunately, this is an extremely tiny
branching ratio, only $0.05\%$ of all $H\to ZZ$ events.  Some studies
consider $jj\ells$ channels, which is a ten-times larger sample, in an
attempt to increase statistics, but this suffers from non-trivial QCD
backgrounds.

%

%%%%%%%%%%%%%%%%%%%%%%%%%%%%%%%%%%%%%%%%%%%%%%%%%%%%%%%%%%%%%%%%%%%%%%%%

\subsubsection{CMMZ technique}
\label{sub:CMMZ}

Ref.~\cite{Choi:2002jk} provides an extension to the Nelson technique
below $ZZ$ threshold.  Its full analysis is far more in-depth,
discussing the angular behavior of the matrix elements for arbitrary
boson spin and parity.  It first demonstrates how objects of odd
normality (spin times parity) can be discriminated via angular
distributions, but for even normality require a further discriminant.
That is, a $J^P=2^+$ boson could mimic a SM Higgs in angular
distribution below $ZZ$ threshold.  (Exotic higher spin states can be
trivially ruled out via the lack of angular correlation between the
beam and the object's flight direction.)

\begin{figure}[hb!]
\includegraphics[scale=0.56]{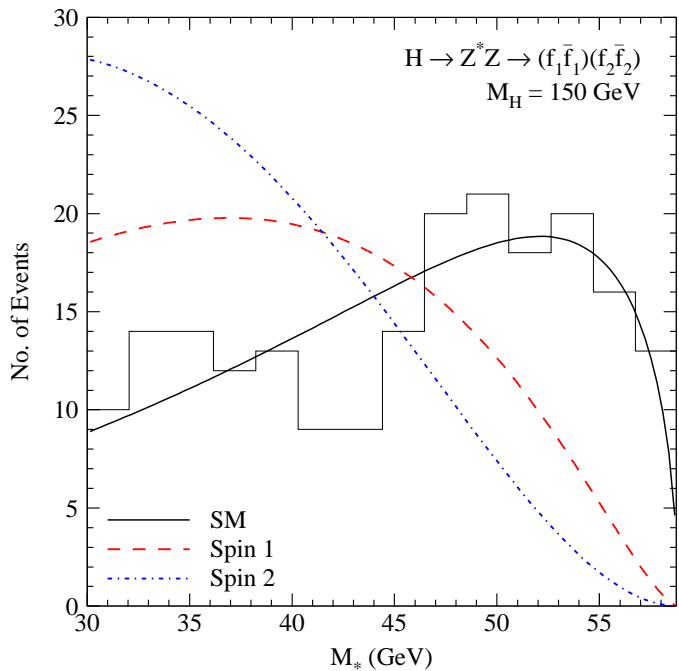}
\vspace*{-2mm}
\caption{Differential decay rate as a function of dilepton invariant 
mass of the off-shell $Z^*$ in $ZZ$ events, for a 150~GeV SM Higgs
v. spin-1 and spin-2 objects of even normality and the same mass.  The
histogram is the SM Higgs case for 300~fb$^{-1}$ of data at LHC.
Figure from Ref.~\protect\cite{Choi:2002jk}.}
\label{fig:dGdM}
\end{figure}

The key discriminant is the differential partial decay rate for the
off-shell $Z$ boson\footnote{Typically only one $Z$ boson is off-shell
for $M_H<2M_Z$, but this ceases to be a good approximation at much
lower (but observable) masses.}.  It depends on the invariant mass of
the final-state lepton pair and is linear in $Z^*$ velocity:
\bq\label{eq:dGdM}
\frac{d\Gamma_H}{dM_*^2} \; \sim \;
\beta \; \sim \; \sqrt{(M_H-M_Z)^2 - M_*^2}
\eq
Fig.~\ref{fig:dGdM} shows the predicted distributions for 150~GeV
spin-0,1,2 even-normality objects as a function of $M_*$, the
off-shellness of the $Z^*\ells$.  The histogram represents about 200
events that a SM Higgs would give in this channel after 300~fb$^{-1}$
of data at LHC.  Unfortunately there are no error bars, although one
can estimate the statistical uncertainty for each bin as $\sqrt{N}$
and observe that the measurement is likely not spectacular.  We can
expect that CMS and ATLAS will eventually get around to quantifying
the discriminating power, but it would not be surprising to learn that
this measurement requires far more data, e.g. at the upgraded
SLHC~\cite{Gianotti:2002xx}.

%

%%%%%%%%%%%%%%%%%%%%%%%%%%%%%%%%%%%%%%%%%%%%%%%%%%%%%%%%%%%%%%%%%%%%%%%%

\subsubsection{CP and gauge vertex structure via WBF}
\label{sub:CP-WBF}

A third technique~\cite{Plehn:2001nj} takes a different approach, but
addressing spin and CP in a slightly different way.  Rather than
examine Higgs decays, it notes that WBF Higgs production is observable
for {\it any} Higgs mass, regardless of decay mode.  Furthermore, the
same $HVV$ vertex appears on the production side for all masses, also
independent of decay.  More precisely, this vertex has the structure
$g^{\mu\nu}HV_{\mu}V_{\nu}$ ($V=W,Z$).  This tensor structure is not
gauge invariant by itself.  It must come from a gauge-invariant
kinetic term $(D_\mu\Phi)^\dagger(D^\mu\Phi)$.  Identifying it in
experiment would go a long way to establishing that the scalar field
is a remnant of spontaneous symmetry breaking.

For a scalar field which couples via higher-dimensional operators to
two gauge bosons, however, we may write down the CP-even and CP-odd
gauge-invariant D6 operators~\cite{Buchmuller:1985jz}:
\bq\label{eq:D6}
{\cal L}_6 \; = \;
\frac{g^2}{2\Lambda_{6,e}} 
     (\Phi^\dagger\Phi) W^+_{\mu\nu} W^{-\mu\nu} \; + \;
\frac{g^2}{2\Lambda_{6,o}} 
     (\Phi^\dagger\Phi) \wt{W}^+_{\mu\nu} W^{-\mu\nu}
\eq
where $\Lambda_6$ is the scale of new physics that is integrated out,
$W^{\mu\nu}$ is the $W$ boson field strength tensor, and
$\wt{W}=\epsilon_{\alpha\beta\mu\nu}W^{\alpha\beta}$ is its dual.
After expanding $\Phi$ with a vev and radial excitation, we obtain two
D5 operators:
\bq\label{eq:D5}
{\cal L}_5 \; = \;
\frac{1}{\Lambda_{5,e}} H W^+_{\mu\nu} W^{-\mu\nu} \; + \;
\frac{1}{\Lambda_{5,o}} H \wt{W}^+_{\mu\nu} W^{-\mu\nu}
\eq
where $\Lambda_5$ are dimensionful but now parameterize both the D6
coefficients and the $\Phi$ vev.

These two D5 operators produce very distinctive matrix element
behavior.  Recalling that the external gauge bosons in WBF are
actually virtual and connect to external fermion currents, the
initial-state scattered quarks, we derive the following approximate
relations for the CP-even operator, using $J_{1,2}$ for the incoming
fermion currents:
\bq\label{eq:D5-eff}
{\cal M}_{e,5}
\propto \frac{1}{\Lambda_{e,5}} J^\mu_1 J^\nu_2
        \biggl[ g_{\mu\nu}(q_1\cdot q_2) - q_{1,\nu}q_{2,\mu} \biggr]
\sim \frac{1}{\Lambda_{e,5}}
[J^0_1 J^0_2 - J^3_1 J^3_2] \; \vec{p}_T^{\;j1} \cdot \vec{p}_T^{\;j2}
\eq
That is, the amplitude is proportional to the tagging jets' transverse
momentum dot product.  This is easy to measure experimentally -- we
just plot the azimuthal angular distribution, i.e. angular separation
in the plane perpendicular to the beam.  It will be minimal, nearly
zero, for $\phi_{jj}=\pi/4$.  In contrast, the $g^{\mu\nu}$ tensor
structure of the SM Higgs mechanism does not correlate the tagging
jets.  The CP-odd D5 operator is different and more complex, but may
be understood by noting that it contains a Levi-Civita tensor
$\epsilon^{\mu\nu\rho\delta}$ connecting the external fermion momenta.
This is non-zero only when the four external momenta are independent,
i.e. not coplanar.  Thus this distribution will be zero for
$\phi_{jj}=0,\pi$.

\begin{figure}[ht!]
\includegraphics[scale=0.8]{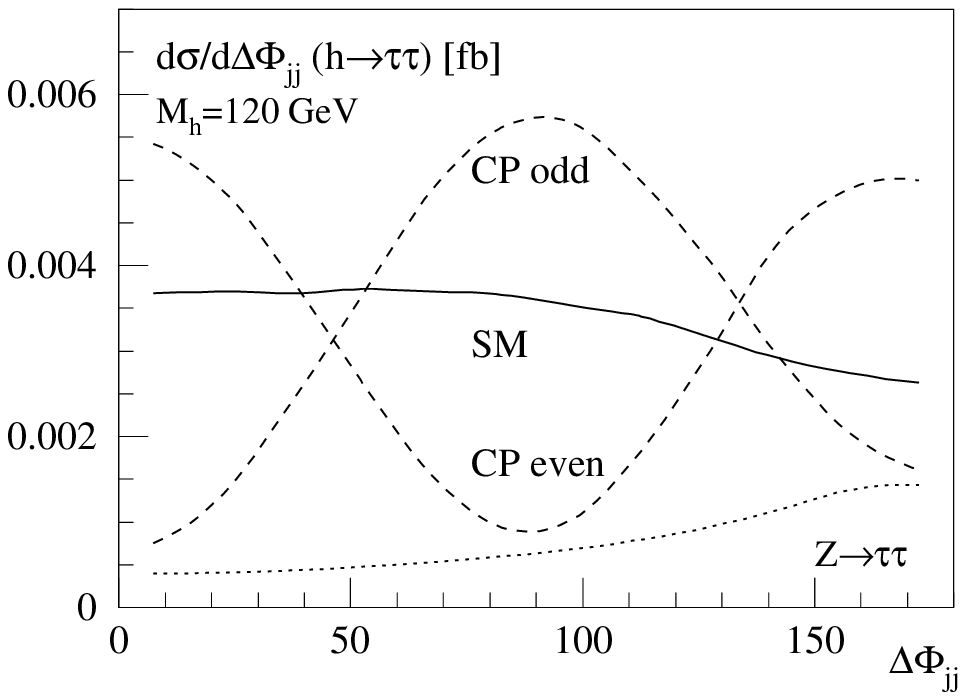}
\includegraphics[scale=0.8]{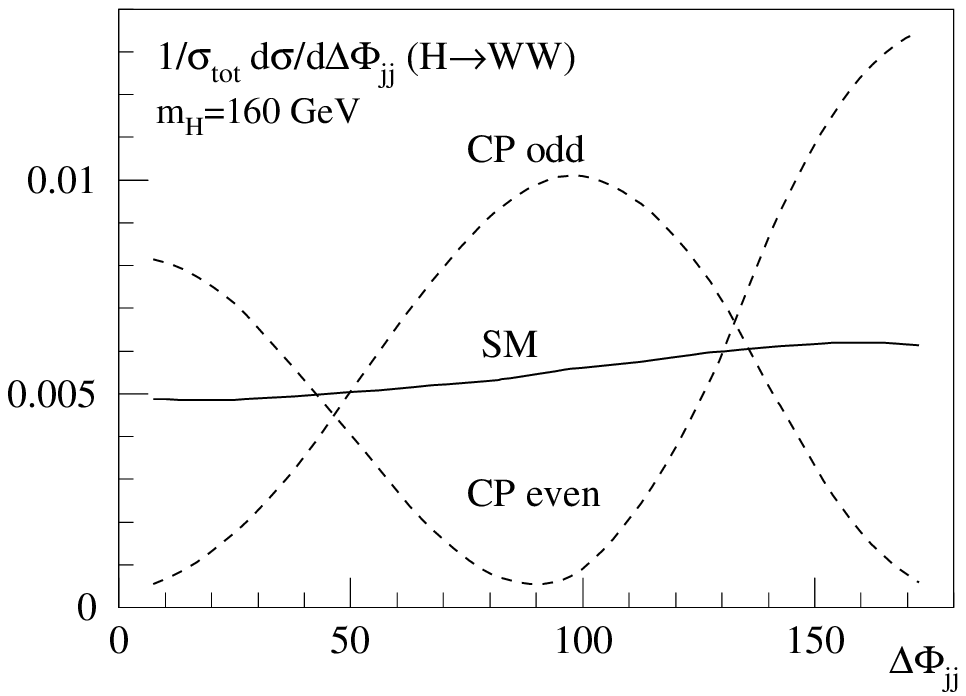}
\vspace*{-2mm}
\caption{Azimuthal angular distributions of the tagging jets in WBF
production of a SM Higgs v. scalar field coupled to weak bosons via
CP-even/odd D6 operators.  The dotted line in the left panel is the SM
background, which is added to the signal curves.  Figures from
Ref.~\protect\cite{Plehn:2001nj}.}
\label{fig:dphijj-1}
\end{figure}

Fig.~\ref{fig:dphijj-1} shows the results of a parton-level simulation
for scalars in both the mass range where decays to taus would be used,
and where $\phi\to\ww$ dominates.  The SM signal curve is not entirely
flat due to kinematic cuts imposed on the final state to ID all
objects.  The D5 operators produce behavior qualitatively distinct
from spontaneous symmetry breaking, with minima for the distributions
exactly where expected, and orthogonal from each other.  It would be
essentially trivial to distinguish the cases from each other shortly
after discovery, regardless of $M_H$ and the particular channel used
to discover the Higgs candidate.  A key requirement for this, of
course, is that the discovery searches don't use this distribution to
separate signal from background.

\begin{figure}[hb!]
\includegraphics[scale=0.8]{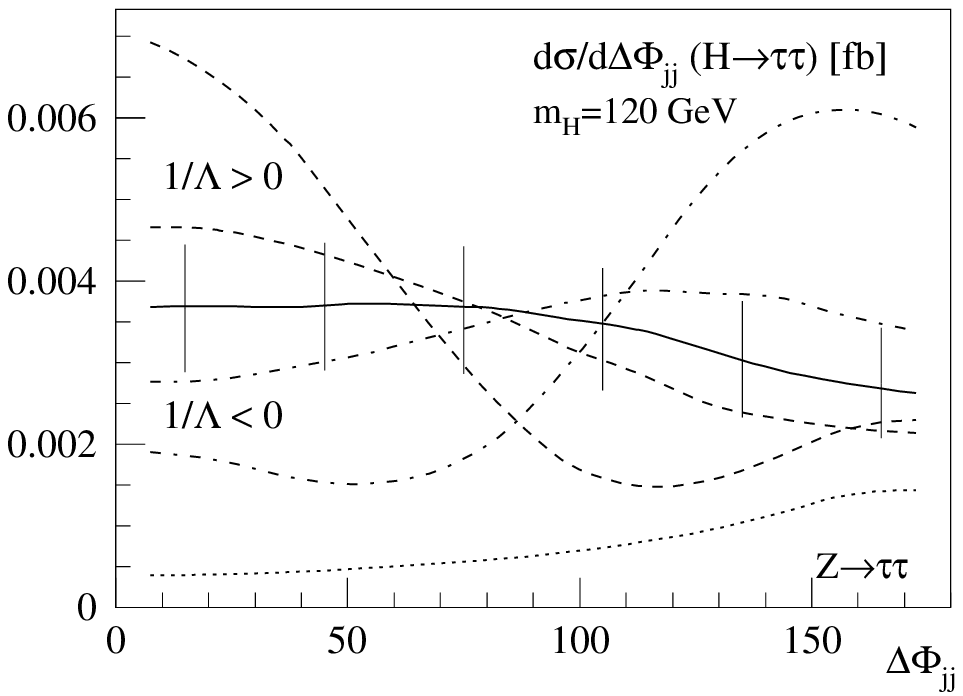}
\includegraphics[scale=0.7]{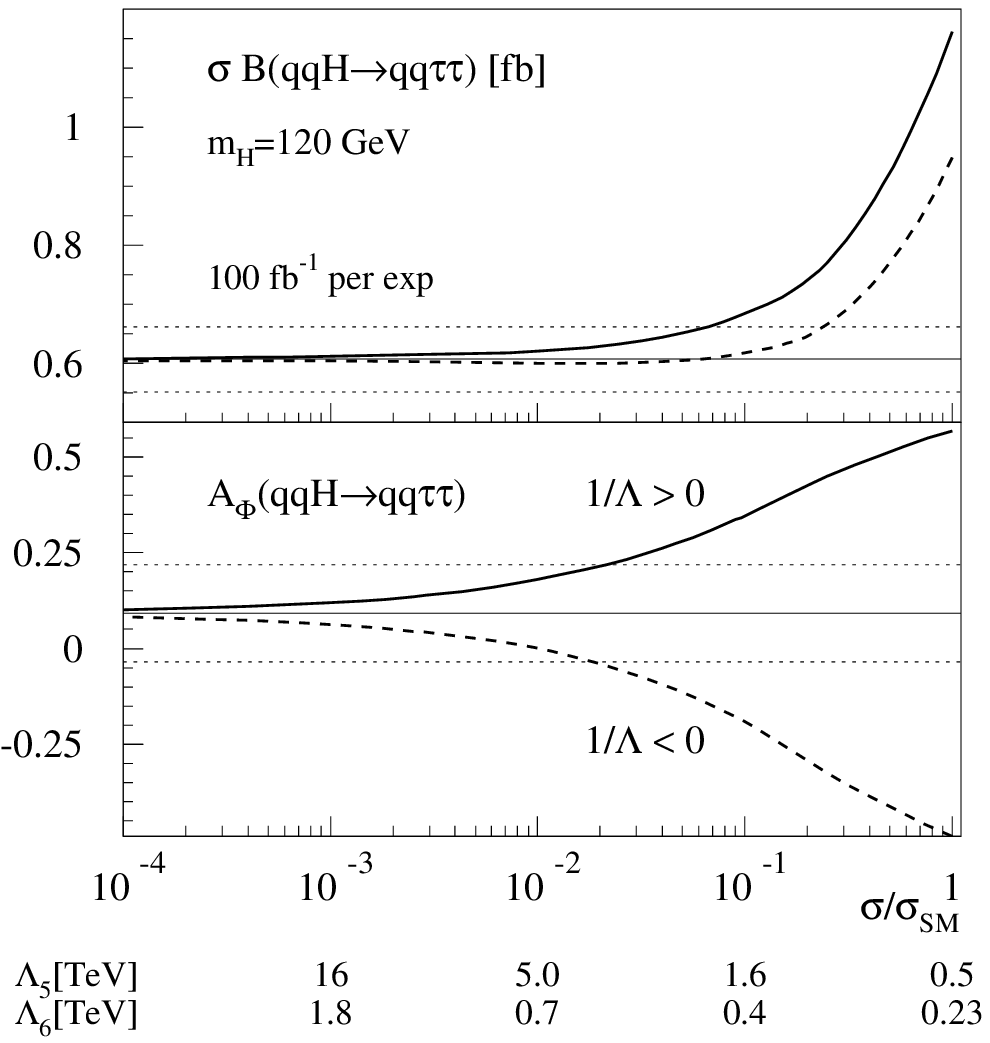}
\vspace*{-2mm}
\caption{Left: As in Fig.~\protect\ref{fig:dphijj-1}, but with 
interference between the SM Higgs and a CP-even D5 operator.  Right:
the effective reach in $\Lambda_{5,e}$ for 100~fb$^{-1}$ at LHC, using
only the rate information (top) or the asymmetry (bottom).  Figures
from Ref.~\protect\cite{Plehn:2001nj}.}
\label{fig:dphijj-2}
\end{figure}

Now, what happens if the Higgs indeed arises from SSB, but new physics
generates sizable D6 operators?  Since $H_{\rm SM}$ is CP-even, a
CP-even D5 operator would interfere with the SM amplitude, while a
CP-odd contribution would remain independent.  This is illustrated in
the left panel of Fig.~\ref{fig:dphijj-2}.  The obvious thing to do is
create an asymmetry observable sensitive to this interference:
\bq\label{WBF-asym}
A_\phi \; = \; \frac
{\sigma(\Delta\phi_{jj} < \pi/2) - \sigma(\Delta\phi_{jj} > \pi/2)}
{\sigma(\Delta\phi_{jj} < \pi/2) + \sigma(\Delta\phi_{jj} > \pi/2)}
\eq
With only 100~fb$^{-1}$ of data at LHC (one experiment), this
asymmetry would have access to $\Lambda_6\sim1$~TeV, which is itself
within the reach of LHC, likely resulting in new physics observation
directly.  One caveat: the study Ref.~\cite{Plehn:2001nj} was done
before the $gg\to Hgg$ contamination~\cite{gg_Hjj} was known, which
will complicate this measurement.

%

%%%%%%%%%%%%%%%%%%%%%%%%%%%%%%%%%%%%%%%%%%%%%%%%%%%%%%%%%%%%%%%%%%%%%%%%

\subsubsection{Spin and CP at an ILC}
\label{sub:CP-ILC}

\begin{figure}[hb!]
\includegraphics[scale=0.65]{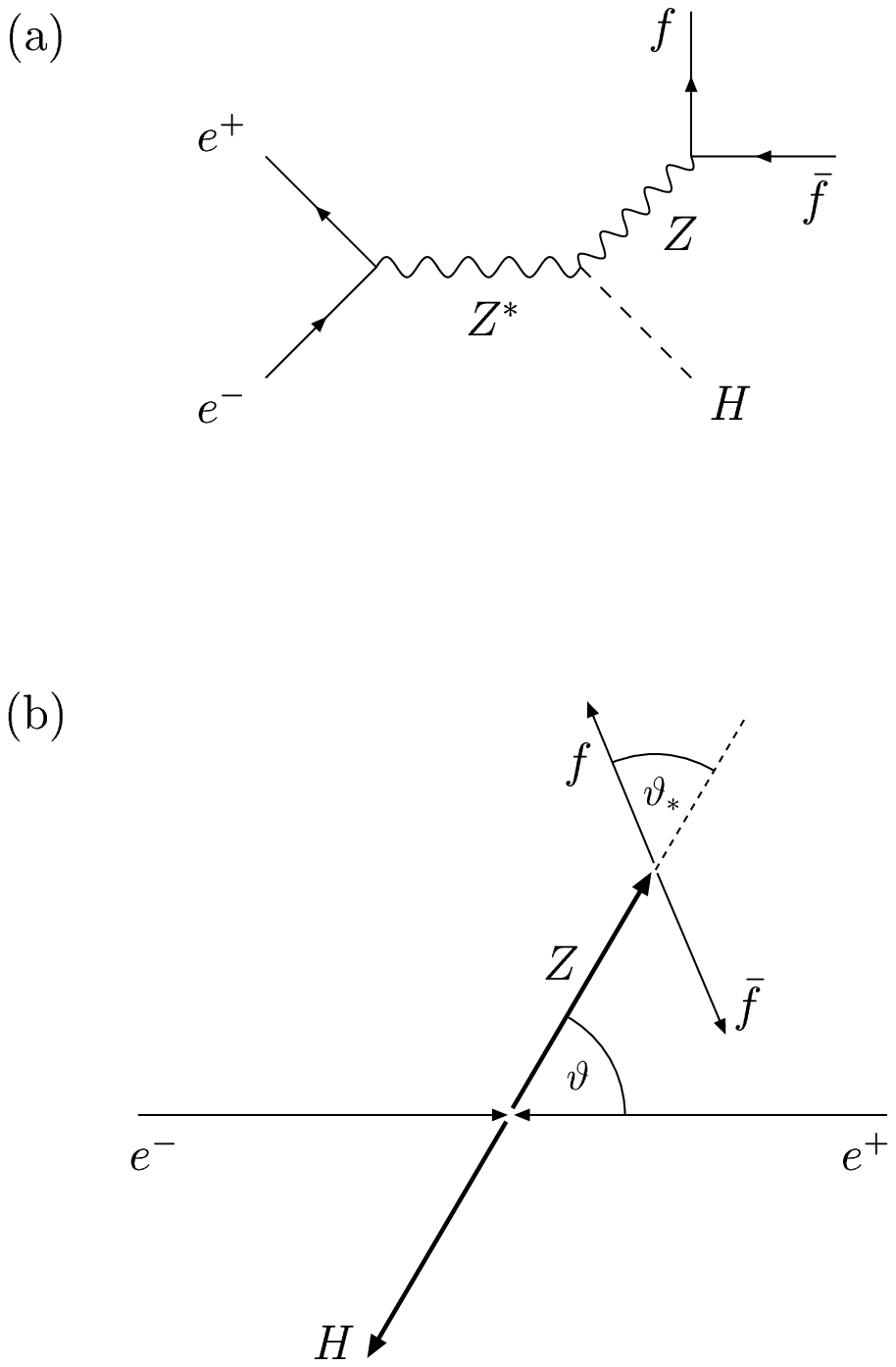}
\includegraphics[scale=0.5]{Higgs.figs/ILC/ILC.H.spin.eps}
\caption{Left: Feynman diagram for $e^+e^-\to ZH$ and 
schematic~\protect\cite{Miller:2001bi} showing the analyzing angles.
Right: curves showing the threshold rate dependence for $J=0,1,2$
states in this channel~\protect\cite{Aguilar-Saavedra:2001rg}.}
\label{fig:ILC-spin}
\end{figure}

The much cleaner, low-background environment of $e^+e^-$ collisions
would be an excellent environment to study a new resonance's spin and
CP properties.  $J^{PC}$ can in fact be determined completely
model-independently.  Recalling the LEP search, the canonical
production mechanism is $e^+e^-\to ZH$.  We would identify the $Z$ via
its decay to leptons, and sum over all Higgs decays (this is possible
using the recoil mass technique, coming up in
Sec.~\ref{sub:ILC-coup}).  $J$ and $P$ are completely determined by a
combination of the cross section rise at threshold and the polar angle
of the $Z$ flight direction in the lab, shown in the left panel of
Fig.~\ref{fig:ILC-spin}.  The differential cross section
is~\cite{Aguilar-Saavedra:2001rg}:
\bq\label{eq:ILC-spin}
\frac{d\sigma}{d\cos\theta_Z} \propto \beta
\bigl[ 
1 + a\beta^2\sin^2\theta_Z + b\eta\beta\cos\theta_Z
  + \eta^2\beta^2(1+\cos^2\theta_Z)
\bigr]
\eq
where $a$ and $b$ depend on the EW couplings and $Z$ boson mass,
$\eta$ is a general pseudoscalar (loop-induced) coupling and $\beta$
is the velocity.  Far more sophisticated analyses techniques exist,
often called ``optimal observable'' analyses~\cite{Hagiwara:2000tk},
but are only for the terminally curious.

If one would have the liberty to perform a threshold scan of $Z\phi$
production at an ILC, distinguishing given-normality $J=0,1,2$ states
is straightforward due to their different $\beta$-dependence.  For
$J=0$ it is linear, but for higher spin is higher-power in
$\beta$~\cite{Miller:2001bi}.  The qualitative behavior is shown in
the right panel of Fig.~\ref{fig:ILC-spin}, complete with error bars
for the SM Higgs case.  However, while the physics is solid,
experiments in the past have generally proved to be a horse race for
highest energy, so there is no guarantee that one would have threshold
scan data available.  The angular distribution fortunately works at
all energies.

%

%%%%%%%%%%%%%%%%%%%%%%%%%%%%%%%%%%%%%%%%%%%%%%%%%%%%%%%%%%%%%%%%%%%%%%%%
%%%%%%%%%%%%%%%%%%%%%%%%%%%%%%%%%%%%%%%%%%%%%%%%%%%%%%%%%%%%%%%%%%%%%%%%

\subsection{Higgs couplings at LHC}
\label{sub:LHC-coup}

Now to something much harder.  It's commonly believed that LHC cannot
measure Higgs couplings, only ratios of BRs~\cite{ATLAS_TDR}.  This is
incorrect, but requires a little explanation to understand why people
previously believed in a limitation.

First, let me state that the LHC doesn't measure couplings or any
other quantum number directly.  It measures {\it rates}.  (This is
true for any particle physics experiment.)  From those we extract
various $\sigma_i\cdot{\rm BR}_j$ by removing detector, soft QCD and
phase space effects, among other things, using Monte Carlo simulations
based on known physics inputs.

Second, we note that for a light Higgs, which has a very small width
(cf. Sec.~\ref{sub:Hdecay}), the Higgs production cross section is
proportional to the partial width for Higgs decay to the initial state
(the Narrow Width Approximation, NWA).  That is, $\sigma_{gg\to
H}\propto\Gamma_{H\to gg}$.  Similarly, $\sigma_{\rm
WBF}\propto\Gamma_{H\to\ww}$.  The student who has never seen this may
easily derive it by recalling the definition of cross section and
partial decay width -- they share the same matrix elements and differ
only by phase space factors\footnote{Well, slightly more than that in
the case of WBF, but the argument holds after careful consideration.}.
Typically we abbreviate these partial widths with a subscript
identifying the final state particle, thus we have $\Gamma_g$,
$\Gamma_\gamma$, $\Gamma_b$, etc.  Since a BR is just the partial
decay width over the total width, we then write:
\vspace*{-2mm}
\bq\label{eq:NWA1}
\bigl( \sigma_H \cdot {\rm BR} \bigr)_i \propto 
\biggl( \frac{\Gamma_p\Gamma_d}{\Gamma_H} \biggr)_i
\eq
where $\Gamma_p$ and $\Gamma_d$ are the ``production'' and decay
widths, respectively.

\begin{figure}[hb!]
\includegraphics[scale=0.95]{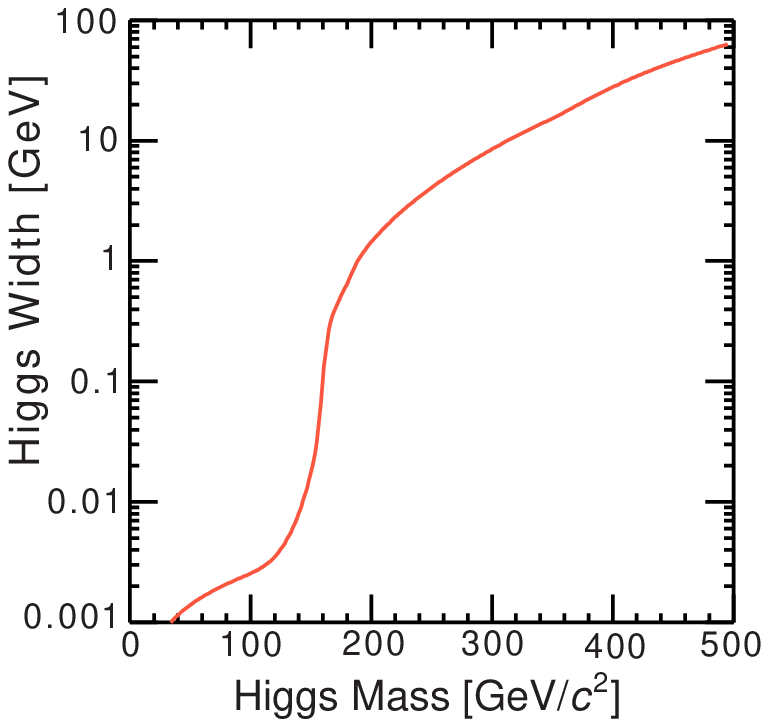}
\hspace*{5mm}
\includegraphics[scale=1]{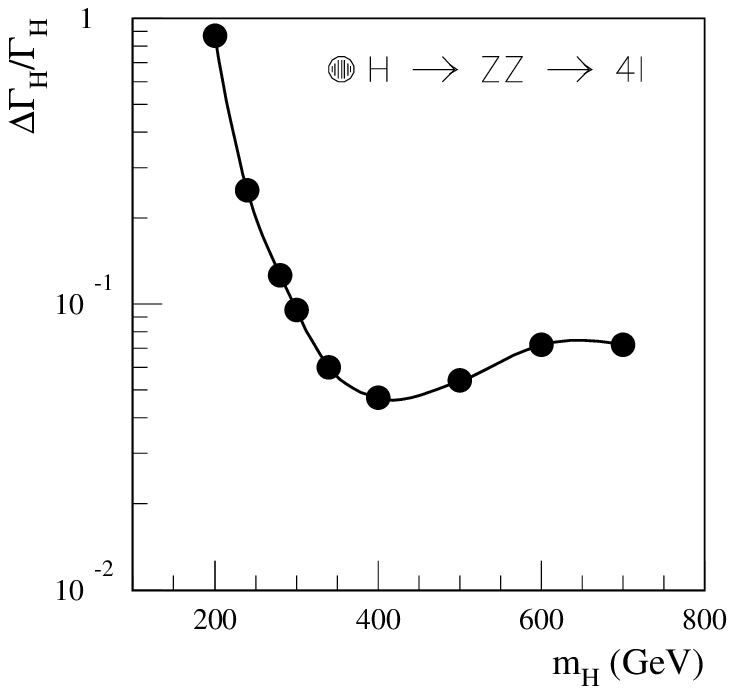}
\caption{Left: Standard Model Higgs total width as a function of $M_H$.
Right: expected experimental precision on $\Gamma_H$ at ATLAS using
the $gg\to H\to ZZ\to 4\ell$ channel~\protect\cite{ATLAS_TDR} (CMS
similar).}
\label{fig:Hwid}
\end{figure}

Third, count up the number of observables we have and measurements we
can make.  Assuming we have a decay channel for each possible Higgs
decay (which we don't), we're still one short: $\Gamma_H$, the total
width.  Now, if the width is large enough, larger than detector
resolution, we can measure it directly.  Fig.~\ref{fig:Hwid} shows
that this can happen only for $M_H\gtrsim 230$~GeV or
so~\cite{ATLAS_TDR}, far above where EW precision data suggests we'll
find the (SM) Higgs.  Below this mass range, we have to think of
something else.

In the SM, we know precisely what $\Gamma_H$ is: the sum of all the
partial widths.  For the moment let's assume we have access to all
possible decays or partial widths via production, ignore the
super-rare decay modes to first- and second-generation fermions.  This
is a mild assumption, because if for some reason the muon or electron
Yukawa were anywhere close to that of taus, where it might contribute
to the total width, it would immediately be observable.  The list of
possible measurements we can form from accessible $(\sigma\cdot{\rm
BR})_{i,exp}$ is:
\bq
X_\gamma, X_\tau, X_W, X_Z, Y_\gamma, Y_W, Y_Z, Z_b, Z_\gamma, Z_W
\eq
where $X_i$ correspond to WBF channels, $Y_i$ are inclusive Higgs
production, and $Z_i$ are top quark associated production\footnote{For
this case, we actually use the Yukawa coupling squared ($y_t^2$)
instead of $\Gamma_t$, because decays to top quarks is kinematically
forbidden.  But this is irrelevant for our argument.}.  We could
easily add measurements like $X_\mu$, $Y_e$, etc. if we wanted,
because measuring zero for any observable is still a measurement -- it
simply places a constraint on that combination of partial widths or
couplings.

In the original implementation of this idea~\cite{Zeppenfeld:2000td},
the authors noted that the $t\bar{t}H,H\to\bb$ channel won't work, so
there is no access at LHC to $\Gamma_b$.  However, there is access to
$\Gamma_\tau$.  In the SM, the $b$ and $\tau$ Yukawa couplings are
related by $r_b=\Gamma_b/\Gamma_\tau=3c_{\rm QCD}m_b^2/m_\tau^2$,
where $c_{\rm QCD}$ contains QCD higher-order corrections and phase
space effects.  $\Gamma_W$ and $\Gamma_Z$ are furthermore related by
$SU(2)_L$, although we don't need to use it.  Now write down the
derived quantity
\bq\label{eq:GamWtilde}
\wt\Gamma_W
\; = \;
X_\tau(1+r_b) + X_W + X_Z + X_\gamma + \wt{X}_g
 \; = \;
\biggl( \sum \Gamma_i \biggr) \frac{\Gamma_W}{\Gamma_H}
\; = \;
(1-\epsilon) \Gamma_W
\eq
where $\wt{X}_g$ is constructed from $X_W$, $X_\gamma$, $Y_W$ and
$Y_\gamma$.  Although $\Gamma_\gamma$ is an infinitesimal contribution
to $\Gamma_H$, it is important as above, and it contains both the top
quark Yukawa and $W$ gauge-Higgs couplings.  Our error is contained in
$\epsilon$ and is typically small.  This provides a good
\underline{lower} bound on $\Gamma_W$ from data.  The total width is
then
\bq\label{eq:totwid}
\Gamma_H \; = \; \frac{\wt\Gamma_W^2}{X_W}
\eq
and the error goes as $(1-\epsilon)^{-2}$.  Assuming systematic
uncertainties of $5\%$ on WBF and $20\%$ on inclusive production, this
would achieve about a $10\%$ measurement of $\Gamma_W$ and $10-20\%$
on the total width for $M_H<200$~GeV.

Voil\`a!  We have circumvented the na\"ive problem of not enough
independent measurements.  The astute observer should immediately
protest, however, and rightly so.  The result is achieved with a
little too much confidence that the SM is correct.  Not only does the
trick rely on a very strong assumption about the $b$ Yukawa coupling,
but there could be funny business in the up-quark sector, giving a
large partial width to e.g. charm quarks, which would not be
observable either via production (too little initial-state charm, and
anyhow unidentifiable) or decay (charm can't be efficiently tagged).
Nevertheless, this was a useful exercise, because a much more
rigorous, model-independent method is closely based on it.

The more sophisticated method is a powerful least-likelihood fit to
data using a more accurate relation than Eq.~\ref{eq:NWA1} between
data and theory~\cite{Duhrssen:2004cv}:
\bq\label{eq:NWA2}
\sigma_H \cdot {\rm BR}(H\to xx) \; = \;
\frac{\sigma^{\rm SM}_H}{\Gamma^{\rm SM}_p}
\cdot
\fbox{$\frac{\Gamma_p\Gamma_d}{\Gamma_H}$}
\eq
where the partial widths in the box are the true values to be
extracted from data, and the $(\sigma/\Gamma)_{\rm SM}$ ratio in front
quantifies all effects shoved into Monte Carlo using SM values: phase
space, QCD corrections, detector, etc.  As before, the ``sum'' of all
channels provides a solid lower bound on $\Gamma_H$, simply because
some rate in each of a number of channels requires some minimum
coupling.  But these are found by a fit, rather than theory
assumptions.  It also properly takes into account all theory and
experimental systematic and statistical uncertainties assigned to each
channel.  We then need only a firm upper bound on $\Gamma_H$ and the
fit then extracts {\it absolute} couplings (transformed from the
partial widths).  This bound comes from unitarity: the gauge-Higgs
coupling can be depressed via mixing in any multi-doublet model, as
well as any number of additional singlets, but it cannot exceed the SM
value, which is strictly defined by unitarity.  Thus
$\Gamma_V\leq\Gamma_V^{\rm SM}$.  (This bound is invalid in triplet
models, but these have other characteristics which should make
themselves apparent in experiment.)  The WBF $H\to\ww$ channel then
provides an upper limit on $\Gamma_H$ via its measurement of
$\Gamma_V^2/\Gamma_H$.

\begin{figure}[hb!]
\includegraphics[scale=0.4]{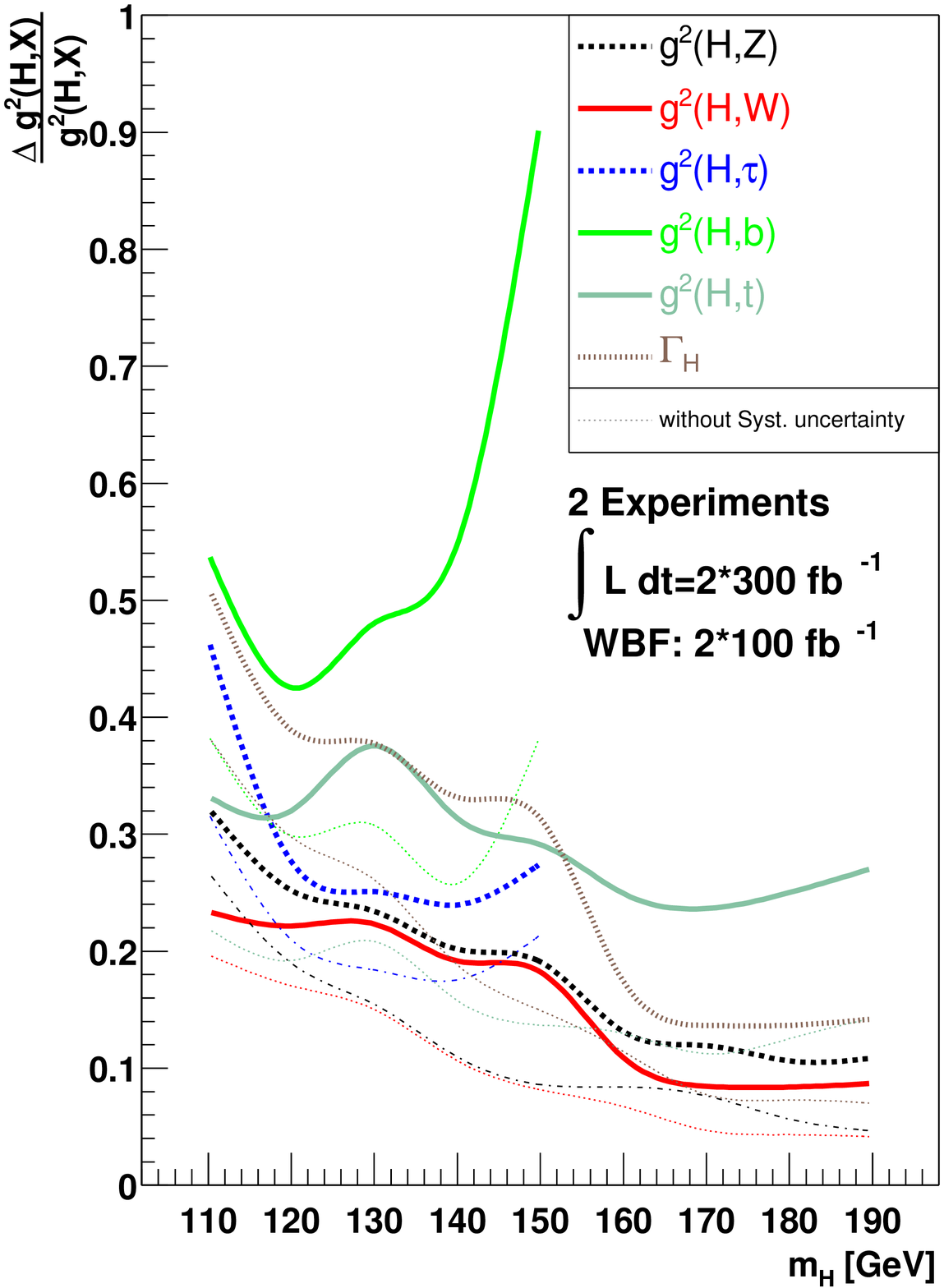}
\hspace*{5mm}
\includegraphics[scale=0.4]{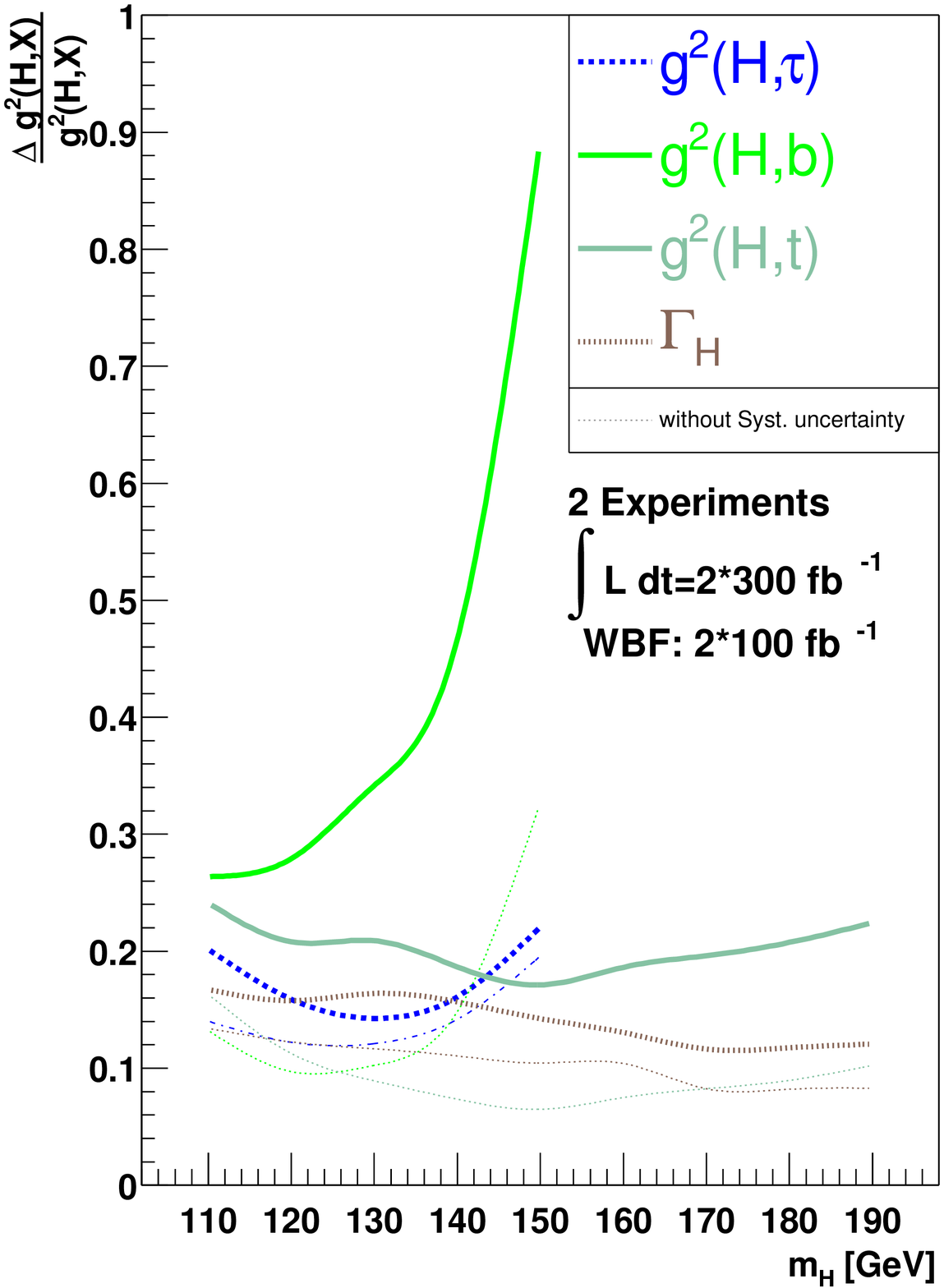}
\caption{Left: a least likelihood general fit on simulated LHC data, 
with no additional assumptions about the Higgs sector.  Right: the fit
assuming no new particles appear in Higgs loop-induced decays, and the
gauge-Higgs coupling fixed exactly to the SM value.  Figures from
Ref.~\protect\cite{Duhrssen:2004cv}.}
\label{fig:Hcoup}
\end{figure}

The method can be further armored against BSM alterations by including
the invisible Higgs channel, allowing additional loop contributions,
and so on.  Of course, the more possible deviations one allows, the
larger the fit uncertainties become.  We see this in the differences
between the left and right panels of
Fig.~\ref{fig:Hcoup}~\cite{Duhrssen:2004cv}.  It is obvious that LHC's
weakness is lack of access to $H\to\bb$.  Nevertheless, LHC can
measure absolute Higgs couplings with useful constraints on BSM
physics.  This is especially true for $M_H\gtrsim 150$~GeV, where LHC
can achieve ${\cal O}(10\%)$ precision on the gauge-Higgs couplings
and the total width.

The fit as implemented in Ref.~\cite{Duhrssen:2004cv} fixes $M_H$.
This is a slight cheat, since for some $M_H$ the BRs change quite
rapidly, and a 1-2~GeV uncertainty can lead to a lot of slop in the
coupling extraction.  This is especially critical for the Higgs sector
of the Minimal Supersymmetric Standard Model (MSSM).  Eventually a fit
to $M_H$ will also have to be included, which will degrade measurement
precision somewhat.

At the same time, there is cause for optimism.  The results of
Fig.~\ref{fig:Hcoup} were based on very conservative, almost
pessimistic assumptions: overly-large systematic errors, WBF not being
possible at all at high-luminosity running, no minijet veto for WBF
(cf. Sec.~\ref{sub:WBF}), and lack of progress in higher-order QCD
calculations for signals and backgrounds.  The reality is that
significant progress has been made regarding QCD corrections, and
we'll see one example shortly.  Also, everyone knows that the minijet
veto is a qualitatively correct aspect of the physics, we just can't
accurately predict its impact.  Early LHC data from $Zjj$ production
should take care of this.  Furthermore, ATLAS and CMS experimentalists
fully expect WBF to work at high-luminosity LHC running, they just
don't have full simulation results for the probable efficiencies.
Also, we may expect far better performance in the WBF $H\to\taus$
channels as discussed in Sec.~\ref{sub:WBF-taus}.  Finally, if new
physics exists up to a few TeV, it will be observable and we can take
it into accounts in Higgs loop-induced decays.

Now to QCD corrections.  Ref.~\cite{Duhrssen:2004cv} used large QCD
uncertainties for $\sigma_{gg\to H}$ and $\Gamma_g$, $20\%$ each,
which is the correct NNLO uncertainty for each by itself.  However,
these two quantities appear as a ratio in our observables formula,
Eq.~\ref{eq:NWA2}.  As pointed out in Ref.~\cite{Anastasiou:2005pd},
most of these uncertainties drop out in the ratio.  The reason for
this is that the QCD corrections to the cross section and partial
width are largely the same:
\ba
\Gamma\sim\alpha^2_s(\mu_R)C_1^2(\mu_R)[1+\alpha_s(\mu_R)X_1+...]
\\
\sigma\sim\alpha^2_s(\mu_R)C_1^2(\mu_R)[1+\alpha_s(\mu_R)Y_1+...]
\ea
The correct uncertainty on the ratio is $5\%$, which will have an
enormous impact on the fits of Fig.~\ref{fig:Hcoup}.  We eagerly await
new results from this and other improvements!

%%%%%%%%%%%%%%%%%%%%%%%%%%%%%%%%%%%%%%%%%%%%%%%%%%%%%%%%%%%%%%%%%%%%%%%%
%%%%%%%%%%%%%%%%%%%%%%%%%%%%%%%%%%%%%%%%%%%%%%%%%%%%%%%%%%%%%%%%%%%%%%%%

\subsection{Higgs couplings at an ILC}
\label{sub:ILC-coup}

Measuring Higgs couplings at an $e^+e^-$ collider would be far more
straightforward and rely on far fewer theoretical assumptions.
Between that and being a colorless collision environment, it would
also involve far fewer systematic uncertainties.  I'll outline the
basic idea.

\begin{figure}[ht!]
\includegraphics[scale=0.5]{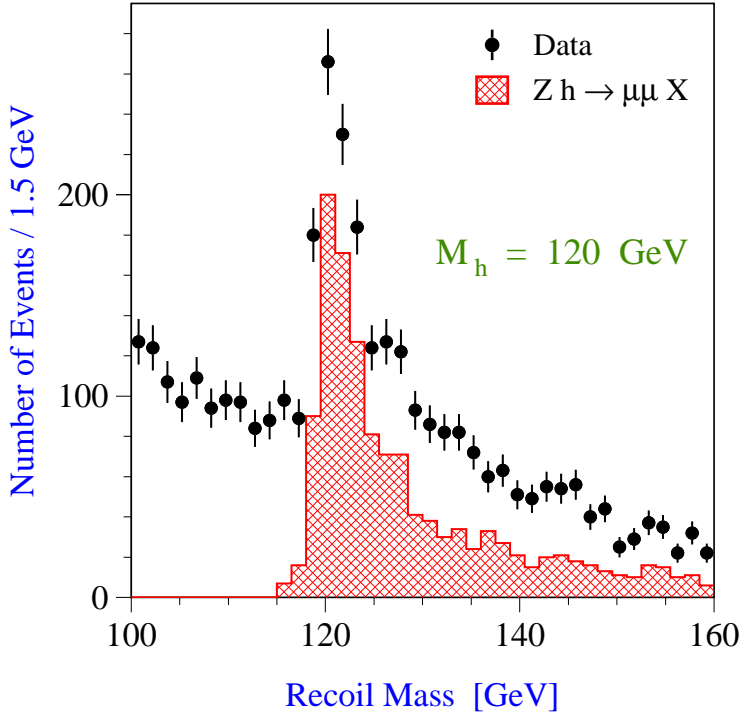}
\caption{Event rate of the recoil mass for $e^+e^-\to\mm+X$ at a
future high-energy linear collider.  $ZH$ production will fall into
this sample, but the Higgs decays are ignored, thus capturing the
total Higgs production rate.  (Figure modified from
Ref.~\protect\cite{Garcia-Abia:2000bk} for a public talk by one of the
authors.)}
\label{fig:recoil}
\end{figure}

In fixed-beam collisions it's possible to measure the {\it total} $ZH$
production rate.  To see this, we just apply a little relativistic
kinematics, rewriting the invariant $M_H^2$:
\bq\label{eq:recoil}
M^2_H = p^2_H = (p_+ + p_- - p_Z)^2 = s + M^2_Z - 2E_Z\sqrt{s}
\eq
We see that observing the Higgs and measuring its total rate boils
down to observing $Z$ bosons via their extremely sharp dimuon peak and
plotting this recoil mass.  Fig.~\ref{fig:recoil} shows what the
resulting event rate looks like in this distribution.  The Higgs peak
is clearly visible and sidebands allow one to subtract the SM
background in the signal region.  This captures all possible Higgs
decays, even though that aren't taggable or even identifiable, simply
by ignoring everything in the event except for the $Z$ dimuons.

Simulations~\cite{Aguilar-Saavedra:2001rg} suggest that the recoil
mass technique would allow for about a $2.5\%$ absolute measurement of
the $ZH$ rate.  Since the cross section depends on the $Z$--Higgs
coupling squared, the coupling uncertainty is then about a percent.

Getting from this one coupling and the total rate to any other
coupling is formulaic:
\begin{itemize}
\vspace{-2mm}
\item[1.] In the total rate, measure the best branching ratios, 
whatever they may be.  Depending on the mass and detector performance,
that's likely one of $\bb$, $\gamma\gamma$ or $\ww$ decays.
\vspace{-2mm}
\item[2.] Now look in WBF Higgs production\footnote{For a linear
collider this is both $e^+e^-\to e^+e^-H$ and $e^+e^-\to\nn H$, since
$e$ and $\nu$ are distinguishable.  Experimentally they become two
different analyses.} with the Higgs decaying to the same best final
state.  This yields the partial width $\Gamma_W$.
\vspace{-2mm}
\item[3.] Calculate the total Higgs width as $\Gamma_W/BR(H\to\ww)$.
\vspace{-2mm}
\item[4.] Any other measured BR now gives that individual partial
width, therefore the relevant coupling (or couplings for some
loop-induced decays).
\end{itemize}
Table~\ref{tab:ILC-coup} enumerates the results of ILC simulation for
select $M_H$~\cite{Abe:2001np}.  (Clearly more thorough work should be
done here.)  There are a few noteworthy features.  First, $H\to\bb$
would be accessible even as a rare BR at larger $M_H$, due to the
nearly QCD-free collision environment.  Second, a weak measurement of
$H\to\cc$ should be possible, for the same reason, and due to the
superior $b$ v. $c$ resolution of the next generation of collider
detectors.  Third, $H\to jj$ is also accessible.  This would be
attributed to $gg$, which is a mild theoretical assumption.  It is in
principle sanity-checkable by the absence of an anomalous high-$x$
Higgs production rate at LHC, which would come from sea or valence
quarks and a non-SM coupling to lighter fermions (which would be
difficult to accommodate theoretically, so not expected).
\begin{table}[ht!]
\begin{tabular}{|c|c|c|c|c|c|c|}
\hline
$M_H$ (GeV) & 120 & 140 & 160 & 180 & 200 & 220 \\
\hline
Decay & 
\multicolumn{6}{|c|}{Relative precision on $\Gamma_i$ ($\%$)} \\ 
\hline
$\bb$     &  1.9 &  2.6 &  6.5 & 12.0 & 17.0 & 28.0 \\
$\cc$     &  8.1 & 19.0 &      &      &      &      \\
$\taus$   &  5.0 &  8.0 &      &      &      &      \\
$gg$      &  4.8 & 14.0 &      &      &      &      \\ 
$\ww$     &  3.6 &  2.5 &  2.1 &      &      &      \\
$ZZ$      &      &      & 16.9 &      &      &      \\
$\gamma\gamma$
          & 23.0 &      &      &      &      &      \\
$Z\gamma$ &      & 27.0 &      &      &      &      \\
\hline
\end{tabular}
\caption{Estimated precision on various SM Higgs partial widths for a
few select values of $M_H$, from measurements at a future $e^+e^-$
collider~\protect\cite{Abe:2001np}.}
\label{tab:ILC-coup}
\end{table}

\begin{figure}[hb!]
\includegraphics[scale=1]{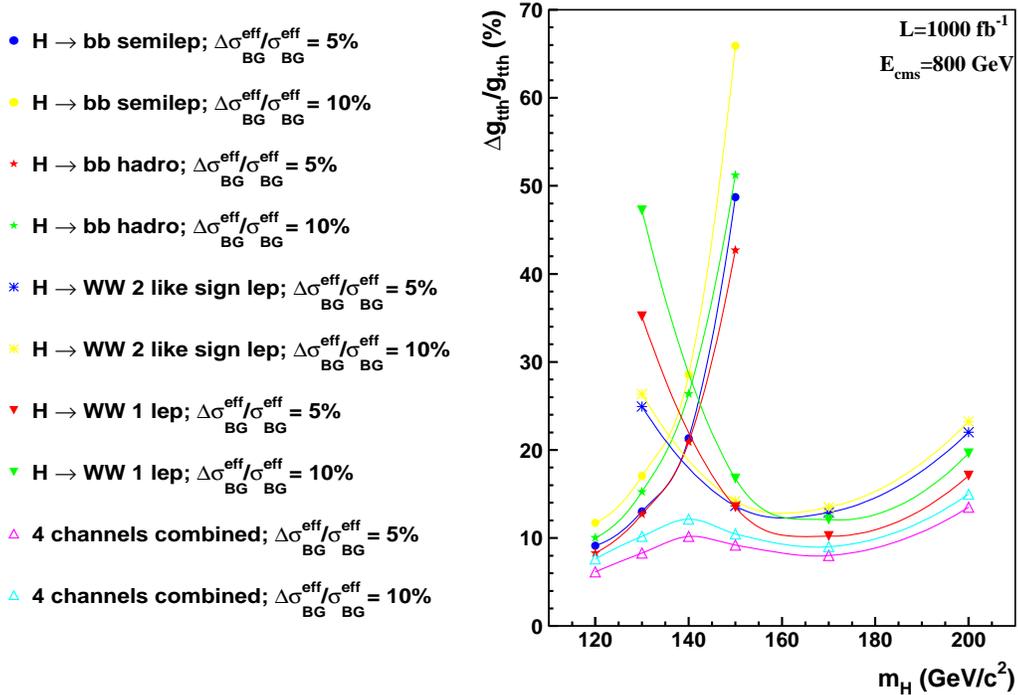}
\caption{Top Yukawa coupling measurement expectations for a future 
800~GeV $e^+e^-$ collider~\protect\cite{Gay:2006vs}.}
\label{fig:ttH}
\end{figure}

But what about the top Yukawa coupling?  Its anticipated value of
approximately one is curious enough to warrant special attention.  A
light Higgs can't decay to top quark pairs, so we'd have to rely on
top quark associated production, as at LHC but without all the nasty
QCD backgrounds.  However, the event rate is far lower than at LHC and
would require an 800~GeV machine collecting
1000~fb$^{-1}$~\cite{Juste:1999af}, the planned lifetime of a
next-generation second-stage machine (justifying my previous statement
about the drive to go to maximum energy and sit there).  One study
combined expected LHC and ILC
results~\cite{Desch:2004kf,Weiglein:2004hn}, and there are more recent
results for ILC, summarized in Fig.~\ref{fig:ttH}~\cite{Gay:2006vs}.
SLHC and an ILC would be complementary, granting superb coverage of
$M_H$ for a $y_t$ measurement at the $10\%$ level.

More sophisticated LC Higgs coupling analyses
exist~\cite{Hagiwara:2000tk}, but aren't often reviewed.  They use a
more complicated ``optimal observables'' (detailed kinematic shape
information, for example) scheme.  It's more powerful, but doesn't
lend itself to the simplistic formulaic approach I just discussed.

I should emphasize that the results I reviewed are relevant only for
the Standard Model.  If the Higgs sector is non-minimal, or any new
physics appears at the weak scale, it could result in altered
couplings (and usually does; see Chapter~\ref{sec:BSM}).  If they're
suppressed, the event rate goes down, resulting in greater
uncertainty.  This is often glossed over or ignored in discussions of
Higgs phenomenology, but is a potential reality and something we'd
just have to lump.  Nevertheless, it should be clear by now that an
ILC would be a spectacular experiment for precision Higgs
measurements.

%%%%%%%%%%%%%%%%%%%%%%%%%%%%%%%%%%%%%%%%%%%%%%%%%%%%%%%%%%%%%%%%%%%%%%%%
%%%%%%%%%%%%%%%%%%%%%%%%%%%%%%%%%%%%%%%%%%%%%%%%%%%%%%%%%%%%%%%%%%%%%%%%

\subsection{Higgs potential}
\label{sub:pot}

Finally we arrive at the most difficult Higgs property to test, the
potential.  This is the hallmark of spontaneous symmetry breaking,
thus ranks at least as high in priority as finding Yukawa couplings
proportional to fermion masses.  To see what's involved, let's review
the SM Higgs potential.  The potential is normally written as:
\bq\label{eq:pot}
V(\Phi) = \mu^2\Phi^\dagger\Phi + \lambda (\Phi^\dagger\Phi )^2
\eq
where $\Phi$ is our $SU(2)_L$ complex doublet of scalar fields.  The
Higgs spontaneous symmetry-breaking mechanism is what happens to the
Lagrangian when $\mu^2<0$ and the field's global minimum shifts to
$v=\sqrt{-\mu^2/\lambda}$.  We then expand $\Phi\to v+H(x)$ (ignoring
the Goldstone modes which you learned about in Sally Dawson's
lectures) where $H(x)$ is the radial excitation, the physical Higgs
boson.  The Higgs mass squared is then $2v^2\lambda$, and is the only
free parameter, although constrained (weakly) by EW precision fits.
The student performing this expansion will also notice $HHH$ and
$HHHH$ Lagrangian terms, which are self-interactions of the Higgs
boson.  The three- and four-point couplings are $-6v\lambda$ and
$-6\lambda$, respectively\footnote{Don't forget the identical-particle
combinatorial factors.}.

To measure the potential is to measure these self-couplings and check
their relation to the measured Higgs mass.  Our phenomenological
approach is to rewire the Higgs potential in terms of independent
parameters and the Higgs candidate field $\eta_H$:
\bq\label{eq:effpot}
V(\eta_H)
\; = \;
\frac{1}{2} \, M^2_H \, \eta_H^2
\, + \,
\lambda \, v \, \eta_H^3
\, + \, 
\frac{1}{4} \, \wt\lambda \, \eta_H^4
\eq
$\lambda$ and $\wt\lambda$ are now free parameters, which we measure
from the direct production rate of $HH$ and $HHH$ events.  This will
ultimately be a voyage of frustration.

%%%%%%%%%%%%%%%%%%%%%%%%%%%%%%%%%%%%%%%%%%%%%%%%%%%%%%%%%%%%%%%%%%%%%%%%

\subsubsection{$HH$ production at LHC}
\label{sub:HH-LHC}

\begin{figure}[ht!]
\includegraphics[scale=0.5]{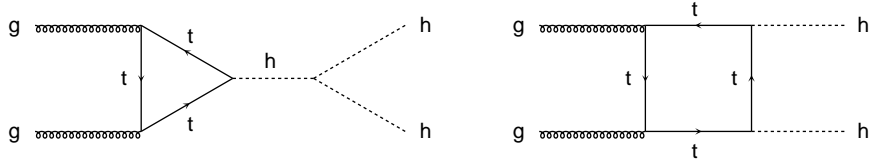}
\caption{Feynman diagrams for the dominant Higgs pair production rate
at LHC, $gg\to HH$.}
\label{fig:gg_HH}
\end{figure}

We begin with Higgs pairs at LHC.  The dominant production mechanism
is gluon fusion, $gg\to
HH$~\cite{Glover:1987nx,Plehn:1996wb,Dawson:1998py}.  The Feynman
diagrams are shown in Fig.~\ref{fig:gg_HH}.  The first diagram is
off-shell single Higgs production which split via the three-point
self-coupling to a pair of on-shell Higgses, which then decay
promptly.  The second diagram is a box (four-point)
\begin{figure}[ht!]
\includegraphics[width=8cm,height=6.3cm]{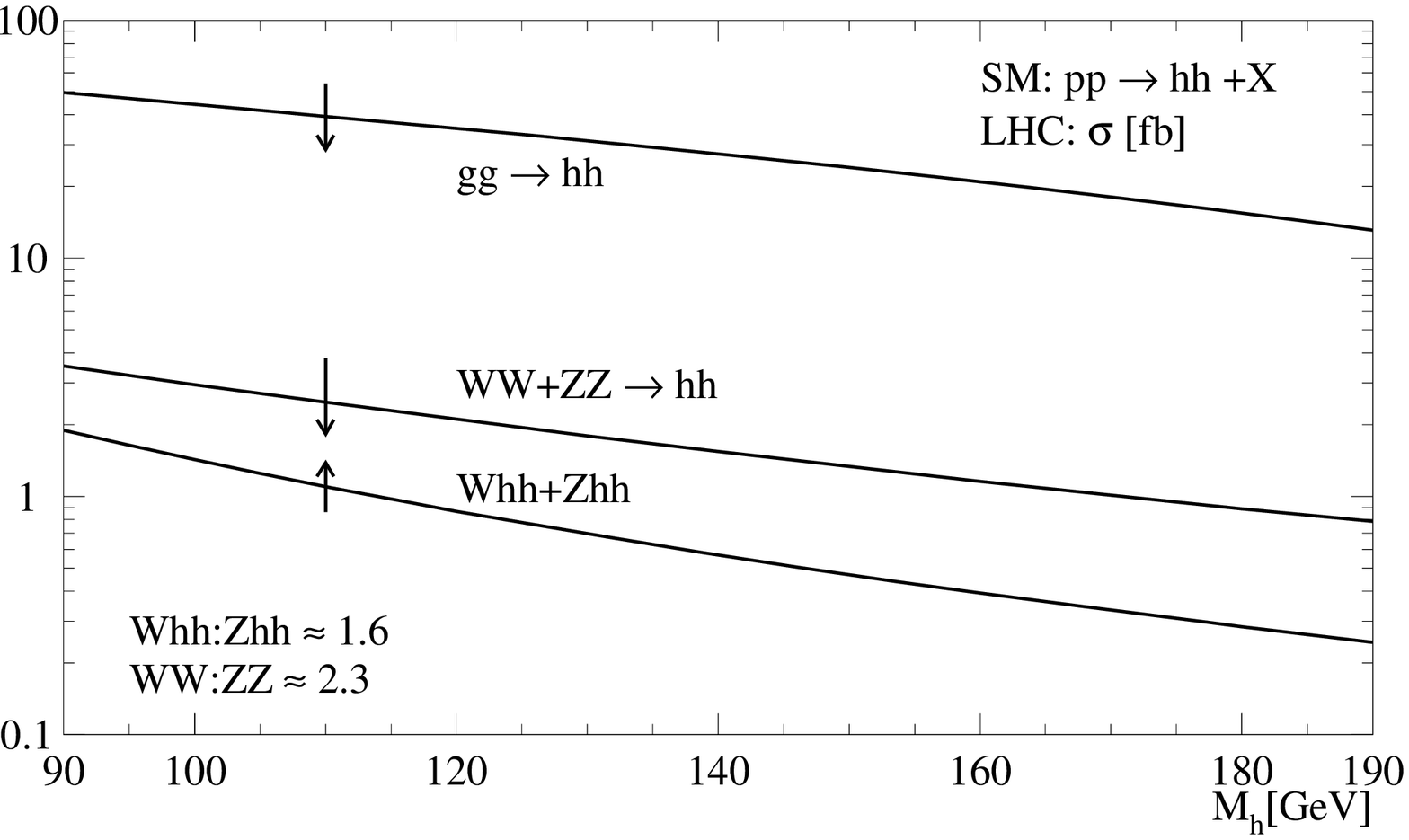}
\includegraphics[scale=0.45]{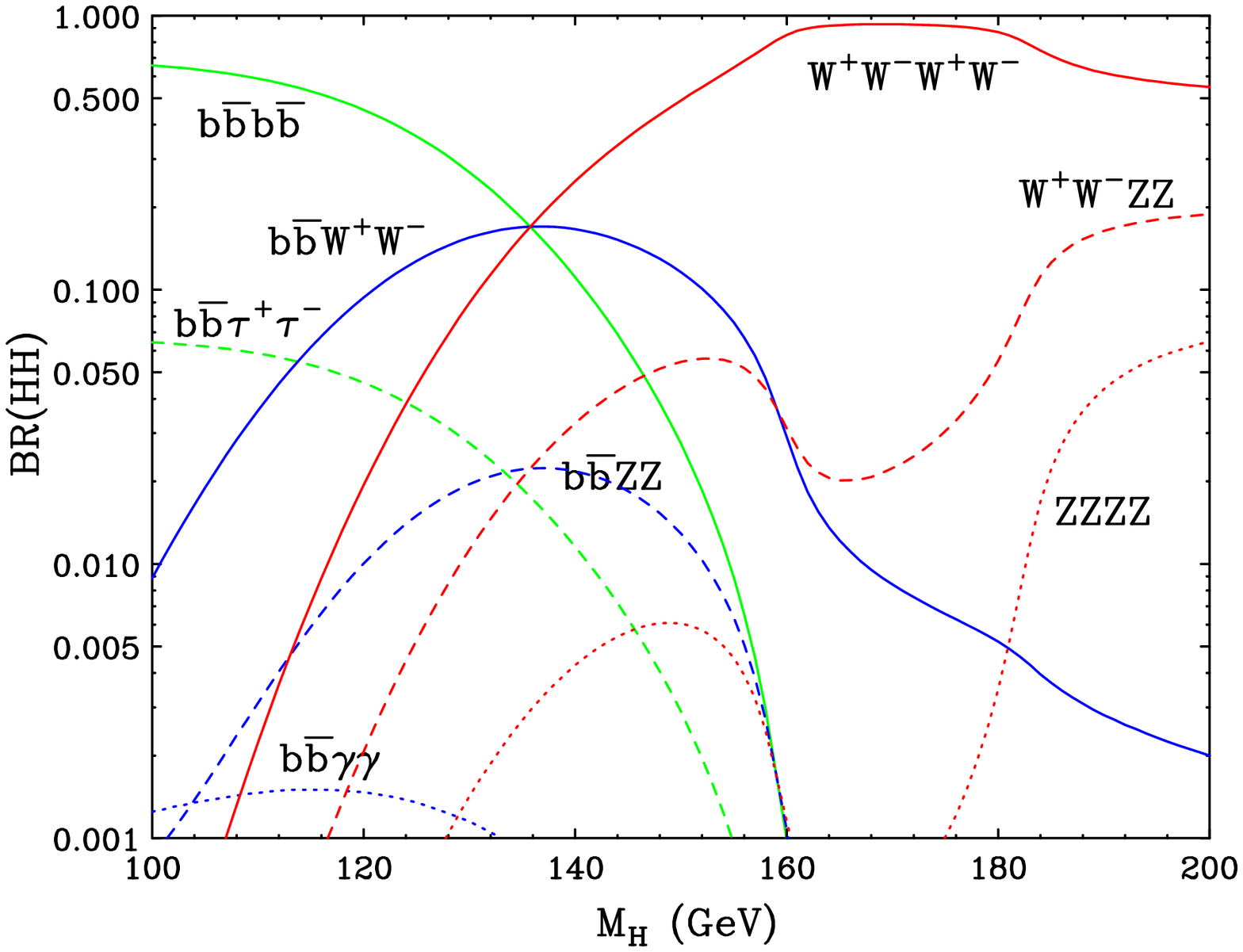}
\caption{Left: Higgs pair production cross sections at LHC as a
function of $M_H$~\protect\cite{Djouadi:1999rc}.  Arrows show the
change of the cross section as $\lambda$ is increase, and the tips are
at one-half and twice the SM value.  Right: Higgs pair branching
ratios as a function of $M_H$, calculated using {\sc
hdecay}~\protect\cite{Djouadi:1997yw}.}
\label{fig:HH-LHC}
\end{figure}
loop contribution which involves only the top quark Yukawa coupling.
Interestingly, the two diagrams interfere destructively and have a
rather large cancellation.  This means the rate is
small~\cite{Djouadi:1999rc}, as shown in Fig.~\ref{fig:HH-LHC}, making
our life difficult with a small statistical sample.  On the other
hand, the destructive interference will turn out to be crucial to
making constructive statements about the self-coupling $\lambda$.

The left panel of Fig.~\ref{fig:HH-LHC} tells us that we can expect
${\cal O}(10k)$ light Higgs pair events per detector over the expected
300~fb$^{-1}$ lifetime of the first LHC run, and ten times that at
SLHC.  That sounds like a lot, but keep in mind that both Higgses have
to decay to a final state we can observe, which will reduce the
captured rate to something much smaller.  Then we have to consider
what backgrounds affect each candidate channel.

The right panel of Fig.~\ref{fig:HH-LHC} shows selected Higgs pair
branching ratios.  At low mass, decays to $b$ pairs dominate, as
expected, while for $M_H\gtrsim135$~GeV mass it's $W$ pairs.  We can
immediately discount the $4b$ final state as hopeless, based on what
we already learned about QCD backgrounds -- but $4W$ is promising for
higher masses.  The next-largest mode from those two is $\bb\ww$,
which unfortunately is the same final state as the far larger top
quark pair cross section.  A few minutes' investigation causes this to
be discarded, even after trying various invariant mass constraints;
$b$ pair mass resolution is just not good enough.  The $\bb\taus$ mode
has very low backgrounds, comparable to the signal, but suffers hugely
from lack of statistics, due to low efficiency for subsequent tau
decays.  However, the rare decay mode $\bb\gamma\gamma$ is extremely
clean and worth further consideration at low masses.

%%%%%%%%%%%%%%%%%%%%%%%%%%%%%%%%%%%%%%%%%%%%%%%%%%%%%%%%%%%%%%%%%%%%%%%%

\bigskip
\underline{$HH\to\ww\ww$ at LHC}
\label{sub:HH-LHC-4W}
\medskip

$HH\to\ww\ww$ has myriad decays, but for triggering purposes and to
get away from QCD background sources of leptons (like top quarks) we
need to select special multilepton final states~\cite{BPR_HH_4W}.  The
most likely accessible channels are same-sign lepton pairs,
$\lpm\lpm+4j$, and three leptons, $\ells\lpm+2j$, since the principal
QCD SM backgrounds can't easily mimic them.  Note that because of
multiple neutrinos departing the detector unobserved, complete
reconstruction is not possible.  The principle backgrounds are
$WWWjj$, $t\bar{t}W$, $t\bar{t}j$, $t\bar{t}Z/\gamma^*$ and $WZ+4j$,
but we also need to consider $\tops\tops$, $4W$, $\ww+4j$, $\ww Zjj$
as well as double parton scattering and overlapping events.  The
calculation of all of these is technical so I won't go into it, rather
simply mention a few noteworthy points.

\begin{figure}[hb!]
\includegraphics[scale=0.72]{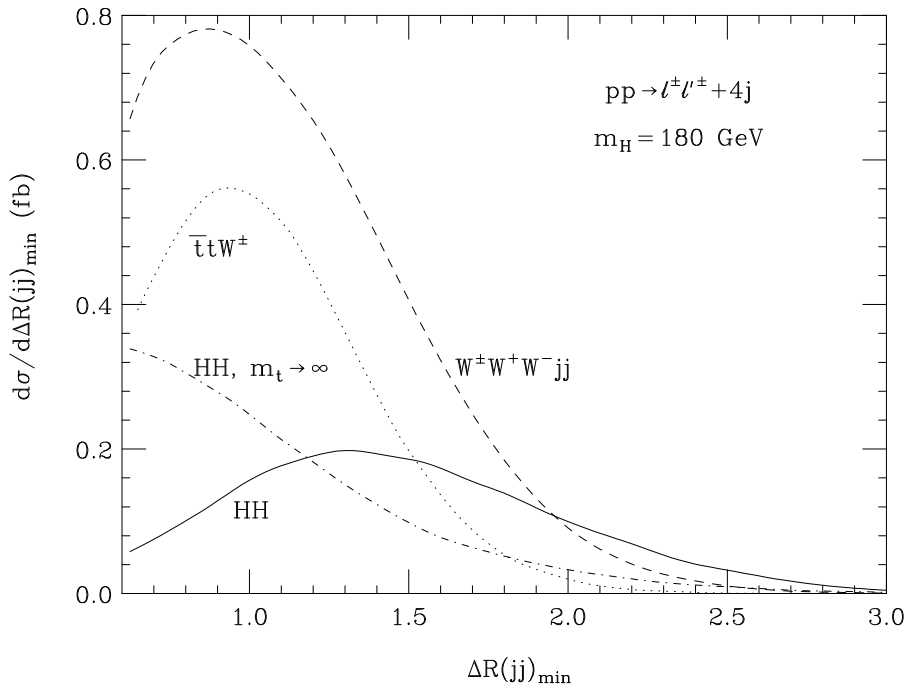}
\caption{Differential cross section as a function of the minimum jet
pair lego plot separation for $\ells+4j$ at events at LHC.  The solid
curve is the correct distribution using exact matrix elements for
$HH$, while the dash-dotted curve comes from effective-Lagrangian
matrix elements where the top quark mass is taken to infinity.  Figure
taken from Ref.~\protect\cite{BPR_HH_4W}.}
\label{fig:HH-dRjj}
\end{figure}

The first is a warning about using the $gg\to HH$ effective Lagrangian
in practical calculations.  It is still a mystery why the leading term
in the $\sqrt{\hat{s}}/m_t$ expansion~\cite{Glover:1987nx} should get
the overall rate so close that of an exact
calculation~\cite{Plehn:1996wb}, but it does.  Because of that, nobody
has ever bothered to calculate higher-order terms in the effective
Lagrangian expansion; in any case, the exact results are available, as
well as NLO in QCD~\cite{Dawson:1998py}.  However, the leading terms
in the expansion cancel too much close to threshold, yielding
incorrect kinematics~\cite{BPR_HH_4W}, as can be seen from
Fig.~\ref{fig:HH-dRjj}.  One should thus use only the exact matrix
element results for practical $gg\to HH$ phenomenology.

The second point is that our main systematic uncertainties will be our
limited knowledge of the top quark Yukawa coupling, which drives the
production rate, and the BR to $\ww$, which drive the decay fraction.
These must be known very precisely for any measurement to be useful.

We will need a discriminating observable to separate signal from
background.  We can speculate that nearly all the signal's kinematic
information is encoded in the invariant mass of the visible final
state particles, so let's construct a new variable, $m_{vis}$:
\bq\label{eq:mvis}
m^2_{vis} \; = \;
\left[ \sum_{i}E_i \right]^2- \left[ \sum_{i}\mathbf p_i \right]^2
\eq
where $i$ are all the leptons and jets in the event.  We suspect a
difference because the signal is a two-body process, which is
threshold-like, while the backgrounds are multi-body processes which
peak at much larger $m_{vis}$ than the sum of their heavy resonances'
masses.

\begin{figure}[hb!]
\includegraphics[scale=0.72]{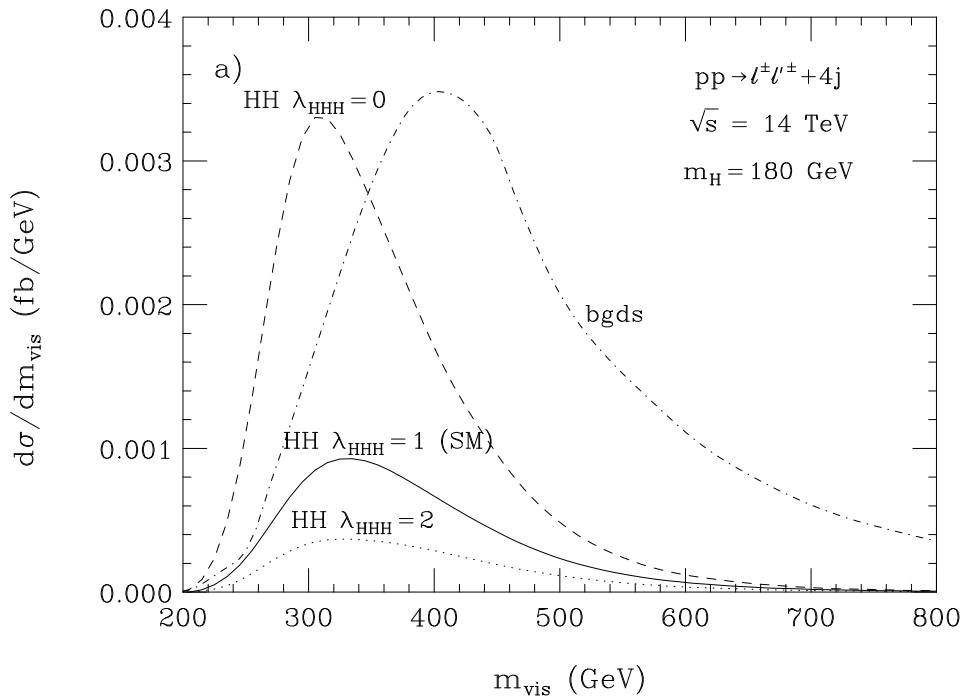}
\caption{Visible invariant mass distribution for same-sign dilepton
plus four jet event at LHC~\protect\cite{BPR_HH_4W}.  All SM
backgrounds are summed into one curve, while the $gg\to HH$ signal is
shown separately, for the SM value of self-coupling $\lambda$, twice
that value, and zero.}
\label{fig:HH-mvis}
\end{figure}

Fig.~\ref{fig:HH-mvis} displays the fruits of parameterizing our
ignorance (or rather, the detector's).  The separation between signal
and background is exactly as expected: the signal peaks much lower,
allowing a $\chi^2$ fit to distinguish it from the backgrounds.  But
the plot also reveals a saving grace in the destructive interference
between triangle and box loop diagrams.  If spontaneous symmetry
breaking isn't the right description and there is no Higgs potential,
then $\lambda=0$ and the lack of destructive interference gives a
wildly larger signal cross section, which is far easier to observe.

\begin{figure}[ht!]
\includegraphics[scale=0.75]{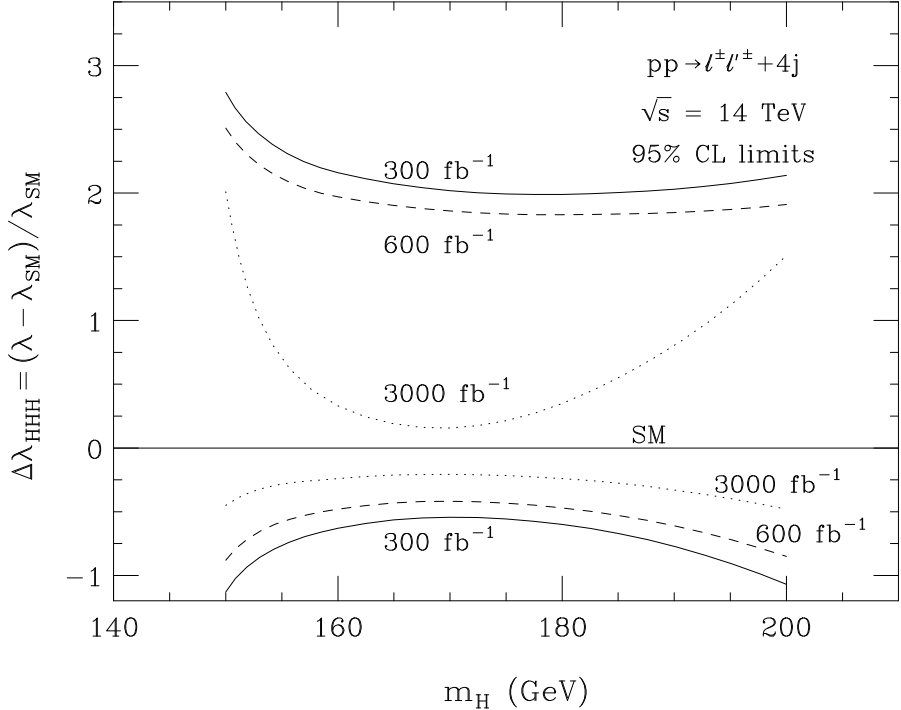}
\caption{$95\%$~CL limits achievable at LHC on the shifted Higgs triple
self-coupling (see text), $\triangle\lambda$, for LHC and SLHC
expected luminosities~\protect\cite{BPR_HH_4W}.}
\label{fig:LHC-HH-4W-fit}
\end{figure}

Fig.~\ref{fig:LHC-HH-4W-fit} summarizes the results of
Ref.~\cite{BPR_HH_4W}.  It plots $95\%$~CL limits on the shifted
self-coupling, $\triangle\lambda=(\lambda-\lambda_{\rm
SM})/\lambda_{\rm SM}$.  This is somewhat easier to understand: zero
is the SM, and -1 corresponds to no self-coupling, or no potential.
For $M_H>150$~GeV, the LHC can exclude $\lambda=0$ at (for some $M_H$
much greater than) $2\sigma$ with only the LHC.  After SLHC running,
this becomes a $20-30\%$ measurement, if other systematics are under
control.  Here, they're assumed to be smaller than the statistical
uncertainty.


Another potential systematics issue is minimum bias, the presence of
extra jets in an event which don't come from the primary hard
scattering.  Here, they could be confused with jets from the $W$
bosons, causing a distortion of $m_{vis}$.  ATLAS has investigated
this and found it to not be a concern -- the shape of $m_{vis}$ for
the signal remains largely unaltered~\cite{Dahlhoff}.

%%%%%%%%%%%%%%%%%%%%%%%%%%%%%%%%%%%%%%%%%%%%%%%%%%%%%%%%%%%%%%%%%%%%%%%%

\bigskip
\underline{$HH\to\bb\gamma\gamma$ at LHC}
\label{sub:HH-LHC-bbaa}
\medskip

We've already ruled out as viable the vast majority of Higgs pair BRs
for $M_H\lesssim150$~GeV due to QCD backgrounds or too-small
efficiencies.  However, the rare decay mode to $\bb\gamma\gamma$ is
worth a closer look~\cite{Baur:2003gp}.  There are many backgrounds to
consider, coming from $b$ or $c$ jets plus photons, or other jets
which fake photons, just as in the single Higgs to photon pairs case.
Table~\ref{tab:eff} highlights the major ID efficiencies and fake
photon rejection factors at LHC and SLHC relevant for us.  The
backgrounds are all calculable at LO, but with significant
uncertainties, probably a factor of two or more.  However, that won't
be a concern as we can identify distributions useful for measuring the
background in the non-signal region.  Note that with this channel we
can completely reconstruct both Higgs bosons.

\begin{table}[hb!]
\begin{tabular}{|c|cccccc|}
\hline
\phantom{i} & $\epsilon_\gamma$ & $\epsilon_\mu$
& $P_{c\to b}$ & $P_{j\to b}$
& $P^{hi}_{j\to\gamma}$ & $P^{lo}_{j\to\gamma}$ \\[1mm]
\hline
LHC  & $80\%$ & $90\%$ & 1/13 & 1/140 & 1/1600 & 1/2500 \\
SLHC & $80\%$ & $90\%$ & 1/13 & 1/23  & 1/1600 & 1/2500 \\
%VLHC & $80\%$ & $90\%$ & 1/13 & 1/140 & 1/1600 & 1/2500 \\
\hline
\end{tabular}
\caption{The major ID efficiencies and fake photon rejection factors
at LHC.  Note the two values for $P_{j\to\gamma}$, which represent the
current uncertainty in detector capability for fake photon rejection.
The true value won't be known until data is collected.  See
Ref.~\protect\cite{Baur:2003gp} for details.}
\label{tab:eff}
\end{table}
\begin{figure}[ht!]
\includegraphics[scale=0.5]{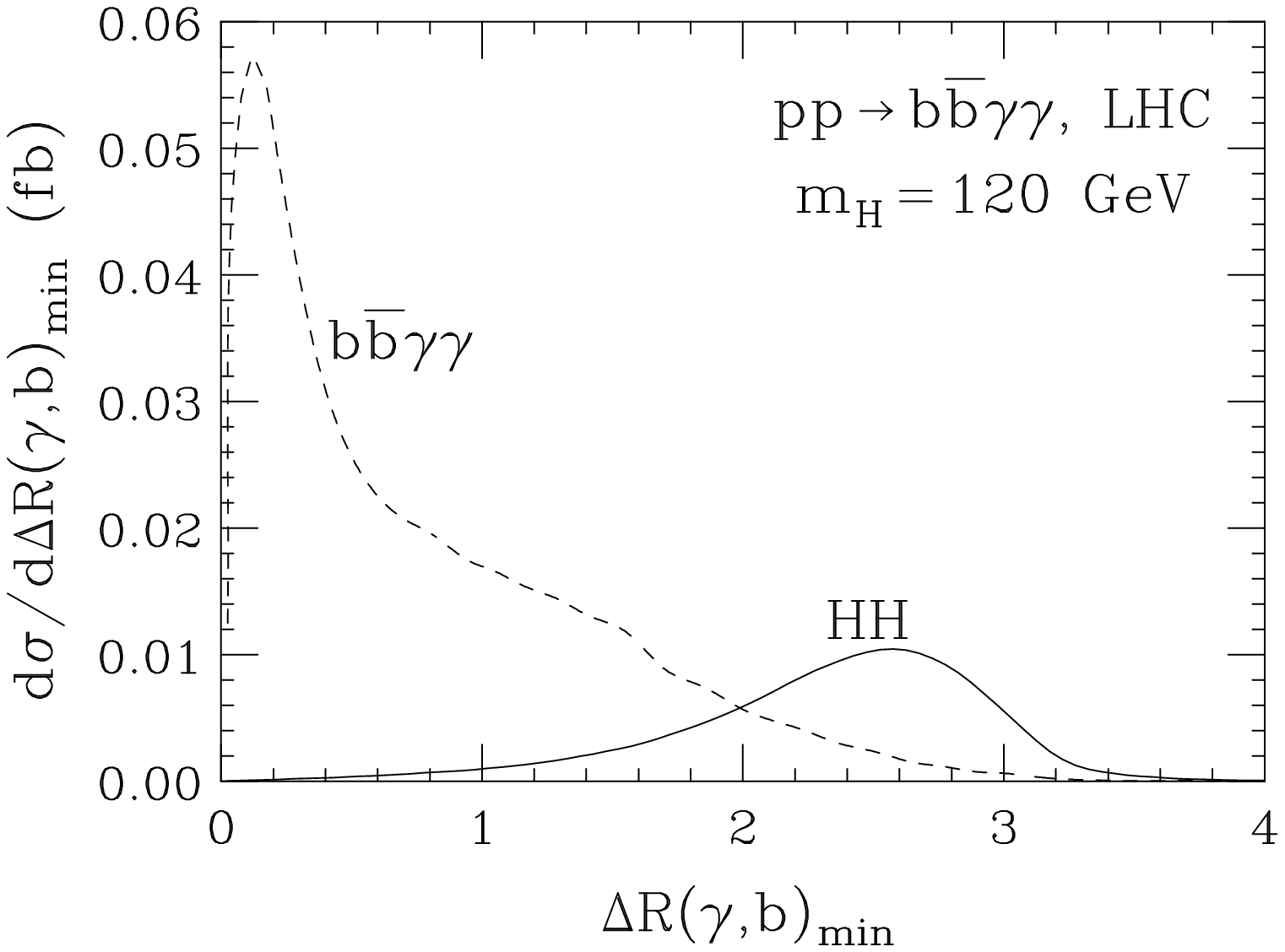}
\hspace*{2mm}
\includegraphics[scale=0.5]{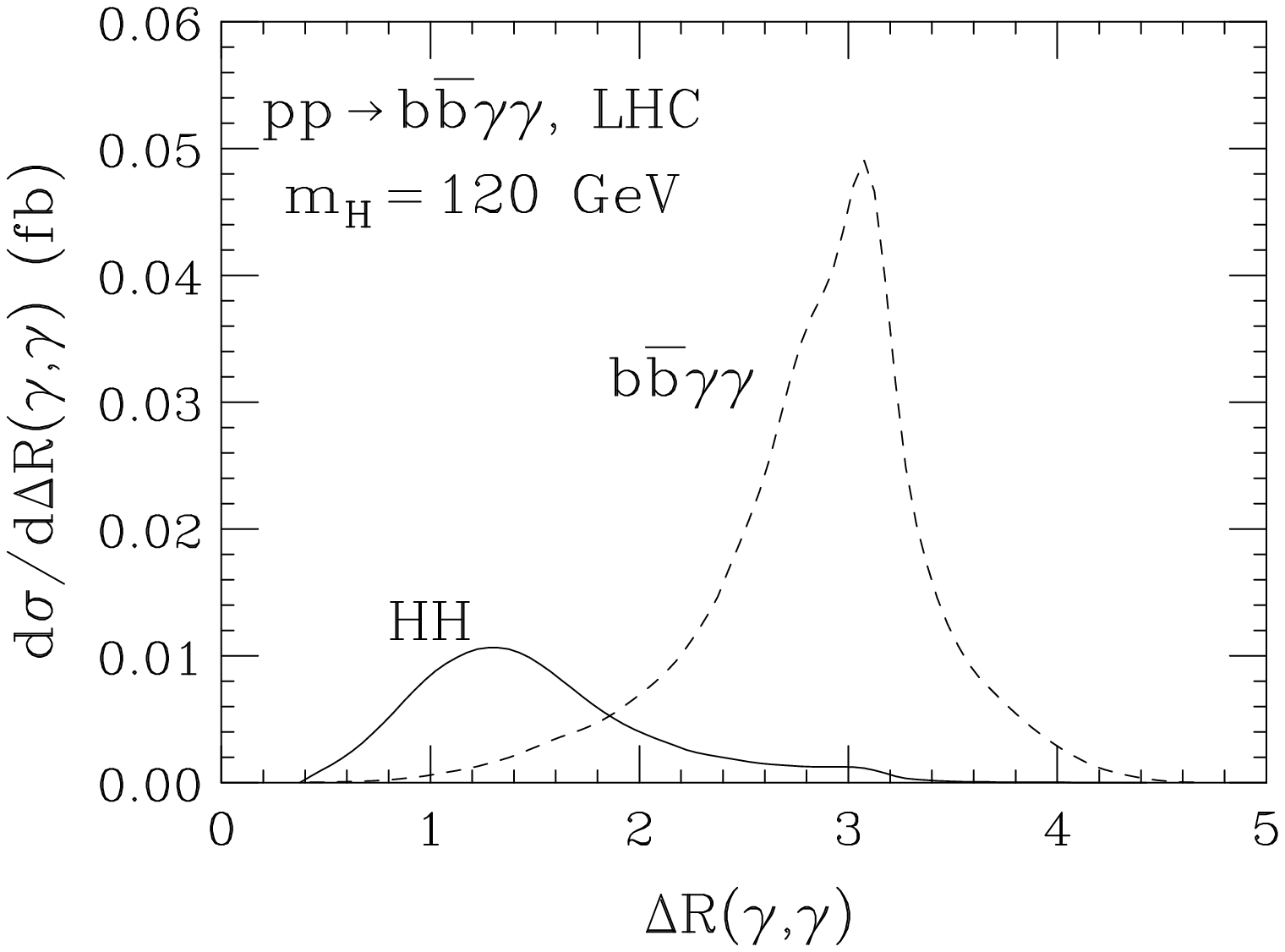}
\vspace*{-2mm}
\caption{Angular separations in the lego plot for $b$ jets and photons
in $gg\to HH\to\bb\gamma\gamma$ signal events and background at the
LHC.  Figures from Ref.~\protect\cite{Baur:2003gp}.}
\label{fig:2dists}
\end{figure}

The background QCD uncertainties have a work-around.  There are two
angular distributions in the lego plot which look very different for
the signal, principally because scalars decay isotropically and thus
are uncorrelated, while the QCD backgrounds have spin correlations.
The two distributions are shown in Fig.~\ref{fig:2dists}.  The
differences are rather dramatic (and even more so in 2-D
distributions).  Tevatron's experiments CDF and D\O\ have used such a
pseudo-sideband analysis for some time to measure a background in a
non-signal region to normalize their Monte Carlo tools, then
extrapolating to the signal region to perform a background
subtraction.  The technique is viable because QCD radiative
corrections {\it in general} do not significantly alter angular
distributions.

Table~\ref{tab:bbaa} summarizes the results of
Ref.~\cite{Baur:2003gp}.  It gives event rates expected with
600(6000)~fb$^{-1}$ of data (two detectors) at LHC(SLHC).  SLHC would
not get ten times as many events because of lower efficiency of having
to tag two $b$ jets instead of only one, to overcome the low fake jet
rejection rate in a high-luminosity environment.  First, note that
fake $b$ jets or fake photons are the largest background: the
measurement would be significantly hampered by detector limitations.
Second, while the $S/B$ ratio is excellent, the overall event rate is
extremely small, definitely in the non-Gaussian statistics regime.

SLHC could make a useful statement about $\lambda$, ultimately
achieving limits on $\triangle\lambda$ of about $\pm0.5$, but this is
not such a strong statement.  It could at best generally confirm the
SM picture of spontaneous symmetry breaking and perhaps rule out
wildly different scenarios, but would never be particularly
satisfying.  On the other hand, it's strong encouragement for ATLAS
and CMS to push the envelope on tagging efficiency and fake rejection,
especially for the detector upgrades necessary for SLHC.  Doing
studies like this well ahead of time is useful for this reason, our
present case being a perfect example.

\begin{table}[hb!]
\begin{tabular}{|c|c|c|c|c|c|c|c|c|c|c|c|c|}
\hline
& $HH$
& $\bb\gamma\gamma$ & $\cc\gamma\gamma$
& $\bb\gamma j$ & $\cc\gamma j$ & $jj\gamma\gamma$
& $\bb jj$ & $\cc jj$ & $\gamma jjj$ & $jjjj$ 
& $\sum ({\rm bkg})$ & S/B \\
\hline
\hline
LHC  &  6 &  2 &  1 &  1 &  0 &  5 &  0 &  0 &  1 &  1 & 11 & 1/2 \\
\hline
\hline
SLHC & 21 &  6 &  0 &  4 &  0 &  6 &  1 &  0 &  1 &  1 & 20 & 1/1 \\
%\hline
%VLHC & 486 & 40 & 70 & 60 & 29 & 137 & 30 & 6 & 36 & 40 & 448 & 1/1\\
\hline
\end{tabular}
\caption{Expected event rates after ID efficiencies and all kinematic
cuts for $\bb\gamma\gamma$ events at LHC (SLHC), two detectors and
600(6000)~fb$^{-1}$ of data~\protect\cite{Baur:2003gp}.  LHC assumes
only one $b$ tag, while SLHC requires two.  Note the increased fake
rate at SLHC.}
\label{tab:bbaa}
\end{table}
%

%%%%%%%%%%%%%%%%%%%%%%%%%%%%%%%%%%%%%%%%%%%%%%%%%%%%%%%%%%%%%%%%%%%%%%%%

\subsubsection{$HH$ production at an ILC}
\label{sub:HH-ILC}

While (S)LHC clearly has access to Higgs pair production and thus
$\lambda$ for $M_H>150$~GeV, it would disappoint at lower masses.  We
should see if a future linear collider could also give a precision
measurement for $\lambda$ as it could for (most) other Higgs
couplings.

\begin{figure}[hb!]
\includegraphics[scale=1]{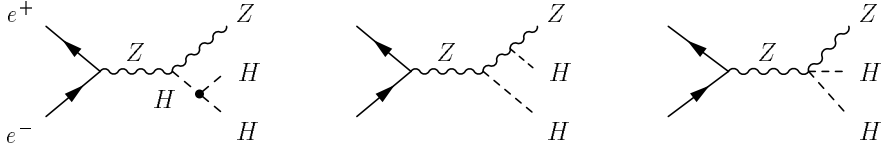}
\caption{Feynman diagrams for double Higgsstrahlung at a future linear
collider, $e^+e^-\to HH$.}
\label{fig:ee_HH}
\end{figure}

\begin{figure}[hb!]
\includegraphics[width=7.5cm,height=6cm]{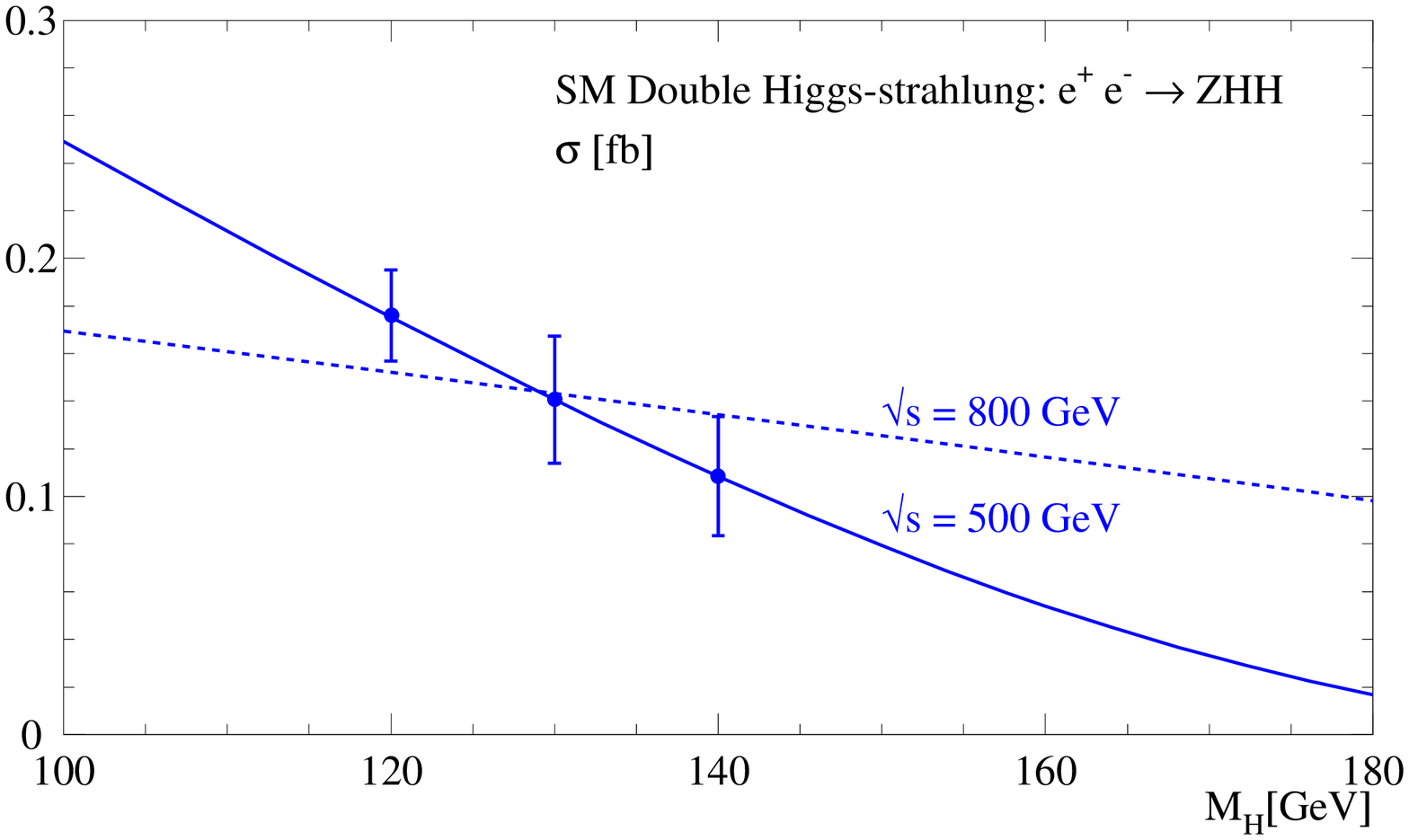}
\includegraphics[scale=0.5]{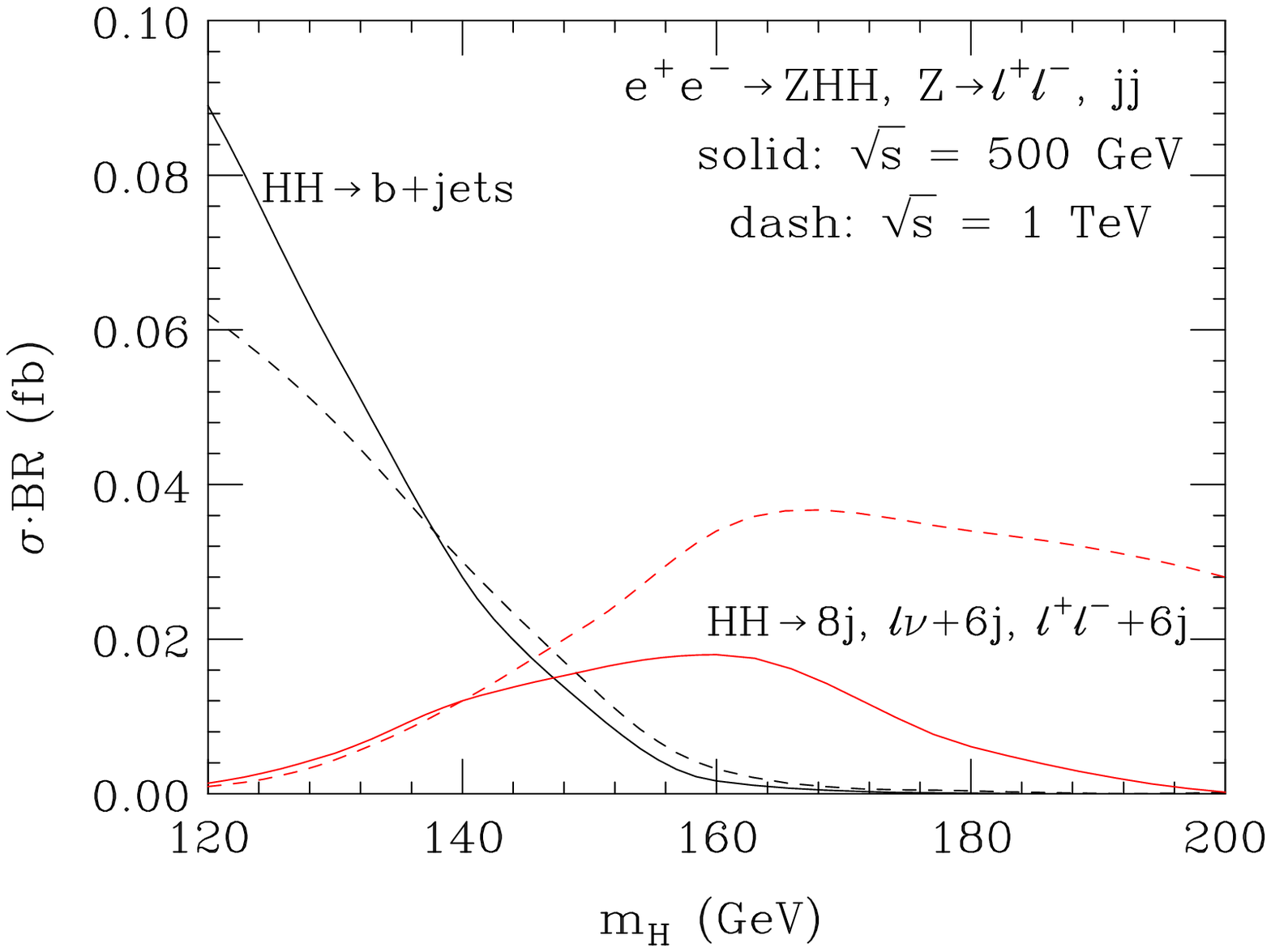}
\caption{Left: the double Higgsstrahlung cross section as a function
of $M_H$ for 500 and 800~GeV $e^+e^-$
collisions~\protect\cite{Djouadi:1999gv}.  Right: the cross section
times BR at 500~GeV and 1~TeV $e^+e^-$ collisions, for the dominant
final state BRs as a function of $M_H$~\protect\cite{Baur:compare}.}
\label{fig:HH-ILC}
\end{figure}

For $e^+e^-$ collisions below about 1~TeV, double Higgsstrahlung is
the largest source of Higgs pairs.  The Feynman diagrams appear in
Fig.~\ref{fig:ee_HH}, while the cross sections as a function of $M_H$
for 500 and 800~GeV collisions~\cite{Djouadi:1999gv} are found in
Fig.~\ref{fig:HH-ILC}, which also shows the cross sections times BRs
for the dominant final states over the range of Higgs masses.
Roughly, this corresponds to $4b$ and $4W$ final state.  The former is
very steeply falling with $M_H$, but the latter is much flatter over
the 100--200~GeV mass region, suggesting broader access if at all
visible.

\begin{figure}[ht!]
\includegraphics[scale=0.75]{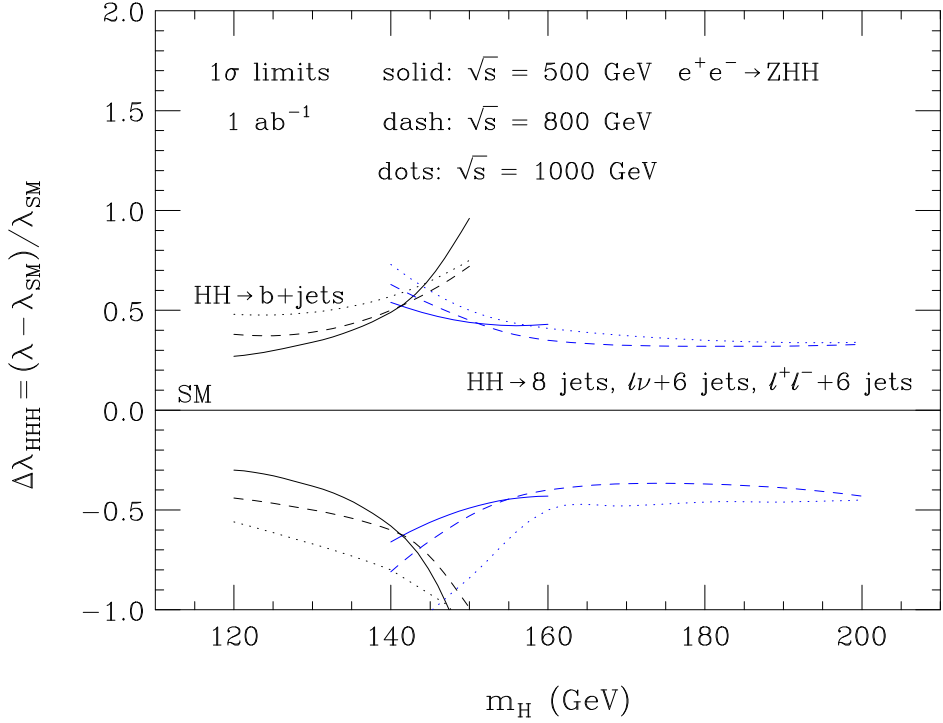}
\caption{Estimated achievable limits in the shifted self-coupling
$\triangle\lambda$ (see Sec.~\protect\ref{sub:HH-LHC}) at future
$e^+e^-$ colliders of various energy, as a function of
$M_H$~\protect\cite{Baur:compare}.}
\label{fig:HH-ILC-limits}
\end{figure}

The parton-level studies performed so far~\cite{Baur:compare} are
fairly encouraging.  As shown in Fig.~\ref{fig:HH-ILC-limits}, an ILC
could achieve about a $20-30\%$ measurement of $\lambda$ over a broad
mass range, with somewhat worse performance around $M_H\sim140$~GeV,
where the $\bb$ and $\ww$ BRs are roughly equal.  Interestingly, for a
lower Higgs mass, the analysis prefers lower machine energy, while the
opposite is true at least to a small degree at higher mass.  This is
largely a phase space effect for the 3-body production mechanism.
Also, SLHC is superior for $M_H\gtrsim150$~GeV (largely due to better
statistics), with an important caveat: controlling systematics in
$gg\to HH\to 4W$ at LHC would require precision input from ILC for the
Higgs couplings and BRs.  This is an excellent example of synergy
between experiments.


\begin{figure}[hb!]
\includegraphics[scale=0.85,angle=270]{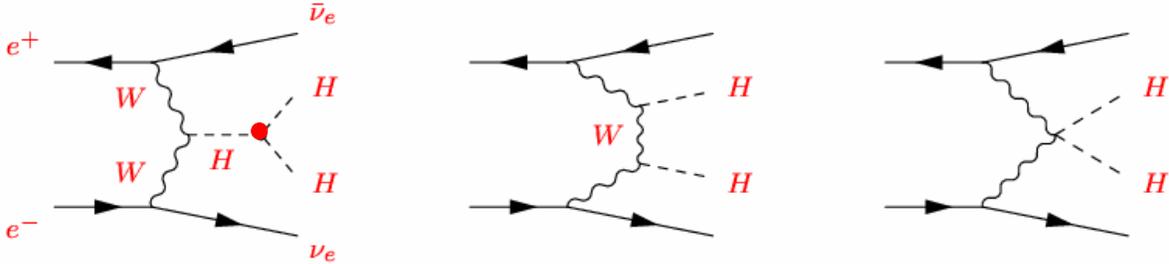}
\caption{Representative Feynman diagrams for the WBF process
$e^+e^-\to\nn HH$.}
\label{fig:ee_vvHH}
\end{figure}

\begin{figure}[ht!]
\includegraphics[scale=0.7]{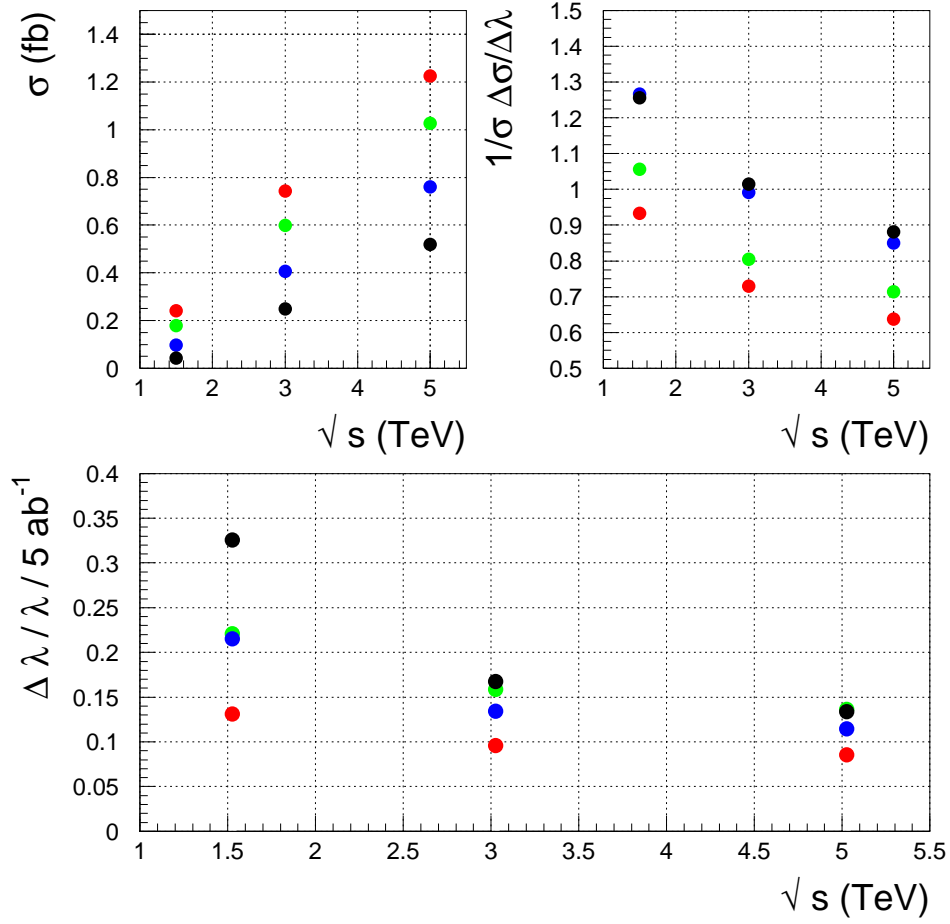}
\vspace*{-2mm}
\caption{The results of Ref.~\protect\cite{DeRoeck:2004zu} for WBF $HH$
production at CLIC, a second-generation multi-TeV $e^+e^-$ collider.
The plot labels are self-explanatory, while the colors are for various
Higgs masses: 120~GeV in red, 140~GeV in blue, 180~GeV in green and
240~GeV in black.}
\label{fig:HH-CLIC}
\end{figure}

Double Higgsstrahlung is not the only source of Higgs pairs at an
$e^+e^-$ collider, however.  In fact, as the energy increases, WBF
Higgs pair production becomes more and more important.  Representative
Feynman diagrams for $e^+e^-\to\nn HH$ are shown in
Fig.~\ref{fig:ee_vvHH}.  A preliminary analysis~\cite{DeRoeck:2004zu}
for CLIC~\cite{Accomando:2004sz}, a second-generation $1-5$~TeV
$e^+e^-$ collider collecting 5000~fb$^{-1}$, found rather interesting
results, summarized graphically in Fig.~\ref{fig:HH-CLIC}.  The
principal finding is that no matter how high the collision energy
goes, and regardless of Higgs mass, the precision on $\lambda$ bottoms
out at $10-15\%$.  This is because the self-coupling has an
$s$-channel suppression, and its contributions becomes washed out as
by other diagrams as $\sqrt{s}$ increases.  A corollary, though, is
that CLIC could potentially achieve better precision than SLHC for
larger $M_H$, although this may be marginal.  Much more detailed work
would be required for both SLHC and CLIC, as well as experience at LHC
and SLHC to determine its true potential, to make conclusive
statements.

%

%%%%%%%%%%%%%%%%%%%%%%%%%%%%%%%%%%%%%%%%%%%%%%%%%%%%%%%%%%%%%%%%%%%%%%%%

\subsubsection{Electroweak corrections to $\lambda$}
\label{sub:HH_EWcorr}

One final word on the trilinear self-coupling $\lambda$:
Ref.~\cite{Kanemura:2002vm} calculated the leading 1-loop top quark EW
corrections to $\lambda_{\rm SM}$.  Their principal SM result is:
\bq\label{eq:L3H_EW}
\lambda^{eff}_{HHH} \; = \; \frac{M^2_H}{2v^2} 
\left[ 1 - \frac{N_C}{3\pi^2}\frac{m^4_t}{v^2M^2_H} + ... \right]
\eq
The correction is $-10\%$($-4\%$) for $M_H=120(180)$~GeV, non-trivial
for smaller Higgs masses, but those are excluded in the SM.  This
correction should obviously be taken into account in any future
analysis, should the Higgs be found.  But it should be clear that
neither (S)LHC nor ILC will be sensitive to it.  Even CLIC would have
only marginal sensitivity, and then only for low $M_H$.

Non-minimal Higgs sectors and new physics effects can tell a very
different story, however, as we'll see, coming up in
Secs.~\ref{sub:HDO}~and~\ref{sub:MSSM-pot}.

%%%%%%%%%%%%%%%%%%%%%%%%%%%%%%%%%%%%%%%%%%%%%%%%%%%%%%%%%%%%%%%%%%%%%%%%

\subsubsection{$HHH$ production anywhere}
\label{sub:HHH}

The trilinear self-coupling $\lambda$ is only part of our
phenomenological Higgs potential of Eq.~\ref{eq:effpot}, though.  We
also need to measure $\wt\lambda$, the quartic self-coupling.  In some
sense this is equally important to measuring $\lambda$.  Recall the
structure of the Higgs potential: $\lambda$ allows the global minimum
to be away from zero, but a non-zero (and positive) $\wt\lambda$ is
required to keep the potential bounded from below.  We can't really
convince ourselves that the potential structure of Eq.~\ref{eq:pot} is
the right picture without a measurement of both these ingredients.
We've just seen that probing $\lambda$ is extremely challenging.  Just
how difficult is this likely to be for $\wt\lambda$?

\begin{figure}[hb!]
\includegraphics[scale=0.35]{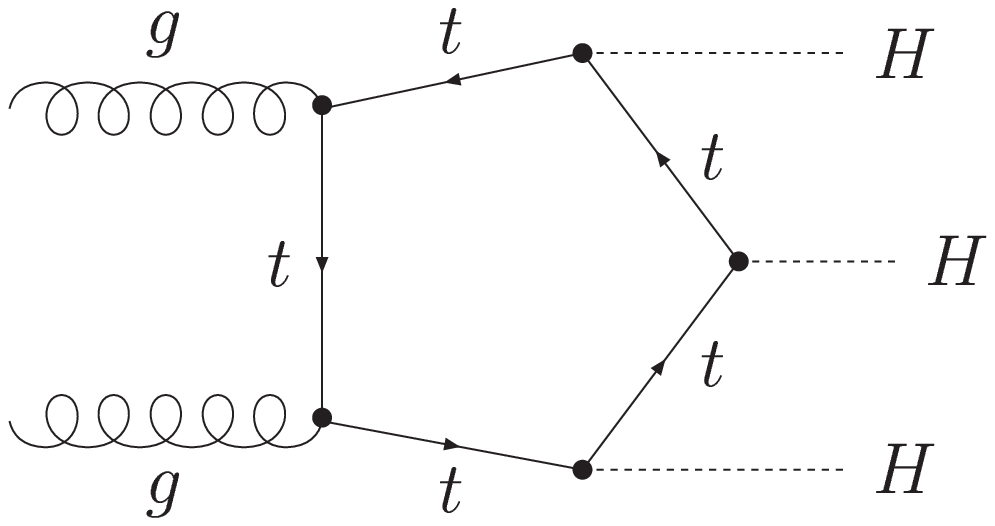}
\hspace*{2mm}
\includegraphics[scale=0.35]{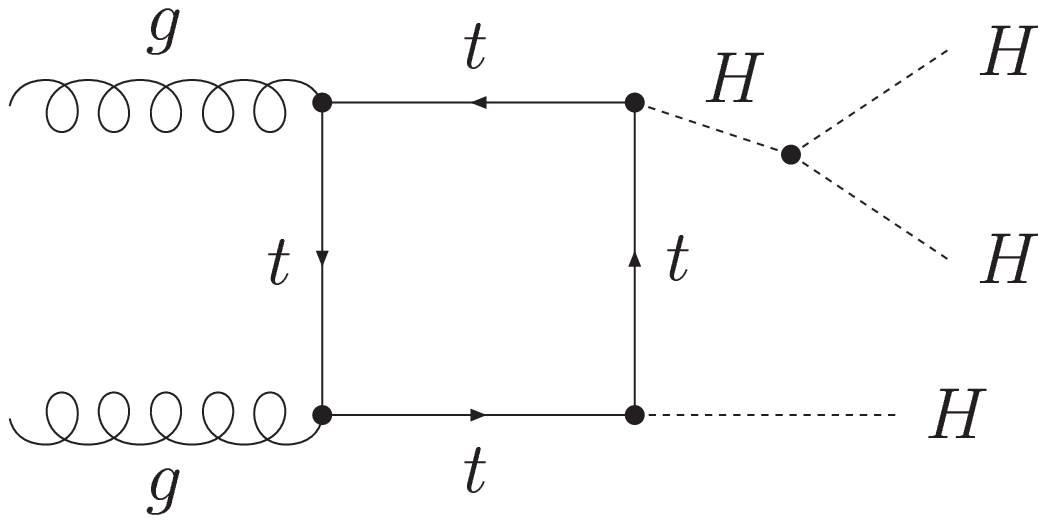}
\hspace*{2mm}
\includegraphics[scale=0.35]{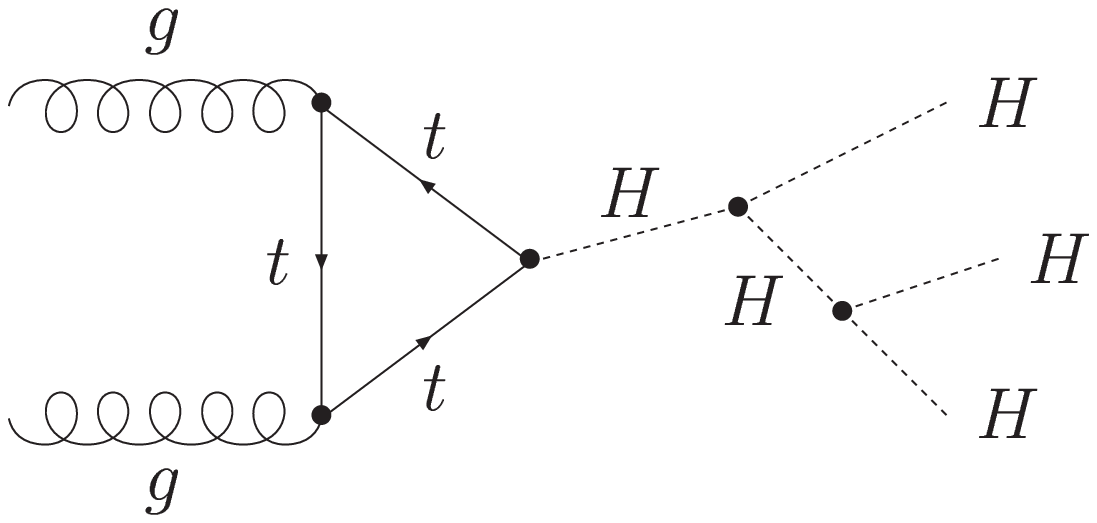}
\hspace*{2mm}
\includegraphics[scale=0.35]{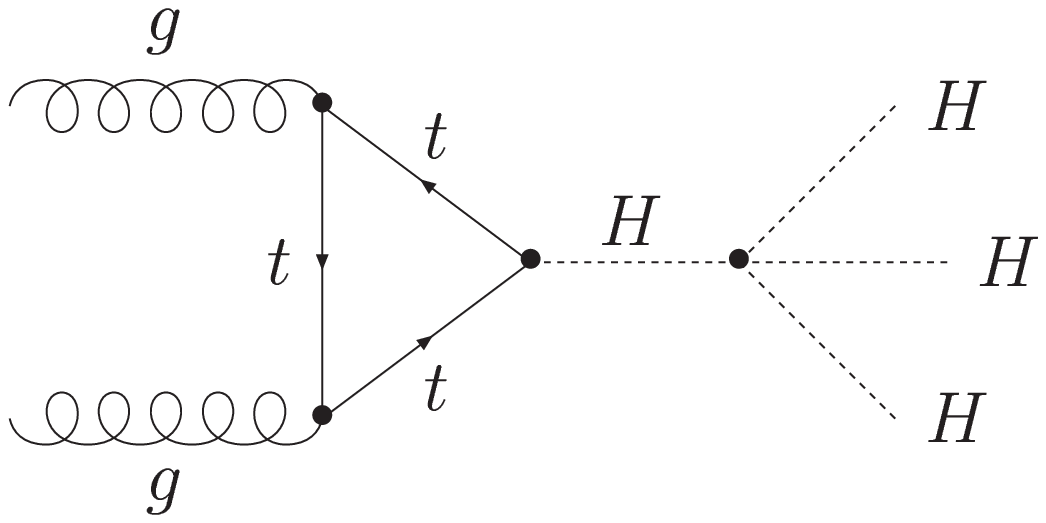}
\vspace*{-2mm}
\caption{Representative Feynman diagrams for $gg\to HHH$.}
\label{fig:gg_HHH}
\end{figure}

For $e^+e^-$ collisions we already know this is hopeless: the $HHH$
rate is both too low and its dependence on $\wt\lambda$ too
weak~\cite{Djouadi:1999gv}.  However, the situation at (S)LHC was only
very recently investigated~\cite{Plehn:2005nk,Binoth:2006ym}.  The
authors calculated the $gg\to HHH$ cross section, which involves
Feynman diagrams like those of Fig.~\ref{fig:gg_HHH}.  Note the
appearance of numerous diagrams dependent on the trilinear
self-coupling, in addition to diagrams dependent only on $y_t$.

\begin{figure}[hb!]
\includegraphics[scale=0.33,angle=270]{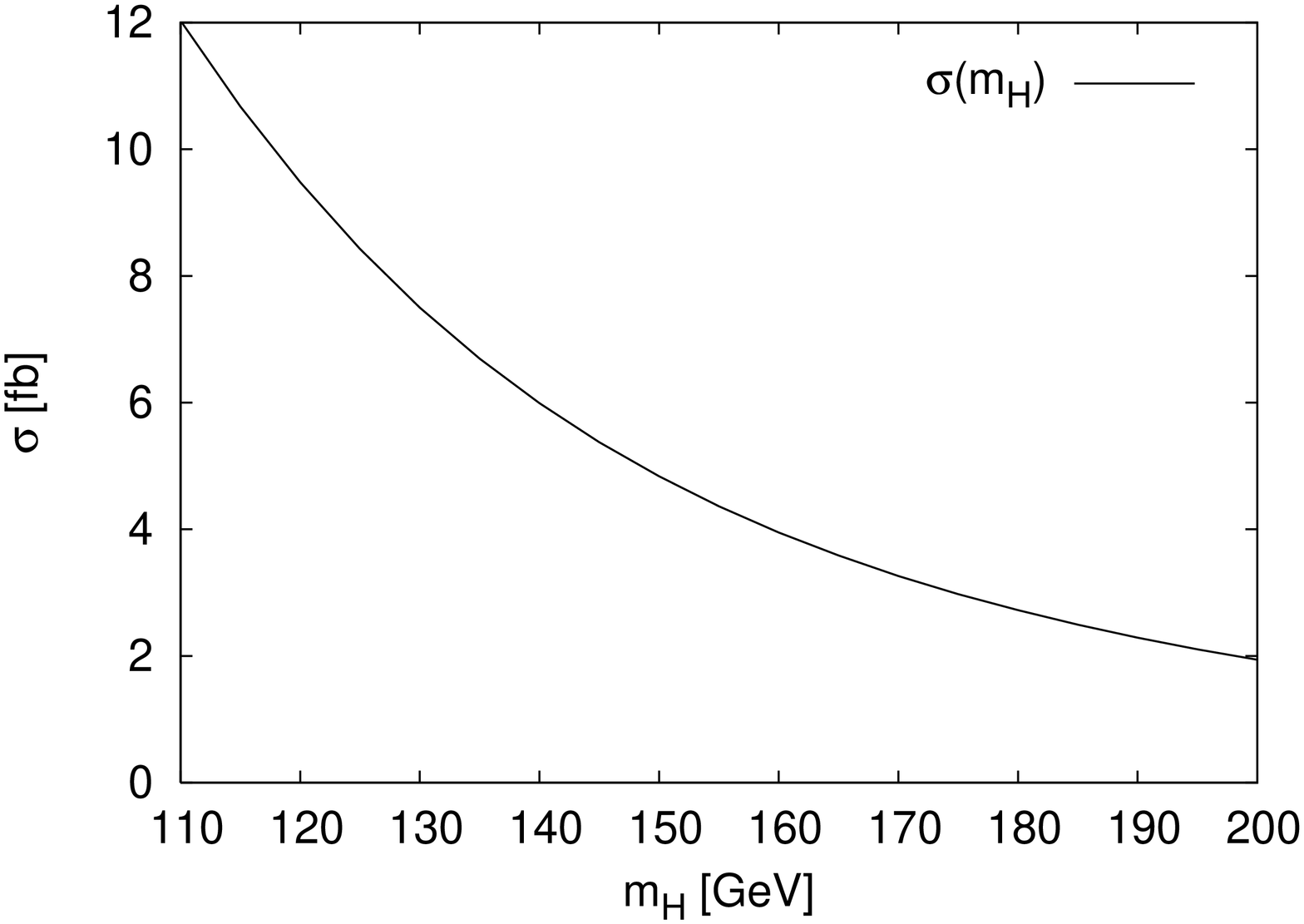}
\includegraphics[scale=0.33,angle=270]{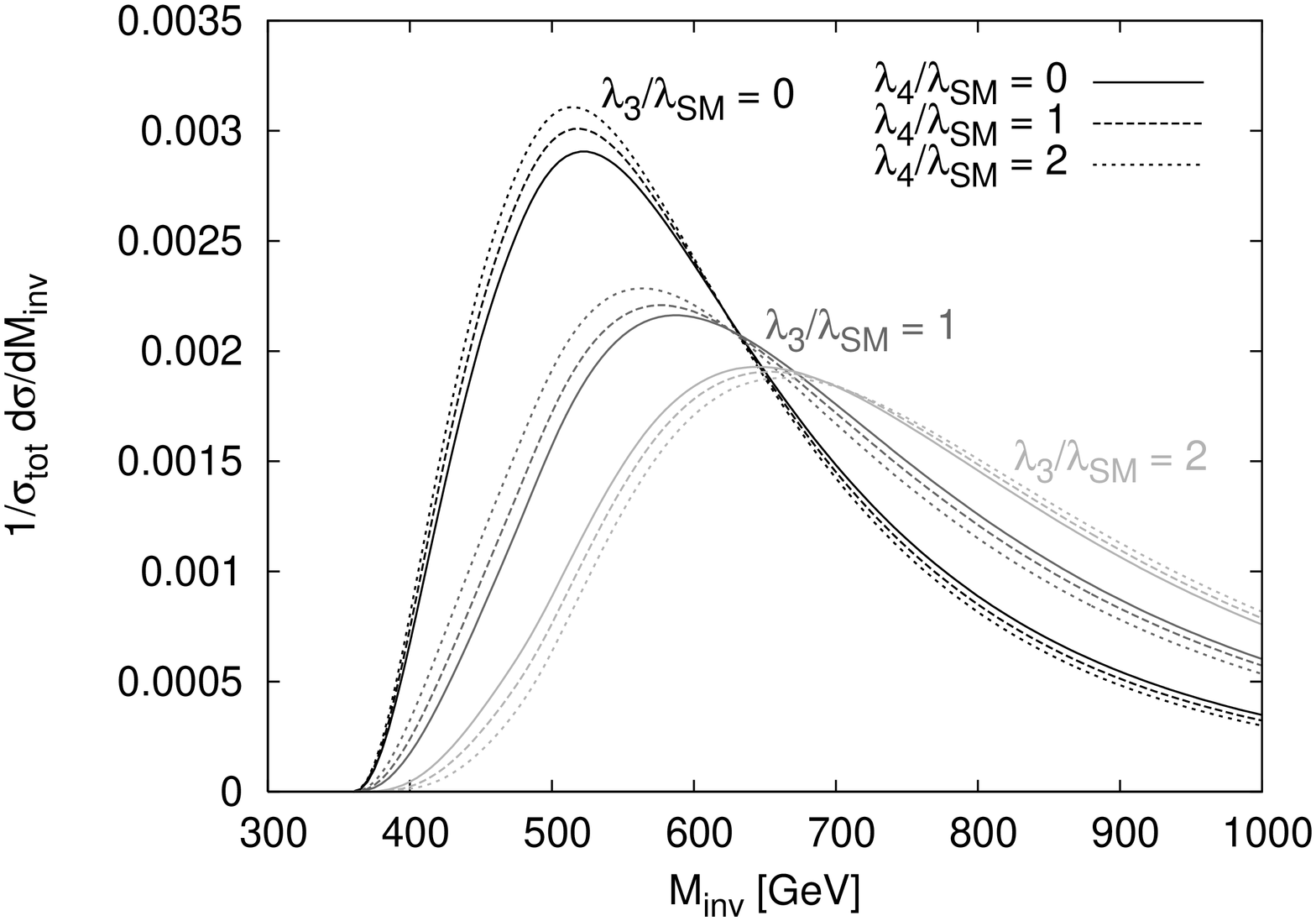}
\caption{Left: 200~TeV VLHC $gg\to HHH$ cross section as a function of
$M_H$.  Right: differential cross section as a function of $M_{HHH}$
for three values each of $\lambda$ and $\wt\lambda$.  Figures from
Ref.~\protect\cite{Plehn:2005nk}.}
\label{fig:LHC-HHH}
\end{figure}

The results of the study are shown in Fig.~\ref{fig:LHC-HHH}, {\it for
a 200~TeV VLHC}.  They're rather deflating because the cross section
is miserably small.  A challenge to the student: find a three-Higgs BR
to a final state that could be observed at a VLHC, where the rate is
not laughable.  Good luck!  In addition, the right panel shows that
any variation of the trilinear coupling $\lambda$ completely swamps
variation of the quartic $\wt\lambda$, whose own variation is already
infinitesimal.

In summary, it appears that we will likely never achieve a complete
picture of the Higgs potential.  This of course applies only to the
Standard Model.  Coming up in Chapter~\ref{sec:BSM} we're going to see
that for BSM physics the situation is even more discouraging.

%

%%%%%%%%%%%%%%%%%%%%%%%%%%%%%%%%%%%%%%%%%%%%%%%%%%%%%%%%%%%%%%%%%%%%%%%%
%%%%%%%%%%%%%%%%%%%%%%%%%%%%%%%%%%%%%%%%%%%%%%%%%%%%%%%%%%%%%%%%%%%%%%%%
%%%%%%%%%%%%%%%%%%%%%%%%%%%%%%%%%%%%%%%%%%%%%%%%%%%%%%%%%%%%%%%%%%%%%%%%

\section{Beyond-the-SM Higgs sectors}
\label{sec:BSM}

Now that we know how the Standard Model Higgs sector works -- how it
could be discovered and measured at LHC -- it's natural to think about
other possibilities for EWSB.  The SM Higgs is elegant in its
simplicity, but as you know from Sally Dawson's SM lectures, it's
probably too minimal -- nagging theoretical questions remain about
Higgs mass stability, flavor (ignoring this is kind of a black eye),
neutrino masses (another black eye), and so on.  Because new physics
that could explain dark matter is likely to also lie at the TeV scale,
most model building makes an attempt to incorporate solutions to some
of these other problems along with EWSB.  The literature is vast, but
let's try to roughly classify some of the major ideas to get a handle
on the variations.

The broadest two categories of classes are weakly-coupled new physics
which can be handled with perturbation theory, and strongly-coupled or
``strong dynamics'' models which are penetrable in some cases, others
not.  These include QCD-inspired theories like
Technicolor~\cite{Weinberg:1975gm,Susskind:1978ms} (or more properly
Extended~\cite{Dimopoulos:1979es,Eichten:1979ah} or
Walking~\cite{Holdom:1981rm} Technicolor, which can handle a top quark
mass very different from the other quark masses) and Topcolor-assisted
Technicolor~\cite{TC2} (``TC2''), which incorporates additional
weakly-coupled gauge structure.  Strong dynamics assumes that some
TeV-scale massive or heavier fermions' attraction became strong at low
energy scales, eventually causing their condensation to mesonic states
(Technipions, Technirho, Technieta, etc.), the neutral scalars of
which can incite EWSB via their $SU(2)_L$ gauge interactions.  Strong
dynamics scenarios are beyond the scope of these lectures, however, so
I leave it for the interested student to study the excellent review
article of Ref.~\cite{Hill:2002ap}.

While strong dynamics theories are Higgsless in some sense, meaning no
fundamental scalar fields, the terms is usually reserved for a new
class of models where the EW symmetry is broken using boundary
conditions on gauge boson wavefunctions propagating in finite extra
dimensions (see e.g. Refs.~\cite{Simmons:2006iw,Cacciapaglia:2004zv}).
We'll also skip these.

There is far more theoretical effort expended on weakly-coupled EWSB,
which is mostly variations on what we can add to the single Higgs
doublet of the SM:
\begin{dingautolist}{192}
\item 1HDM + invisible (high-scale) new physics, 
      hidden from direct detection
\vspace{-1mm}
\item CP-conserving 2HDM: 4 types
      (minimal supersymmetry, MSSM, is Type II)
\vspace{-1mm}
\item CP-violating 2HDM
\vspace{-1mm}
\item Higgs singlet(s)
      (\eg next-to-minimal supersymmetry, NMSSM)
\vspace{-1mm}
\item Higgs triplets
      (often appear in Grand Unified Theories)
\vspace{-1mm}
\item Little Higgs models: 
      $SU(2)_L\times U(1)_Y$ is part of larger gauge and global group
\end{dingautolist}
The first item, new high-scale physics hidden from direct detection,
sounds like a cheat.  It actually involves an important aspect of
phenomenology: effective Lagrangians from higher-dimensional
operators.  We'll come back to these in a moment.  Two Higgs doublets
instead of one is an idea with multiple sources.  For instance, one
doublet could give mass to the leptons and the other to the quarks, or
one to the up-type fermions and the other to the down-type, etc.
We'll return to these after effective operators.  Additional Higgs
singlets likewise have a variety of reasons for being written down,
but usually it's just ``we can do it, so we will''.  We'll skip these.
Higgs triplets originated from natural appearance in left-right
symmetric GUTs.  They're a bit exotic and typically have issues with
precision EW data, but are interesting in that they predict the
existence of doubly-charged Higgs states $H^{\pm\pm}$, and a
tree-level $H^\pm W^\mp Z$ coupling, which must be zero in most
Higgs-doublet models.  It would therefore stand out experimentally.
For all these cases I don't have time to cover, the Higgs Hunter's
Guide is the best place to start to learn more~\cite{HHG}.

Little Higgs theories, on the other hand, are different in that they
necessarily involve new scalar {\it and} gauge structure arising from
an enlarged global symmetry from which the SM emerges, as well as
additional matter content.  Interestingly, in these models the Higgs
looks very much like the SM Higgs, but with ${\cal O}(v^2/F^2)$
corrections, where $F$ is typically a few TeV, parametrically $4\pi$
larger than the EW scale.  The smallness of $v^2/F^2$ could make it
very difficult to measure Little Higgs corrections to Higgs
observables.  These models are probably ultimately strongly-coupled at
a scale $\Lambda\sim 4\pi F$, but this is an open question.  If nature
chose this course, the most interesting physics is the new gauge boson
and matter fields that appears at a scale $F$.
Refs.~\cite{LH-reviews} provide nice overviews and simple explanations
of the two primary Little Higgs mechanisms.

%%%%%%%%%%%%%%%%%%%%%%%%%%%%%%%%%%%%%%%%%%%%%%%%%%%%%%%%%%%%%%%%%%%%%%%%
%%%%%%%%%%%%%%%%%%%%%%%%%%%%%%%%%%%%%%%%%%%%%%%%%%%%%%%%%%%%%%%%%%%%%%%%

\subsection{Higher-dimensional operators}
\label{sub:HDO}

The new physics responsible for dark matter, flavor, neutrino masses,
etc., might very well be too massive to produce directly at colliders.
This the dreaded SM-Higgs-only scenario, where LHC sees nothing new.
It would really be an invitation to take a more rigorous look at all
data -- new physics effects might still appear as small deviations in
precision observables.

The standard way of parameterizing this is to write down all the
possible Lagrangian operators with the heavy fields integrated out
which preserve $SU(3)_c\times SU(2)_L\times U(1)_Y$ gauge invariance.
This was done over two decades ago for operators up to dimension
six~\cite{Buchmuller:1985jz}.  Although not often emphasized in
today's phenomenology, I consider this paper a must-read for all
students.

Let's begin by considering the possible operators involving only the
SM Higgs doublet.  There are two, of dimension six:
\bq\label{eq:O1O2}
{\cal O}_1 = \frac{1}{2}
\,\partial_\mu\,(\Phi^\dagger\Phi)\,\partial^\mu(\Phi^\dagger\Phi)
\qquad \& \qquad
{\cal O}_2 = -\frac{1}{3}\,(\Phi^\dagger\Phi)^3
\eq
for the effective Lagrangian contribution
\bq\label{eq:eff-L}
{\cal L}_{6D,\Phi} \; = \;
\sum\limits^2_{i=1} \frac{f_i}{\Lambda^2}{\cal O}_i \; , \;\; f_i>0
\eq
$\Lambda$ must be at least a couple TeV, otherwise we'd likely observe
it directly at LHC.  If you've somewhere seen an alternative effective
theory for the Higgs potential written as
\bq\label{eq:alt-eff-pot}
V_{\rm eff} \; = \; 
\sum\limits_{n=0} \frac{\lambda_n}{\Lambda^{2n}}
\biggl( |\Phi|^2 - \frac{v^2}{2} \biggr)^{2+n}
\eq
the operators written above correspond to the $n=1$ term in this
expansion.

${\cal O}_1$ modifies the Higgs kinetic term, while ${\cal O}_2$
modifies the EW vev, $v$:
\bq\label{eq:O1O2-mods}
{\cal L}_{\rm kin} \; = \;
\frac{1}{2} \, \partial_\mu\phi \, \partial^\mu\phi
 +
\frac{1}{2}f_1\frac{v^2}{\Lambda^2}
\, \partial_\mu\phi \, \partial^\mu\phi
\; , \qquad
\frac{v^2}{2} \; \approx \;
\frac{v^2_0}{2}
\bigl( 1-\frac{f_2}{4\lambda}\frac{v^2_0}{\Lambda^2} \bigr)
\eq
where $v$ is what $G_F$ measures.  We must also canonically normalize
the physical Higgs field: $\phi=NH$ with
$N=1/(1+f_1\frac{v^2}{\Lambda^2})$.

This results in a number of alterations to masses and
couplings~\cite{Barger:2003rs}.  First, the Higgs mass itself receives
corrections from the expected value, given $\lambda$:
\bq\label{eq:O1O2-M_H}
M_H^2 \; = \; 2\lambda v^2
\bigl( 1 - f_1\frac{v^2}{\Lambda^2}
         + \frac{f_2}{2\lambda}\frac{v^2}{\Lambda^2} \bigr)
\eq
where the $f_2$ term is independent of $\lambda$.  Next, Higgs gauge
couplings receive $v^2/\Lambda^2$ shifts:
\ba\label{eq:O1O2-VVH}
\frac{1}{2}g^2v \biggl( 
 1-\frac{f_1}{2}\frac{v^2}{\Lambda^2} \biggr) H W^+_\mu W^{-\mu}
\;\; & \;\;
\frac{1}{4}g^2 \bigl( 
 1-f_1\frac{v^2}{\Lambda^2} \bigr) HH W^+_\mu W^{-\mu}
\\ \notag
\frac{1}{2}\frac{g^2}{c_W}v \biggl( 
 1-\frac{f_1}{2}\frac{v^2}{\Lambda^2} \biggr) H Z_\mu Z^\mu
\;\; & \;\;
\frac{1}{4}\frac{g^2}{c_W} \bigl( 
 1-f_1\frac{v^2}{\Lambda^2} \bigr) HH Z_\mu Z^\mu
\ea
Finally, the Higgs boson self-couplings are (phases vary with Feynman
rule convention):
\ba\label{eq:O1O2-Hself}
|\lambda_{3H}| = \frac{3m^2_H}{v} \left[ 
\left( 
1 - \frac{f_1}{2}\frac{v^2}{\Lambda^2} 
  + \frac{2f_2}{3}\frac{v^2}{M^2_H}\frac{v^2}{\Lambda^2}
\right) 
+ \frac{2f_1}{3M^2_H}\frac{v^2}{\Lambda^2}\sum_{i<j}^3 p_i\cdot p_j \right]
\\
|\lambda_{4H}| = \frac{3m^2_H}{v^2} \left[ 
\left( 
1 - f_1\frac{v^2}{\Lambda^2} 
  + 4f_2\frac{v^2}{M^2_H}\frac{v^2}{\Lambda^2}
\right) 
+ \frac{2f_1}{3M^2_H}\frac{v^2}{\Lambda^2}\sum_{i<j}^4 p_i\cdot p_j \right]
\ea
Note that ${\cal O}_1$ and ${\cal O}_2$ both enter here, but more
importantly there are momentum-dependent terms, which are typical of
higher-dimensional operators.  The effect of these terms would be
anomalous high-$p_T$ Higgses in pair production.

Only one phenomenological analysis exists for these effects, and only
for precision experiments at a future ILC and
CLIC~\cite{Barger:2003rs}.  In this study, measurements are expressed
in terms of $a_i=f_i\,v^2/\Lambda^2$, since $f_i$ and $\Lambda$ can't
be easily separated from so few measurements.  Higgsstrahlung, double
Higgsstrahlung and WBF Higgs pair production together measure a
combination of $a_1$ and $a_2$.

\begin{figure}[hb!]
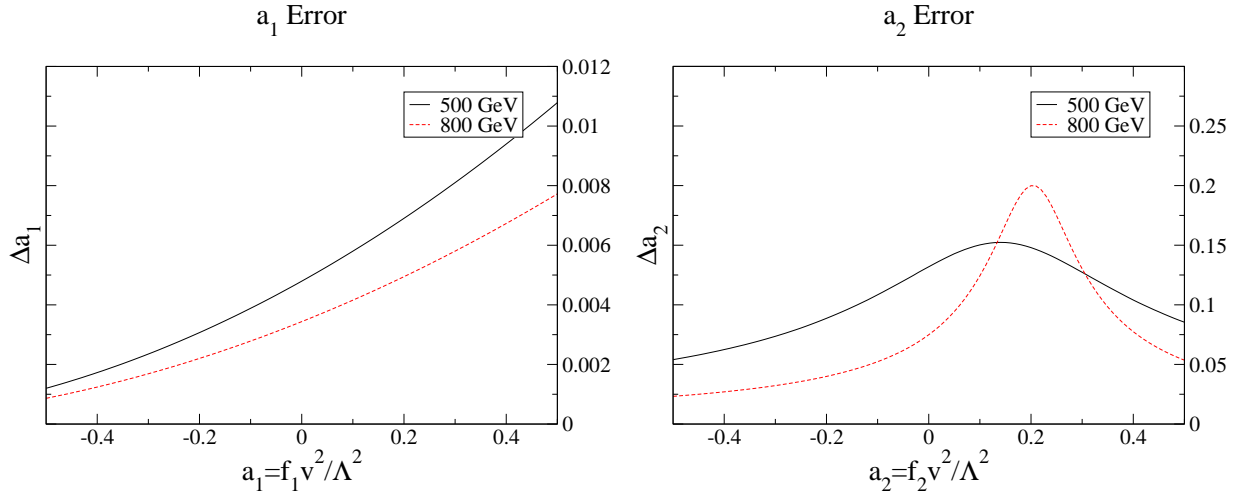

\includegraphics[scale=0.315]{Higgs.figs/ILC/err_h_a1_LC.eps}
\hspace*{18mm}
\includegraphics[scale=0.315]{HH.figs/err_hh_a2_LC.eps}
\vspace*{-4mm}
\caption{Achievable uncertainty on measurements of the $a_1$ (left) 
and $a_2$ (right) coefficients of Eq.~\protect\ref{eq:O1O2} (also see
text) at a future ILC for 500 and 800~GeV
running~\protect\cite{Barger:2003rs}.}
\label{fig:a1a2-lim}
\end{figure}

Fig.~\ref{fig:a1a2-lim} shows the expected achievable uncertainties
(not limits!) on $a_1$ and $a_2$ at a future ILC.  For $f_1=1$, this
corresponds to a reach in $\Lambda$ of about 4~TeV, possibly out of
the reach of LHC depending on what might be directly produced.  For
$f_2=1$, however, this corresponds to only about $\Lambda\sim
0.8$~TeV, easily accessible at LHC.  Put another way, an ILC could
have access to new high-scale physics via altered Higgs--gauge boson
couplings, but not via Higgs self-couplings.  This is in line with
what we'd come to expect, as $HH$ production is much smaller.  The
shapes of the uncertainty curves in the figures depend on what values
of the operator coefficients add to or subtract from the signal, with
the added feature that the momentum dependence of the Higgs
self-couplings that the ${\cal O}_1$ operator introduces changes to
kinematic distributions.

In addition to the Higgs-only D6 operators, there are a handful of
operators involving the Higgs and gauge boson fields
together~\cite{Buchmuller:1985jz}:

\medskip
%\begin{table}[h!]
\begin{tabular}{cl}
\qquad & $O_{WW} \;=\; 
         (\phi^\dagger\phi)\bigl[W^+_{\mu\nu}W^{-\mu\nu}
                                 +\frac{1}{2}W^3_{\mu\nu}W^{3\mu\nu}\bigr]$ \\
\qquad & $O_{BB\phantom{h}} \;=\; 
         (\phi^\dagger\phi) B_{\mu\nu} B^{\mu\nu}$ \\
\qquad & $O_{BW\phantom{i}} \;=\; 
         B^{\mu\nu}\bigl[(\phi^\dagger\sigma^3\phi)W^3_{\mu\nu}
                         +\sqrt{2}\bigl[(\phi^\dagger T^+\phi)W^+_{\mu\nu}
                         +(\phi^\dagger T^-\phi)W^-_{\mu\nu}\bigr]\bigr]$ \\
\qquad & $O_{B\phantom{Bh}} \;=\; 
         (D^\mu\phi)^\dagger(D^\nu\phi) B_{\mu\nu}$ \\
\qquad & $O_{W\phantom{Bi}} \;=\; 
         (D^\mu\phi)^\dagger\bigl[\sigma^3(D^\nu\phi)W^3_{\mu\nu}
                                  +\sqrt{2}\bigl[T^+(D^\nu\phi)W^+_{\mu\nu}
                                  +T^-(D^\nu\phi)W^-_{\mu\nu}\bigr]\bigr]$ \\
\qquad & $O_{\Phi,1\phantom{w}} \;=\; 
         (D_\mu\phi)^\dagger\phi\phi^\dagger(D^\mu\phi)$
\end{tabular}
%\notag
%\end{table}
\medskip

\noindent
These induce momentum-dependent $HHVV$ vertices, so could be studied
at an ILC or CLIC in the same manner as the Higgs-only couplings, as
well as with rare Higgs decays~\cite{Hagiwara:1993qt}, but in general
they're highly constrained by EW precisions observables ($S$, $\rho$,
$g_{VVV}$)~\cite{Hagiwara:1995vp}.  Interestingly, it appears there
has not been an update of the EW constraints on these operators since
1997~\cite{Szalapski:1997rf}, although there are predictions for
limits at an ILC~\cite{Beyer:2006hx}.  There is, however, a new
analysis for WBF Higgs at LHC includes the effects of some of these
operators and finds that they would be encoded in the tagging jet
azimuthal separation~\cite{Hankele:2006ma}.

There is also a set of D6 operators involving the Higgs, fermion and
gauge boson fields~\cite{Buchmuller:1985jz}:

\medskip
\begin{tabular}{clll}

\qquad & $O_{d\phi} \;\, = \; (\phi^\dagger\phi)(\bar{q}d\phi)$
\qquad & $O_{\phi d} \;\, = \; i(\phi^\dagger D_\mu\phi)(\bar{d}\gamma^\mu d)$ \qquad
\qquad & $O_{\bar{D}d} \; = \; (D_\mu\bar{q}d)D^\mu\phi$ \\

\qquad & $O^{(1)}_{\phi q}\;=\;i(\phi^\dagger D_\mu\phi)(\bar{q}\gamma^\mu q)$
\qquad & $O_{\phi\phi} \:\, = \; i(\phi^\dagger\epsilon D_\mu\phi)(\bar{u}\gamma^\mu d)$ \qquad
\qquad & $O_{dW} \, = \; (\bar{q}\sigma^{\mu\nu}\sigma^i d)\phi W^i_{\mu\nu}$\\

\qquad & $O^{(3)}_{\phi q}\;=\;i(\phi^\dagger D_\mu\sigma^i\phi)(\bar{q}\gamma^\mu\sigma^iq)$
\qquad & $O_{Dd} \; = \; (\bar{q} D_\mu d)D^\mu\phi$ \qquad
\qquad & $O_{dB} \:\, = \; (\bar{q}\sigma^{\mu\nu} d)\phi B_{\mu\nu}$ \\

\end{tabular}
\medskip

\noindent
Some of these are constrained by precise LEP measurements of $Z\bb$,
$\gamma\bb$ couplings, but not severely.  They would give interesting
rare Higgs decays like $H\to\bb Z,\bb\gamma$.  Their phenomenology for
LHC and even ILC is not really studied.  Thus, I can't say to what 
scale they might be sensitive given a SM Higgs discovery with nothing
else observed.

%%%%%%%%%%%%%%%%%%%%%%%%%%%%%%%%%%%%%%%%%%%%%%%%%%%%%%%%%%%%%%%%%%%%%%%%
%%%%%%%%%%%%%%%%%%%%%%%%%%%%%%%%%%%%%%%%%%%%%%%%%%%%%%%%%%%%%%%%%%%%%%%%

\subsection{Two-Higgs doublet models (2HDMs)}
\label{sub:2HDM}

The most-often studied extension to the SM Higgs sector is the
two-Higgs doublet model (2HDM)~\cite{Lee:1973iz,HHG}.  That is, we add
one additional $SU(2)_L$ doublet.  Both of the doublets acquire a vev.
For now let's assume CP conservation and work with in the real-vev
basis.  Counting degrees of freedom, four per complex doublet, and
knowing that three modes are ``eaten'' to give the $W^\pm$ and $Z$
their masses, after SSB there must be five physical states.  Two of
them will necessarily be charged ($H^\pm$) regardless of how we
assigned hypercharge to each doublet, leaving the other three neutral.
Of those, two ($h$, $H$) will be CP-even and one will be CP-odd ($A$),
the last of which won't couple to the weak bosons at tree level.  The
general 2HDM potential is quite messy~\cite{CTHill,HHG} , so we'll not
discuss it.

Recall the primary role of the Higgs sector: to restore unitarity to
weak boson scattering.  This requires the gauge coupling to $WW$ to be
exactly $\frac{1}{2}g_W^2v$, where $v$ is what we measure with $G_F$.
In the amplitude, then, the coupling squared is $\frac{1}{4}g_W^4v^2$.
With two vevs, there is the automatic constraint $v_1^2+v_2^2\equiv
v^2$~\cite{Huffel:1980sk}.  The ratio is
$\tan\beta\equiv\frac{v_2}{v_1}$.  The CP-even mass eigenstates, which
couple to the weak bosons, thus boil down to simply mixing:
\ba\label{eq:hH-mix}
h \;\; = \; \sqrt{2} \bigl[ 
-({\rm Re}\phi^0_1-v_1)\sin\alpha+({\rm Re}\phi^0_2-v_2)\cos\alpha
\bigr]
\\
H \; = \; \sqrt{2} \bigl[
\phantom{-}
 ({\rm Re}\phi^0_1-v_1)\cos\alpha+({\rm Re}\phi^0_2-v_2)\sin\alpha
\bigr]
\ea
where $\alpha$ is the angle which diagonalizes the $2\times2$ mixing
matrix.  The Higgs sector is typically defined by $\alpha$,
$\tan\beta$ and the potential parameters which govern the
self-couplings.  Some models are defined instead by $M_A$ and $M_Z$.

Let's pause for a moment to reflect on what would happen if we
introduced CP violation~\cite{Lee:1973iz}.  This is a well-motivated
exercise since there isn't enough CP violation in the SM model to
account for baryogenesis in the early universe.  The most immediate
impact is that $h$, $H$ and $A$ now mix.  $M_A$ is supposed to
parameterize the pseudoscalar pole, but it's now mixed into three
physical states, so it becomes ill-defined.  Instead, we typically use
the charged Higgs mass.  It would be logical to use $M_{H^\pm}$ for
CP-conserving scenarios as well, but this is one of those historical
accidents that has too much momentum to change.

Regarding the fermions, we can apportion the two doublets in four
general ways~\cite{HHG}:
\begin{itemize}
\vspace{-2mm}
\item[I] \; only $\Phi_2$ couples to fermions
\vspace{-2mm}
\item[II] \; $\Phi_1$ couples to down-type, $\Phi_2$ to up-type fermions
\vspace{-2mm}
\item[III] \; $\Phi_1$ couples to down quarks, 
              $\Phi_2$ to up quarks and down leptons
\vspace{-2mm}
\item[IV] \; $\Phi_1$ couples to quarks, $\Phi_2$ to leptons
\end{itemize}
Types III and IV induce flavor-changing neutral currents (FCNCs),
which are highly constrained, thus these models are not much studied
any more.  Types I and II are qualitatively different and worth a
quick look at the differences in their couplings, shown in
Table~\ref{tab:I-II}\footnote{Note that various references use
different phase conventions for the Lagrangian.  The important
distinction is the phase between Higgs couplings, and a reference SM
coupling such as $ee\gamma$.  I use positive terms in the covariant
derivative and drop the overall superfluous factor of $i$ typical of
most Lagrangians.}.  Because of which doublet gives the down-type
fermions their masses, those Yukawa couplings to $h$ and $H$ are
swapped between models, with a phase factor from mixing.  Similarly,
the $Af\bar{f}$ coupling is inverted and changes sign:
$\cot\beta\to-\tan\beta$.  The gauge coupling for $h$ and $H$, of
course, are unaffected by the Yukawa couplings and are fixed to
$\sin(\beta-\alpha)$ and $\cos(\beta-\alpha)$.  (The sum of their
squares in the amplitude must equal 1!)

The charged Higgs Yukawa couplings are slightly different yet.  The
left-handed coupling is proportional to the up-type Yukawa coupling,
and the right-handed coupling the down-type Yukawa, {\it for an
out-flowing $H^-$}.  The reverse is true for an outflowing $H^+$.  We
have:
\ba\label{eq:H+-_1}
g_{H^-D\bar{U}}
\; = \; \frac{g}{2\sqrt{2}M_W}
\bigl[ m_U\cot\beta(1+\gamma_5) - m_D\cot\beta(1-\gamma_5) \bigr]
\\\label{eq:H+-_2}
g_{H^-D\bar{U}} \; = \; \frac{g}{2\sqrt{2}M_W}
\bigl[ m_U\cot\beta(1+\gamma_5) + m_D\tan\beta(1-\gamma_5) \bigr]
\ea
where $H^-$ flows out, $D$ is incoming and $\bar{U}$ is outgoing.

\begin{table}[ht!]
\begin{large}
\begin{tabular}{|c|c|c|c|c|c|}
\hline
$\Phi$ & $\frac{g_{\Phi\uu}}{g_f}$ & $\frac{g_{\Phi\dd}}{g_f}$ 
       & $\frac{g_{\Phi VV}}{g_V}$ & $\frac{g_{\Phi ZA}}{g_V}$ \\
\hline
$h$ & $-\frac{\cos\alpha}{\sin\beta}$ & $-\frac{\cos\alpha}{\sin\beta}$ 
    & $\sin(\beta-\alpha)$            & $-\frac{1}{2}i\cos(\beta-\alpha)$ \\
$H$ & $-\frac{\sin\alpha}{\sin\beta}$ & $-\frac{\sin\alpha}{\sin\beta}$ 
    & $\cos(\beta-\alpha)$            & $\phantom{-}\frac{1}{2}i\sin(\beta-\alpha)$ \\
$A$ & $-i\gamma_5\cot\beta$           & $i\gamma_5\cot\beta$
    & 0                               & 0                               \\
\hline
$h$ & $-\frac{\cos\alpha}{\sin\beta}$ & $\phantom{-}\frac{\sin\alpha}{\cos\beta}$ 
    & $\sin(\beta-\alpha)$            & $-\frac{1}{2}i\cos(\beta-\alpha)$ \\
$H$ & $-\frac{\sin\alpha}{\sin\beta}$ & $-\frac{\cos\alpha}{\cos\beta}$ 
    & $\cos(\beta-\alpha)$            & $\phantom{-}\frac{1}{2}i\sin(\beta-\alpha)$ \\
$A$ & $-i\gamma_5\cot\beta$           & $-i\gamma_5\tan\beta$
    & 0                               & 0                               \\
\hline
\end{tabular}
\end{large}
\caption{Fermion and gauge boson couplings in Type I (upper) and II
(lower) 2HDMs.}
\label{tab:I-II}
\end{table}
%

%%%%%%%%%%%%%%%%%%%%%%%%%%%%%%%%%%%%%%%%%%%%%%%%%%%%%%%%%%%%%%%%%%%%%%%%
%%%%%%%%%%%%%%%%%%%%%%%%%%%%%%%%%%%%%%%%%%%%%%%%%%%%%%%%%%%%%%%%%%%%%%%%

\subsection{Type~II 2HDM in the MSSM}
\label{sub:MSSM-2HDM}

At this point we should focus on the Type~II 2HDM, because that's the
one required to appear in the MSSM\footnote{A superpotential can't be
constructed from conjugate fields, else the supersymmetry
transformations aren't preserved.  For an excellent SUSY tutorial, see
Ref.~\cite{Martin:1997ns}.} (see Ref.~\cite{JackHowie_MSSM_H} for a
detailed description).  Model~I will have similar features, modulo the
couplings swaps given in Table~\ref{tab:I-II}, so is understandable by
analogy.  We'll spend the remaining portion discussing only SUSY Higgs
phenomenology, and specifically minimal SUSY, the MSSM.  However, by
the end it should be apparent that extended Higgs sectors may often be
treated as variations on a theme, with much of the phenomenology based
on the same collider signatures.

The MSSM imposes tree-level constraints on the Higgs potential which
require the various $\lambda$ to be gauge parameters (MSSM extensions
add non-gauge terms).  We'll come back to what the potential looks
like in Sec.~\ref{sub:MSSM-pot} and study its phenomenology, and for
now simply examine the implication of this structure on the mass
spectrum.  Because we consider only the CP-conserving case here, we
can get away with using $M_A$ as an input.  The others will be
$\tan\beta$ as discussed before, the average top squark mass $M_S$,
and an encoded trilinear mixing parameter for the top sector, $X_t$.
This last one is important because of the large top Yukawa corrections
the MSSM Higgs sector receives.  The values $0$ and $\sqrt{6}\,M_S$
are referred to as ``no mixing'' and ``maximal mixing'', because they
extremize the loop corrections.  The $h-H$ mixing angle is
\bq\label{eq:alpha}
\alpha \; = \; \frac{1}{2}\tan^{-1}
\biggl[ \tan 2\beta \frac{M_A^2+M_Z^2}{M_A^2-M_Z^2} \biggr]
\; , \qquad -\frac{\pi}{2} \leq \alpha \leq 0
\eq
to first order.  The CP-even masses are given by:
\ba\label{eq:hH-masses}
M^2_{H,h} \; = \; &
\frac{1}{2}\biggl(  M^2_A + M^2_Z 
                   \pm \sqrt{(M_A^2 + M_Z^2)^2+4M_A^2M_Z^2\sin^2(2\beta)}
           \biggr) \\[-1mm] \nonumber
& +\frac{3}{8\pi^2}\cos^2\alpha \: y_t^2 m_t^2
   \biggl[ \log\frac{M_S^2}{m_t^2} 
    + \frac{X_t^2}{M_S^2}\biggl(1-\frac{X_t^2}{12M_S^2} \biggr) \biggr]
\;\;\;\; {\rm for} \, M_h \, {\rm only}
\ea
where the top Yukawa correction can be significant, a couple tens of
GeV.  The charged Higgs mass is rather more simple:
\bq\label{eq:xH-mass}
M_{H^\pm}^2=M_A^2+M_W^2
\eq
\begin{figure}[ht!]
\includegraphics[scale=0.75]{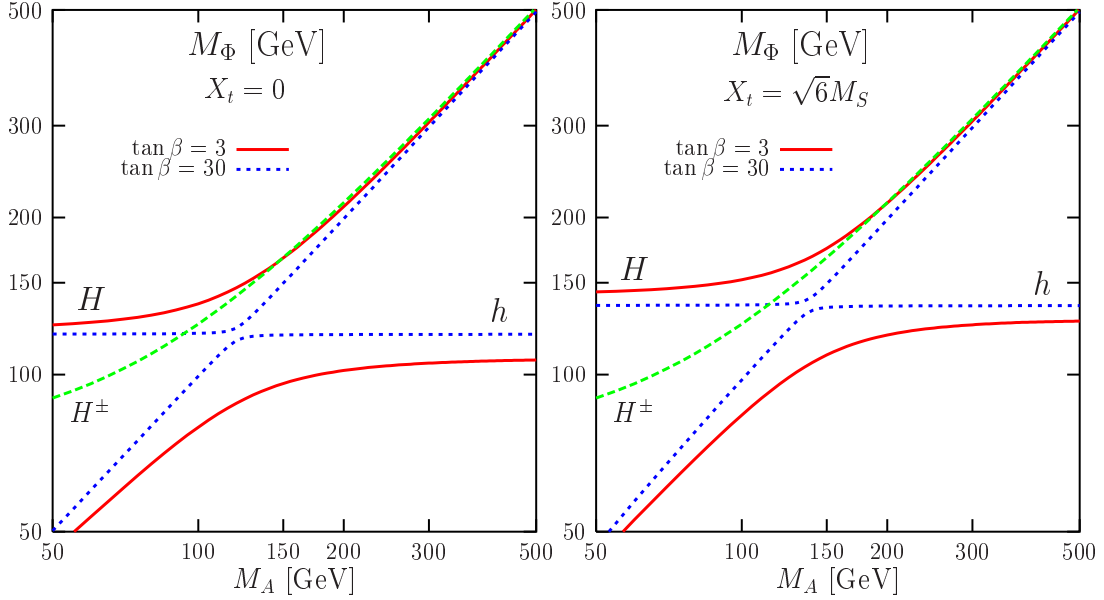}
\vspace*{-4mm}
\caption{MSSM Higgs boson masses as a function of pseudoscalar mass
$M_A$ and two choices of $\tan\beta$, for no (left) and maximal
(right) mixing ($X_t$ parameter; see text).  Figures from
Ref.~\protect\cite{Djouadi}.}
\label{fig:MSSM-masses}
\end{figure}
These equations exhibit the interesting property of $h$ decoupling
with increasing pseudoscalar mass: for large $M_A$ the heavy states
$H$, $A$ and $H^\pm$ tend to be closely degenerate, and the light $h$
has an asymptotic maximum mass which depends mostly on $\tan\beta$.
We see this behavior, along with a plateau effect for $M_h$ and $M_H$,
in Fig.~\ref{fig:MSSM-masses}.  There is always at least one CP-even
Higgs boson in the mass region $90\lesssim M_\phi\lesssim 145$~GeV,
assuming perturbativity to high scales.  For large $M_A$, toward the
decoupling region, it is the lighter state, $h$, but at low $M_A$ it
is the heavier state, $H$.  The transition region is sharper for
larger $\tan\beta$.

\begin{figure}[ht!]
\includegraphics[scale=0.77]{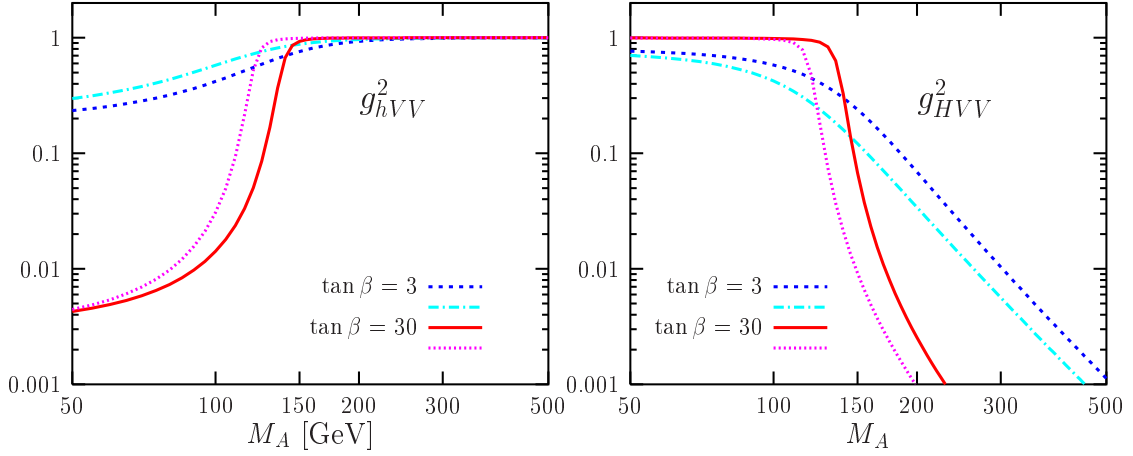}
\vspace*{-2mm}
\caption{MSSM CP-even Higgs boson couplings to the weak gauge bosons
as a function of $M_A$ and for two choices of $\tan\beta$, and for no
mixing (darker colors) and maximal mixing (lighter colors).
Figures from Ref.~\protect\cite{Djouadi}.}
\label{fig:MSSM-Vcoup}
\end{figure}

The mass spectrum is not the only feature to exhibit the decoupling
and transition behavior, however.  Both the gauge and Yukawa couplings
do the same.  The $VV\phi$ couplings are shown in
Fig.~\ref{fig:MSSM-Vcoup}.  By comparison with
Fig.~\ref{fig:MSSM-masses}, we easily see that when either $h$ or $H$
is in its plateau mass region, it holds most of the gauge coupling;
$\sin(\beta-\alpha)\to1$ or $\cos(\beta-\alpha)\to1$.  In the
transition region, the two states share the gauge coupling, and both
are of comparable importance in unitarity cancellation.  As with the
mass spectrum, and by now as anticipated, the transition region is
sharper for larger $\tan\beta$.  Hold these two figures in your mind,
as they are going to play an extremely important phenomenological role
shortly.

\begin{figure}[hb!]
\includegraphics[scale=0.77]{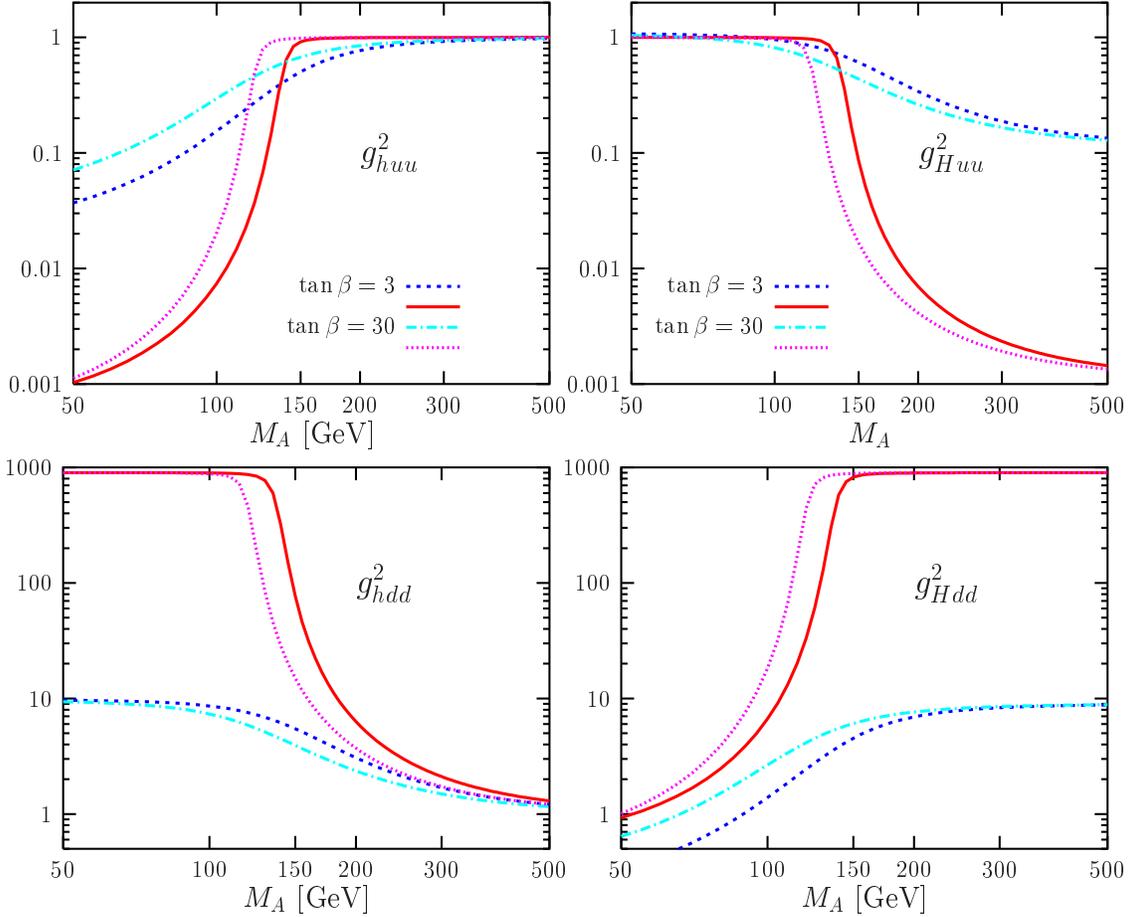}
\vspace*{-1mm}
\caption{MSSM CP-even Higgs boson couplings to fermions as a function
of $M_A$ and for two choices of $\tan\beta$, and for no mixing (darker
colors) and maximal mixing (lighter colors).  Figures from
Ref.~\protect\cite{Djouadi}.}
\label{fig:MSSM-fcoup}
\end{figure}

Using just trigonometry, let's rewrite the Yukawa couplings of
Table~\ref{tab:I-II} to see better how they depend on $M_A$ and
$\tan\beta$:
\ba\label{eq:Yuk-coup}
g_{h\uu} = &  -\frac{\cos\alpha}{\sin\beta} \, Y_u
  & = \; - [ \sin(\beta-\alpha) + \cot\beta\cos(\beta-\alpha) ] \: Y_u
\\ \nonumber
g_{h\dd} = &  \phantom{-}\frac{\sin\alpha}{\cos\beta} \, Y_d
  & = \; - [ \sin(\beta-\alpha) - \tan\beta\cos(\beta-\alpha) ] \: Y_d
\\ \nonumber
g_{H\uu} = &  -\frac{\sin\alpha}{\sin\beta} \, Y_u
  & = \; - [ \cos(\beta-\alpha) - \cot\beta\sin(\beta-\alpha) ] \: Y_u
\\ \nonumber
g_{H\dd} = &  -\frac{\cos\alpha}{\cos\beta} \, Y_d
  & = \; - [ \cos(\beta-\alpha) + \tan\beta\sin(\beta-\alpha) ] \: Y_d
\ea
This is a far more convenient form, since $\tan\beta$ is an input and
$\sin(\beta-\alpha)/\cos(\beta-\alpha)$ is the reduced $h/H$ gauge
coupling.  These are both natural, convenient parameters to describe
production cross sections and decay partial widths (thus branching
ratios), rather than the CP-even mixing angle and $\sin\beta$ or
$\cos\beta$, or their inverses.  Check Fig.~\ref{fig:MSSM-fcoup} to
see if you agree.


These are the most salient features of the MSSM Higgs sector,
sufficient to understand the bulk of MSSM Higgs phenomenology.  For a
more in-depth discussion, especially of why SUSY imposes these
constraints, and for more detailed formulae, see
Refs.~\cite{Djouadi,JackHowie_MSSM_H}.

Now that we know the couplings, we can obtain cross sections for $h$
and $H$ production simply as correction factors to the SM channels of
equal mass.  There is no WBF or $W/Z$-associated pseudoscalar
production, but there is both $gg\to A$ inclusive and top quark
associated production, $t\bar{t}A$, which are easily obtained if one
inserts the $\gamma_5$ factor into the loop derivation for $gg\to
A$~\cite{Djouadi}.  The charged Higgs is a special case as there is no
SM analogue; we'll discuss this in Sec.~\ref{sub:MSSM-search} in the
context of searches.  For the moment, let's examine the neutral
states' branching ratios, just to get an idea of how they behave.
It's easy to suffer plot overload about now, so don't try to absorb
every last detail; focus on the general behavior, which you already
should be able to guess from the couplings plots.

\begin{figure}[hb!]
\includegraphics[width=14cm,height=6.5cm]{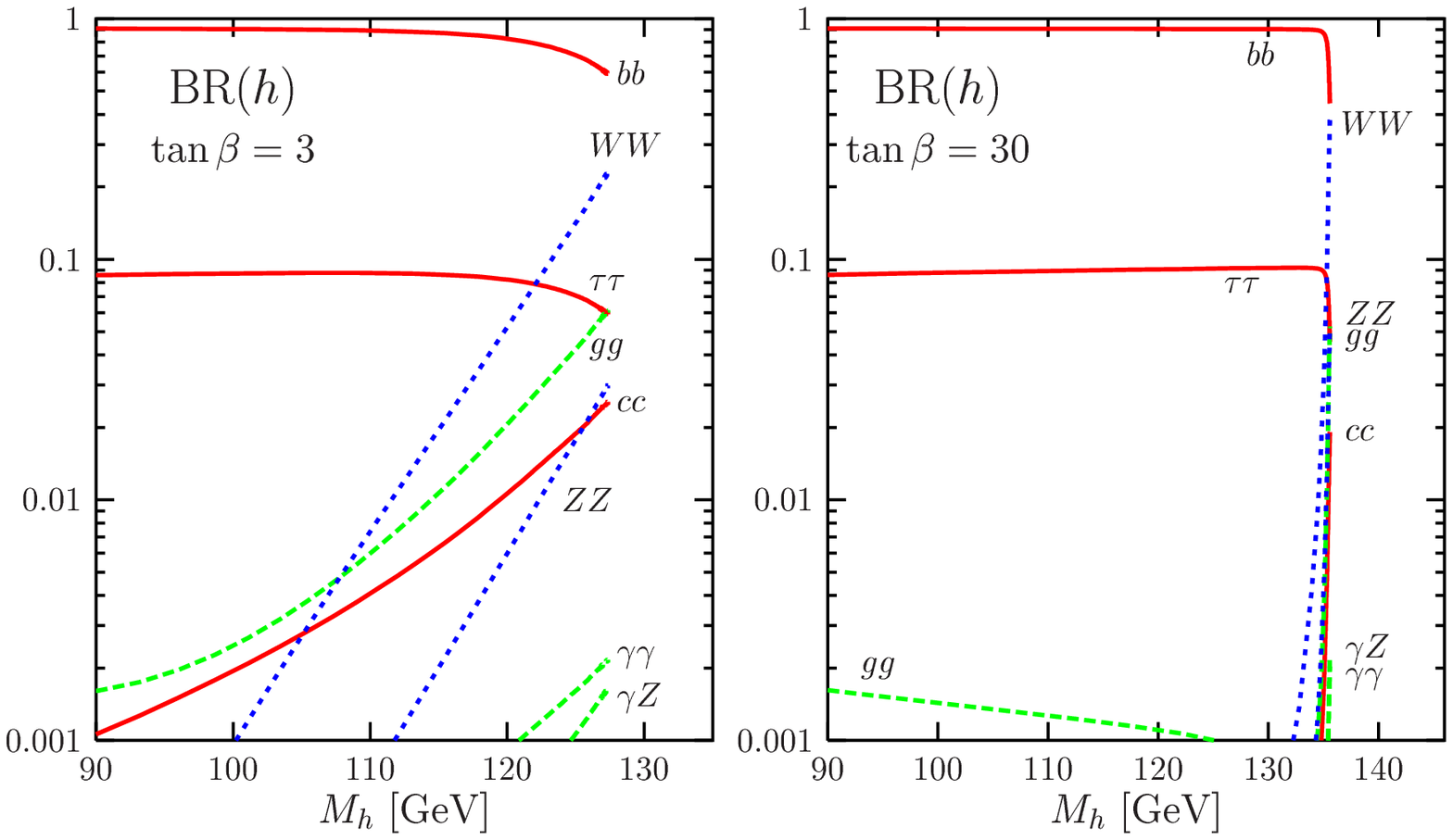}
\includegraphics[width=14cm,height=6.5cm]{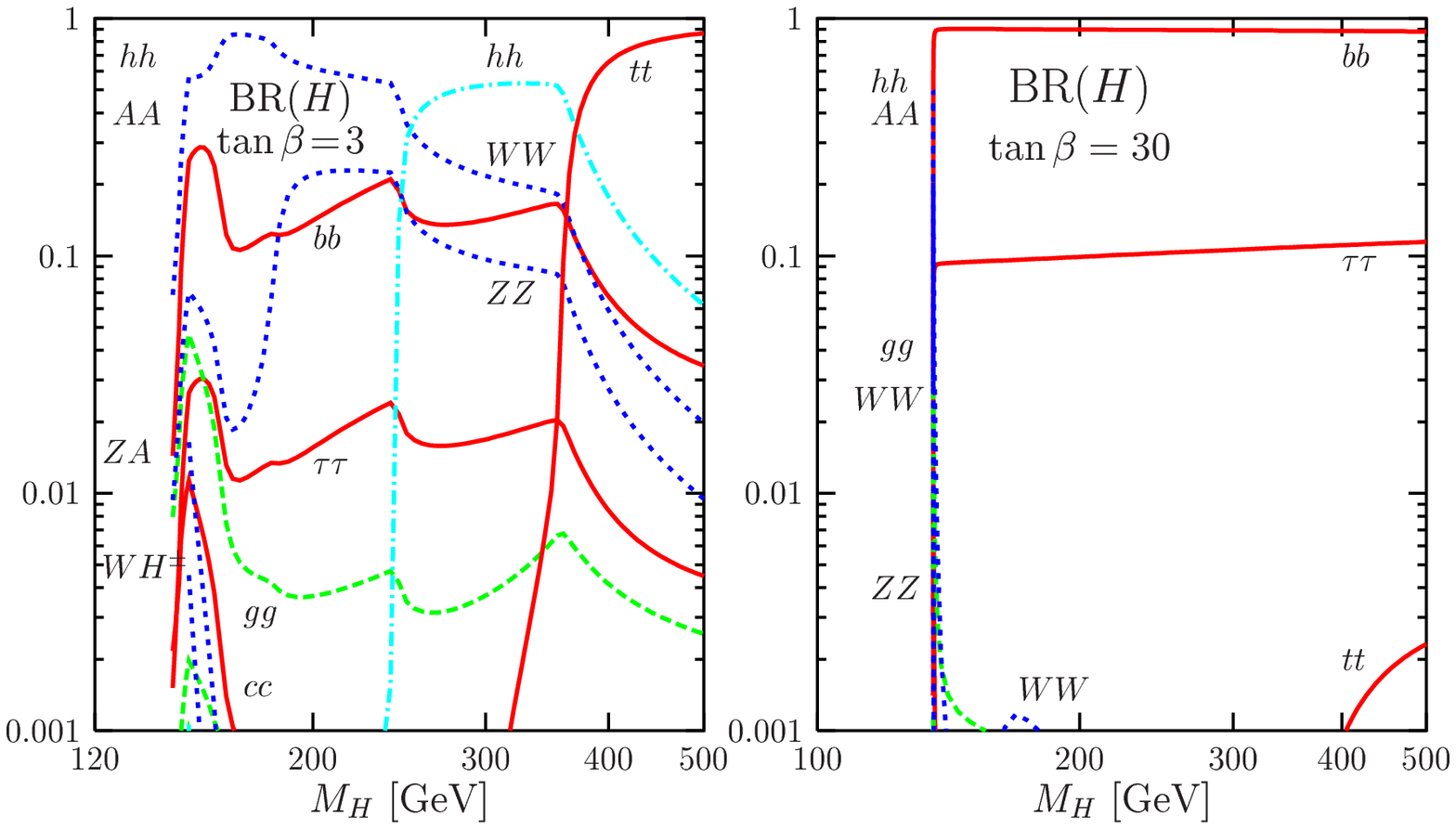}
\vspace*{-2mm}
\caption{MSSM CP-even Higgs boson branching ratios as a function of 
$M_A$ for $\tan\beta=3,30$.  Figures from
Ref.~\protect\cite{Djouadi}.}
\label{fig:MSSM-BR-hH}
\end{figure}

Fig.~\ref{fig:MSSM-BR-hH} shows the BRs for the CP-even states $h$ and
$H$, cut off at the mass plateaus.  They're basically what we would
expect: both $h$ and $H$ behave like a SM Higgs of equal mass, except
that the various couplings are dialed up or down.  $M_h$ can never be
above $\sim145$~GeV, so it almost never has a significant BR to gauge
bosons.  Because the fermionic partial widths can be enhanced by a
factor of $\tan^2\beta$, the rare modes like $\phi\to\gamma\gamma,gg$
\begin{figure}[ht!]
\includegraphics[width=14cm,height=6.3cm]{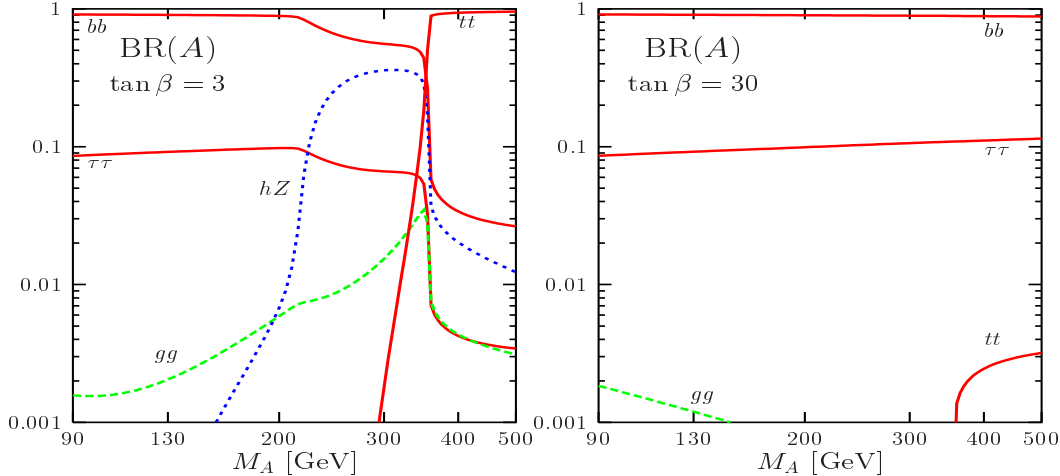}
\vspace*{-2mm}
\caption{MSSM CP-odd Higgs boson branching ratios as a function of 
$M_A$ for two choices of $\tan\beta$.  Figures from
Ref.~\protect\cite{Djouadi}.}
\label{fig:MSSM-BR-A}
\end{figure}
tend to be suppressed (and the top quark loop can better cancel the
$W$ loop for some parameter choices, suppressing the partial width).
The only new features are $H\to hh,AA$ decays, possible for limited
parameter choices but making for interesting additional channels.

The pseudoscalar BRs behave similarly, as shown in
Fig.~\ref{fig:MSSM-BR-A}.  The new feature here is at small
$\tan\beta$, where decays $A\to hZ$ are possible.  But otherwise $A$
prefers to decay $\sim90\%$ to $\bb$ and $\sim10\%$ to $\taus$, unless
it is heavy enough to produce top quark pairs.  That dominates only at
small $\tan\beta$ (large $\cot\beta$), where the up-type coupling
dominates.  At large $\tan\beta$, $\bb$ and $\taus$ both still win by
a considerable margin.

There are similar plots for $H^\pm$, but they're not particularly
enlightening as its decay patterns are drastically simpler: as far as
phenomenology is concerned, it's BR$\sim1$ to $tb$ when kinematically
accessible, $\tau\nu$ if lighter.  For low $\tan\beta$ there is a rare
BR to $hW^\pm$, but that is predicted to always be difficult to
observe.

All Higgs bosons can decay to SUSY particle pairs if they're light
enough, but this is not a very common occurrence across parameter
space (especially since so much of it is ruled out already by LEP SUSY
searches), so we'll bypass that discussion here.

%%%%%%%%%%%%%%%%%%%%%%%%%%%%%%%%%%%%%%%%%%%%%%%%%%%%%%%%%%%%%%%%%%%%%%%%
%%%%%%%%%%%%%%%%%%%%%%%%%%%%%%%%%%%%%%%%%%%%%%%%%%%%%%%%%%%%%%%%%%%%%%%%

\subsection{MSSM Higgs searches}
\label{sub:MSSM-search}

\begin{figure}[ht!]
\includegraphics[scale=0.46]{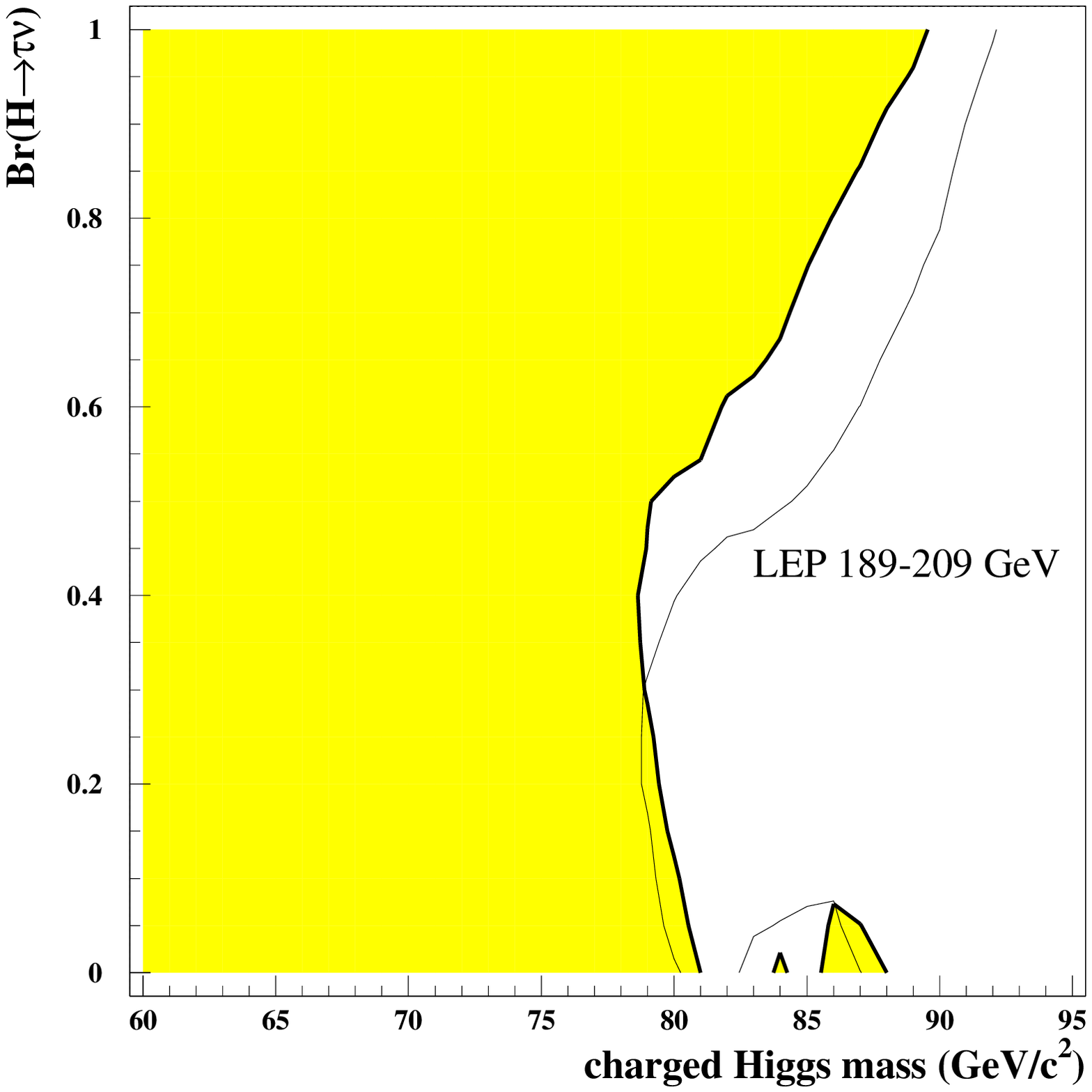}
\includegraphics[scale=0.4]{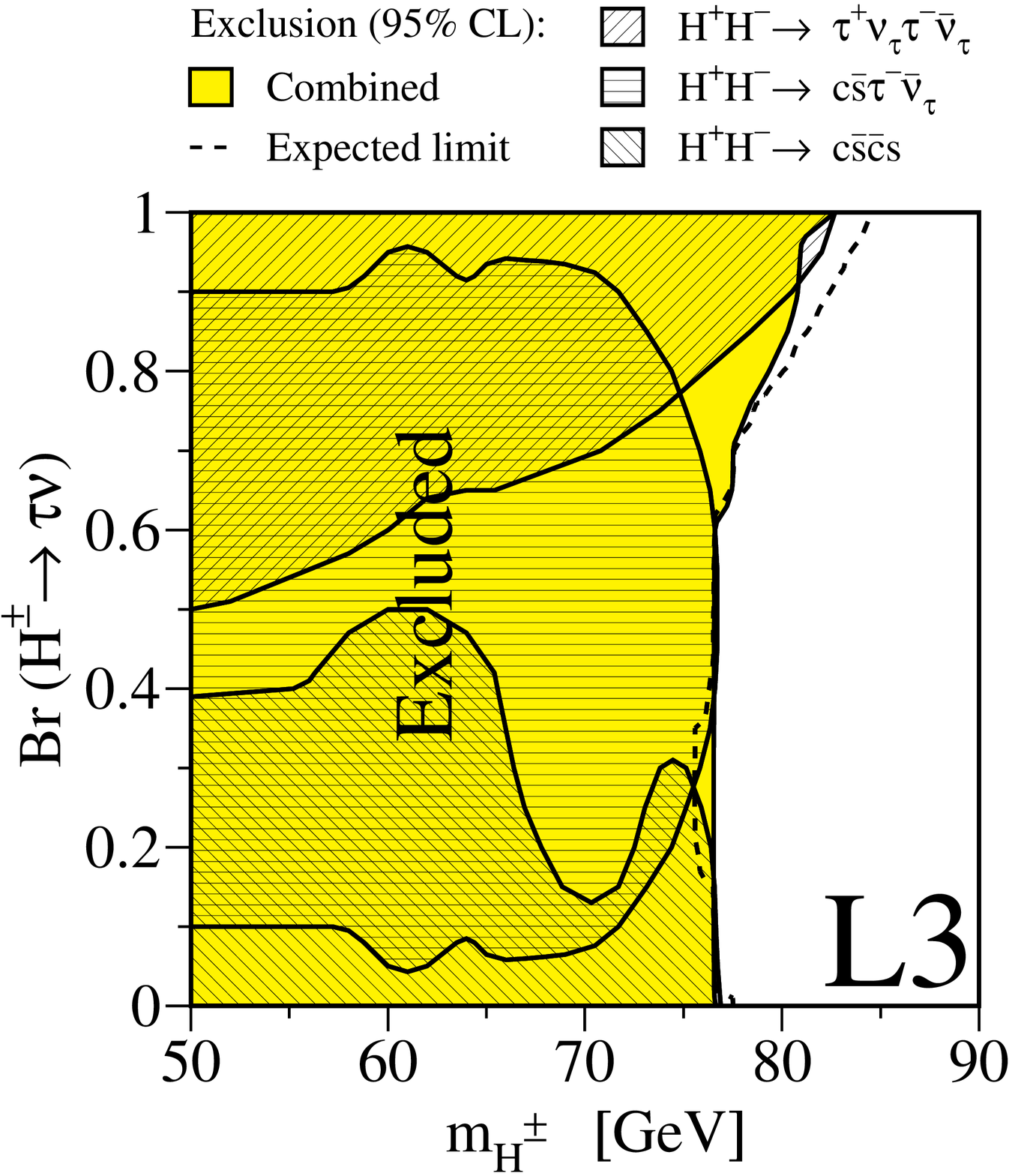}
\caption{Left: LEP preliminary combined-experiment charged Higgs search
$95\%$CL limits (2001), from Ref.~\protect\cite{unknown:2001xy}.
Right: L3 published limits from 2003, illustrating where each of the
three decay channels discussed in the text contributes to the overall
limit~\protect\cite{Achard:2003gt}.  There is no final LEP combined
limit, but judging from each of the individual
limits~\protect\cite{Heister:2002ev,Abdallah:2003wd,Achard:2003gt,Horvath:2003kd},
it does not change significantly from the preliminary results.}
\label{fig:LEP-xHiggs}
\end{figure}

For MSSM Higgs searches past, we start again with LEP.  It didn't find
anything, but placed various limits.  Let's begin with the charged
Higgs search, because it's the simplest.  This proceeded via $H^+H^-$
pair production (the only mechanism accessible at LEP) and decay to
$\tau\nu$ or $cs$, as there was never kinematic room for $tb$.  Thus,
the search had three channels: dual taus, mixed tau plus hadronic
decays, and an all-hadronic mode~\cite{Okpara:2001ef}.  Because the
production mechanism depends on only gauge-fixed couplings, the MSSM
charged Higgs search is usually presented as a more general 2HDM
search, with limits presented in the $M_{H^\pm}$
v. BR($H^\pm\to\tau^\pm\nu$) plane.  Fig.~\ref{fig:LEP-xHiggs}
summarizes the obtained limits.  To translate the general search
limits to the MSSM Higgs sector inputs, recall Eq.~\ref{eq:xH-mass},
$M_{H^\pm}^2=M_A^2+M_W^2$.  The difficulty of this search was the low
ID efficiency for taus and charm quarks.  Unfortunately, there is no
final combined limit, but each of the collaborations has published
final independent
limits~\cite{Heister:2002ev,Abdallah:2003wd,Achard:2003gt,Horvath:2003kd}.
Watch the LEP-Higgs web page for updates~\cite{LEP-HWG}.

The basic neutral Higgs boson search channels are exactly the same as
in the SM for each of $h$ and $H$, to which we add $e^+e^-\to Z^*\to
hA/HA$ production via the additional couplings of
Table~\ref{tab:I-II}.  Each of the four LEP collaborations presented a
multitude of MSSM $h/H/A$ search limits, and there are combined LEP
results with CP-conservation~\cite{Schael:2006cr} and CP-violation
(CPX)~\cite{Schael:2006cr,Bechtle:2006iw}.  However, one should be
somewhat wary of what precisely is presented.  The results are usually
shown as shaded exclusion blobs in either $M_A$-$\tan\beta$ space (for
a very specific set of additional assumptions) or
$M_{h_i}$-$\tan\beta$ space, also given some assumptions.  There are
literally dozens of pages of exclusion plots, depending on what one
chooses for the mixing parameter $X_t$, top quark mass (recall the
strong $M_h$ dependence on $m_t$), stop masses, $\mu$, and so on.
This is far too much to show here, because the exclusion contours
change so much from assumption to assumption -- it's impossible even
to select a representative sample without misleading the uninitiated.
See e.g. Ref.~\cite{Sopczak:2006vn}.

The curious student should flip through the plots in
Refs.~\cite{Schael:2006cr,Sopczak:2006vn} simply to get a feel for how
wild this variation is.  Observe how much the contours change
depending on the top quark mass -- it is obviously still fairly poorly
measured, as far as fits to supersymmetry go.  Note also that the
plots are always logarithmic in $\tan\beta$, which compresses the
unexcluded large-$\tan\beta$ region, making it appear that parameter
space is vastly ruled out in many cases.  This simply isn't true.
Finally, I should comment that the ``theoretically inaccessible''
disallowed blobs are even more grossly misleading.  All one has to do
is move the stop masses up slightly and these retreat dramatically.
Perhaps a more logical approach is the model-independent $h/H/A$
search of OPAL~\cite{Abbiendi:2004gn}.

\begin{figure}[hb!]
\includegraphics[scale=0.75]{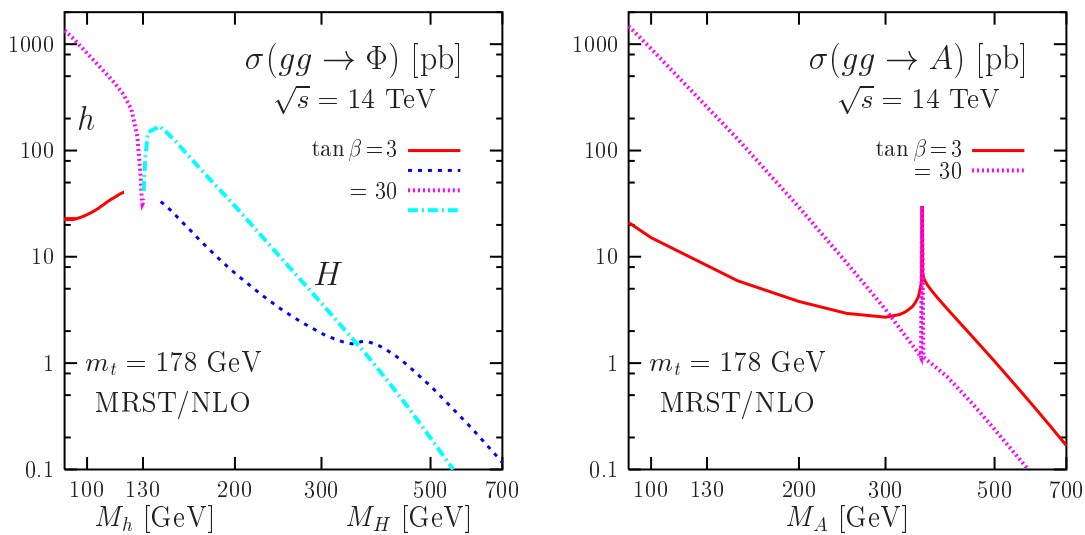}
\vspace*{-3mm}
\caption{Gluon fusion MSSM Higgs production cross sections at LHC for
the CP-even states $h$ and $H$ (left) and the pseudoscalar $A$
(right), for two values of $\tan\beta$.  Figures from
Ref.~\protect\cite{Djouadi}.}
\label{fig:LHC-MSSM-gg_H}
\end{figure}

MSSM Higgs Searches at LHC are also mostly variants on the SM search
channels, the exceptions being charged Higgses, rare (SUSY or Higgs
pair) decay modes, and one new production channel, $\bb\phi$, which is
important at large $\tan\beta$ where the coupling is enhanced to
top-quark Yukawa strength.  $t\bar{t}\phi$ rates tend to be about the
same as the SM for equal mass, or slightly suppressed.  WBF $h$ or $H$
rates can only be suppressed relative to the SM, due to the appearance
of $\sin^2(\beta-\alpha)$ or $\cos^2(\beta-\alpha)$, respectively.
Inclusive rates can change rather dramatically, however, because the
$b$ loop can be extremely important.  Fig.~\ref{fig:LHC-MSSM-gg_H}
shows the cross sections for $gg\to\phi$ as a function of the physical
masses, for small and large $\tan\beta$.  These may be compared with
the SM cross sections of Fig.~\ref{fig:LHC-xsecs}.

%
%\begin{figure}[ht!]
%\includegraphics[scale=0.42]{Higgs.figs/LEP/LEP-con1.eps}
%\includegraphics[scale=0.42]{Higgs.figs/LEP/LEP-con2.eps}
%\includegraphics[scale=0.42]{Higgs.figs/LEP/LEP-con3.eps}
%\includegraphics[scale=0.42]{Higgs.figs/LEP/LEP-con4.eps}
%\vspace*{-3mm}
%\caption{LEP CP-even Higgs $h$ search $95\%$~CL mass limits as a 
%function of $M_h$ (left) or $M_A$ (right), and for the maximal mixing
%(top) and no-mixing scenarios (bottom).}
%\label{fig:LEP-Higgs}
%\end{figure}
%

\begin{figure}[ht!]
\includegraphics[scale=0.65]{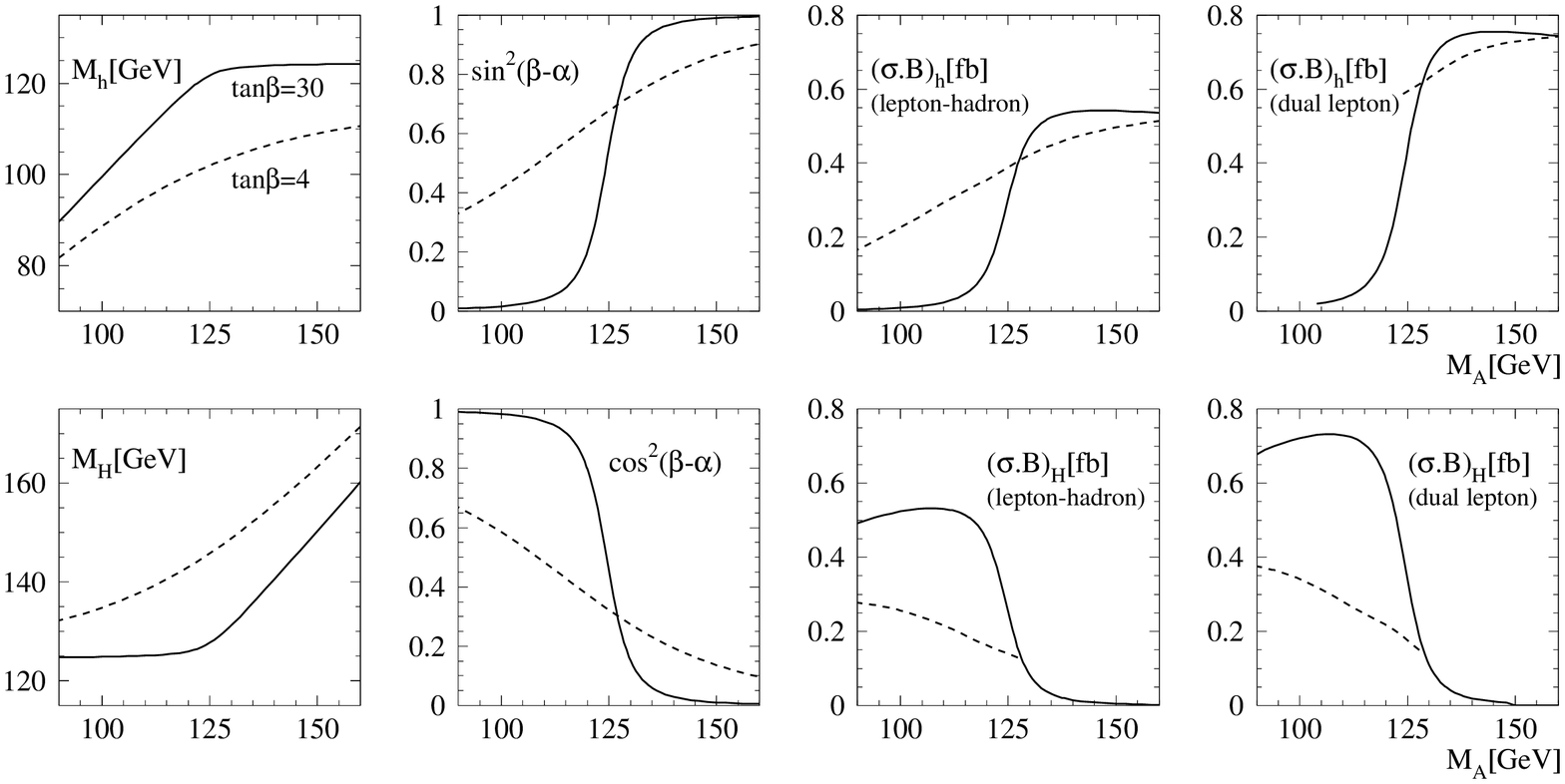}
\vspace*{-5mm}
\caption{From left to right, each plot as a function of $M_A$: $h/H$
mass, gauge coupling suppression factor squared, WBF cross section
times BR to taus to the lepton-hadron final state, and the same for
the dual lepton mode.  The upper (lower) row is for $h$($H$).
Fig. from Ref.~\protect\cite{Plehn:1999xi}.}
\label{fig:hH-versus}
\end{figure}

Let's concentrate on the WBF modes, however, as they turn out to be
the most interesting.  Recall the plateau behavior of $h$ and $H$
masses as a function of $M_A$ (cf. Fig.~\ref{fig:MSSM-masses}), and
simultaneously the $h$ and $H$ gauge coupling behavior
(cf. Fig.~\ref{fig:MSSM-Vcoup}).  The astute student will realize that
this implies that WBF Higgs production in an accessible mass region
probably always occurs at a good rate, somewhat suppressed but never
much so.  Fig.~\ref{fig:hH-versus} summarizes some of this previous
information and goes on to show the cross section times BR to tau
pairs (in the two accessible tau decay modes), also as a function of
$M_A$~\cite{Plehn:1999nw}.  Indeed, eyeballing the upper and lower
rows, it appears that between $h$ and $H$, there's always a signal in
WBF.  It may be slightly suppressed, but we know from SM WBF Higgs
studies (cf. Sec.~\ref{sub:WBF-taus}) that since so little data is
required to make an observation, the signal could be suppressed by a
factor of several and be detectable.  The reason is that in the MSSM
the $h$ and $H$ plateau mass ranges are in the ``good'' region of WBF
Higgs observability.  Actually, quite a large mass region is
observable, but if the MSSM predicted Higgs masses closer to the $Z$
pole, there could be trouble (but LEP would already have discovered
such a Higgs).

\begin{figure}[ht!]
\includegraphics[scale=0.73]{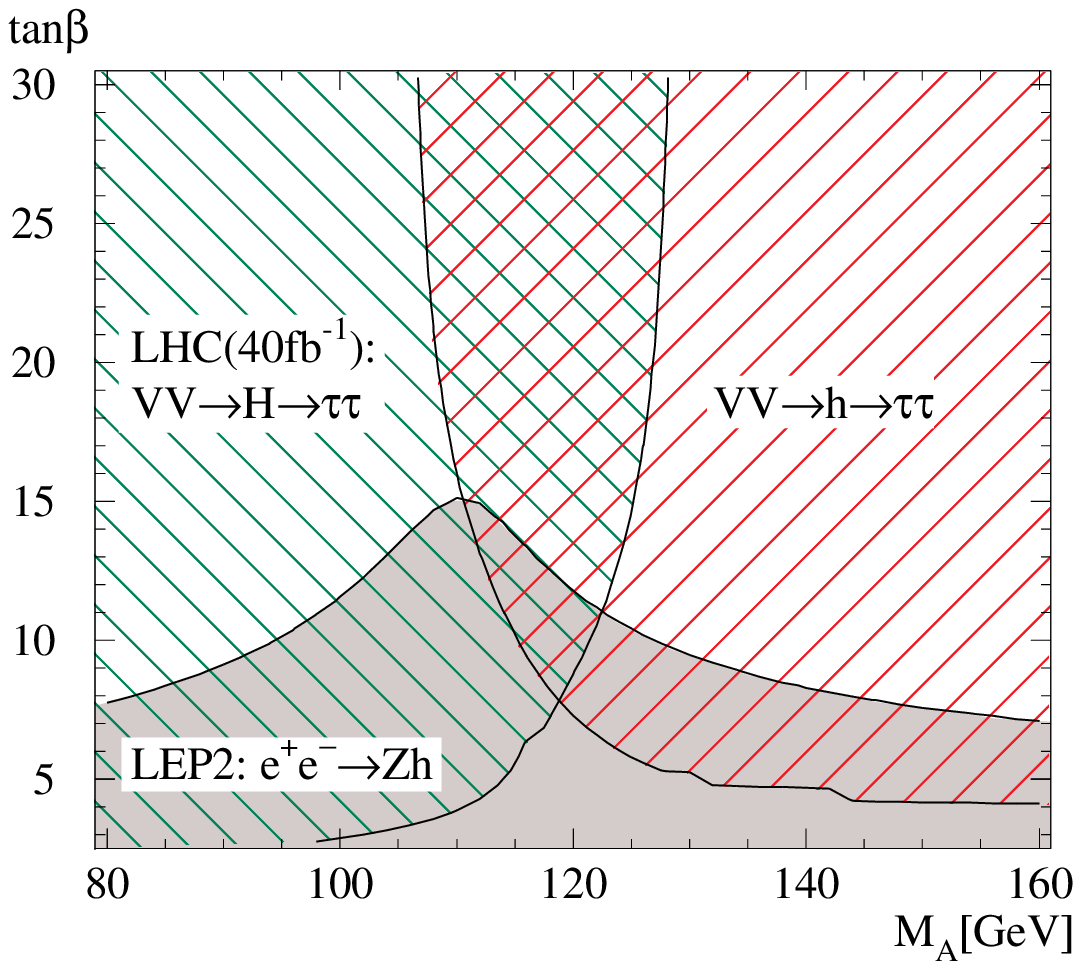}
\includegraphics[scale=0.73]{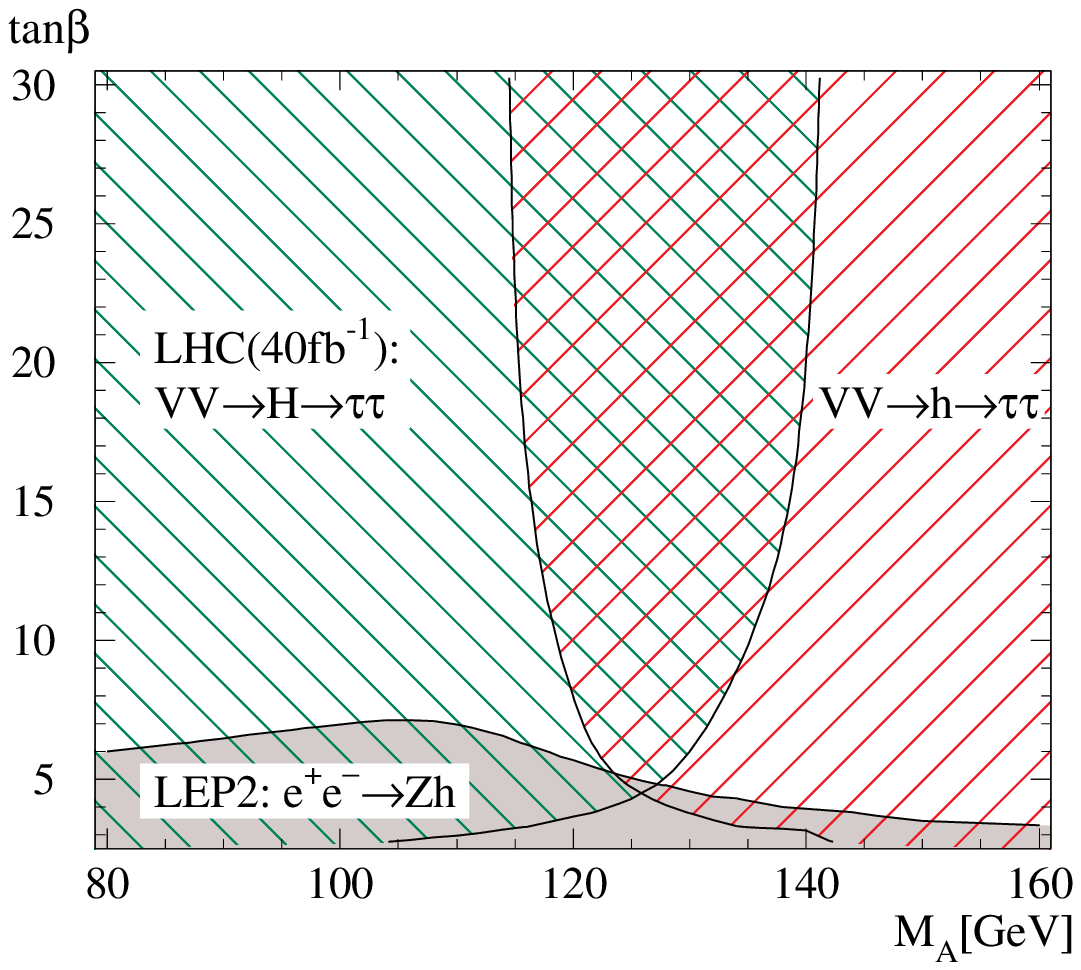}
\vspace*{-2mm}
\caption{MSSM parameter space coverage of WBF $h/H\to\taus$ for the
no-mixing ($A_t=0$, left) and maximal mixing
($X_t=\sqrt{6}\,M_{SUSY}$, right) cases~\protect\cite{Plehn:1999nw}.}
\label{fig:LHC-WBF-hH-coverage}
\end{figure}

This bit of luck forms the basis of the MSSM Higgs No-Lose Theorem: at
least one of the CP-even Higgs states, $h$ or $H$, is guaranteed to be
observable in WBF at LHC~\cite{Plehn:1999nw,Plehn:1999xi}.  The
original parton-level studies have since been confirmed with full
ATLAS detector simulation, and actually
improved~\cite{Schumacher:2004da}.  The parton-level coverage plots
shown in Fig.~\ref{fig:LHC-WBF-hH-coverage}, however, are simpler to
grasp.  Very little data would be required for discovery, and for some
$M_A$ it would be possible to observe both $h$ and $H$ simultaneously.

One caveat: the final state $\taus$ is not always
accessible!\footnote{There's always fine print...}  It's possible to
zero out the MSSM down-type fermion coupling at tree level -- an
interesting exercise for the student.  If this happens,
$h/H\to\gamma\gamma$ and $h/H\to\ww$ are ``large'' partial widths, so
their BRs take up the coverage
slack~\cite{Plehn:1999nw,Schumacher:2004da}, saving the No-Lose
Theorem.  There's been some work on an NMSSM No-Lose
Theorem~\cite{Ellwanger:2001iw,Ellwanger:2003jt,Ellwanger:2005uu,Moretti:2006hq},
which extends the Higgs sector by a complex singlet~\cite{HHG}.  The
outlook for LHC is promising, but not obviously rock-solid.

The No-Lose Theorem is great for the CP-even states, but what about
the other Higgses?  I'll gloss over the bulk of searches, since
they're mostly variants on the SM ones, and move on to the special
case of heavy $H/A$ (towards decoupling) and this new channel
$\bb\phi$.  The Feynman diagrams appear in Fig.~\ref{fig:bbphi-Feyn}.
Recall that $H$ has a $\tan\beta$ enhancement to down-type quarks in
the decoupling region, and $A$ always has this enhancement.  We
already know that means that $H$ and $A$ prefer to decay $90\%$ of the
time to $\bb$ and $10\%$ to $\taus$, but it would be impossible to
observe either of those final states in inclusive production, and WBF
production is zilch for $H$ in the decoupling region.  However, the
LHC being essentially a gluon collider, the initial state can create
high-energy $b$ pairs, which can then Brem a Higgs, either $H$ or $A$,
which are essentially degenerate (but do not interfere due to the
$\gamma_5$ coupling).  Since the $b$ jets are produced at high-$p_T$,
the $H/A$ must recoil against them, so it also produced with a
transverse boost.  It's decay products are then not back-to-back,
allowing for tau pair reconstruction; $H/A\to\mm$ may also be used,
but is a rare mode.  The final state is then $\bb\taus$ (or $\bb\mm$),
which is taggable and distinguishable from mixed QCD-EW backgrounds
because the tau pair invariant mass is in the several-hundred GeV
region.

\begin{figure}[ht!]
\includegraphics[scale=0.43]{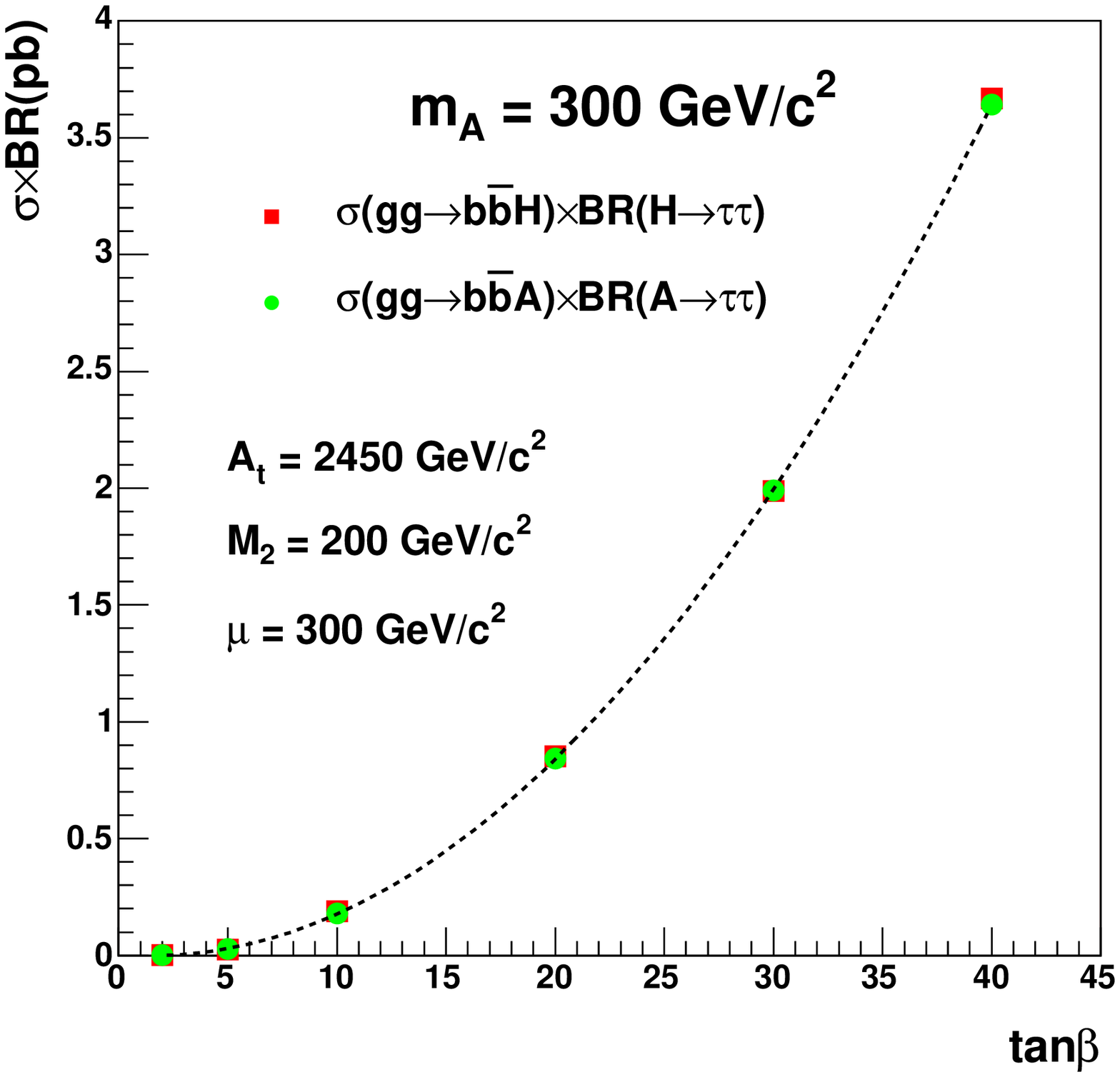}
\hspace*{2mm}
\includegraphics[scale=0.44]{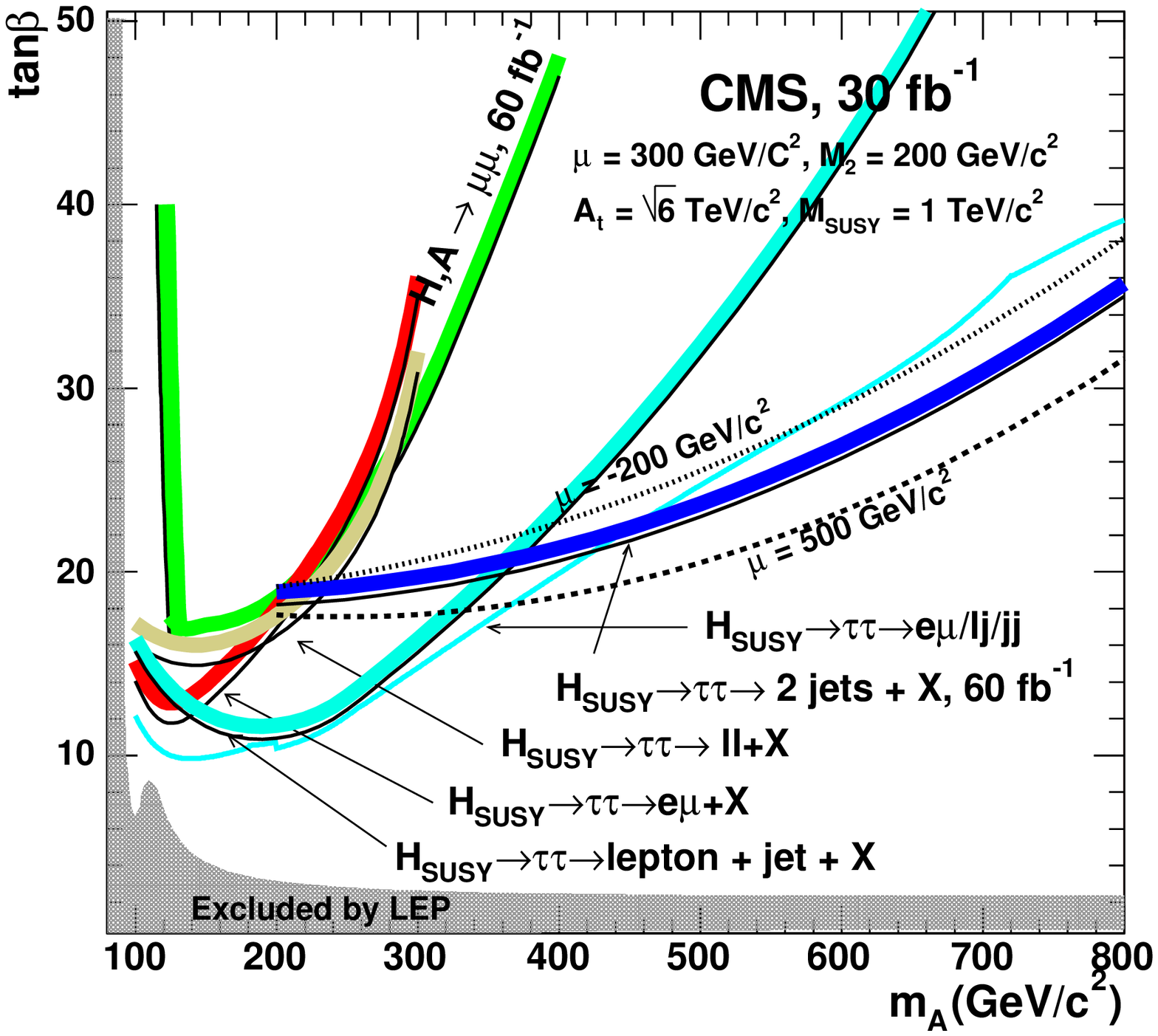}
\caption{Left: $\bb\phi$ production cross section at LHC times the BR
to tau pairs, as a function of $\tan\beta$ for $M_A=300$~GeV.  Right:
expected CMS reach using only 30~fb$^{-1}$ of data for $\bb
H/A\to\taus,\mm$ as a function of $M_A$.  Figures from
Ref.~\protect\cite{Kinnunen:2004ji}.}
\label{fig:bbphi-CMS}
\end{figure}

Fig.~\ref{fig:bbphi-CMS} shows the cross section times BR to tau pairs
for 300~GeV Higgs bosons as a function of $\tan\beta$, and also the
CMS expected discovery reach for various final states in tau or muon
pairs, with only 30~fb$^{-1}$ of luminosity, or about 1/10 of the
total LHC data expected.  Coverage is not complete, because this mode
doesn't produce enough rate at low $\tan\beta$ where there is little
coupling enhancement, but is still a significant search tool.  The
mass resolution achievable for $H$ and $A$ using taus in this mode is
even pretty good, on the order of a couple tens of GeV, possibly
better.  Of course, if the decay to muons is accessible (at very large
$\tan\beta$, then mass resolution would be on the order of a GeV.

This would determine $M_A$ quite well, good enough for comparison with
theory (at least at first), but what about the other major Higgs
parameter, $\tan\beta$?  The $\bb\phi$ production rate is directly
proportional to $\tan^2\beta$, so we can measure it using the overall
rate, with the mild (but not rock solid) assumption that the ratio of
$\bb$ and $\taus$ BRs is the ratio of the $b$ and $\tau$ squared
masses, i.e. that
BR($H/A\to\taus$)\,$\sim10\%$~\cite{Kinnunen:2004ji}.  The major
sources of uncertainty are this assumption, the machine luminosity
uncertainty of $5-10\%$, PDF uncertainties of probably about $5\%$,
and higher-order QCD corrections to the production process of probably
about $20\%$~\cite{Dittmaier:2003ej,Dawson:2005vi}.

\begin{figure}[ht!]
\includegraphics[scale=0.43]{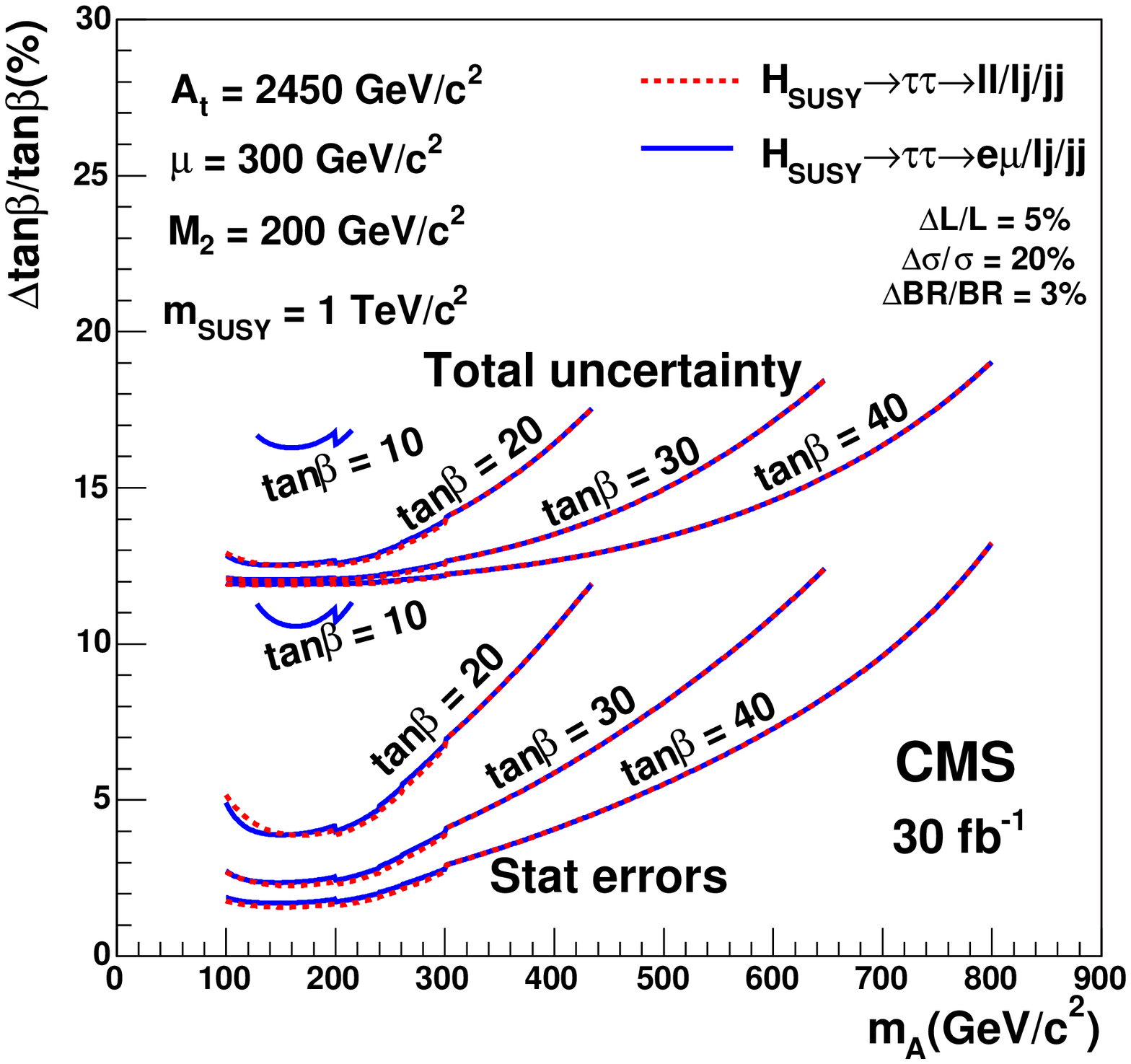}
\includegraphics[scale=0.43]{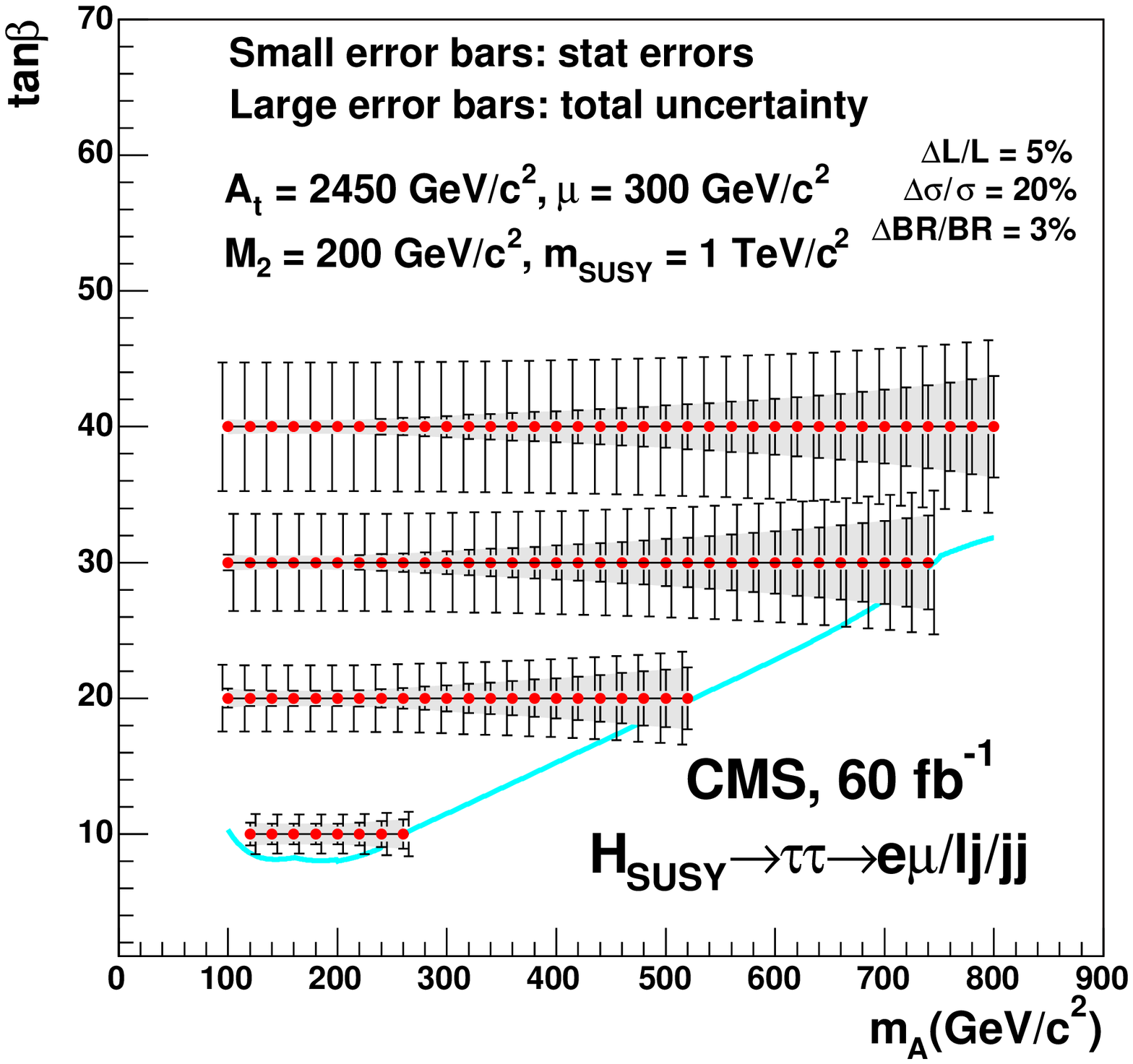}
\caption{CMS expected precision on $\tan\beta$ at LHC using $\bb\phi$ 
production as described in the text.  Figures from
Ref.~\protect\cite{Kinnunen:2004ji}.}
\label{fig:bbphi-meas}
\end{figure}

Fig.~\ref{fig:bbphi-meas} shows the CMS expected uncertainty on
$\tan\beta$ using this method, as a function of $M_A$ and for 30 or
60~fb$^{-1}$ of data.  In general, $10-20\%$ appears achieveable.
This is not spectacular, but would be a significant first step toward
sorting out the new Higgs sector and presumably comparing to other
SUSY discovery measurements.  Clearly the higher-order QCD
uncertainties dominate, which could probably be improved with better
theoretical calculations over the next decade.  This will be done if
heavy Higgses are discovered.

\begin{figure}[hb!]
\includegraphics[scale=1.4]{Feynman/MSSM_bbH-bbA.eps}
\caption{Feynman diagrams for $gg\to\bb\phi$ production at LHC.}
\label{fig:bbphi-Feyn}
\end{figure}

Now, what about charged Higgs discovery?  We know nothing about its
phenomenology, because there is no SM analogue.  All we do know is the
very important fact that, {\it despite everything else we may see at
Tevatron or LHC, the only way to prove the existence of two Higgs
doublets is to directly observe the charged Higgs states}.  I cannot
emphasize this enough.  For all we know, an extra neutral state might
simply be the residue of an extra Higgs singlet; there could be more
to the flavor sector that confuses us when we try to measure Yukawa
couplings or $\tan\beta$.  Thus, observing the $H^\pm$ states would be
a huge qualitative step toward understanding what the Higgs sector is.
How would this proceed experimentally?

\begin{figure}[ht!]
\includegraphics[scale=0.75]{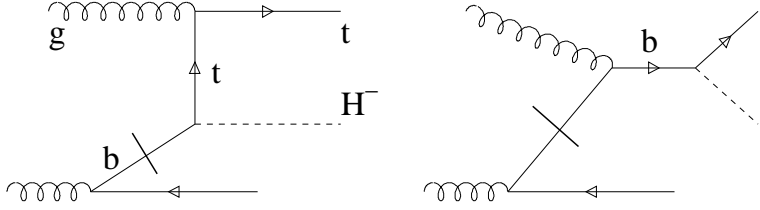}
\caption{Feynman diagrams for charged Higgs production at hadron 
colliders.  The short line breaking the $b$ quark propagator
represents how the process may also be regarded as initiated by a $b$
parton in the proton, rather than from gluon splitting to a $b$ quark
pair.}
\label{fig:xH-Feyn}
\end{figure}

\begin{figure}[hb!]
\includegraphics[scale=0.74]{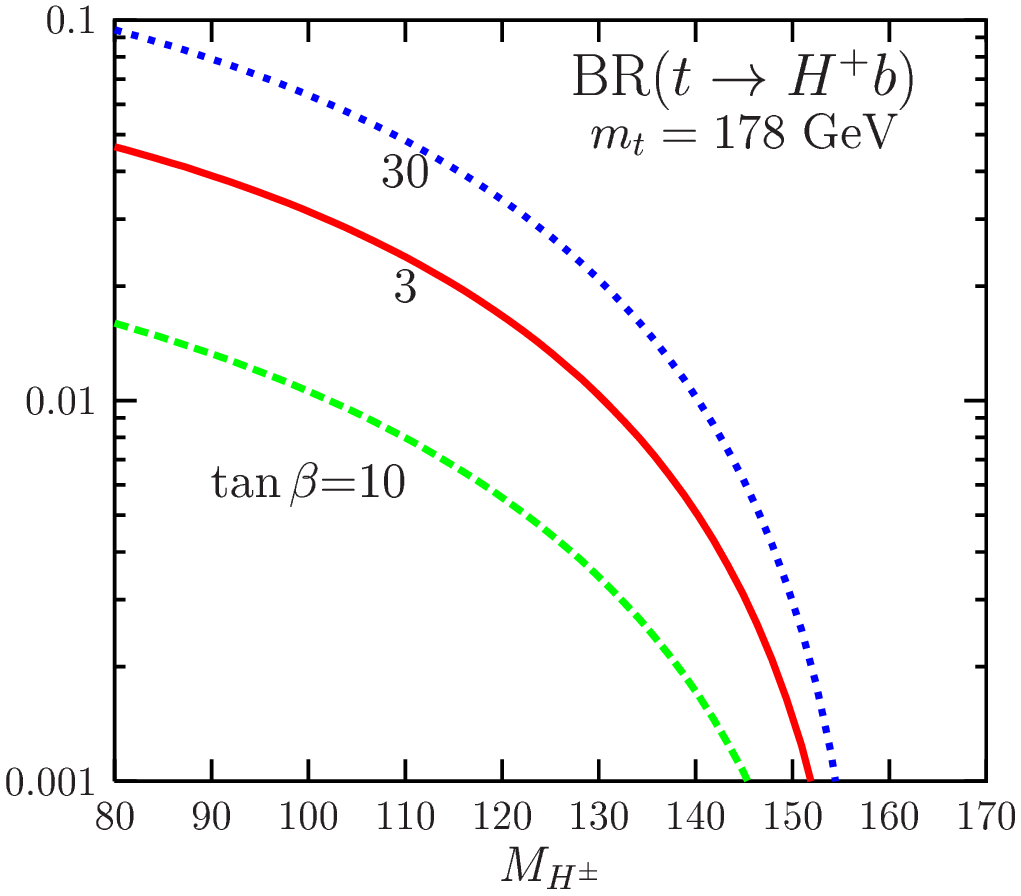}
\hspace*{5mm}
\includegraphics[scale=0.68]{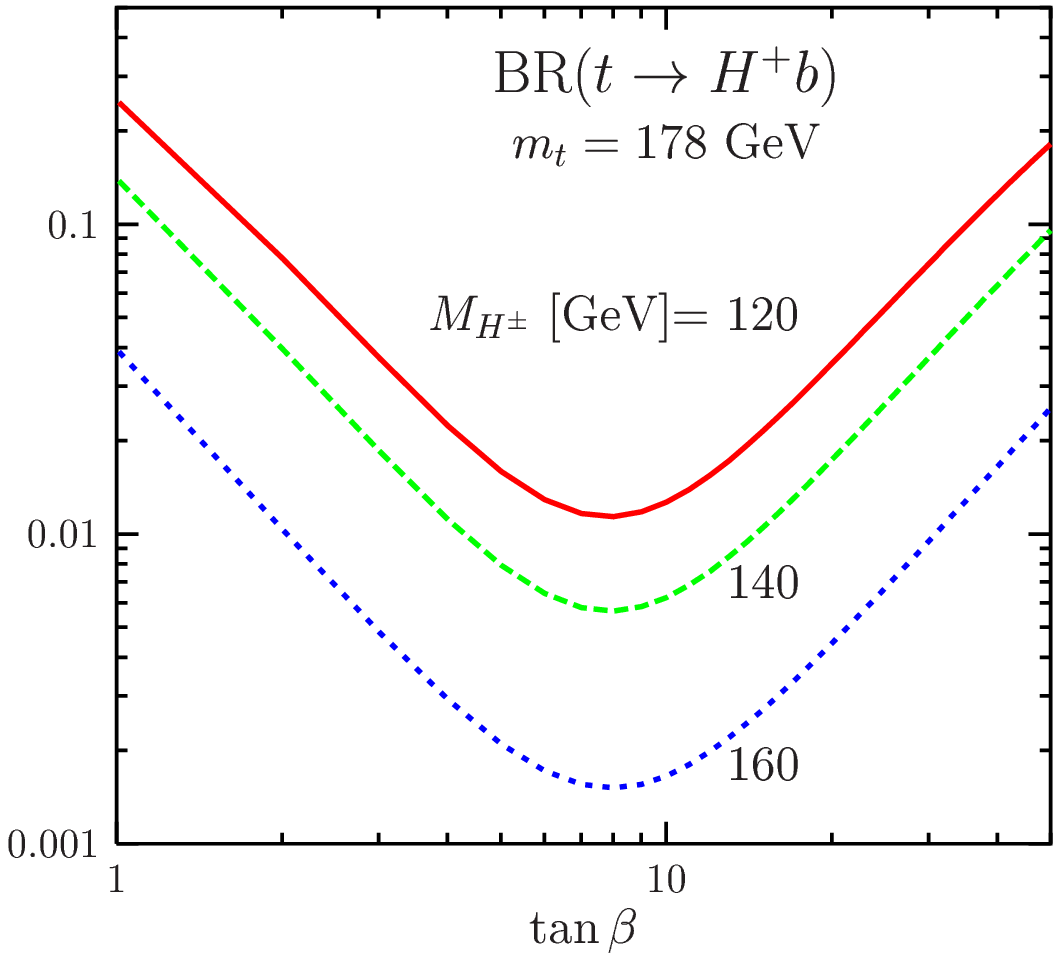}
\vspace*{-1mm}
\caption{Branching ratio for top quark to bottom quark plus charged
Higgs boson, as a function of $M_{H^\pm}$ for a few select values of
$\tan\beta$ (left) and as a function of $\tan\beta$ for
$M_{H^\pm}=120$~GeV (right).  Figures from
Ref.~\protect\cite{Djouadi}.}
\label{fig:t_bH}
\end{figure}

At Tevatron there is very little energy available for direct charged
Higgs production, since it must be produced in association with a top
quark (large coupling), as shown in the Feynman diagrams of
Fig.~\ref{fig:xH-Feyn}.  However, if $M_{H^\pm}$ is small enough, the
top quark can decay to $bH^\pm$ followed by $H^\pm\to\tau\nu$ if
$\tan\beta>1$, and equally to $bc$ and $cs$ if $\tan\beta<1$; if
$M_{H^\pm}\gtrsim120$~GeV, then the BR to $W^\pm\bb$ via a top quark
loop becomes significant.  Fig.~\ref{fig:t_bH} shows the $t\to bH^\pm$
BR as a function of $M_{H^\pm}$ for a few select $\tan\beta$, and as a
function of $\tan\beta$ for $M_{H^\pm}=120$~GeV.  At low $\tan\beta$,
the partial width is driven mainly by the top quark Yukawa, while at
large $\tan\beta$ it's primarily the bottom quark.  Weakness of both
Yukawas in the intermediate-$\tan\beta$ regime results in a
comparatively reduced top quark partial width (recall
Eqs.~(\ref{eq:H+-_1},\ref{eq:H+-_2})).  For fixed $M_{H^\pm}$, the
partial width is symmetric in $\log(\tan\beta)$ about a minimum at
$\tan\beta=\sqrt{m_t/m_b}$.  Charged Higgs decays to $hW^\pm$ or
$AW^\pm$ are generally disallowed in the MSSM from LEP mass limits on
$h$ and $A$.

\begin{figure}[ht!]
\includegraphics[scale=0.4]{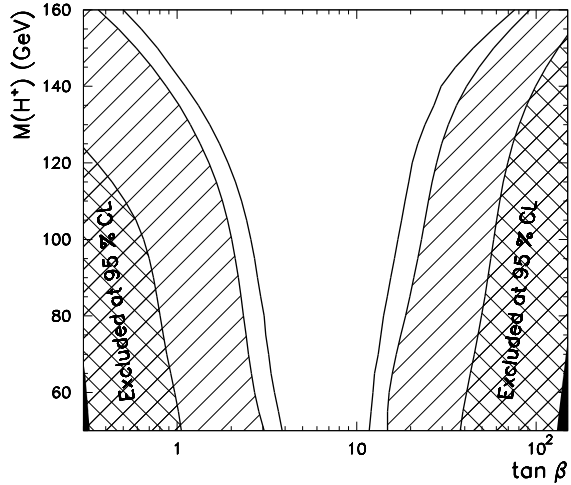}
\vspace*{-3mm}
\caption{Tevatron Run~I $95\%$~CL charged Higgs mass limits (double
hatched lines) as a function of $\tan\beta$ from searches for top
quark decays to bottom quark plus charged Higgs, and expected limits
achievable in Run~II (single hatched lines for 2~fb$^{-1}$, unhatched
curves for 10~fb$^{-1}$).  Fig. from
Ref.~\protect\cite{Chakraborty:2003iw}.}
\label{fig:Tev2-xH}
\end{figure}

The Tevatron search proceeds both as appearance (i.e. looking directly
for $H^\pm$ in the top quark sample) and disappearance, or missing
rate for top quark to $bW^\pm$.  Fig.~\ref{fig:Tev2-xH} goes on to
show the expected $95\%$~CL limits in the $M_{H^\pm}-\tan\beta$ plane
that Tevatron Run~I achieved, and Run~II might reach depending on how
much data it ultimately records.  The very slight change between 2 and
10~fb$^{-1}$ reveals that the experiments there are
statistics-limited, but not by a great margin.


\begin{figure}[hb!]
\includegraphics[scale=0.55]{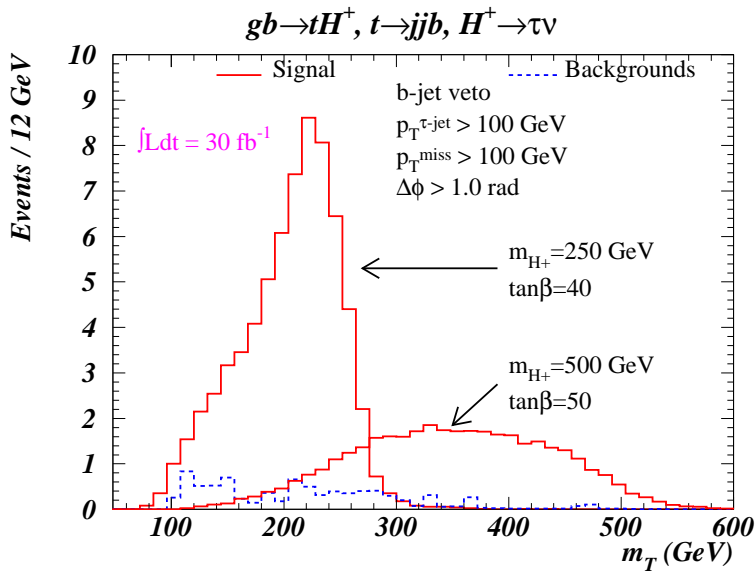}
\vspace*{-3mm}
\caption{Expected transverse mass distributions for light and heavy
$H^\pm\to\tau\nu$ at ATLAS~\protect\cite{Assamagan:2002ne}.}
\label{fig:LHC-xH}
\end{figure}

LHC will search for $tH^\pm$ direct production
(Fig.~\ref{fig:xH-Feyn}), covering the mass range $M_{H^\pm}>m_t$.
Due to nasty QCD backgrounds, the $tb$ decay will be
inaccessible~\cite{Assamagan:2004mu}, leaving $\tau\nu$ with
BR$\sim10\%$.  This is very difficult due to a subtlety of tau decays.
Left-handed taus decay to soft leptons~\cite{Hagiwara:1989fn}.  Since
neutrinos are left-handed, helicity conservation in scalar decay means
all taus are as well.  We need a lepton to trigger the event, and it
must come from $H^\pm$ instead of $t$, so that there is only one
source of missing transverse momentum and we can fully reconstruct
$t$, and $H^\pm$ transversely.  Only a small fraction of the small
rate could pass the necessary detector kinematic cuts to be recorded.
This limits the search to large $\tan\beta$ or small $M_{H^\pm}$,
where the production rate is largest.  Fig.~\ref{fig:LHC-xH} shows
ATLAS's expected transverse mass distributions for a fairly light and
a heavy $H^\pm$.

\begin{figure}[hb!]
\includegraphics[scale=0.8]{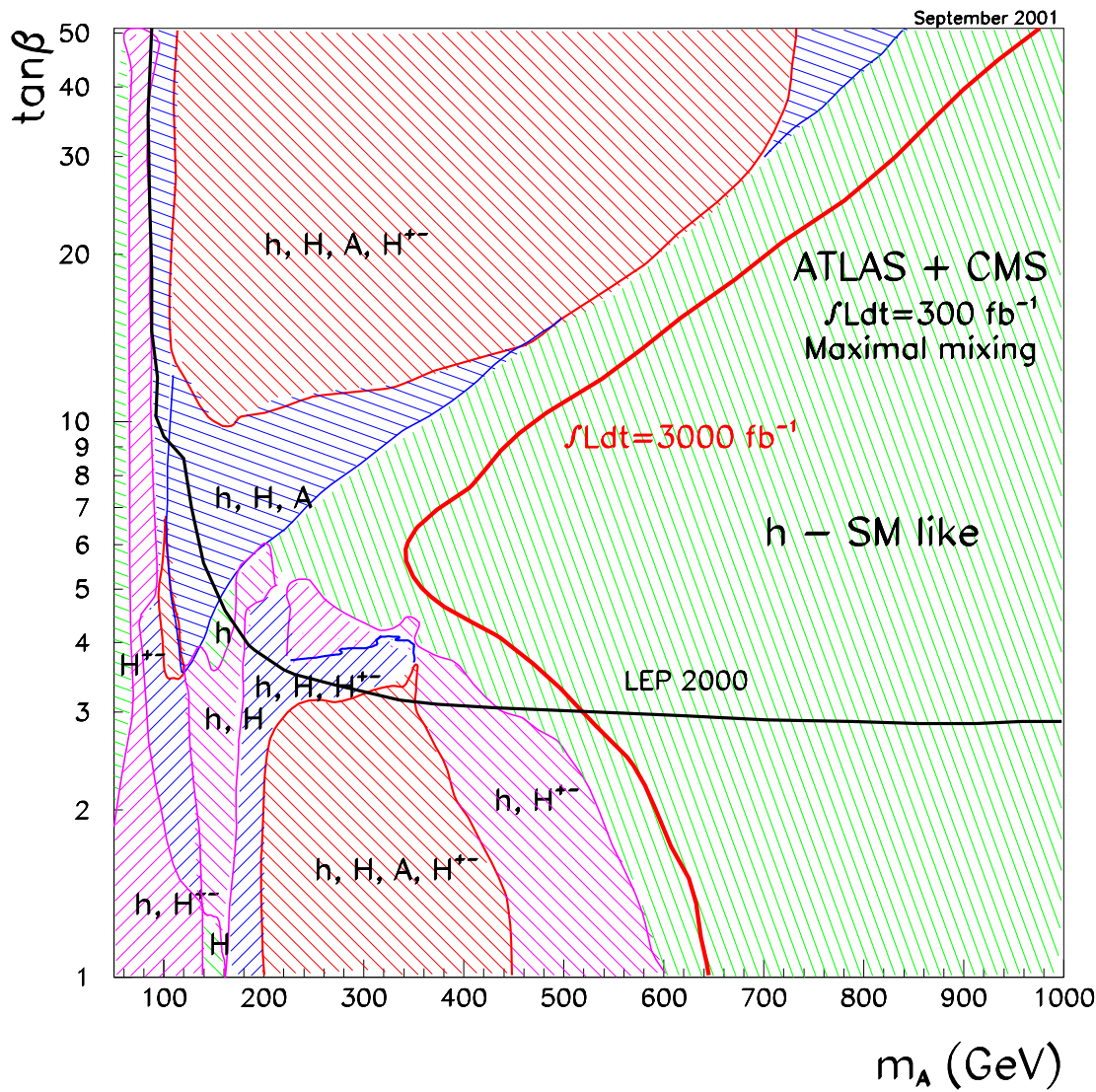}
\caption{Summary of MSSM Higgs boson discovery reaches at LHC (and
extended to SLHC via the solid red line), combining ATLAS and CMS, in
the $\tan\beta-M_A$ plane in the maximal mixing scenario.  The reach
is defined as $5\sigma$ discovery in at least one production and decay
channel.  Below the solid black curve is the region excluded by LEP.
Figure from Ref.~\protect\cite{Gianotti:2002xx}.}
\label{fig:LHC-sum}
\end{figure}

Finally, we come to the overall picture of MSSM Higgs phenomenology at
LHC.  Primarily we're concerned with discovering all the states, but
especially the charged Higgs as it's the key to confirming the
existence of two Higgs doublets.  That turns out to be extraordinarily
difficult due to a combination of factors, from overwhelming QCD
backgrounds to characteristics of left-handed tau decays.
Fig.~\ref{fig:LHC-sum} summarizes the reach for $h$, $H$, $A$ and
$H^\pm$~\cite{Gianotti:2002xx}.  It's reassuring that the No-Lose
Theorem holds and we're guaranteed to find at least one of the CP-even
states, $h$ or $H$.  However, moderate $\tan\beta$ and the decoupling
limit (large $M_A$) both present significant gaps in coverage to
observe any of the additional states.  This is especially more
apparent once one realizes that the region below the solid black curve
is already excluded by LEP, so those LHC access regions don't matter.
The figure is from 2001 and needs updating -- some significant
positive changes exist -- but the general picture remains.

%

%%%%%%%%%%%%%%%%%%%%%%%%%%%%%%%%%%%%%%%%%%%%%%%%%%%%%%%%%%%%%%%%%%%%%%%%
%%%%%%%%%%%%%%%%%%%%%%%%%%%%%%%%%%%%%%%%%%%%%%%%%%%%%%%%%%%%%%%%%%%%%%%%

\subsection{MSSM Higgs potential}
\label{sub:MSSM-pot}

I've touched on the bits of Higgs gauge and Yukawa couplings in the
MSSM that are qualitatively different that the SM: $M_A$ and
$\tan\beta$.  But we should look at self-couplings more closely,
because in a general 2HDM (or the subset MSSM) they are radically
different.  First, because there are more Higgs bosons, there are more
self-couplings -- six for the neutral states alone, to be precise:
$\lambda_{hhh}$, $\lambda_{Hhh}$, $\lambda_{HHh}$, $\lambda_{HHH}$,
$\lambda_{hAA}$, $\lambda_{HAA}$.  In the MSSM these are all equal to
$M_Z^2/v$ times various mixing angles (which aren't particularly
enlightening so I don't show them) plus additional shifts from top
quark Yukawa loop corrections.  That is, they are all (mostly) gauge
parameters.  However, in the large-$M_A$ decoupling limit which
recovers the SM, $\lambda_{hhh}\to\lambda_{\rm SM}$.

\begin{figure}[hb!]
\includegraphics[scale=1]{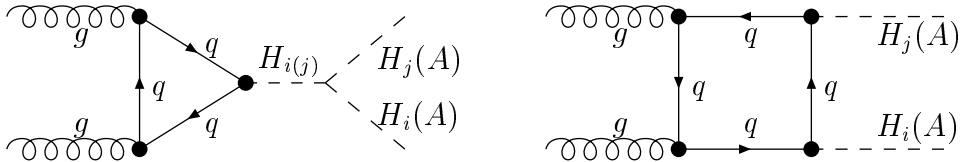}
\vspace*{-3mm}
\caption{Feynman diagrams for Higgs pair production in a 2HDM like the
MSSM.  The loops include both top and bottom quarks, and there are six
possible processes (see text).}
\label{fig:gg_phiphi}
\end{figure}

If we discover SUSY, we'd start by assuming it's the MSSM.  To measure
the MSSM potential in that case, we'd have to observe at least six
different Higgs pair production modes to measure the six
self-couplings.  (Note that I'm leaving out the possible
self-couplings involving charged Higgses.)  Inclusive Higgs pair
production looks generally like it does in the SM, $gg\to\phi_1\phi_2$
via triangle and box loop diagrams as shown in
Fig.~\ref{fig:gg_phiphi}, but the $b$ quark loops become important and
must be included.

\begin{figure}[hb!]
\includegraphics[scale=0.72]{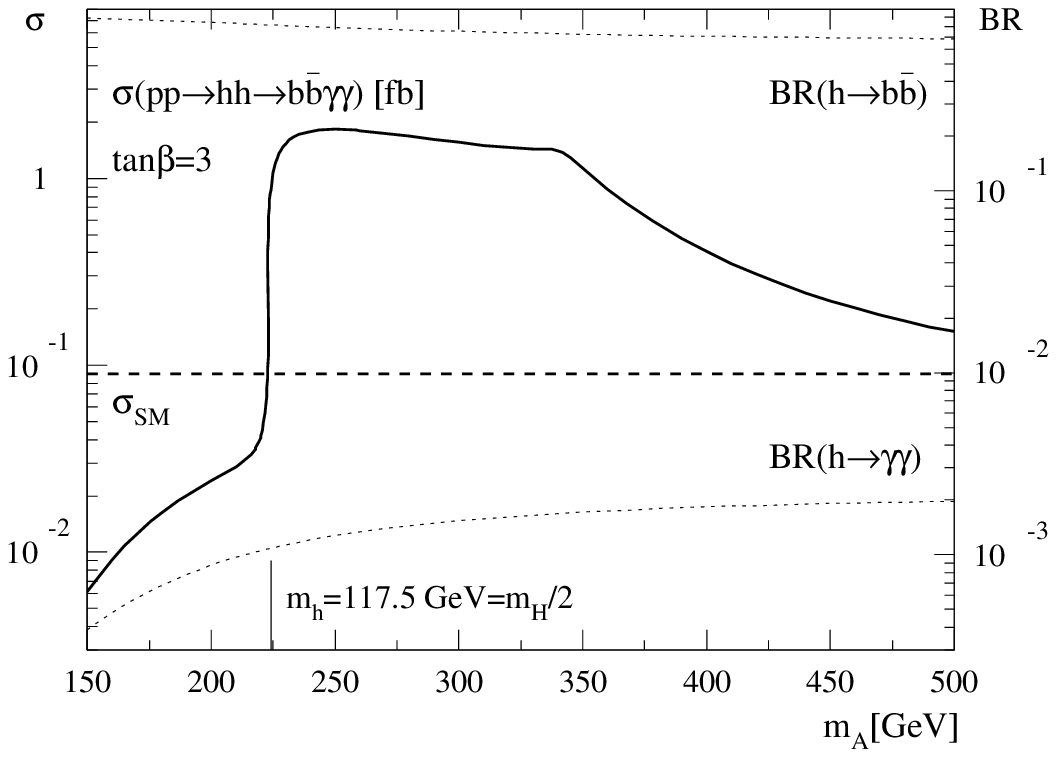}
\includegraphics[scale=0.45]{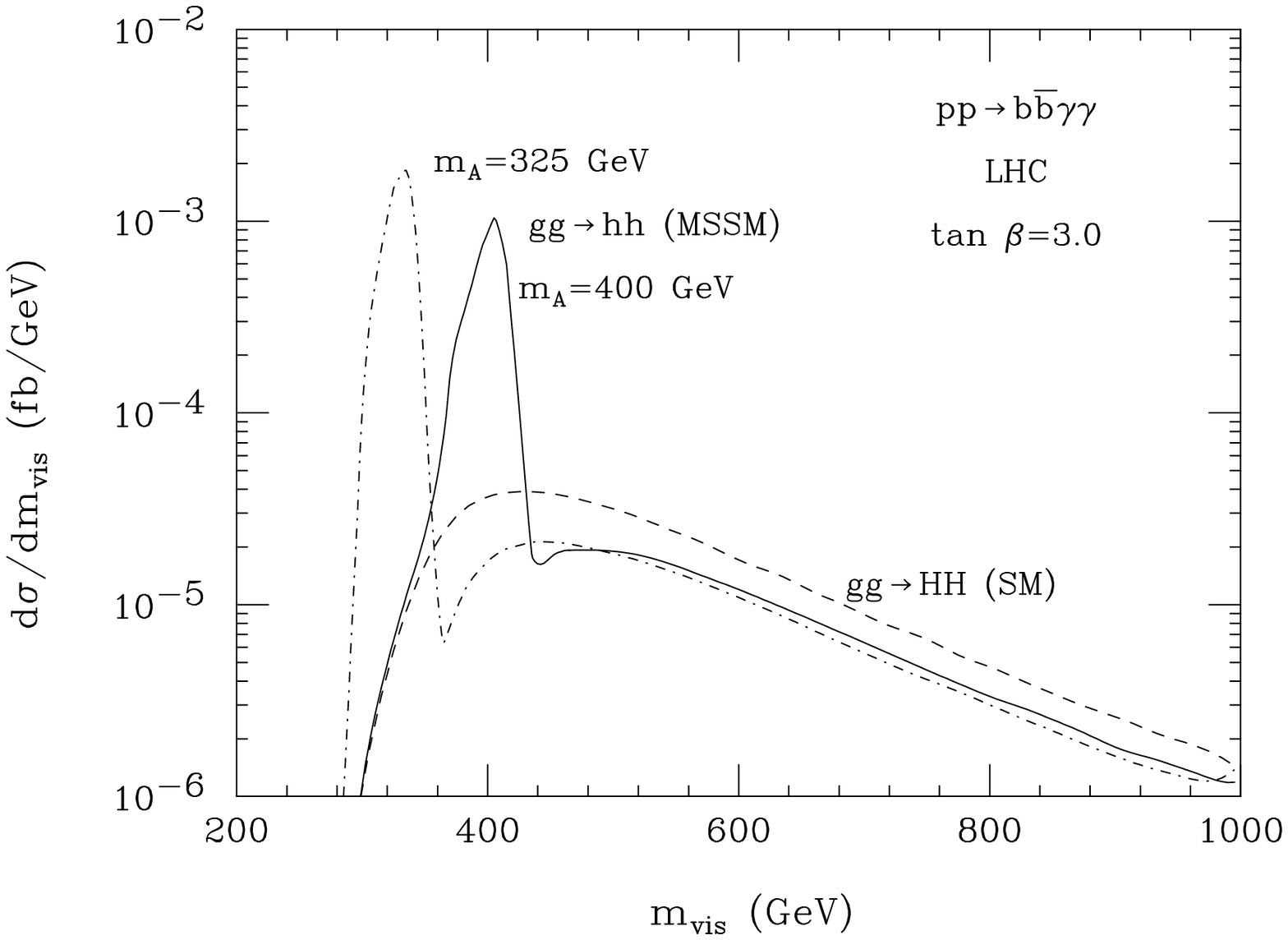}
\vspace*{-3mm}
\caption{Resonant MSSM Higgs pair production at LHC and decay to
$\bb\gamma\gamma$ final states~\protect\cite{Baur:2003gp}.}
\label{fig:LHC-phiphi-res}
\end{figure}

Unfortunately, the box diagram totally swamps the one containing the
self-coupling we care about by a factor $\tan^2\beta$, and in any case
backgrounds from $H/A\bb$ production appear to be
overwhelming~\cite{Baur:2003gp}: very generally, LHC would not obtain
any $\lambda$ measurements at all.  The one very limited exception is
that LHC could clearly observe Higgs pair production if it came from
resonant heavy Higgs decay, $H/A\to hh$.  An example peak is shown in
Fig.~\ref{fig:LHC-phiphi-res}.  However, this would measure only a BR,
at best, not an absolute coupling.  Sadly, exactly the same situation
exists for Higgs pairs at a future ILC~\cite{Djouadi:1999gv}.

%

%%%%%%%%%%%%%%%%%%%%%%%%%%%%%%%%%%%%%%%%%%%%%%%%%%%%%%%%%%%%%%%%%%%%%%%%
%%%%%%%%%%%%%%%%%%%%%%%%%%%%%%%%%%%%%%%%%%%%%%%%%%%%%%%%%%%%%%%%%%%%%%%%
%%%%%%%%%%%%%%%%%%%%%%%%%%%%%%%%%%%%%%%%%%%%%%%%%%%%%%%%%%%%%%%%%%%%%%%%

\section{Conclusions}
\label{sec:conc}

The purpose of these lectures has not been to provide exhaustive
coverage of all aspects of collider Higgs phenomenology.  Rather, it's
a solid introduction, focusing on the basics.  This includes SM
production and decay, mostly at LHC, where we're confident we could
discovery a SM-like Higgs, and many non-SM-like variants.  I focused
on the most important channels which guarantee discovery, and
especially in weak boson fusion (WBF) as those are the most powerful
(best $S/B$, distinctive) search channels, covering the broadest range
of Higgs mass.  I emphasized that our understanding of LHC Higgs
physics has changed dramatically from the days of the ATLAS TDR, for
example, which is now quite obsolete.  However, ATLAS has produced a
plethora of Notes and summaries of Notes to cover the changes, and CMS
published a fresh TDR~\cite{CMS_TDR} in 2006 which covers the changes
as well.

We now understand the LHC to be such a spectacular Higgs factory that
not only can it discover any mass of SM-like Higgs boson, it can also
do an impressive job of measuring all its quantum properties.
Granted, Higgs couplings measurements won't be precision-level if the
Higgs is light, as expected from EW precision data, but they would
nonetheless be absolute couplings measurements.  The LHC can even make
significant steps toward measuring the SM Higgs potential, at least
the Higgs trilinear self-coupling, although depending on $M_h$ it may
require precision gauge and Yukawa couplings input from a future
$e^+e^-$ collider (an ILC) to control the major systematic
uncertainties.  I also highlighted where an ILC could make
improvements to the LHC's measurements, and where it would be vital to
filling in gaps in LHC results.

The final third of the lectures discussed BSM Higgs sectors, but only
the 2HDM MSSM Higgs sector in any detail.  Many SM Higgs sector
extensions are rather simple variants on SM phenomenology, involving
factorizable changes in production and decay rates (couplings), mostly
arising from mixing angles.  This is not general, however, and there
are plenty of ``exotic'' models -- Higgs triplets, for example --
which would be qualitatively different, but therefore simultaneously
distinctive.  The popular focus on the MSSM 2HDM is because of several
other outstanding questions in particle physics, like dark matter or
the theoretical dirty laundry of the SM Higgs sector, which strongly
motivate the other new physics.

Students who wish to engage in Higgs phenomenology research should
definitely take the time to expand their scope beyond the SM and the
MSSM.  Other extensions are equally well-motivated, such as Little
Higgs, not to mention strong dynamics.  But the two well-studied basic
models I covered here give one a strong foundation for other BSM Higgs
phenomenology by analogy.  Happy Higgs hunting!

%%%%%%%%%%%%%%%%%%%%%%%%%%%%%%%%%%%%%%%%%%%%%%%%%%%%%%%%%%%%%%%%%%%%%%%%

\begin{acknowledgments}
I would like to thank Sally Dawson and the TASI 2006 organizers for
the opportunity to give these lectures, and for an extremely pleasant
experience at the summer school.  Gracious thanks also go to Dan
Berdine, John Boersma, Fabio Maltoni, Tilman Plehn, J\"urgen Reuter,
and especially Steve Martin for proofreading contributions above and
beyond the call of duty.
\end{acknowledgments}

%%%%%%%%%%%%%%%%%%%%%%%%%%%%%%%%%%%%%%%%%%%%%%%%%%%%%%%%%%%%%%%%%%%%%%%%
%%% References
%%%%%%%%%%%%%%%%%%%%%%%%%%%%%%%%%%%%%%%%%%%%%%%%%%%%%%%%%%%%%%%%%%%%%%%%

\end{document}